\documentclass[twocolumn,showpacs,preprintnumbers,amsmath,superscriptaddress,amssymb,10pt]{revtex4}

\usepackage{graphicx}
\usepackage{dcolumn}
\usepackage{bm}
\usepackage{hyperref}

\def\l#1{\left#1}
\def\r#1{\right#1}
\def\ds{\displaystyle}
\def\eref#1{(\ref{#1})}

\def\simlt{\lower.5ex\hbox{$\; \buildrel < \over \sim \;$}}
\def\simgt{\lower.5ex\hbox{$\; \buildrel > \over \sim \;$}}

\begin{document}

\title{Polarized CMB power spectrum estimation using the pure pseudo cross-spectrum approach}


\author{J. Grain}
\email{julien.grain@ias.u-psud.fr}
\affiliation{CNRS, Laboratoire AstroParticule \& Cosmologie, Universit\'e Paris 7, Denis Diderot\\ 10, rue Alice Domon et L\'eonie Duquet, 75205 Paris Cedex 13, France}
\affiliation{CNRS, Institut d'Astrophysique Spatiale, Universit\'e Paris-Sud 11\\ B\^atiments 120-121, 91405 Orsay Cedex, France}

\author{M. Tristram}
\email{tristram@lal.in2p3.fr}
\affiliation{CNRS, Laboratoire de l'Acc\'el\'erateur Lin\'eaire, Universit\'e Paris-Sud 11\\
B\^atiment 200, 91898 Orsay Cedex, France}

\author{R. Stompor}
\email{radek@apc.univ-paris7.fr}
\affiliation{CNRS, Laboratoire AstroParticule \& Cosmologie, Universit\'e Paris 7, Denis Diderot\\ 10, rue Alice Domon et L\'eonie Duquet, 75205 Paris Cedex 13, France}

\date{\today}

\begin{abstract}
We extend the pure pseudo-power-spectrum formalism proposed recently in the context of the Cosmic Microwave 
Background polarized power spectra estimation by Smith (2006) to incorporate cross-spectra computed for multiple
maps of the same sky area. We present an implementation of such a technique, paying particular attention to a
calculation of the relevant window functions and mixing (mode-coupling) matrices. We discuss the relevance and
treatment of the residual $E/B$ leakage for a number of considered sky apodizations as well as compromises and
assumptions involved in an optimization of the resulting power spectrum uncertainty. In particular, we investigate the importance of a pixelization scheme, patch geometry, and sky signal priors used in apodization optimization
procedures. In addition, we also present results derived for more realistic sky scans as motivated by the proposed 
balloon borne experiment EBEX.
We conclude that the presented formalism thanks to its speed and efficiency
can provide an interesting alternative to the CMB polarized power spectra estimators based on the optimal methods at least on angular scales smaller than $\sim$10 degrees.
In this regime, we find that it 
is capable of suppressing the total variance of the estimated $B$-mode spectrum to within a factor
of $\sim2$ of the variance due to only the sampling and noise uncertainty of the $B$-modes alone,
as derived from the Fisher matrix approach.
\end{abstract}


\maketitle

\section{Introduction}
\label{sect:intro}
The reliable characterization and scientific exploitation of the polarized Cosmic Microwave 
Background (CMB) Anisotropy signal is one of the main challenges facing the CMB research at the 
present. The challenge is aimed at by an entire slew of the experimental efforts which are currently 
on-going~\cite{bicep_website}, being deployed~\cite{planck_website,polarbear_website,quiet_website,clover_website,ebex_website,spider_website}
or planned~\cite{brain_website, bpol_website}. The majority, if not all, of these experiments will effectively 
observe only part of the sky. This is either due to observational/hardware limitations, particularly for ground-based and balloon-borne experiments, or the presence of foregrounds, or both. 
The resulting so-called cut-sky effects will therefore have to be taken into account in the data analysis 
of most of these experiments. 

The focus of this paper is on the CMB power spectrum estimation via the pseudo-spectrum technique
~\cite{hauser_peebles_1973}. The power spectrum is the most fundamental statistics considered in the
context of the CMB fluctuations, which are thought of as being Gaussian or nearly Gaussian distributed. The pseudo
spectrum approach provides a computationally quick and flexible framework for estimating the power
spectra, which has been extensively and successfully employed in past analyses of the CMB 
data sets (e.g., \cite{hivon_etal_2002, hinshaw_etal_2003}).
However, 
it has been long recognized~\cite{bunn_etal_2003} that a straightforward application of the  pseudo
spectrum technique to cut-sky polarized CMB maps leads to the so-called "$E$-to-$B$" leakage, or 
power aliasing, as a consequence of which the cosmologically important information contained in the CMB 
$B$-modes is overwhelmed by the statistical uncertainty of the (much larger) $E$-modes. To date two techniques 
alleviating the problem have been proposed: one correcting the leakage on the correlation function 
level~\cite{chon_etal_2004} and the other attempting to do so directly on the maps~\cite{smith_2006}. 
The latter approach relies on a suitably chosen sky apodization to remove from the map harmonic 
modes which are neither solely $E$  nor $B$. In a recent paper~\cite{smith_zaldarriaga_2007} it has 
been suggested that the apodization can be appropriately optimized to bring down the resulting 
$B$-mode spectrum uncertainty in-line with the optimal methods, building up a strong case in favor
of this approach. In 
this paper we extend the original formalism of~\cite{smith_2006} to incorporate it within 
the framework 
of the cross-spectra computations and discuss a number of implementation issues, related to 
derivative computation and apodization choices in actual applications. Later, in this context we discuss 
the performance of the approach.

We note here that other approaches to the polarized power spectrum, which do not suffer from
the leakage problem, exist but are however hampered either by numerical efficiency
~\cite{tegmark_costa_2001} or convergence~\cite{larson_etal_2007} issues in particular for the high 
resolution observations (but see~\cite{jewell_etal_2008} for a potential resolution of this problem). Consequently, the pure pseudo-spectrum approach stands out as an important 
alternative potentially capable of addressing successfully the needs of the future CMB B-mode polarization 
experiments.

The outline of this paper is as follows. In Sect.~\ref{sect:formalism} we start from presenting the standard 
pseudo-power-spectrum technique and the $E/B$-leakage problem. We also summarize the original 
approach of Smith (2006)~\cite{smith_2006} and describe its specific variant as proposed in this paper 
and which invokes harmonic domain derivatives and the cross-spectrum idea. We conclude
that Section with an overview of the numerical implementation of the method. 
In Sect.~\ref{sect:apodizations} we 
elaborate on the issues related to the proper sky apodizations. In Sect.~\ref{sect:applications} we consider 
the performance of the formalism focusing on the role of the apodization and its impact on the quality of 
the final results, as expressed by the level of the remaining residual $E/B$ leakage and the uncertainty of 
the estimated power spectra. We summarize the main results of this work in Sect.~\ref{sect:concl}.

For definiteness, hereafter, we will reserve the term mask to a binary, pixelized map of a sky
assuming only two values 1 and 0, corresponding to an observed or unobserved (or just not to be taken into 
the analysis) sky pixel respectively. The  pixel weights which can assume any nonzero value 
will be referred to as either apodizations or windows. In this language a mask just defines an 
observed sky. 
We will also refer to aliasing involving different spin harmonics as leakage, reserving the
term aliasing solely to mixing between different modes of the same type of the same basis
functions.
Though not strictly precise such differentiations will be useful in the following discussion.

\section{Notation and Formalism}

\label{sect:formalism}

\subsection{Harmonic decomposition}
\label{subsec:HarmDec}

The CMB (linear) polarization field is completely described by two Stokes parameters, $Q$ and 
$U$. Those can be combined into two spin-2 and spin-(-2) fields defined as,
\begin{equation}
	P_{\pm2} \equiv Q\pm iU.
\end{equation}
For the full sky, the spin fields can be expressed in the harmonic space making use of the spin-weighted 
spherical harmonic basis
\begin{equation}
	P_{\pm2}=\displaystyle\sum_{\ell m} {}_{\pm2}a_{\ell m} {}_{\pm2}Y_{\ell m}.
	\label{eqn:spinalm}
\end{equation}
Alternatively, the polarization field can be decomposed into a gradient, $E$, and a curl, $B$, part, both of which are coordinate independent.
In the harmonic space, these two approaches are related via the following relations,
\begin{eqnarray}
	a^E_{\ell m}&=&-\frac{1}{2}\left( {}_{2}a_{\ell m}+ {}_{-2}a_{\ell m}\right), \\
	a^B_{\ell m}&=&\frac{i}{2}\left( {}_{2}a_{\ell m}- {}_{-2}a_{\ell m}\right).
\end{eqnarray}
From a point of view of physics, the $E/B$ decomposition of CMB polarization may seem more natural 
as it is  directly linked to the primordial cosmological perturbations leading to the presently observed CMB 
anisotropies. In particular, only primordial gravity waves, and not density perturbations, can create 
$B$-type polarization~\cite{zaldarriaga_seljak_1997}. As a consequence constraining the amplitude 
of such modes could allow us to study the character of the primordial cosmological perturbations. This 
simple picture is spoilt somewhat by a presence of an additional $B$-mode type contribution generated by 
the gravitational lensing of the $E$-mode polarization~\cite{zaldarriaga_seljak_1998} and which needs to 
be accounted for if the primordial component is to be recovered.

A computation of polarized angular power spectra makes use of spin-weighted spherical harmonics. 
Those functions, a natural harmonic basis for spin-$s$ fields on the sphere, are derived from the standard 
(or spin-0) spherical harmonics by applying the spin-raising ($\eth$) or spin-lowering ($\bar\eth$) 
operators~\cite{zaldarriaga_seljak_1997},
\begin{equation}
\begin{array}{l c l}
{}_{s}Y_{\ell m}&=&{\ds \sqrt{\frac{(\ell-s)!}{(\ell+s)!}}\,\eth^sY_{\ell m}}, \\
{}_{-s}Y_{\ell m}&=&{\ds (-1)^s\sqrt{\frac{(\ell-s)!}{(\ell+s)!}}\,\bar\eth^sY_{\ell m}},
\end{array}
\end{equation}
where hereafter $s\geq0$. In the following, we will always assume that $s$ is nonnegative. The main properties of the spin-weighted 
spherical harmonics under different transformation can be found elsewhere, e.g.,~\cite{vrashalovich_etal_1988, zaldarriaga_seljak_1997, kamionkowski_etal_1997}.

Any complex, spin-$s$ field, with ${}_s\phi^\dag={}_{-s}\phi$ is completely characterized by its projection on 
the spin-$s$ spherical harmonics, defining its harmonic representation,
\begin{equation}
{}_{\pm s}a_{\ell m}=\displaystyle\int\,d^2\Omega\;_{\pm s}\phi\;{}_{\pm s}Y^{\dag}_{\ell m}.
\end{equation}
Just as is the case for spin-2 polarization fields a spin-$s$ field 
can be also fully described
by its $E/B$ harmonic decomposition. Adopting a vectorial notation,
\begin{equation}
\mathbf{\Phi}=
\left(
\begin{array}{c}
\mathrm{Re}\left[{}_{s}\phi\right] \\
\mathrm{Im}\left[{}_{s}\phi\right]
\end{array}
\right),
\end{equation}
we can define the $E$ and $B$ spherical harmonic basis as,
\begin{eqnarray}
{}_E\mathbf{Y}_{s,\ell m}&\equiv&\mathbf{D}_s^{E} \, Y_{\ell m}
\label{eqn:eYsDef}
\\
&=&\frac{1}{2}\sqrt{\frac{(\ell-s)!}{(\ell+s)!}}
\left(\begin{array}{c}
\eth^s+(-1)^s\bar\eth^s \\
-i(\eth^s-(-1)^s\bar\eth^s)
\end{array}\right)Y_{\ell m} \nonumber \\
&=&\frac{1}{2}\left(\begin{array}{c}
{}_{s}Y_{\ell m}+(-1)^s{}_{-s}Y_{\ell m} \\
-i({}_{s}Y_{\ell m}-(-1)^s{}_{-s}Y_{\ell m})
\end{array}\right),\nonumber
\end{eqnarray}
and
\begin{eqnarray}
{}_B\mathbf{Y}_{s,\ell m}&\equiv&\mathbf{D}_s^{B} \, Y_{\ell m}
\label{eqn:bYsDef}	
\\
&=&\frac{1}{2}\sqrt{\frac{(\ell-s)!}{(\ell+s)!}}\left(\begin{array}{c}
i(\eth^s-(-1)^s\bar\eth^s) \\
\eth^s+(-1)^s\bar\eth^s
\end{array}\right)Y_{\ell m} \nonumber \\
&=&\frac{1}{2}\left(\begin{array}{c}
i({}_{s}Y_{\ell m}-(-1)^s{}_{-s}Y_{\ell m}) \\
{}_{s}Y_{\ell m}+(-1)^s{}_{-s}Y_{\ell m}
\end{array}\right).\nonumber
\end{eqnarray}
Here we have defined two differential operators $\mathbf{D}^{E(B)}_s$ which generalize
to arbitrary spin the operators used in \cite{bunn_etal_2003}. The harmonic representation of 
the field $\mathbf{\Phi}$ in the $E/B$ subspace then reads,
\begin{equation}
{}_{X}a_{s,\ell m}=\displaystyle\int\,d^2\Omega\,W\left(\Omega\right)\,\mathbf{\Phi}\cdot{}_{X}\mathbf{Y}^\dag_{s,\ell m},
\label{scal-prod}
\end{equation}
where $X = E$ or $B$ and we have introduced an arbitrary (scalar) weight, $W$, which we 
will later need to account
either for the noise present in the data, $\mathbf{\Phi}$, or the cut-sky effects. The
dot denotes a standard componentwise dot product. In the special case of CMB polarization, 
the vector $\mathbf{\Phi}$ is simply the polarization vector defined as,
\begin{displaymath}
\mathbf{P}=\left(\begin{array}{c}
Q \\
U
\end{array}\right).
\end{displaymath}
It can be easily verified that introducing the polarization vector, $\mathbf{P}$ into Eq.~\eref{scal-prod} 
and setting $s=2$ and $W=1$ over the full sphere, leads to the standard definition of the $E$ and $B$ 
multipoles.

\subsection{Pseudo power spectra}

The pseudo-$C_\ell$ estimator, {\it e.g.}~\cite{hivon_etal_2002, hinshaw_etal_2003, tristram_etal_2005}, in a given 
bandpower $\alpha$ with a width $\Delta\ell$, is then based on the multipole decomposition Eq.~\eref{scal-prod} 
and defined as,
\begin{equation}
\tilde{\mathcal{C}}^{X,\;\alpha}_{\l(i,j\r)}=\displaystyle\sum_{\ell\in\alpha}\frac{\ell(\ell+1)}{2\pi\Delta\ell}\displaystyle\sum_m\frac{{}_X\tilde{a}^{\l(i\r)}_{2,\ell{m}}\,{}_X\tilde{a}^{\l(j\r)\;\dag}_{2,\ell{m}}}{2\ell+1},
\label{eqn:pseudocldef}
\end{equation}
where $X$ stands for $E$ or $B$ and $i$ and $j$ distinguish two sets of measurements (maps) of the sky 
for which the harmonic coefficients have been computed. Throughout this paper, we use $\ell(\ell+1)/2\pi$ as $\ell$-weighting 
while computing the bandpowers. Though, this may not be the most natural choice from the $B$ mode power spectrum perspective, the results
presented hereafter are expected to be largely independent on it, given the relatively 
narrow $\ell$--bins used in this work and  the fact that the $B$-mode power spectrum is nearly structure free. For definiteness we will be assuming hereafter 
that the two different sets of data always refer to the same sky area, though we will allow for different (nonzero)
weights assigned to each of them. The pseudo-spectrum formalism can be used to estimate $EE$ and 
$BB$ as well as $EB$ spectra. Hereafter, we will use a calligraphic typeface, ${\cal C}$, for 
the estimated and a roman one, $C$, for the true underlying CMB power spectrum.

Because of the limited sky coverage, and nonuniform, pixel-dependent weights, the above estimator is biased and 
its average over CMB realizations, $\l<\tilde{\mathcal{C}}^{X,\; \alpha}_{\l(i,j\r)}\r>$, involves a mixing between 
different $\ell$ modes (or bins) and polarization states. The latter can be described by a so-called {\em mixing kernel} which we will 
denote as $M^{XX'}_{\alpha\alpha'}$. The unbiased estimator $\mathcal{C}^{X,\; \alpha}_{\l(i,j\r)}$ is thus obtained 
by inverting the following linear system,
\begin{equation}
\left(\begin{array}{cc}
\medskip
M^{diag}_{\alpha\alpha'} & M^{off}_{\alpha\alpha'} \\
M^{off}_{\alpha\alpha'}  & M^{diag}_{\alpha\alpha'}
\end{array}\right)
\left(\begin{array}{c}
\medskip
\mathcal{C}^{E,\;\alpha'}_{\l(i,j\r)} \\
\mathcal{C}^{B,\;\alpha'}_{\l(i,j\r)}
\end{array}\right)=\left(\begin{array}{c}
\medskip
\tilde{\mathcal{C}}^{E,\;\alpha}_{\l(i,j\r)}-N^{E,\;\alpha}_{\l(i,j\r)} \\
\tilde{\mathcal{C}}^{B,\;\alpha}_{\l(i,j\r)}-N^{B,\;\alpha}_{\l(i,j\r)}
\end{array}\right).
\label{unbiased}
\end{equation}
Here $N^{X',\;\alpha}_{\l(i,j\r)}$ is the noise contribution to the estimated pseudo-spectra and vanishes, whenever the noise in the two data sets is not correlated, as for example, in a 
case of two data sets produced by two different experiments or two uncorelated detectors of a single experiment. 
In fact, we will always assume that 
$N^{X',\;\alpha'}_{\l(i,j\r)} = 0$, whenever $i \ne j$, and refer to the latter cases as pseudo cross-spectra. We will 
call a pseudo-spectrum computed from a single data set, i.e., $i = j$, a pseudo autospectrum. We note that in 
this case the noise term is usually nonvanishing and has to be estimated from the data and/or our knowledge 
of the experiment to allow the estimator in Eq.~\eref{eqn:pseudocldef} to be correctly unbiased. This emphasizes one 
of the biggest advantages of the cross-spectrum based estimators, which do not need such a correction.
However, the mixing kernel, $M^{XX'}_{\alpha\alpha'}$ will depend on the window functions applied 
to each of the data sets and therefore be somewhat more cumbersome to calculate in the cross-spectrum case. 
Moreover, as we will discuss that later on, some of the most appealing choices for the window function will often 
require the knowledge of the noise in the data. Nevertheless, as the latter will be used only for the power spectrum 
error estimation and not for the computations of the spectral estimates, lower precision in modeling the noise may 
be sufficient. In the very least, in a case of real data the cross-spectra provide a useful and handy diagnostic in 
the analysis of any data set~\cite{polenta_etal_2006}, while for a discussion as the one presented here the 
cross-spectra allow to simplify the analysis by avoiding any explicit treatment of the noise bias, while retaining all 
the other aspects essentially unchanged. 

The mixing matrix can be computed as,
\begin{widetext}
\begin{equation}
M^{XX'}_{\alpha\alpha'}=\displaystyle\sum_{\ell\in\alpha}\sum_{\ell'\in\alpha'}\frac{\ell(\ell+1)}{\ell'(\ell'+1)\Delta\ell}\displaystyle\sum_{m}\frac{1}{2\ell+1}\sum_{ij}\tilde{\mathbf{Y}}^\dag_{X,\ell m}(x_i)\mathbf{S}^{XX'}_{\ell'}(x_i-x_j)\tilde{\mathbf{Y}}_{X,\ell m}(x_j),
\label{mix-kernel-pixel}
\end{equation}
\end{widetext}
where $x_i$ denotes a position of an $i$-th pixel.
The above expression is fully general, as $\tilde{\mathbf{Y}}_{X,\ell m}$ 
can be any function used to project the Stokes parameters maps into the harmonic $E$ and $B$ subspaces. In particular, for the standard spectra the windowing of the data with a window $W$ can be taken into account by setting,
\begin{equation}
\tilde{\mathbf{Y}}_{X,\ell m}\,\equiv\,W\,\mathbf{D}^X_2\,Y_{\ell m}.
\label{eqn:YtildeDefStd}
\end{equation}
The specific expressions for $M^{XX'}_{\alpha\alpha'}$ relevant for the case of standard pseudo spectra can be found elsewhere \cite{hinshaw_etal_2003}.

The $\mathbf{S}^{XX'}_\ell$ matrix is a convolution kernel relating the pixel-domain correlation functions 
of the $Q$ and $U$ Stokes parameters to the $E$ and $B$ angular power spectrum (the explicit link between 
two-point correlation functions and power spectra for polarization fields can be found in the Appendix 
of~\cite{tegmark_costa_2001}). We point out that in some applications considered in the following the window, 
$W$, and therefore also the functions $\tilde{\mathbf{Y}}_{X,\ell m}$ will all depend on the bin number, $\alpha$, 
(see Eq.~\eref{unbiased}).

In  general, the kernel, $M_{\alpha\alpha'}^{X,X'}$, will mix different $\ell$-modes as well as $E$ and $B$ 
polarization types. The consequences of this fact and the conditions, which ensure the diagonality of the kernel 
in the polarization states are discussed in the following Sections.

\subsection{The $E/B$ leakage problem}

The operators defined in Sect. \ref{subsec:HarmDec}, valid for any spin-$s$ complex field, define two orthogonal subspaces as,
\begin{displaymath}
{\mathbf{D}^{E}_s}^\dag\cdot\mathbf{D}^{B}_s=0.
\end{displaymath}
However, the $E/B$ subspaces defined by ${}_{E/B}\mathbf{Y}_{s \ell m}$ are orthogonal (in the sense of 
Eq.~\eref{scal-prod}) only on the full sky.  This means that if the polarization field is known on a limited part 
of the celestial sphere, the reconstruction of $E$ and $B$ modes by projecting the measured polarization on the  
${}_{E/B}\mathbf{Y}_{2 \ell m}$ basis is ``nonpure'', i.e., the estimated $E$ or $B$ multipoles receive a contribution 
from both $E$ and $B$ modes. This applies in particular to the pseudo-power spectra, $\tilde{\mathcal{C}}^{EE}$ 
and $\tilde{\mathcal{C}}^{BB}$, as defined in Eq.~\eref{eqn:pseudocldef}, each of which will include a contribution 
from the other mode.
Though such a mixing can be removed on average, the leaked modes will still contribute to the variance of the estimated power spectra. Therefore, if the power contained in one of the polarization states is much higher than 
in the other, as is the case of, for example, the CMB anisotropies, the error bars estimated for the mode with 
lower power will be drastically exaggerated due to sample variance of the leaked contribution.

The same conclusion can be arrived at by starting from Eq.~\eref{scal-prod} which can be viewed as a harmonic decomposition of the $(W\mathbf{P})$ field. Multiplication by the window function, $W$, is a convolution not only 
in the harmonic space but also in the $E/B$ subspace, turning thus $\tilde{a}^B_{\ell{m}}$ into a mixture of the true 
$B$ and $E$ CMB multipoles.

This last observation highlights the fact that the standard pseudo-$C_\ell$ technique may be in practice unable 
to provide good estimates of the $B$-mode angular power spectra in particular from data sets coming from 
small-scale CMB experiments. Moreover, even for satellite experiments for which angular power spectra can be
estimated from a significant fraction of the sky, the discussed effects may be significant enough not to be negligible.

We can look at the $E/B$ leakage problem also from the mixing matrix perspective. The lack of orthogonality of $E$ and $B$ type of spherical harmonics results in the coupling between the $E$ and $B$ pseudo power spectra. For this reason the inversion in Eq.~\eref{unbiased} needs to be performed simultaneously for both $E$ and $B$ spectra
to allow for unscrambling the modes with the help of nonvanishing off-diagonal $EB$ and $BE$ blocks of the mixing
matrix.  The resulting expression for the power spectrum estimate, derived in the case of the $B$-mode spectrum 
reads now,
\begin{eqnarray}
\mathcal{C}^{B,\;\alpha}_{\l(i,j\r)} & = &
\l(M^{diag}_{\alpha\alpha'} -  M^{off}_{\alpha\alpha'}\, {M^{diag}_{\alpha\alpha'}}^{-1} \, M^{off}_{\alpha\alpha'}\r)^{-1} \nonumber\\
&\times& \l[\l(\tilde{\mathcal{C}}^{B,\;\alpha'}_{\l(i,j\r)}-N^{B,\;\alpha'}_{\l(i,j\r)} \r)\r. \\
&-& \l. M^{off}_{\alpha\alpha'}\, {M^{diag}_{\alpha\alpha'}}^{-1} \, \l(\tilde{\mathcal{C}}^{E,\alpha'}_{\l(i,j\r)}-N^{E,\;\alpha'}_{\l(i,j\r)}\r)\r]. \nonumber
\end{eqnarray}
In the ensemble average sense the above expression is unbiased as a result of a subtle cancellation of the $E$ 
mode power present in the pseudo- $B$ and $E$ spectra. Such a cancellation does not however apply to 
the variance of the estimator and as a result the variance of the spectra of one type will include a contribution from 
the other. 

In the following we will quantify the $E/B$ leakage using both these perspectives,  i.e., either looking at an excess variance of the estimated $B$-mode power spectrum or at the magnitude of nonvanishing, off-diagonal  blocks of the 
mixing matrix.

\subsection{Pure pseudo-$C_\ell$ estimators for cross-spectrum}

\label{subsect:purespec}

\subsubsection{Definition}

Pseudo-$C_\ell$ estimators of the polarization power spectra which do not mix $E$ and $B$ modes can be 
constructed in projecting the polarization fields on the ``pure'' $E$ and $B$ subspaces~\cite{bunn_etal_2003}. 
Pure $E$ and $B$ multipoles on a partial sky are defined as follows (the vector $\mathbf{P}$ now represents 
the polarization signal as measured on the sky) \cite{smith_2006}:
\begin{eqnarray}
{}_{E}\tilde{a}_{\ell{m}}&=&\displaystyle\int{dx}\,\left[\mathbf{D}^E_2\left(W(x)Y_{\ell{m}}(x)\right)\right]^\dag\cdot\mathbf{P}(x), \label{aetilde} \\
{}_B\tilde{a}^B_{\ell{m}}&=&\displaystyle\int{dx}\,\left[\mathbf{D}^B_2\left(W(x)Y_{\ell{m}}(x)\right)\right]^\dag\cdot\mathbf{P}(x).
\label{abtilde}
\end{eqnarray}
with $W$ a spin-0 window function satisfying the Dirichlet and Neumann conditions on the boundary of the 
observed sky region. Such conditions on the window function are mandatory for the estimated multipoles to be 
free of a $E/B$ leakage due to partial sky. This can be easily shown in the pixel-domain. On applying  an integration 
by parts twice to Eqs.~\eref{aetilde} \& \eref{abtilde} and taking into account the boundary properties of $W$, we 
can rewrite the estimated multipoles as~\cite{smith_2006,smith_zaldarriaga_2007},
\begin{eqnarray}
{}_E\tilde{a}_{\ell{m}}&=&\displaystyle\int{dx}\,W(x)Y_{\ell{m}}(x)\,{\mathbf{D}^E_2}^\dag\cdot\mathbf{P}(x), 
\label{eqn:chiEdef}
\\
{}_B\tilde{a}_{\ell{m}}&=&\displaystyle\int{dx}\,W(x)Y_{\ell{m}}(x)\,{\mathbf{D}^B_2}^\dag\cdot\mathbf{P}(x).
\label{eqn:chiBdef}
\end{eqnarray}
Given that the operator $\mathbf{D}^E_2~(\mathbf{D}^B_2)$ filters out all the $B~(E)$ modes, the multipole decomposition defined above indeed should be free of any $E/B$ mixing. (We point out that the above harmonic
decomposition corresponds to the decomposition on a partial sky of the so-called $\chi$ fields 
defined 
in~\cite{smith_zaldarriaga_2007}.) A simple way to understand this point is to notice that applying 
$\mathbf{D}^B_2$ to the polarization field corresponds to a local filtering of the $E$ modes whereas projecting 
on the ${}_B\mathbf{Y}^\dag_{2,\ell{m}}$ basis corresponds to a global filtering. Because of this difference, 
the latter of these two techniques suffers from partial sky effects whereas the former one does not. In other words, 
if the window, $W$, satisfies the Dirichlet and Neumann boundary conditions, the multipoles estimated in 
Eqs.~\eref{aetilde} and \eref{abtilde} are pure and the mixing kernel, $M^{XX'}_{\alpha\alpha'}$, is a block 
diagonal matrix in the $E/B$ subspace, i.e.,
\begin{equation}
M^{XX'}_{\alpha\alpha'}=M^{XX}_{\alpha\alpha'}\delta_{X,X'}.
\label{eqn:blockMix}
\end{equation}
This last condition, intuitively obvious given our previous discussion, can be formally derived from 
Eq.~\eref{mix-kernel-pixel} defining,
\begin{equation}
\tilde{\mathbf{Y}}_{X,\ell m}\equiv\mathbf{D}^X_2WY_{\ell m}.
\label{eqn:tildeYdefPure}
\end{equation}
A more detailed discussion of the mixing matrix and related issues is presented in the next Section and 
Appendix~\ref{app:mixKernel}.

\subsubsection{Numerical implementation}
\label{subsubsect:NumImp}

\paragraph{Pure multipoles.}

In the actual implementation,  we do not make a direct use of Eqs.~\eref{eqn:chiEdef} \&~\eref{eqn:chiBdef}
as this would require taking numerical derivatives of noisy maps. We rather calculate the pure multipoles using 
Eqs.~\eref{aetilde} \& \eref{abtilde}. We refer hereafter to this way of performing calculations as the spin-weighted approach. Consequently, $\mathbf{D}^{E/B}_2$ operators need to be applied directly only to products 
of the spherical harmonics times the window function. This is an approach also adopted in \cite{smith_2006,smith_zaldarriaga_2007}. Our numerical implementation proceeds in two steps. 
First, we define two spin-weighted windows,
\begin{eqnarray}
W_1=\eth W & \mathrm{and} &W_2=\eth^2 W,
\label{deriv-cond}
\end{eqnarray}
and construct three new fields,
\begin{eqnarray}
\tilde{P}_2&=&W(Q+iU), \nonumber\\
\tilde{P}_1&=&W^\dag_1(Q+iU), \label{eqn:psfields}\\
\tilde{P}_0&=&W^\dag_2(Q+iU).\nonumber
\end{eqnarray}
$\tilde{P}_s$ is a spin-$s$ field, since $W^\dag_s=W_{-s}$ has spin $(-s)$ and $(Q+iU)$ -- spin 2. Then, on the second step, we calculate the pure multipoles as linear combinations of the $E$ and $B$ multipoles of the three new fields. 
The latter can be derived from Eqs.~\eref{aetilde} \&~\eref{abtilde} with the help of Eqs.~\eref{eqn:eYsDef} 
\&~\eref{eqn:bYsDef}. For instance, in the case of the $B$-modes, the pure estimated multipoles read,
\begin{equation}
{}_B\tilde{a}_{\ell m} =  B_{2,\ell m}+2\sqrt{\frac{(\ell-2)!(\ell+1)!}{(\ell+2)!(\ell-1)!}}B_{1,\ell m}+\sqrt{\frac{(\ell-2)!}{(\ell+2)!}}B_{0,\ell m},
\label{counterterm}
\end{equation}
with $B_{s,\ell m}$ denoting the $B$-type multipoles of the $\tilde{P}_s$ field as in Eq.~\eref{scal-prod}. Numerically the entire computation comes down to an efficient calculation of the spin spherical harmonic transforms. We comment on those in more detail  later in this Section. Equation \eref{counterterm} offers another insight into the pure estimator formalism. The first term on the {\em rhs} of this equation corresponds to the standard estimator, while the two following ones act as counter-terms removing the leaked $E$-mode. Indeed we have found that they are anticorrelated with the first term as expected in such an interpretation.

\paragraph{Numerical derivatives.}

The knowledge of the three spin window functions is needed for numerically computing the pure multipoles. As will be clear in the next Section, there are cases where only the spin-0 window is known.  Therefore, we need an efficient way to derive the spin-1 and spin-2 windows from the spin-0 one, which in turn requires a numerical calculation of window derivatives.
 Hereafter we will perform this computation in the harmonic space,
\begin{equation}
W_s=\eth^sW \rightarrow w_{s,\ell m}=\sqrt{\frac{(\ell+s)!}{(\ell-s)!}}w_{\ell m}.
\label{eqn:wells}
\end{equation}
The final spin-1 and spin-2 windows are then obtained by transforming back to the pixel-domain. This
technique allows us to use simple analytic window functions on complicated sky coverage, for which 
only the spin-0 part can be easily computed. It also ensures that the conditions given in Eq.~\eref{deriv-cond} 
are fulfilled. By doing this computation in harmonic space, we may introduce some ringing due to the pixelization 
(or gridding) of the sphere.  Indeed, we have found some minor artefacts of this sort  when comparing numerically 
computed windows with the respective analytic expressions in cases when the latter 
could be calculated, e.g.,  in cases with apodizations as in Eqs.~\eref{eqn:smithWinDef} \&~\eref{eqn:grainWinDef}
and for simple observed sky patches.
It appears, however, that such ringing does not seem to affect our ability to control the $E/B$ leakage at scales where this control is mandatory. For instance, for a $20\times20$ degree square patch pixelized with HEALPix pixels with $N_{side}=512$, the residual, leaked $E$-modes calculated
using the numerical spin windows reproduce closely those obtained analytically all the way up 
to $\ell\sim400$. For higher values of $\ell$, the use of numerical computation leads to a slight increase of the 
residual leakage.  We expect that to be always the case, independently of the apodization length as long as it is sufficiently greater than the pixel size.

In general the discrepancy between the exact and numerical spin windows depends on patch geometry. For example, the discrepancy is reduced for a spherical cap as compared to a square patch. Its magnitude will also in general depend on the assumed pixelization and in particular may be greatly enhanced if the considered region is in a particular position and/or orientation with respect to the pixelization scheme. 
However, whenever necessary, such a discrepancy can be reduced or at least shifted to higher $\ell$-values 
by going to a higher resolution. We demonstrate some of these effects in Sect.~\ref{sect:applications}.

\paragraph{Mixing kernel.}

\label{sect:mixKernelPure}

The mixing kernel can be expressed as a sum over the $\ell$-modes
in the harmonic space of products of the harmonic coefficients of the spin window functions and $3j$ Wigner 
coefficients~\cite{hivon_etal_2002}. This is done by using Eq.~\eref{eqn:tildeYdefPure} in Eq.~\eref{mix-kernel-pixel} 
and subsequently expanding the window function in harmonic space. We present 
relevant formulae in Appendix \ref{app:mixKernel} (see also \cite{smith_2006} for an alternative derivation of those kernels). They 
generalize the standard expressions for the ordinary pseudo spectra \cite{kogut_etal_2003, hansen_gorski_2003}.
In our numerical computations
we implement these formulae and perform all the required calculations directly in the harmonic domain. This is in contrast to 
the previous study of \cite{smith_2006}, which strives to make computations predominantly in the pixel-domain.
\begin{figure}[ht]
\includegraphics[scale=0.246]{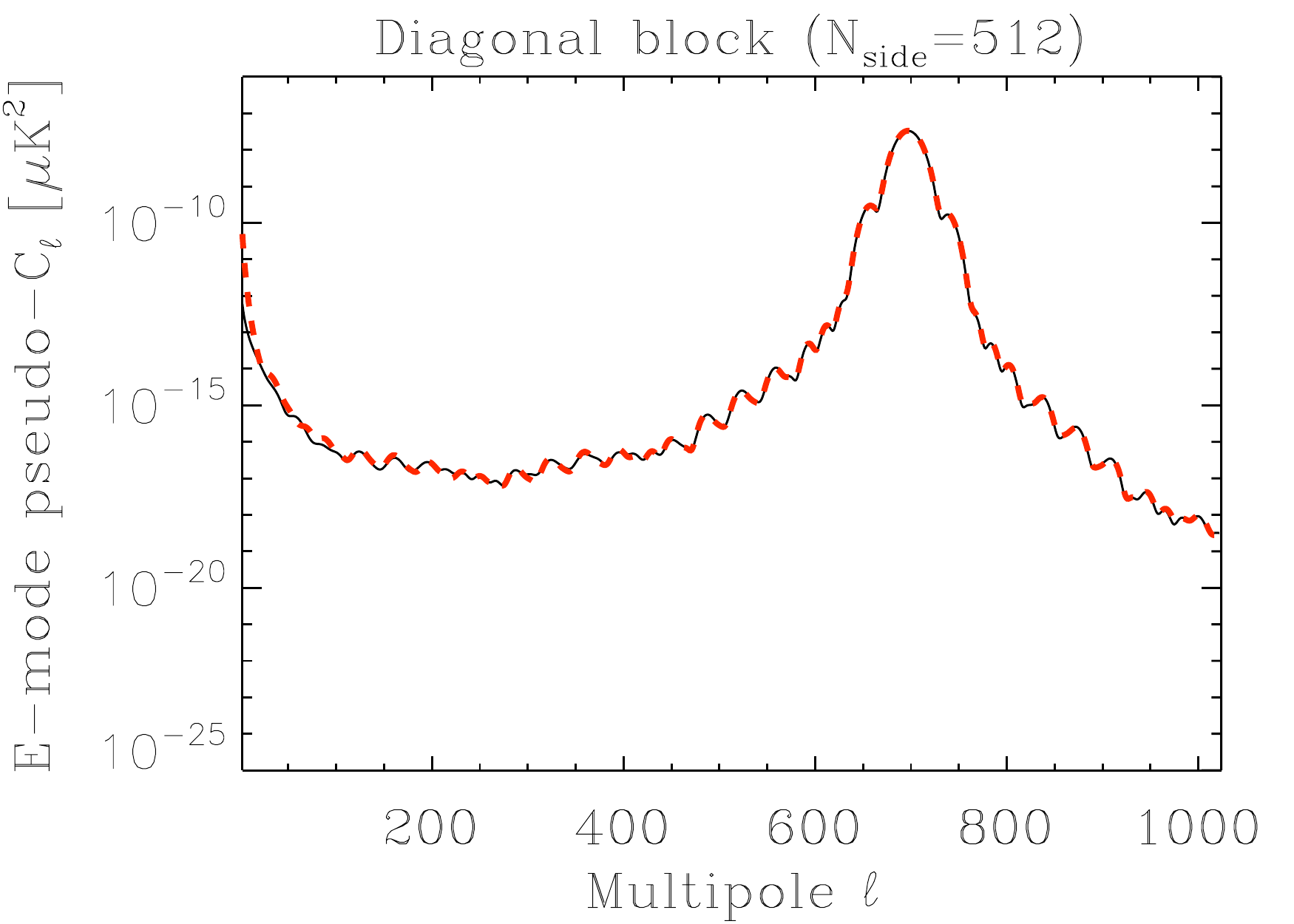} \includegraphics[scale=0.246]{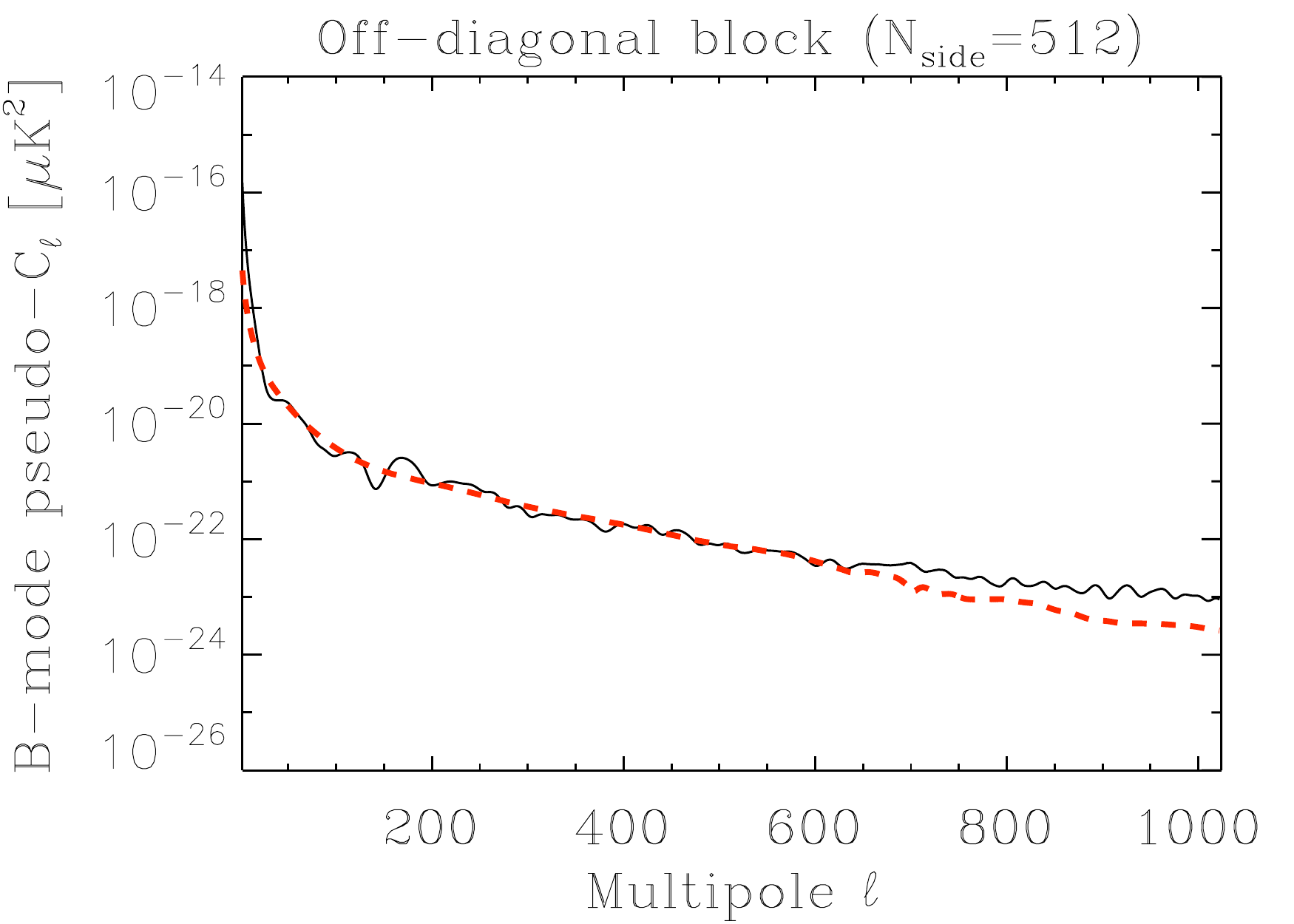} \\
\includegraphics[scale=0.246]{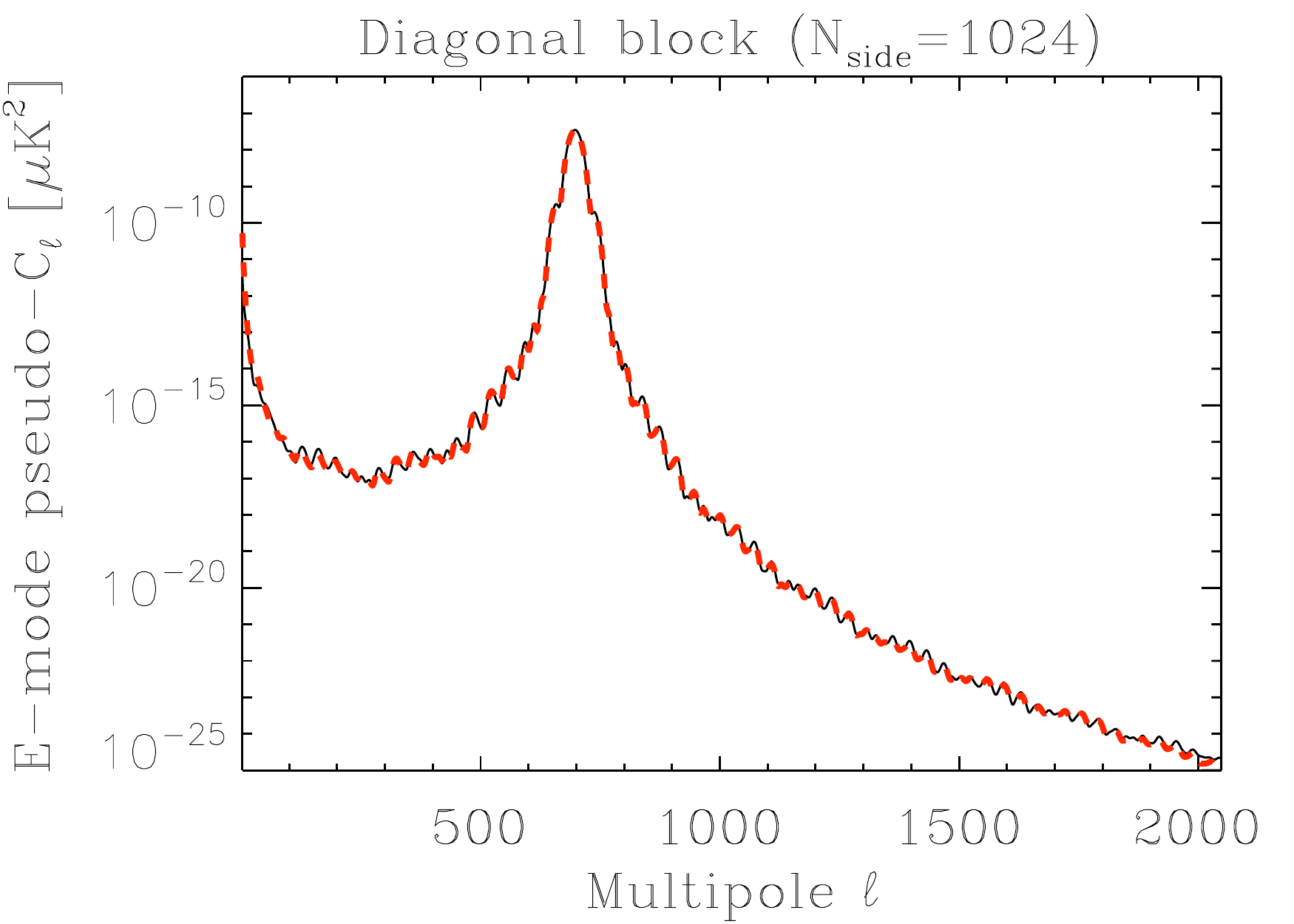} \includegraphics[scale=0.246]{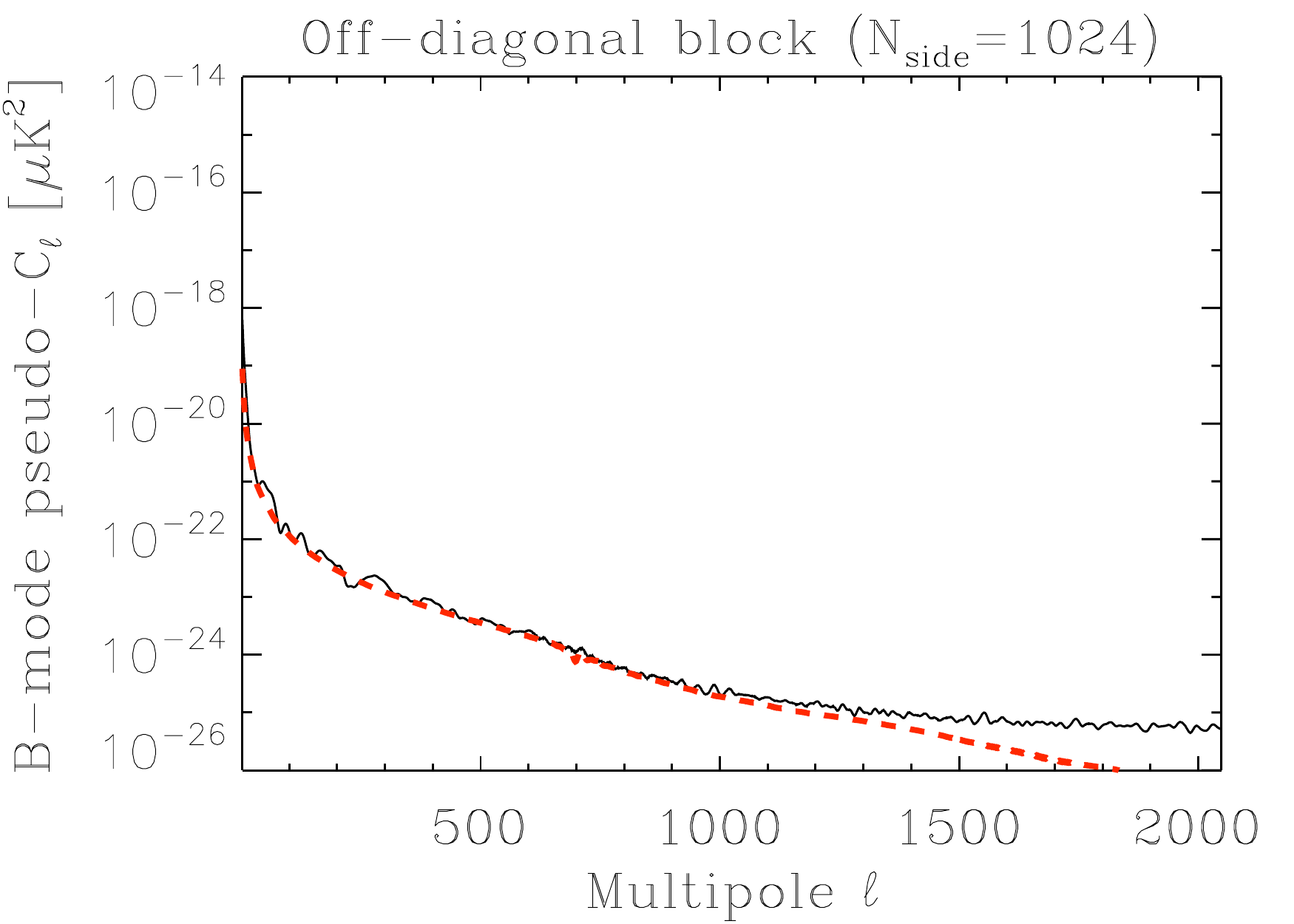}
\caption{Dashed (red) curves show a single column of the mixing kernel at $\ell_0=700$ computed using our numerical 
approach as described in the text, for the diagonal, left, and off-diagonal blocks, right, and two different resolutions: $n_{side}=512$, top row, and $n_{side} = 1024$, bottom. The overlapping solid (black) lines depict the $EE$, left, and $BB$,
right,  pseudo-spectra computed from $Q$ and $U$ maps containing only $E$-mode power at $\ell_0=700$ and apodized with analytic window functions satisfying the boundary condition and constrains in 
Eq.~\eref{deriv-cond}. The apodization length is set to 7 degrees.
A very good agreement seen in all the cases validates the proposed numerical approach. The off-diagonal elements in the
left panels are due to the cut-sky mode mixing, while the nonzero values in the right panels are due to a residual
pixel-induced $E/B$ leakage.
}
\label{fig:oneEmode}
\end{figure}

As we mentioned before for any scalar window function we can calculate the harmonic coefficients of the spin windows by using the relation in Eq.~\eref{eqn:wells}. If we then use these in the computation of the mixing kernel, Eqs.~\eref{eqn:mlldiaganal} \&~\eref{eqn:mlloffanal}, we find that the off-diagonal blocks of the mixing matrices vanish within the numerical precision. This simply reflects the fact that the pure estimators are indeed "pure" and do not mix $E$ and $B$ modes.
This is however also not realistic. 
In practice, we expect that some residual $E/B$ leakage will be present due to the sky discretization -- an effect not corrected by the pure pseudo spectra formalism and not characterized properly
in the above mentioned calculation of the mixing matrix.
 To account for the pixelization effects we propose to compute the mixing kernel using coefficients of each of the spin windows calculated via numerical spherical
harmonic transforms (SHT) of the pixel-domain representation, Eqs.~\eref{eqn:mlldiaganalspin} \&~\eref{eqn:mlloffanalspin}. 
We do so even in the cases, when the window derivatives are computed numerically and thus are directly available in the harmonic domain. In such cases, our
procedure
corresponds simply to taking first an inverse SHT of the spin windows as given in the harmonic domain. This produces their pixel-domain representation, which we
transform back to harmonic domain via a direct SHT.
Because of sampling effects such an operation does not reproduce precisely the input coefficients. The new harmonic coefficients of the windows, when used now
to calculate the mixing matrix, lead the higher values of the off-diagonal block
elements, which, though still small when compared with the diagonal elements of the diagonal blocks, are not any more numerically zero.
This can be understood from Eq.~\eref{counterterm}: because of gridding effects during the SHT, 
the multipoles involved in the pure estimators are not computed exactly and the precise cancellation of the leaked 
$E$-mode is not perfect.

As those are the pixel-domain representation of the windows, which are used in the pure pseudo-spectra computation, we may expect
that the observed change by explicitly taking the SHT of each spin windows while computing the mixing kernel, goes in the right direction, properly
reflecting the correlation pattern between the modes of the pure spectra.
 We can test this expectation with the help of numerical experiments. We first note that we can compute any of the columns
of the mixing matrix by computing the pure $E$ and $B$ pseudo-spectra 
of the sky signal containing power only in one selected harmonic mode
$\ell=\ell_0$. Examples of such computations are shown in Fig~\ref{fig:oneEmode}.
There we have allowed  for nonzero power only in a single $\ell_0=700$ mode of the $E$ polarization. The derived pure $E$ and $B$ pseudo spectra are contrasted with the corresponding columns of the $EE$ and $BE$ blocks of $M_{\ell\ell'}$, both computed using the procedure outlined before.

For a square patch assumed in the calculation visualized in Fig. ~\ref{fig:oneEmode}, the overall agreement between the pseudo spectra and the column of the kernel is  indeed very good.
For the off-diagonal block some differences can be seen only at the high-$\ell$ end with the mixing kernel underestimating the residual leakage. The discrepancy appears mostly benign as it affects the multipoles for which the leaked signal is found to be negligible.

We thus conclude that the simple way of incorporating the pixel effects as proposed here works indeed very well everywhere, where the leakage is significant and thus requires a sufficiently accurate treatment. This also demonstrates that the method of computing the
mixing matrix advocated here adequately describes the actual
correlations between the $\ell$-modes of the computed polarized pseudo-spectra.
For more relevant examples of such a comparison see Sect.~\ref{sect:optMasks} and Fig.~\ref{fig:mixMatPCGebex}.

We note that in general different window functions can be used for the estimation of the power in different $\ell$-bins.
Indeed, as we discuss that later on, it is a common occurrence in practical applications in which we usually strive 
for the smallest  attainable uncertainty on a final power spectrum. 
In such cases, we calculate the kernel row-by-row, i.e. bin-by-bin, using an appropriate window assigned to
each of them. The kernel computed in this way is later used to solve the inverse problem in Eq.~\eref{unbiased}.

\paragraph{Spherical harmonic transforms.}
\label{sect:s2hat}

At the core of the numerical implementation of the formalism presented in the paper is a calculation of the spin spherical harmonic transforms. For this purpose we utilize here a publicly available,  Scalable Spherical Harmonic Transforms (S$^2$HAT) library~\cite{s2hat}. This parallel library allows for an efficient computation of arbitrary spin transforms for a general class of sky griddings and pixelization in which the pixels (grid points) are distributed along the constant declination lines. These include therefore the most frequently used in the CMB work, e.g., HEALPix~\cite{gorski_etal_2005}, GLESP~\cite{doroshkievich_etal_2005}, ECP~\cite{mucciacia_etal_1997} as well as some others.
The spin spherical harmonics are computed using the recurrence relations as listed in~\cite{lewis_2005}, which are appropriately rescaled to avoid numerical under- and overflows. The library is designed to treat optimally the cases with only a partial sky coverage and is therefore
particularly suitable for our purposes here. It is parallelized using the Message Passing Interface (MPI)  and distributes over the processors both the pixel (maps) and harmonic (harmonic coefficients) domain objects permitting to achieve high resolutions with modest memory resources available per single processor.

\section{Sky apodizations}

\label{sect:apodizations}

Performance and in a result applicability of the pure (cross) spectra formalism is clearly
dependent on our ability to compute a suitable apodization for any specific CMB polarization 
experiment.
In this Section we introduce a number of proposal functions, which could be used as apodizations
in calculations of the pure (cross) spectra. The discussed functions range from ones derived as 
a result of optimization procedures and therefore often numerically involved, to ones proposed 
somewhat {\em ad hoc} but quick and easy to calculate. We will discuss their relative merit only
briefly here, postponing a more thorough comparison of their performance to the following 
sections.

\subsection{Variance-optimized apodizations}

\label{sect:optMasks}

In realistic applications we would like to have an apodization function which would ensure that the total (i.e., noise plus sample variance) uncertainty of the recovered power spectra is (nearly) minimal. 
In the case of the auto pure pseudo spectra a general ansatz for such optimized
weighting has been proposed recently by Smith and Zaldarriaga~\cite{smith_zaldarriaga_2007}. It is derived as a condition on the apodization window required
to minimize the difference between the auto pseudo-spectrum of a map and its 
optimal counterpart.
Those authors have obtained a following relation for the (nearly) optimal weights,
\begin{equation}
\displaystyle\sum^{N_{obs}}_{i=1}\mathbf{C}_{ij}\mathbf{P}^{(\alpha)}_{ij}W_i=1,~~~~~\mathrm{for~all}~j.
\label{equ-sz}
\end{equation}
Here the summation is made over the $N_{obs}$ observed pixels. The matrix $\mathbf{C}$ is the covariance matrix of the data and $\mathbf{P}$ is a geometrical matrix which comes from the projection on harmonic space (for example the Legendre polynomials matrix in the case of temperature).
The optimized weight function can be then obtained by a numerical inversion of the above linear system for each bin $\alpha$. 
It can be shown~\cite{smith_zaldarriaga_2007} that such optimized windows converge towards inverse square-noise weighting in the noise-dominated case. This is a regime usually reached at the high-$\ell$ end of the estimated power spectrum.
We refer the reader for
more details about the derivation of this relation to~\cite{smith_zaldarriaga_2007}.

For the computation of the cross-spectrum we need to compute an apodization for each of the available maps. The approach of~\cite{smith_zaldarriaga_2007} has to be therefore extended to accommodate such cases. This can be done
in a number of ways. For instance, using Eq.~\eref{equ-sz} we can compute an apodization for each map considering them separately or together. Alternately, we 
could apply a reasoning similar to that of~\cite{smith_zaldarriaga_2007} but in the context of the cross-spectrum to derive a relation analogous to Eq.~\eref{equ-sz} but applicable to the cross-spectra. Though this last approach may seem the most appropriate for the case at hand it unfortunately results in an expression, Eq.~\eref{x-opt3}, which is computationally rather involved.
The two first options mentioned above are numerically more tractable, though they produce windows, which minimize the difference between the pure pseudo-spectrum estimator and its optimal counter-part on the auto spectrum level. 
The first of these two approaches uses for that purpose the auto spectrum of the map for which the apodization is sought, while the second one -- the single auto spectrum of all the maps considered. Clearly, which one of the two methods is more appropriate will depend on the specific goals of the entire estimation procedure. We note for example that in the case of two maps with the same noise properties the difference between the two optimization approaches is limited only to a different overall factor 
determining the level of the noise with respect to the sky signal (see Appendix~\ref{app:CrossOptWin} for more details).
Hereafter, we will utilize the first option and solve Eq.~\eref{equ-sz}
as in the pure pseudo auto spectrum case for each of the two maps. We will support 
this choice in the following Sections with the help of numerical experiments.

Equation \eref{equ-sz} can be specialized to a particular case of polarization when the input data used are composed of the Stokes $Q$ and $U$ maps.
The geometrical matrix $\mathbf{P}^{(\alpha)}_{ij}$ needs to be then extended into the spin subspaces and nonzero spin windows introduced. Their implicit definition consequently reads,
\begin{equation}
\displaystyle\sum^{N_{obs}}_{i=1}\mathbf{C}_{ij}\sum_{s=0}^2\mathbf{P}^{(\alpha)}_{i+N_{obs}s,j+N_{obs}s'}W_{i+N_{obs}s}=\mathbf{e}_{j+N_{obs}s'},
\label{equ-sz-spin}
\end{equation}
where an additional summation over spins of the windows is added. The geometrical matrix,  $\mathbf{P}^{(\alpha)}$, describing the projection on spin-weighted spherical harmonics, 
is now explicitly dependent on the spin value, $s$, and in general is not diagonal in the spin subspace.
The vector $\mathbf{e}$ is a $5N_{obs}$ vector whose first $N_{obs}$ entries are equal to one, corresponding to the normalization of the spin-0 component, and all the others are set to zero~\cite{smith_zaldarriaga_2007}.

The approach can be also looked at as an attempt to minimize the total, signal plus noise, power aliased to a given bin either from the other adjacent bins or the other modes (e.g., $E/B$ leakage).
This viewpoint is referred to as the variational interpretation in~\cite{smith_zaldarriaga_2007}. Consequently, in general there is always a different window function calculated for each bin as a result of this procedure.
This has two consequences. First of all, one has to deal with a set of windows optimized for 
each bin. Second, this means that one has to define each bin parameters before performing the 
optimization procedure.

Below we discuss implementation of this kind of optimization procedure in both pixel and harmonic domains.

\subsubsection{Pixel-domain implementation}

The matrix operations  in Eq.~\eref{equ-sz-spin} should be done preferably in a pixel-domain. 
That allows to consider a full potential complexity of the map noise properties as well 
as arbitrary masks with complicated boundaries and/or holes due to point-sources extraction, 
rendering such an implementation very flexible. 

\begin{figure}
\includegraphics[scale=0.246]{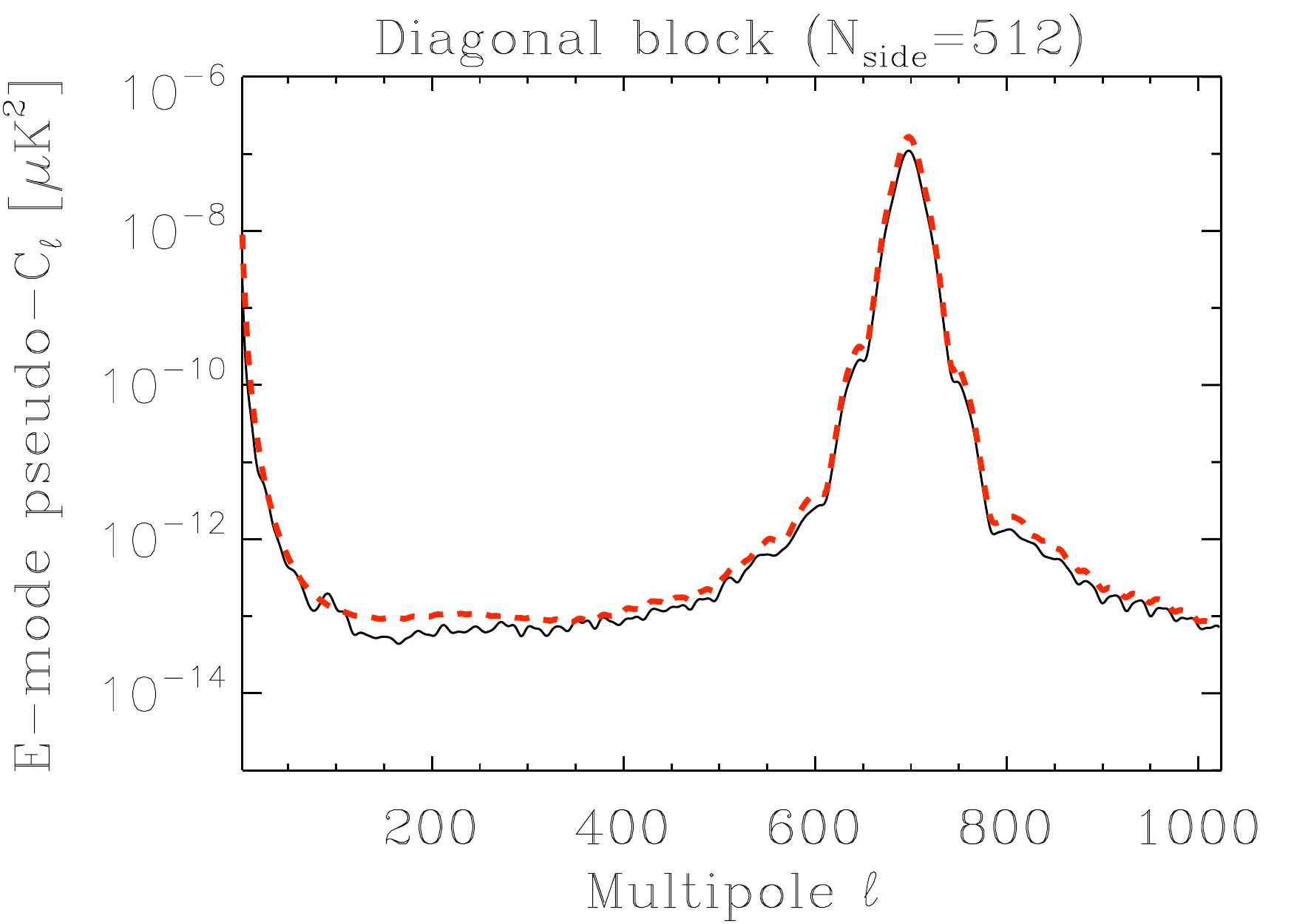} \includegraphics[scale=0.246]{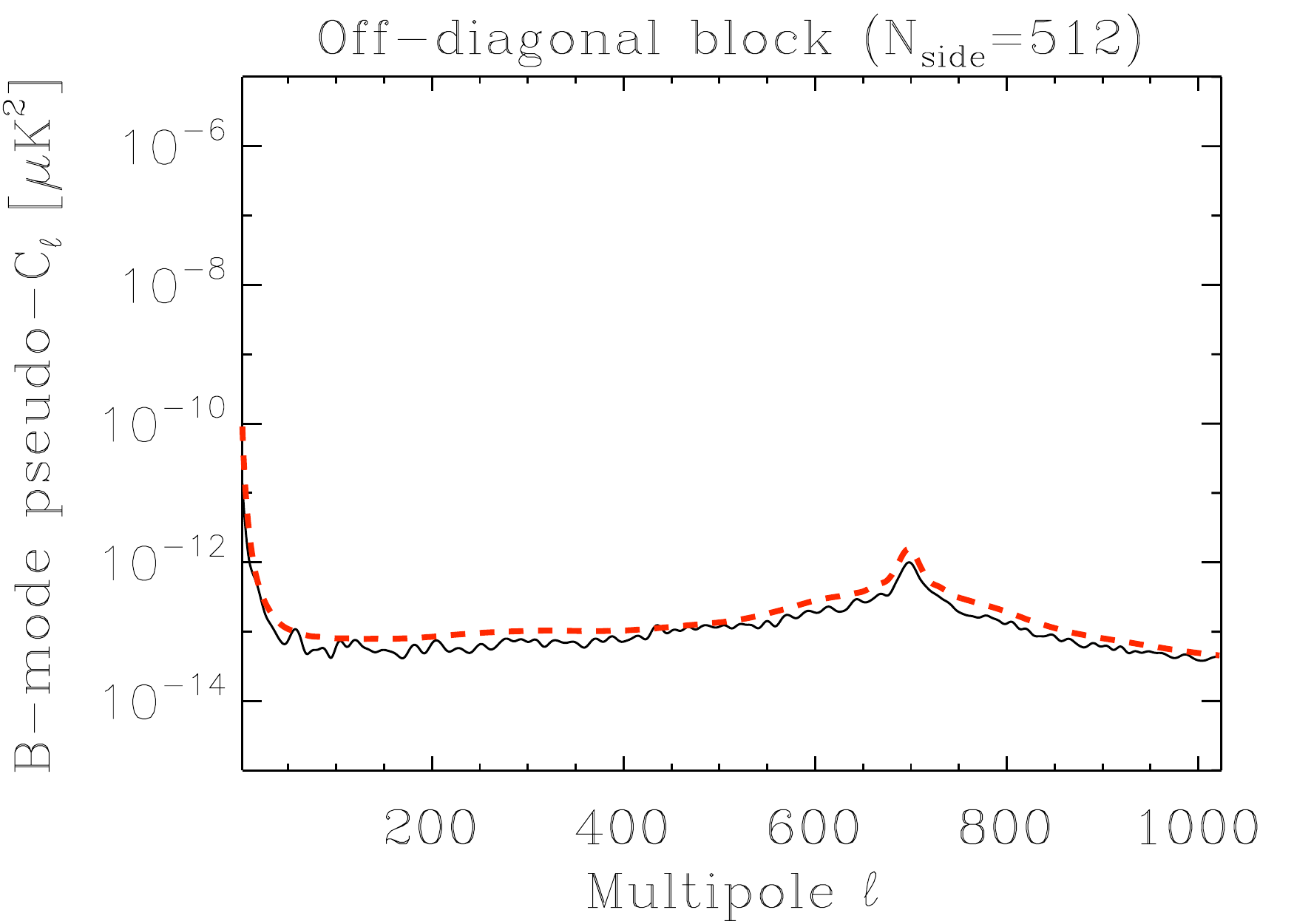}
\caption{As figure~\ref{fig:oneEmode}, but computed assuming numerically optimized pixel-domain
window functions and
a more realistic sky patch as could be  observed by CMB B-mode balloon-borne experiments, 
such EBEX, (see Fig.~\ref{nhit} and Sect.~\ref{sect:applications} for more details).}
\label{fig:mixMatPCGebex}
\end{figure}

In the context of the pure pseudo-spectrum estimators, one should in principle ensure that the spin window 
functions fulfill the relation in  Eq.~\eref{deriv-cond} as well as boundary conditions. However, hereafter 
following \cite{smith_zaldarriaga_2007}, we relax these constraints in the numerical implementation of the 
procedure. This allows for greater flexibility and therefore enables finding window functions better suited for our needs. 
Nevertheless, this simplification has a major consequence: the pure pseudo-$C_\ell$ estimator is not perfectly 
pure and will unavoidably introduce some $E/B$ leakage in addition to the one induced by the pixelization itself. 
This residual leakage may become problematic in the low-$\ell$ region where sampling variance is the major 
contribution to the error bars. We note however that in the high signal-to-noise ratio regime the resulting windows 
are expected to nearly fulfill all the constraints, even if those are not explicitly imposed. This is because the total 
variance minimization approach used in~\cite{smith_zaldarriaga_2007} to derive Eq.~\eref{equ-sz} implies that 
the solution for which the counter-terms in Eq.~\eref{counterterm} nearly cancel the leaked signal included in the 
first term is preferred over those for which three terms are independent, which unavoidably boosts the variance.
In the noise-dominated regime, the leaked $E/B$ signal is quickly becoming subdominant and therefore it will be the 
noise variance, not the leaked $E$ signal, which will be the major target of the optimization. In this case, the perfect 
apodization of the window, i.e., as required by the pure formalism, may not be the best solution, which should
gradually, with a decreasing signal-to-noise ratio, approach the standard inverse noise weighting.

We point out here that the window function satisfying Eq.~\eref{equ-sz} ensures only that the uncertainty is 
minimized on the pseudo-spectrum not on the power spectrum level. In particular, if the mixing matrix is not 
block diagonal, i.e., some residual $E/B$ leakage allowed, the final power spectrum 
uncertainty may not be strictly at its minimum. Nevertheless, though some $E/B$ leakage may be indeed present, 
it will be correctly accounted for in our calculation of the mixing matrix, and therefore consistently treated in the 
presented method. We show in Fig.~\ref{fig:mixMatPCGebex} that the computational approach to the calculation 
of the mixing matrix proposed here, and already discussed in the case of a simple sky patch and analytic window, 
fares indeed very well also in the case with the nearly optimized window functions.

The fact that  Eq.~\eref{equ-sz-spin} is solved relaxing the derivative constraints between the spin-$s$ and spin-0 components
does not reduce it to a set of three independent equations -- one 
for each spin-weighted window. This is because the geometrical matrix $\mathbf{P}^{(\alpha)}_{ij}$ is not diagonal 
in the spin subspace. Consequently all three windows have to be solved for simultaneously. In order to do 
so, following \cite{smith_zaldarriaga_2007} we have implemented an iterative Preconditioned Conjugate Gradient (PCG) solver~\cite{golub_vanloan_1987}. The numerical cost of such an operation is high and on order of 
${\cal O}\l(n_{iter}\,n_{pix}^2\r)$, if no symmetries are present. A number of the iterations, $n_{iter}$, is usually 
as high as a few hundreds and it is strongly dependent on a dynamic range spanned by the pixel noise 
amplitudes. Nonetheless, for small-scale experiments with the help of a proper thresholding the computational 
load is clearly within the limits of available presentday (super)computers. We note that for the high signal-to-noise 
cases imposing explicitly the constraints on the window functions could reduce the number of iterations required 
for the procedure to converge. We have not however attempted to include such an optimization in the current 
version of the software in a general case (see however the homogenous noise discussion in the next Section).

\begin{figure}[ht!]
\includegraphics[scale=0.175]{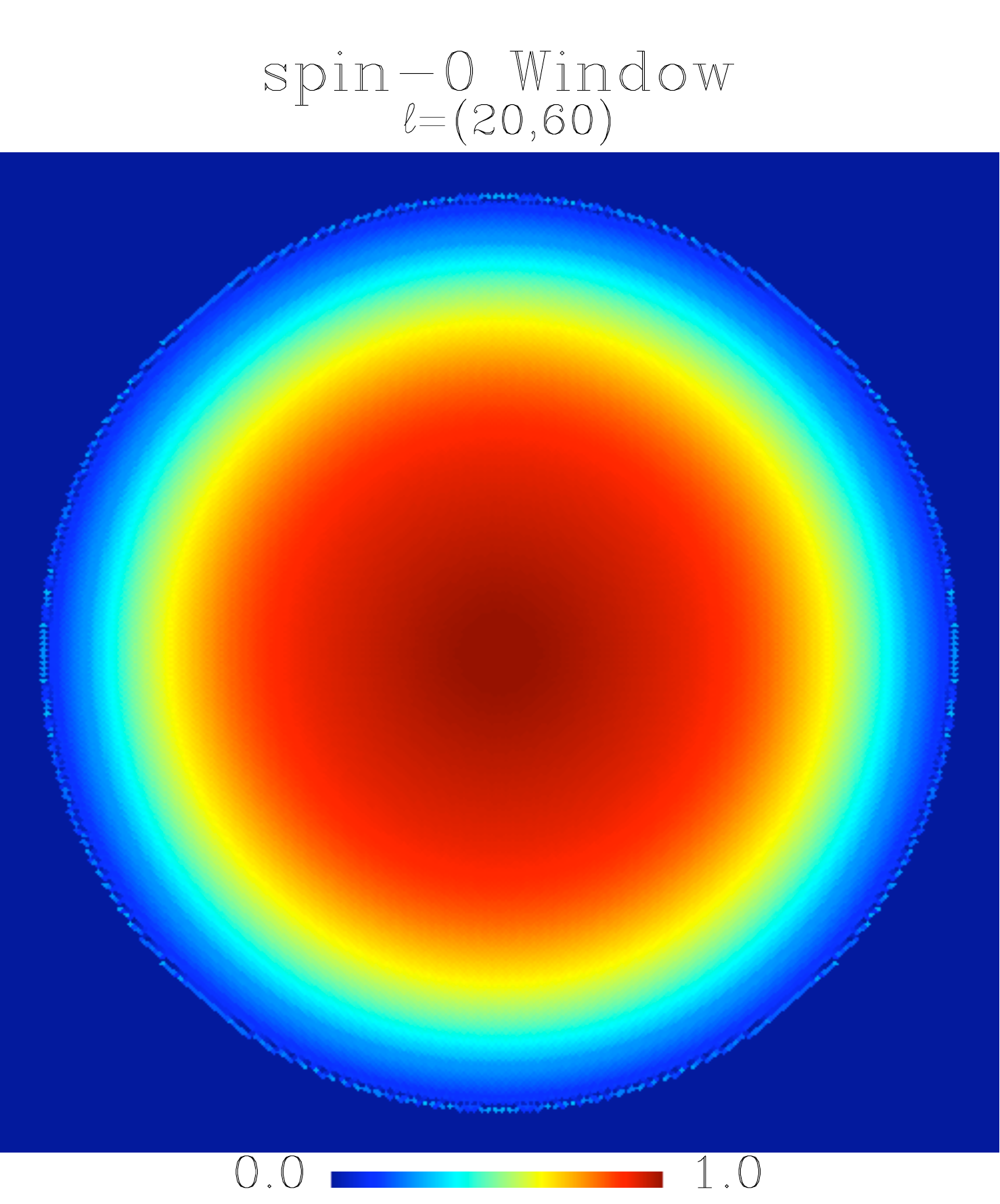} \includegraphics[scale=0.175]{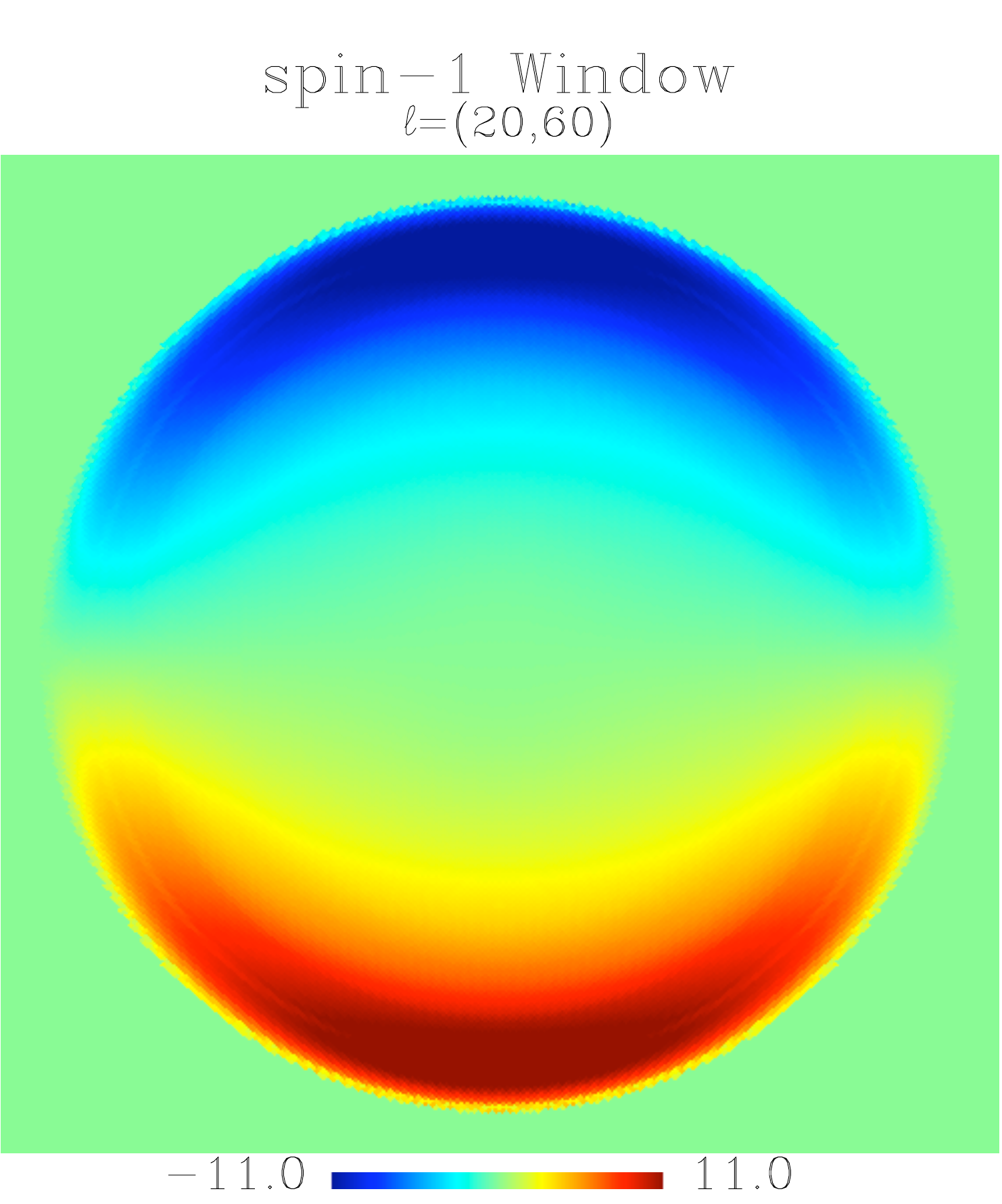} \includegraphics[scale=0.175]{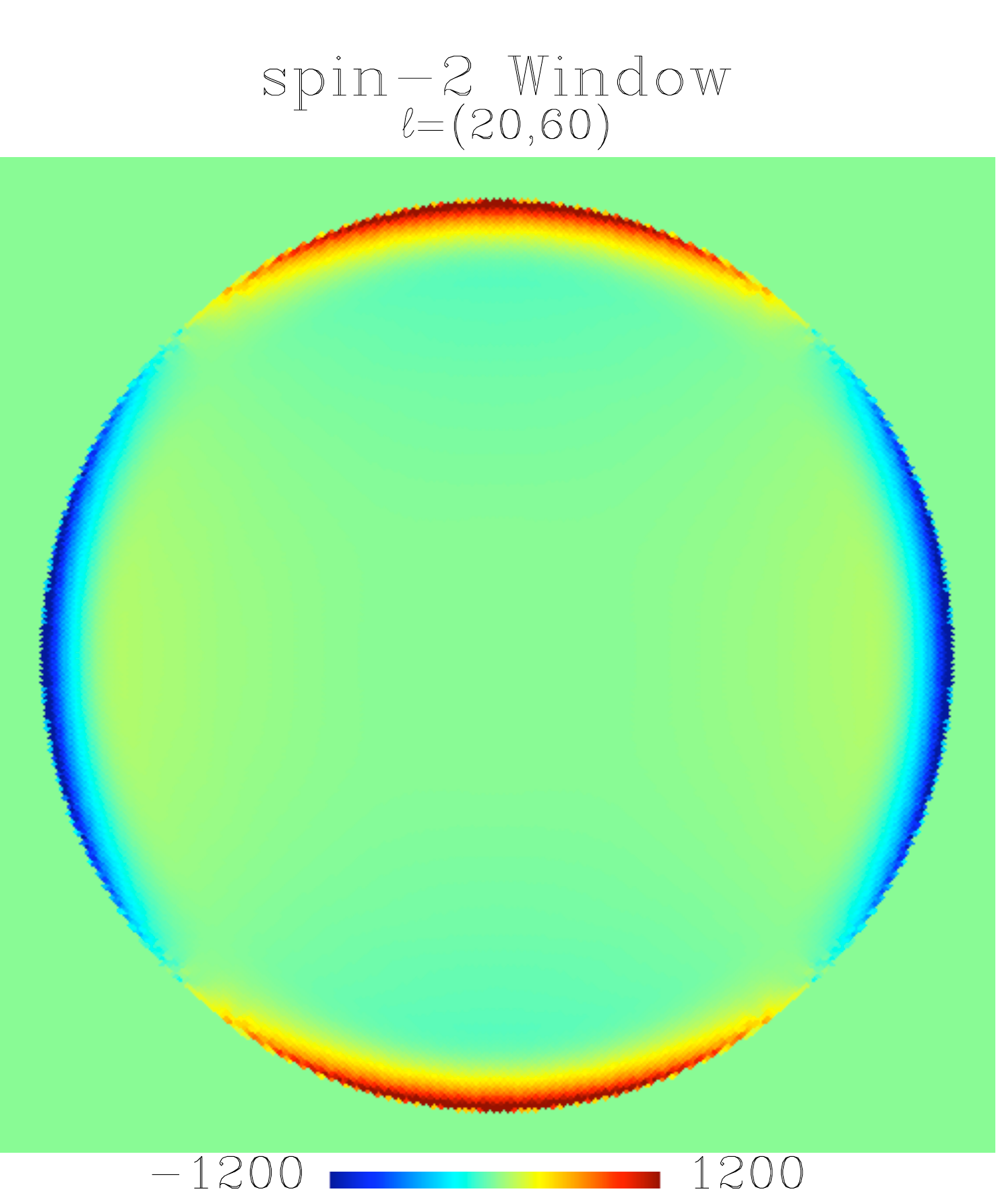} \\
\includegraphics[scale=0.175]{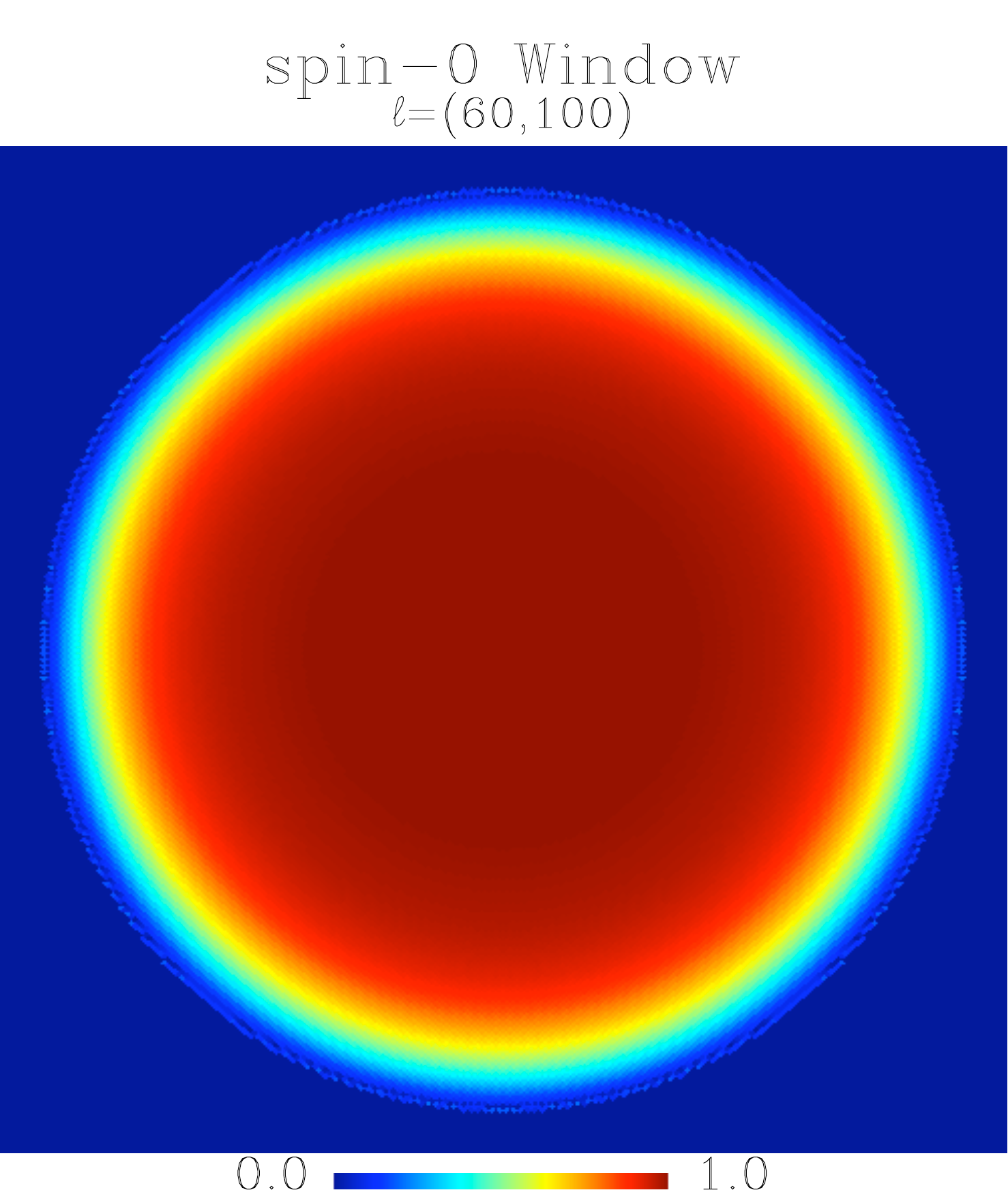} \includegraphics[scale=0.175]{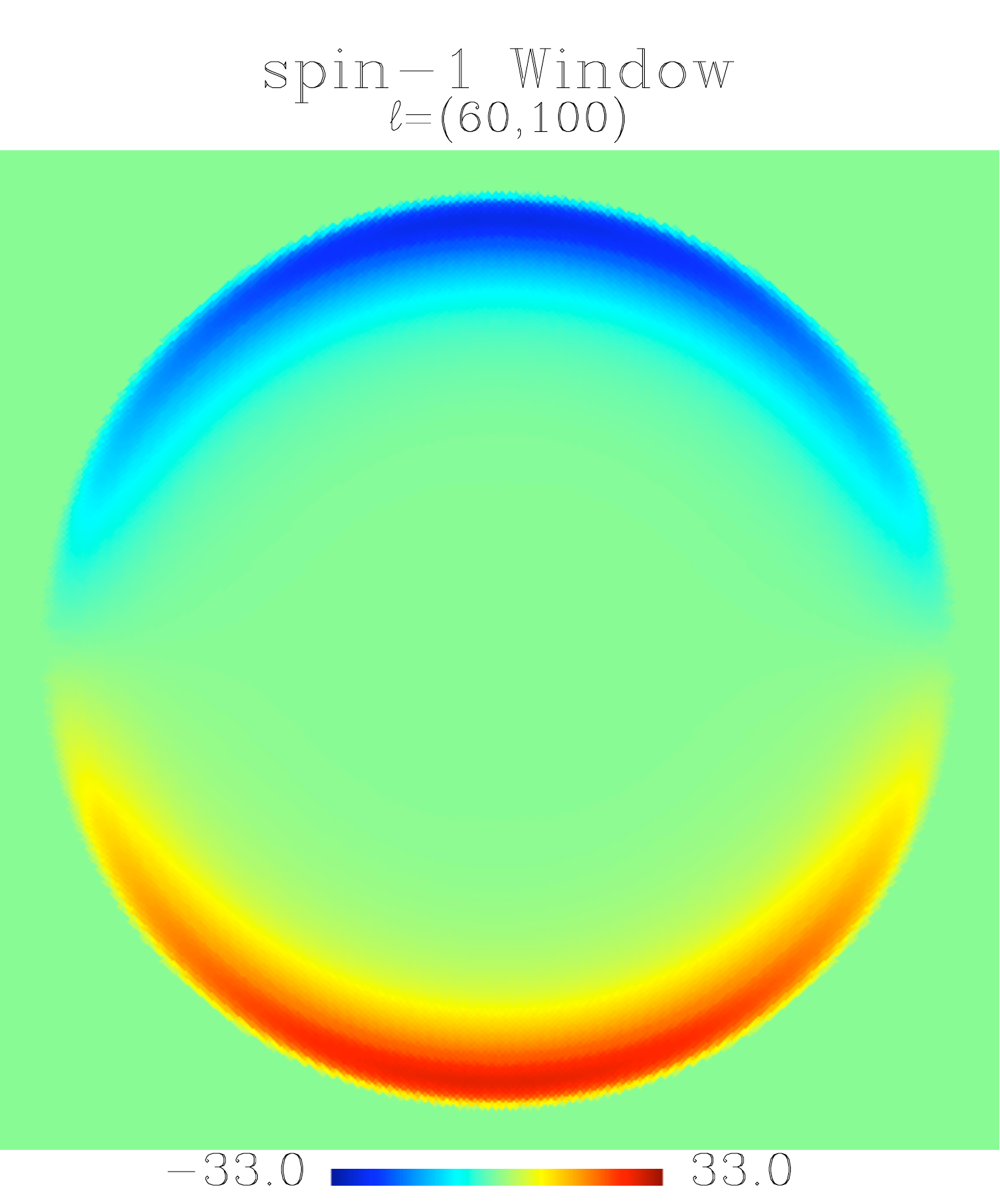} \includegraphics[scale=0.175]{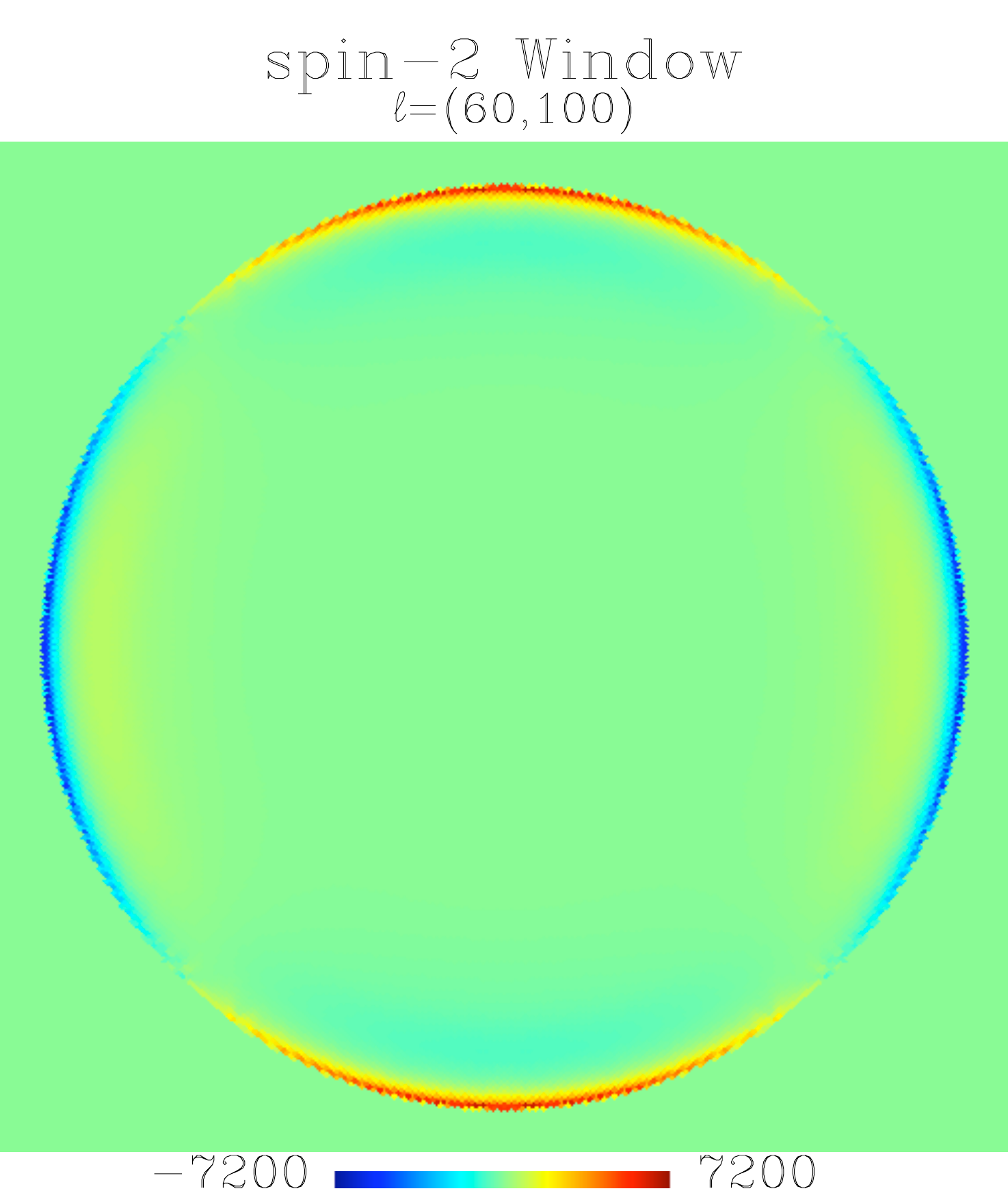}
\caption{From left to right: spin-0, spin-1 \&\ spin-2 spin-weighted optimized windows for two bins 
$\ell\in(20,60)$ and $\ell\in(60,100)$ are shown in the upper and the lower panels respectively. In all cases 
only the real parts of the spin-1 and spin-2 windows are displayed. The noise level is set to $5.75~\mu$K-arcmin, 
the $E$-mode signal is  the WMAP 5-year best fit~\cite{dunkley_etal_2008} and the $B$-mode signal includes 
lensing-induced and primordial $B-$mode ($T/S=0.05$).  
The sky patch is a cap of a radius of 11~degree leading thus to a significant amount of $E$-to-$B$ leakage.
Note that in the middle (left) panel, the color stretch is 3 (6) times bigger for the case with $\ell\in(60,100)$ 
than for $\ell\in(20,60)$.}
\label{pcg1}
\end{figure}

To illustrate the results of our numerical optimization,  in Fig.~\ref{pcg1} we show three windows computed for a spherical cap with homogeneous noise and two different $\ell$-bins. The noise level is set to $5.75~\mu$K-arcmin, the 
$E$-mode signal is obtained with the WMAP 5-year cosmological parameters~\cite{dunkley_etal_2008} and the 
$B$-mode signal includes lensing-induced and primordial $B$-mode ($T/S=0.05$). The radius of the cap is 
roughly 11~degrees, leading to a significant level of $E$-to-$B$ leakage in the represented bins. The resulting 
window functions are well but not perfectly apodized as they do not precisely vanish at the edge of the cap. 
However, as expected, the apodization length, defined as a width of the boundary layer, see Eq.~\eref{eqn:smithWinDef}, decreases for bins with higher $\ell$. This is in agreement with 
the discussion of \cite{smith_2006} concerning the choice of the apodization length for homogeneous noise cases.

\subsubsection{Harmonic domain implementation}

If the noise in the pixel-domain is uncorrelated and homogeneous over the observed sky patch, the overall 
optimization procedure can be simplified by performing the calculations in the harmonic space. That can be
done most straightforwardly starting from Eqs.~\eref{eqn:chiEdef} \&~\eref{eqn:chiBdef} rather than 
Eqs.~\eref{aetilde} \& \eref{abtilde} and therefore results in windows for which the boundary conditions, 
Dirichlet~+~Neumann, are explicitly imposed.

We begin deriving the average $B$-mode pseudo-spectrum (see Appendix~\ref{app:harmOptWin}),
\begin{equation}
\left<\tilde{C}_\ell\right>=\displaystyle\sum_{\ell''m''}{C}^\chi_{\ell,\ell''}w_{0,\ell''m''}w^*_{0,\ell''m''},
\label{equ-chi}
\end{equation}
where,
\begin{eqnarray}
{C}^\chi_{\ell,\ell''}&=&\displaystyle\sum_{\ell'}\frac{2\ell'+1}{4\pi}{\frac{(\ell-2)!(\ell'+2)!}{(\ell+2)!(\ell'-2)!}}
\left(\begin{array}{ccc}
\ell & \ell' & \ell'' \\
0 & 0 & 0
\end{array}\right)^2\nonumber \\
&\times& \left(C^B_{\ell'}B^2_{\ell'}+\sigma^2\right).\nonumber
\end{eqnarray} 
Here $\sigma^2$ denotes the noise power spectrum, $B_\ell$ -- the effect of an azimuthally symmetric beam.
The power spectrum ${C}^\chi_{\ell,\ell''}$ represents the total aliased signal and noise power leaking from the $\ell'$ bin into the $\ell$ one.

The optimized apodization is then given by applying the variational principle to $w_{0,\;\ell''m''}$ with three external 
constrains, which have to be imposed: the window has to be normalized within the observed region and together with its first derivative, they have to vanish at the contour of the observed region. The computation 
then consists of solving a differential equation with the appropriate Dirichlet~+~Neumann
boundary conditions thus leading to the final expression for the optimized window given by,
\begin{equation}
w_{0,\ell''m''}=\frac{1}{{C}^\chi_{\ell,\ell''}}\left(\lambda \mathcal{M}_{\ell''m''}+\mu \mathcal{P}_{\ell''m''}\right).
\label{opt-sht}
\end{equation}
Here $\mathcal{M}_{\ell''m''}$ and $\mathcal{P}_{\ell''m''}$ are the harmonic representation of the mask and the 
contour and $\lambda$ and $\mu$ are two constants given in Appendix~\ref{app:harmOptWin},  which  both 
depend on the signal and noise power.

In practice the optimization proceeds in two steps. First, Eq.~\eref{opt-sht} is used to provide the scalar component of the optimized window. This requires taking the spherical harmonic transform of the mask and of its contour and calculating the weights $C^\chi_{\ell,\ell''}$. Subsequently,  the spin-1 and spin-2 windows are derived from the scalar one in the harmonic space using the numerical tools.
This procedure requires taking the SHT of the binary mask and the one of the contour of this mask. Because this last SHT is prone to a numerical error (as it corresponds to calculating an integral of a pixelized line), the resulting scalar window function is not perfectly apodized. As a consequence, the window derived in this way does not ensure that the estimator is perfectly (or sufficiently) pure and may therefore reintroduce $E/B$ leakage.  We point out that unlike the pixel-domain implementation for which the level of leakage is by design smaller than the noise level, and therefore under control, the extra-leakage introduced is here not controlled and thus may or may not be smaller than the noise. To cure this problem, we add an extra apodization to the scalar window, $W_0$, before computing its first and second derivatives. This extra-apodization is obtained by multiplying the spin-0 window by  the $C^2$-window with a small apodization length (see the following Section
 ). This makes our window optimal throughout the entire mask (its shape is driven by the optimization procedure) except very close to the boundary, where its shape is driven by the extra apodization.

\begin{figure}
\includegraphics[scale=0.175]{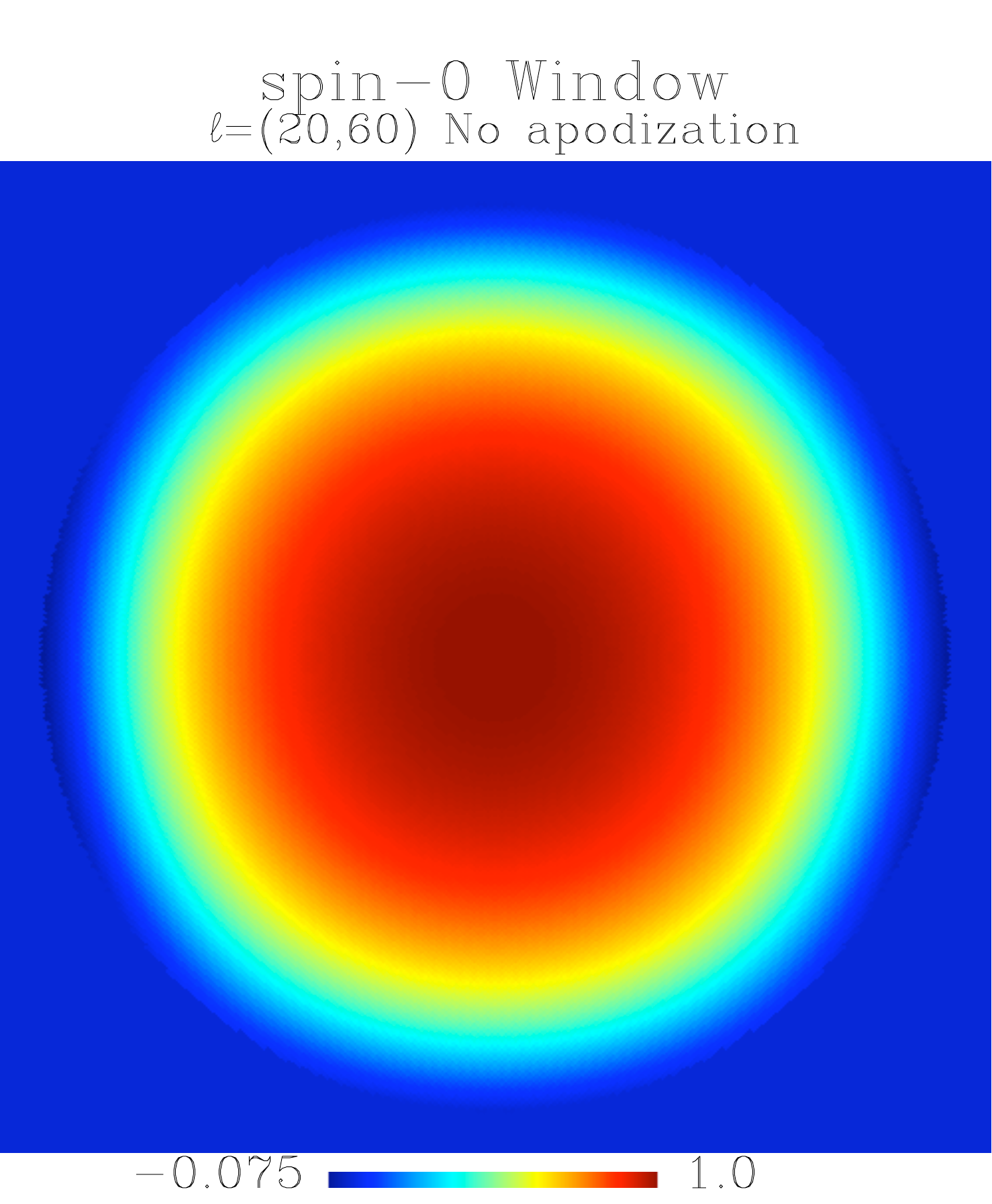} \includegraphics[scale=0.175]{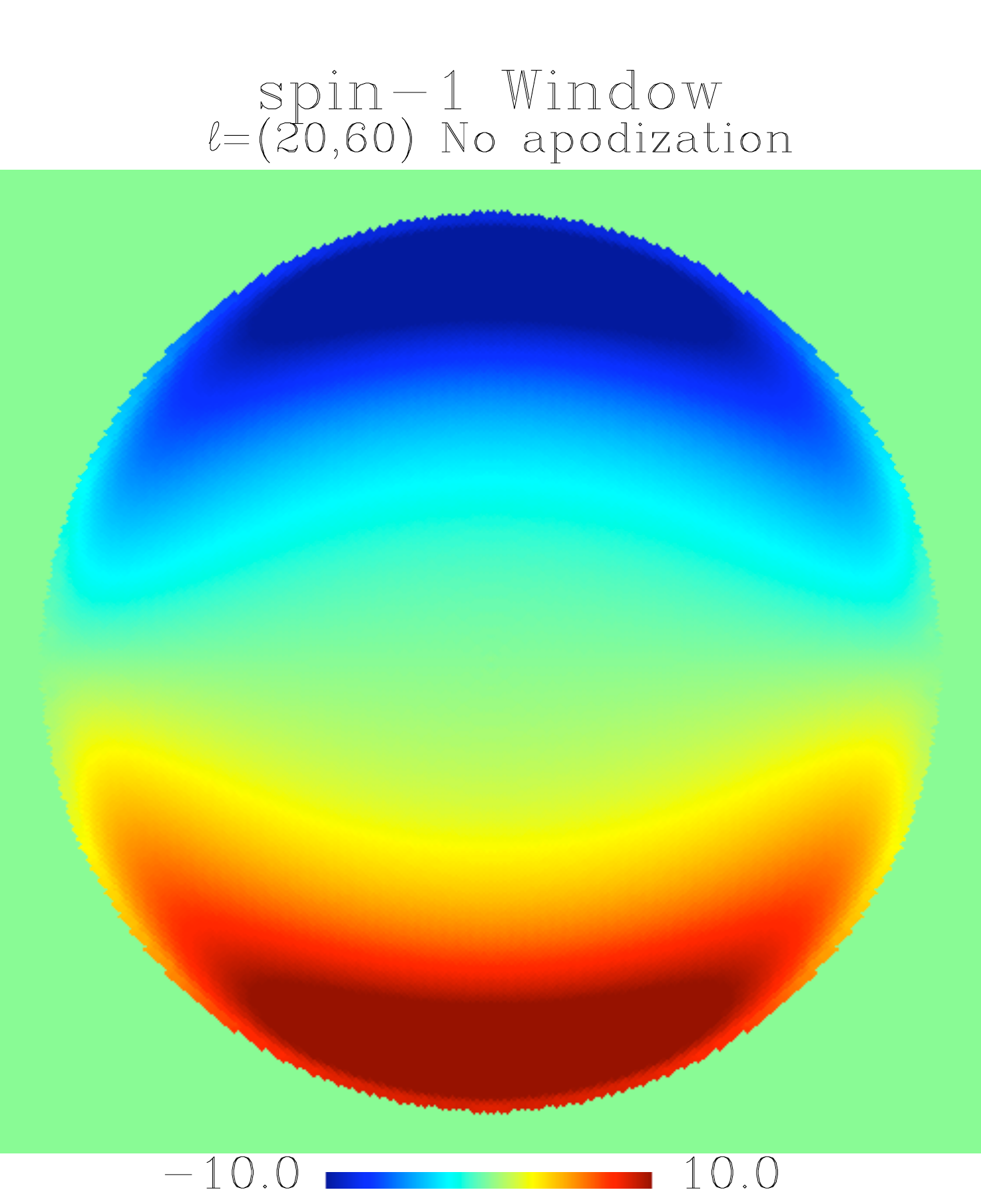} \includegraphics[scale=0.175]{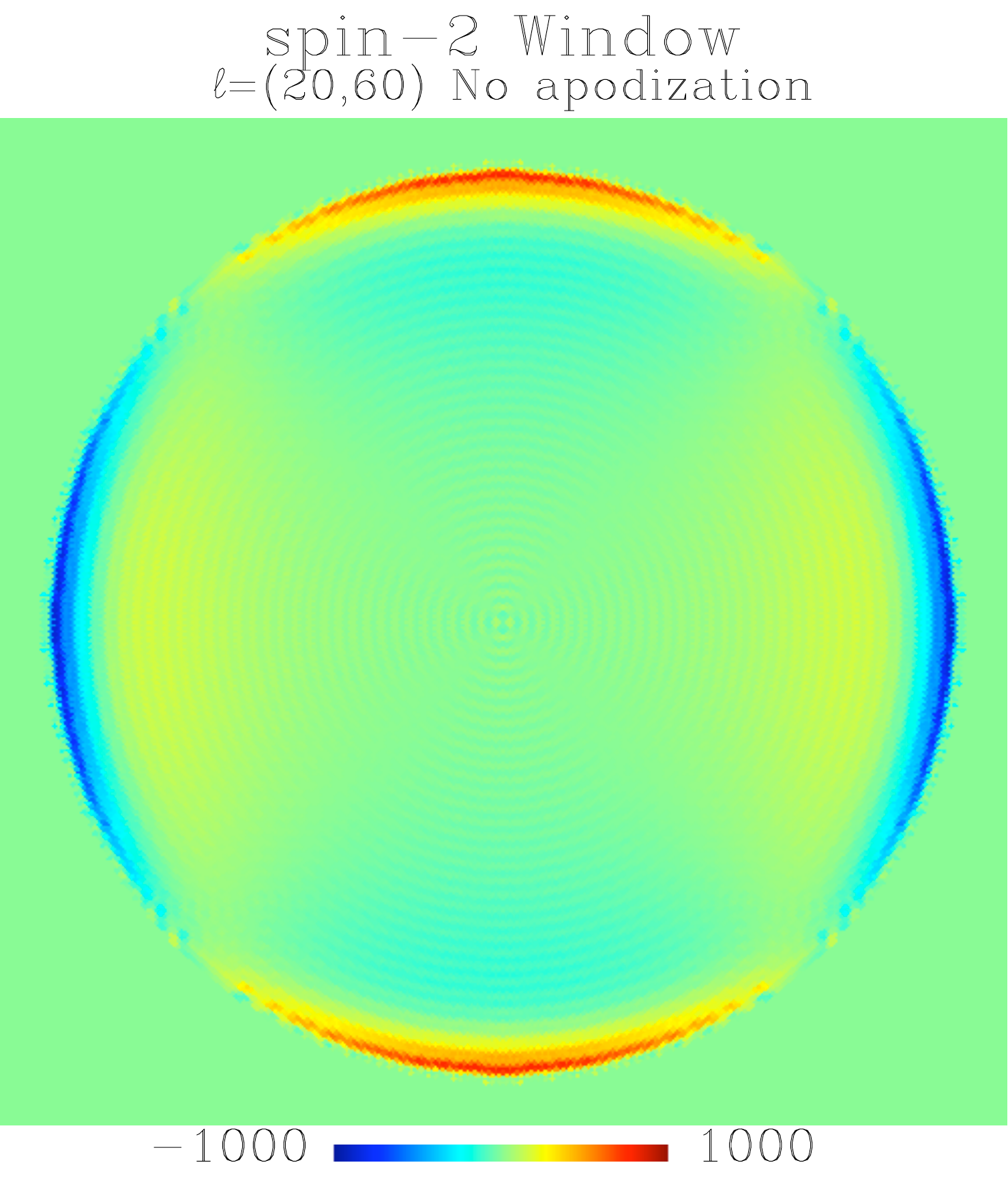} \\
\includegraphics[scale=0.175]{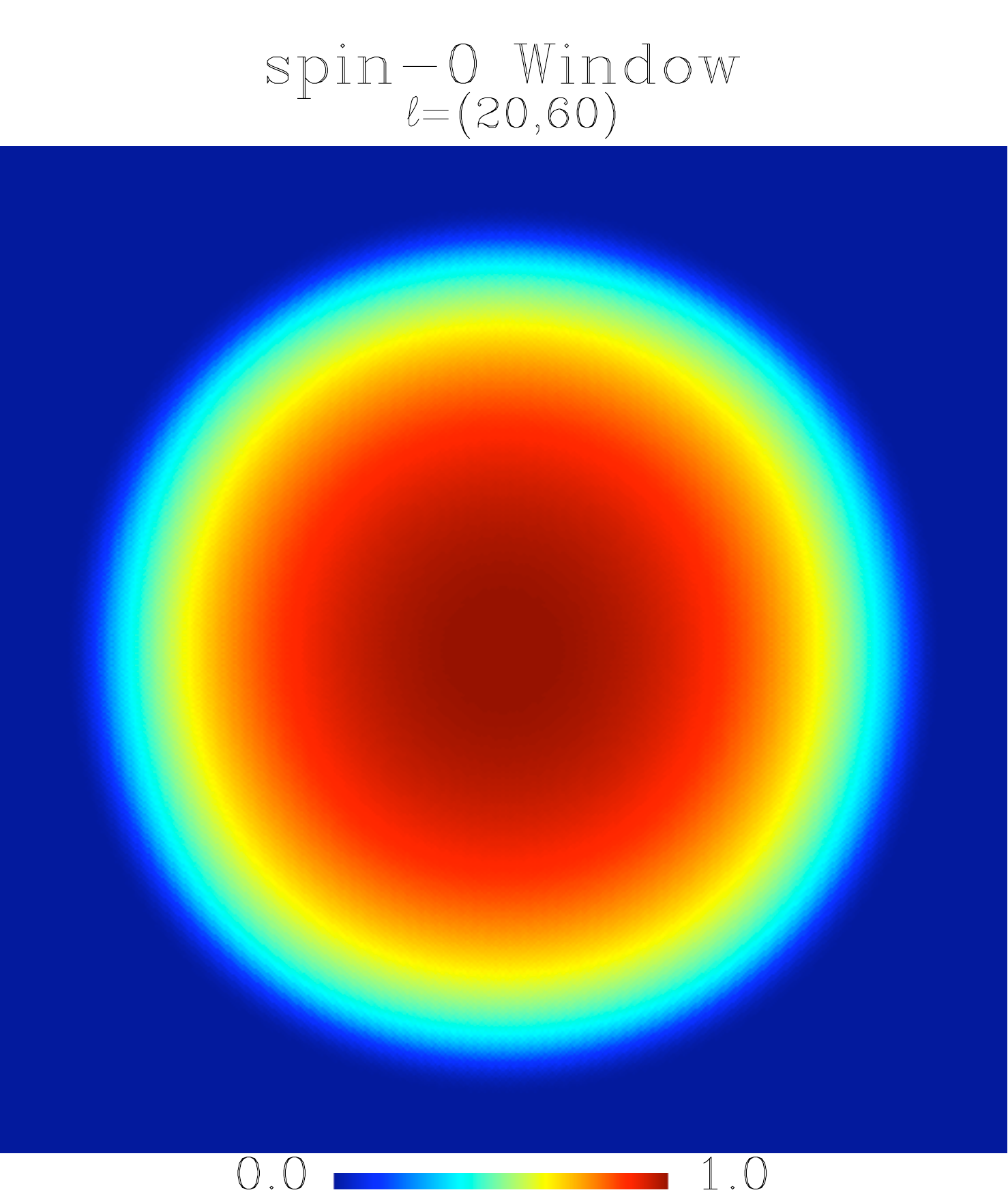} \includegraphics[scale=0.175]{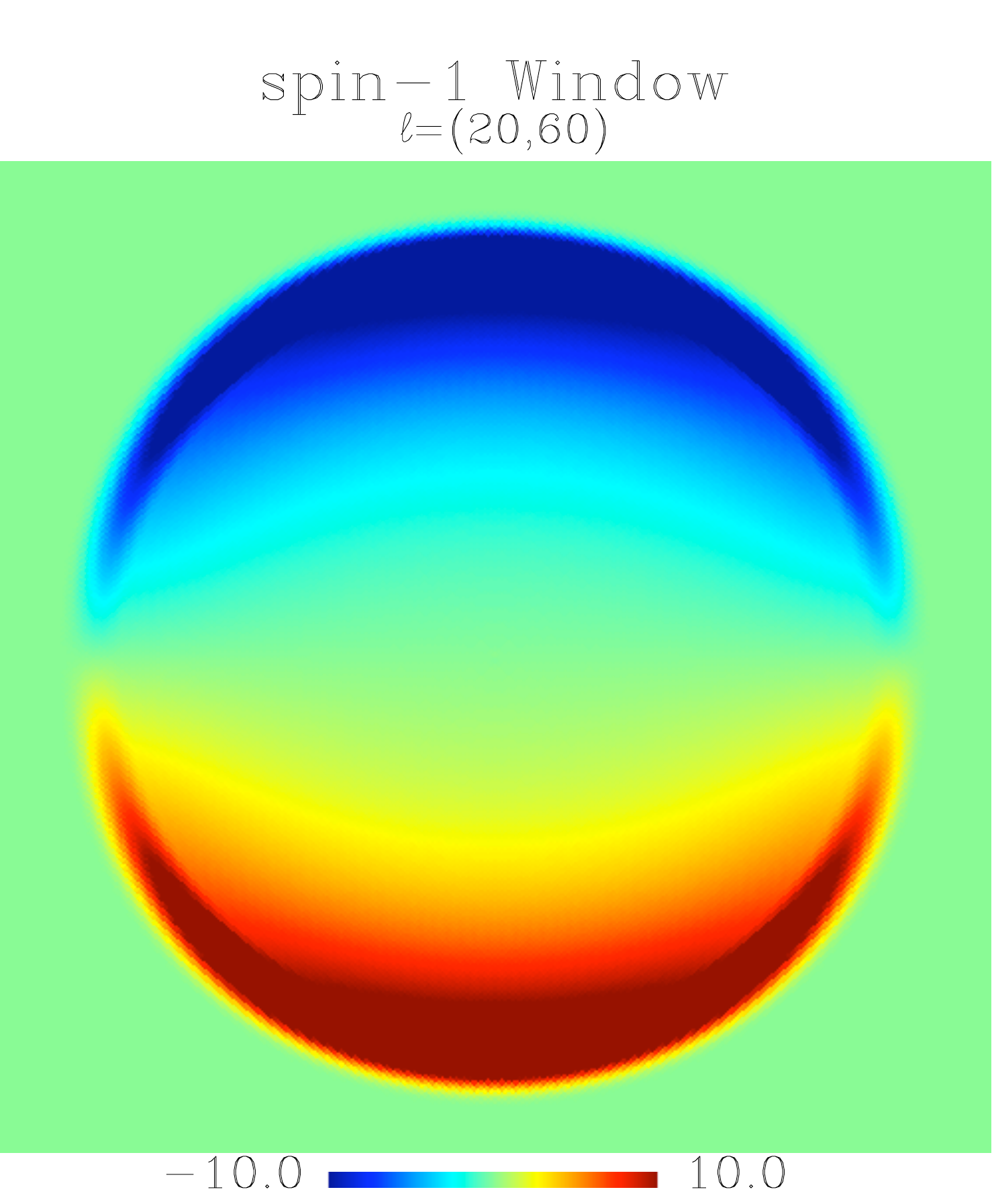} \includegraphics[scale=0.175]{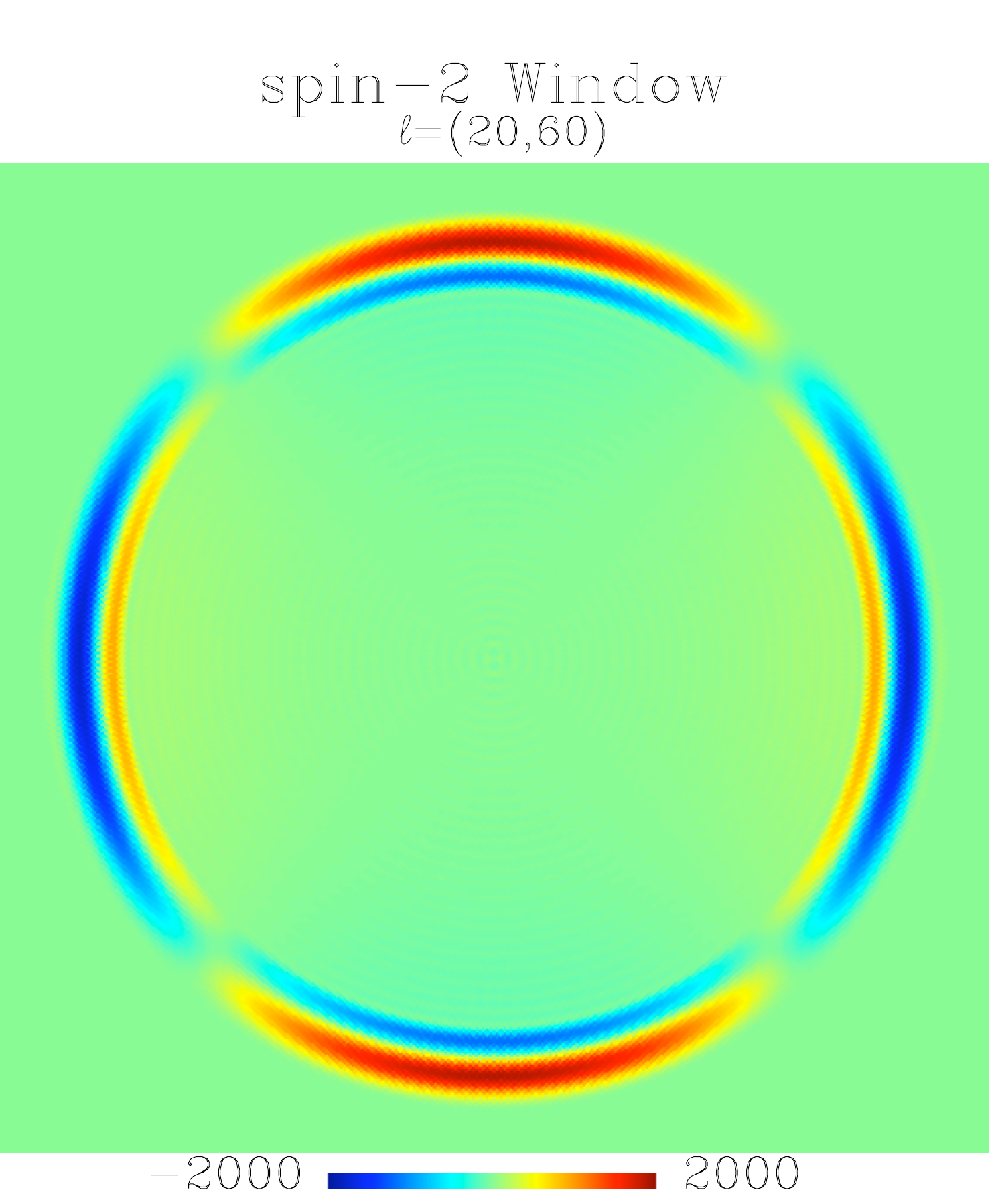} \\
\includegraphics[scale=0.175]{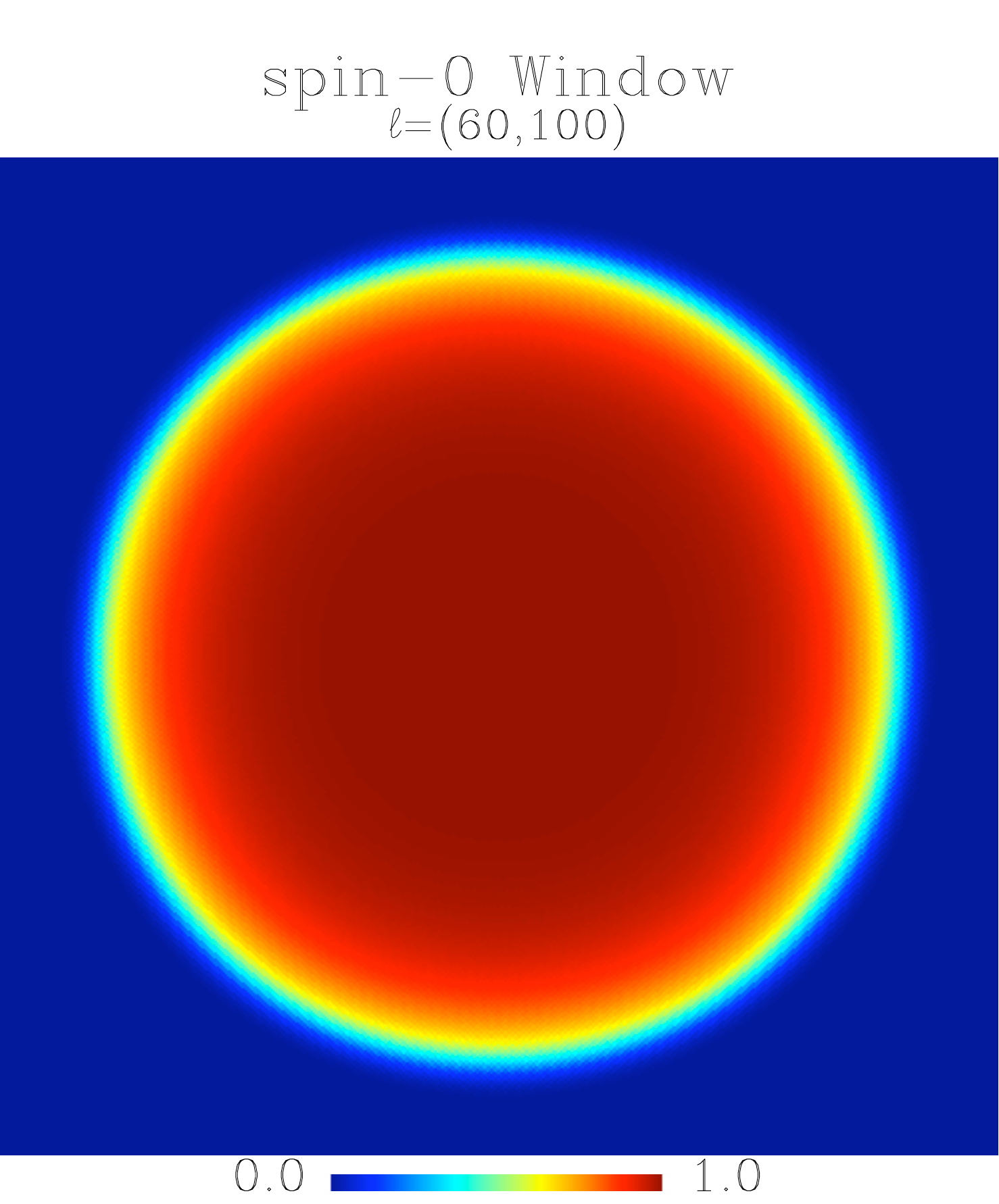} \includegraphics[scale=0.175]{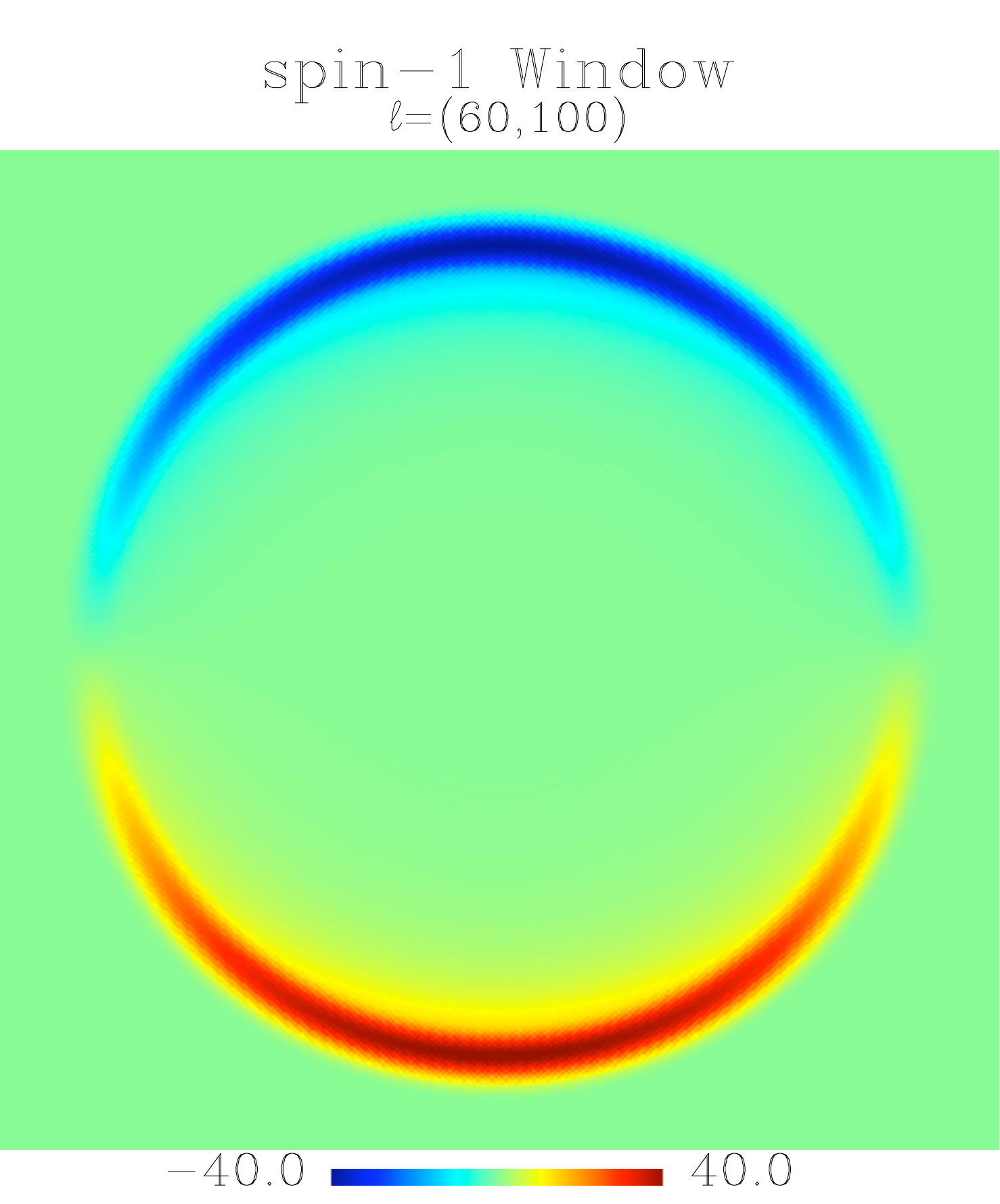} \includegraphics[scale=0.175]{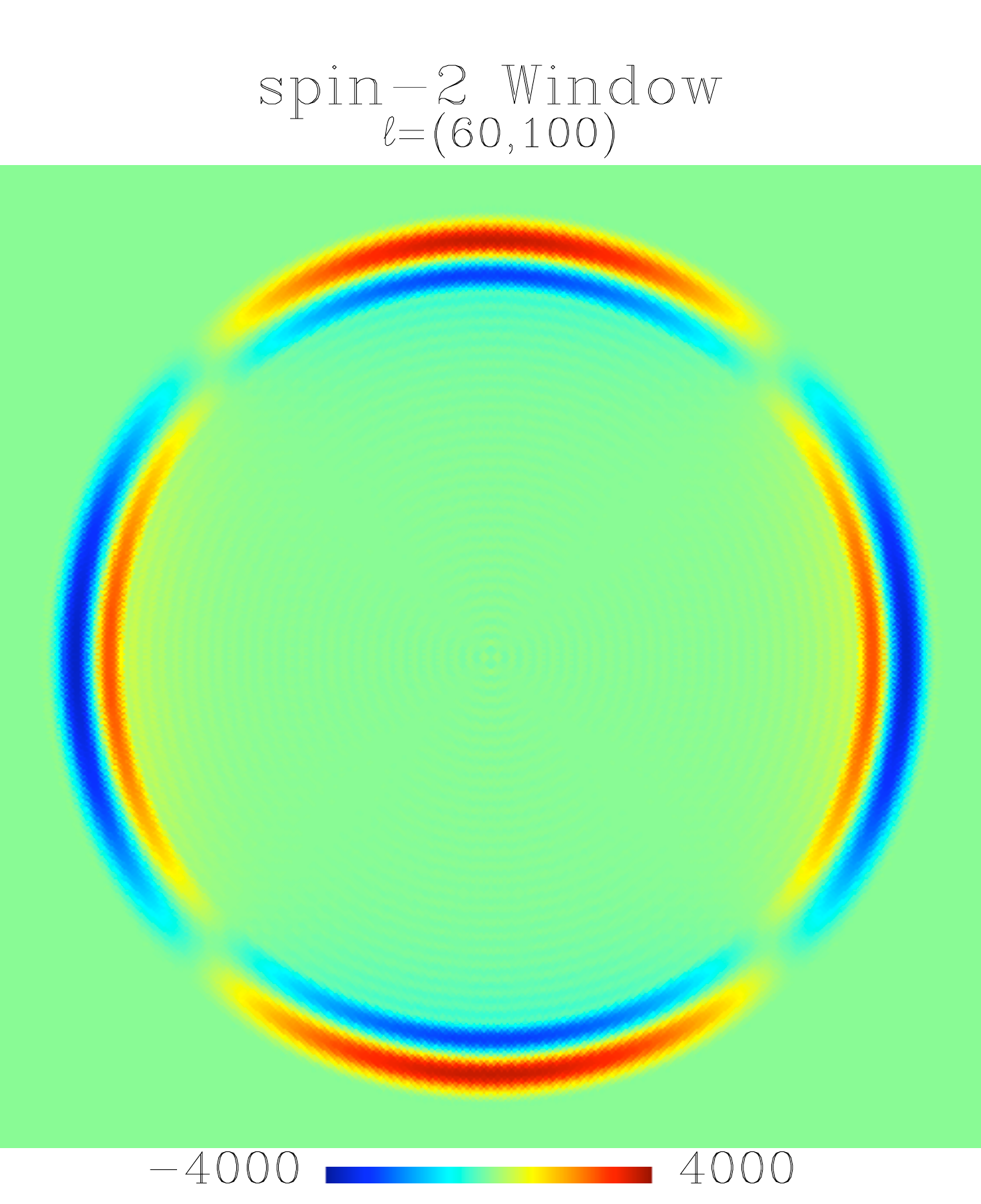} \\
\caption{Left to right: spin-0, spin-1, \& spin-2, spin-weighted optimized windows computed in the harmonic space 
for $\ell\in(20,60)$ without, {\it top row}, and with, {\it middle row}, the extra-apodization as discussed in the text. We 
show here only the real parts of the spin-1 and spin-2 windows.  The color stretch of the right panel is 2 times bigger from top to middle row. {\it Bottom:} The same as middle row but optimized 
for a bin $\ell\in(60,100)$. In the middle and right panels, the color 
stretch is set to be 4 (2) times bigger, respectively, as compared to the corresponding panels of the middle row. In all the shown cases the $B$-mode signal includes lensing and primordial contribution 
for $T/S=0.05$ and all the other parameters are set as in Fig.~\ref{pcg1}.}
\label{sht1}
\end{figure}

The results of the numerical implementation for a spherical cap with homogeneous noise are shown in Fig.~\ref{sht1}. These windows show the same behavior as the windows obtained from the PCG with respect to bins and signal level. In particular, we find that the apodization length decreases for higher bins.  
The shapes of spin-1 and spin-2 windows computed in the pixel- and harmonic domains are 
however significantly different. 
Because of the extra-apodization and numerical issues, the windows obtained in 
the harmonic space in a way proposed here are not any more an exact solution of the optimization problem 
fulfilling the boundary constraints. Nevertheless they provide an easy and quick-to-compute apodization, which
one may hope will result in spectral variances not substantially larger than the one produced by the computationally 
heavy pixel-domain optimized windows described below. We will test these expectations against simulations in 
Sect.~\ref{sect:applications}.

\begin{figure}
\includegraphics[scale=0.15]{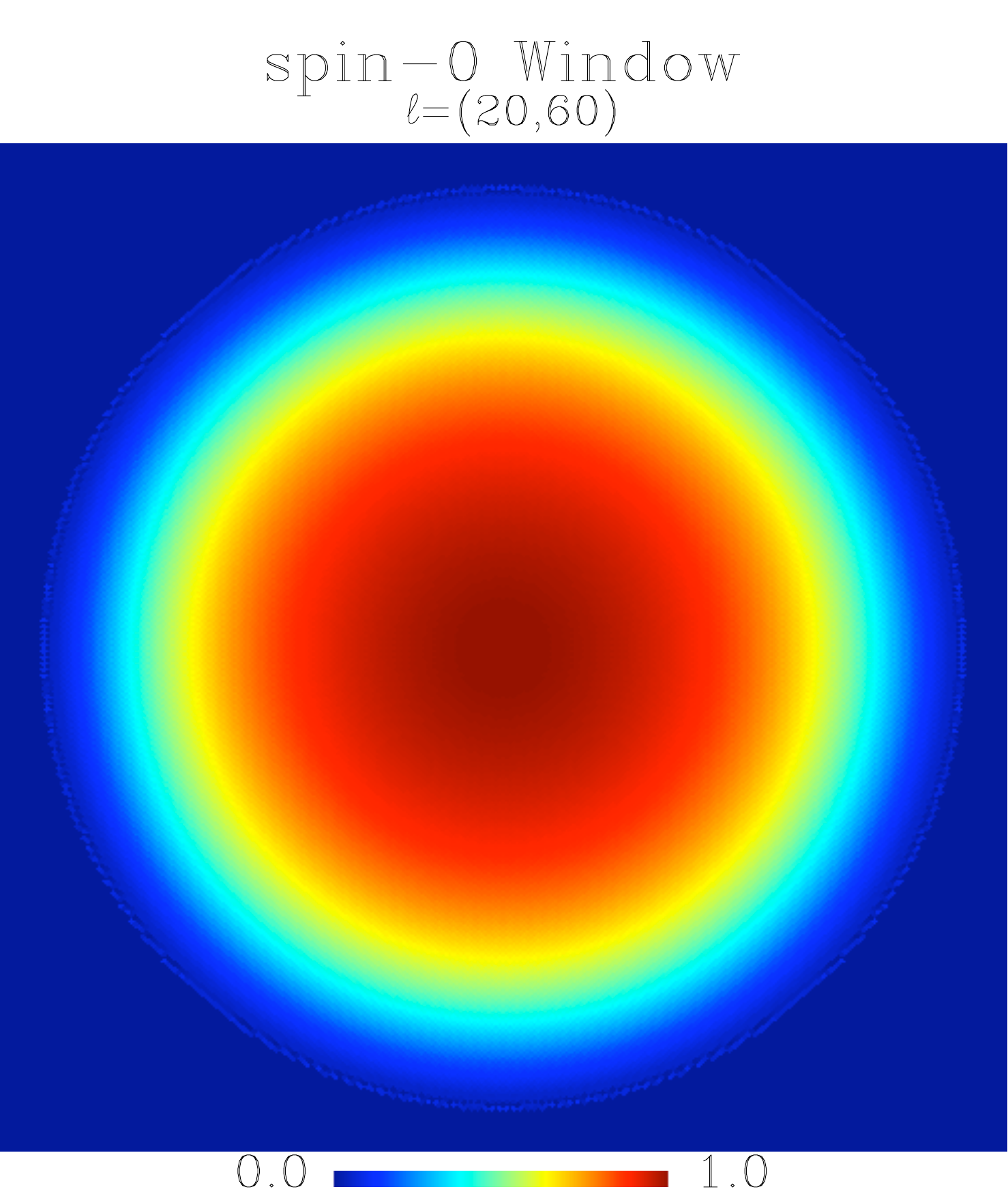}  \includegraphics[scale=0.15]{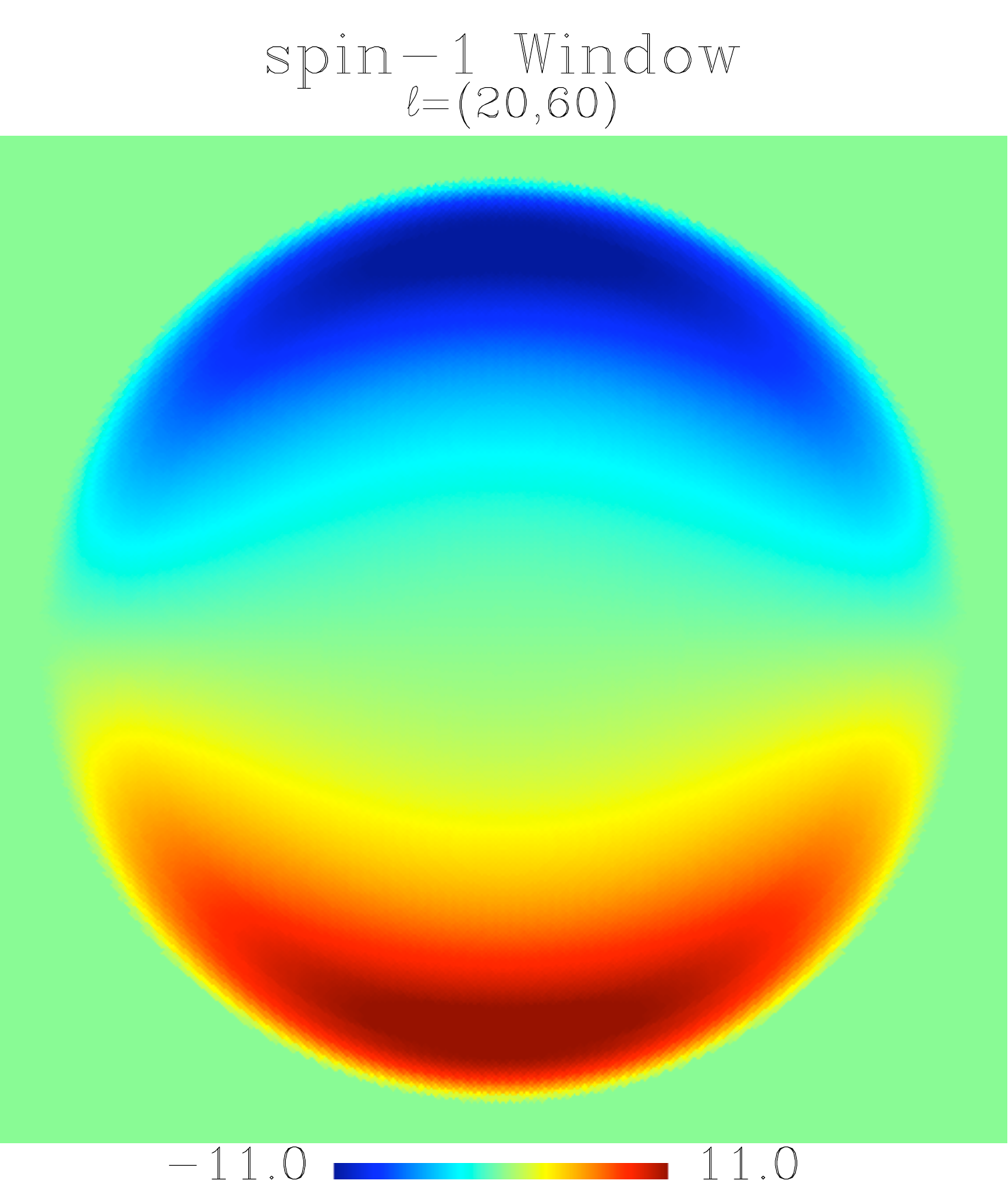} \includegraphics[scale=0.15]{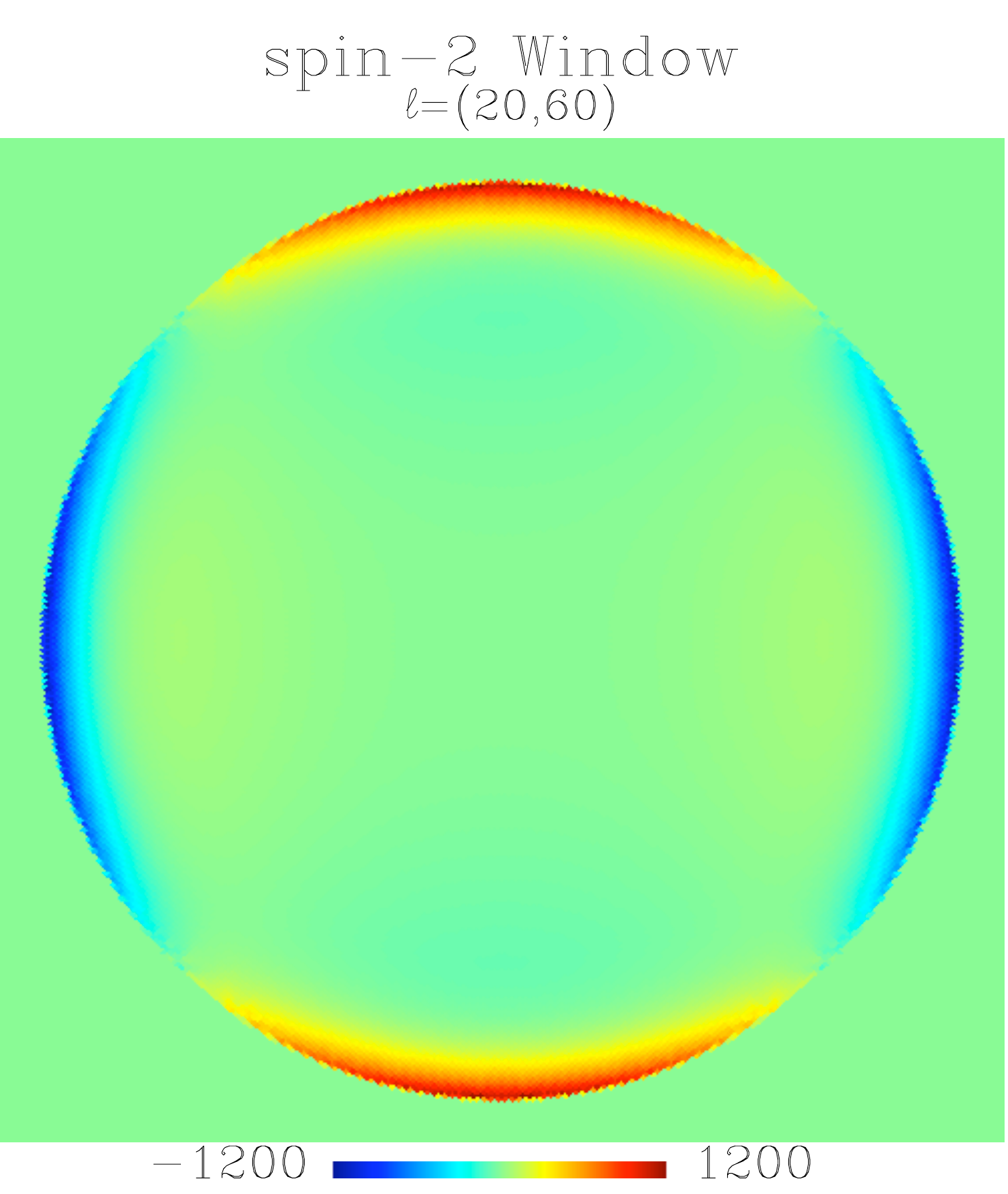}\\
\includegraphics[scale=0.15]{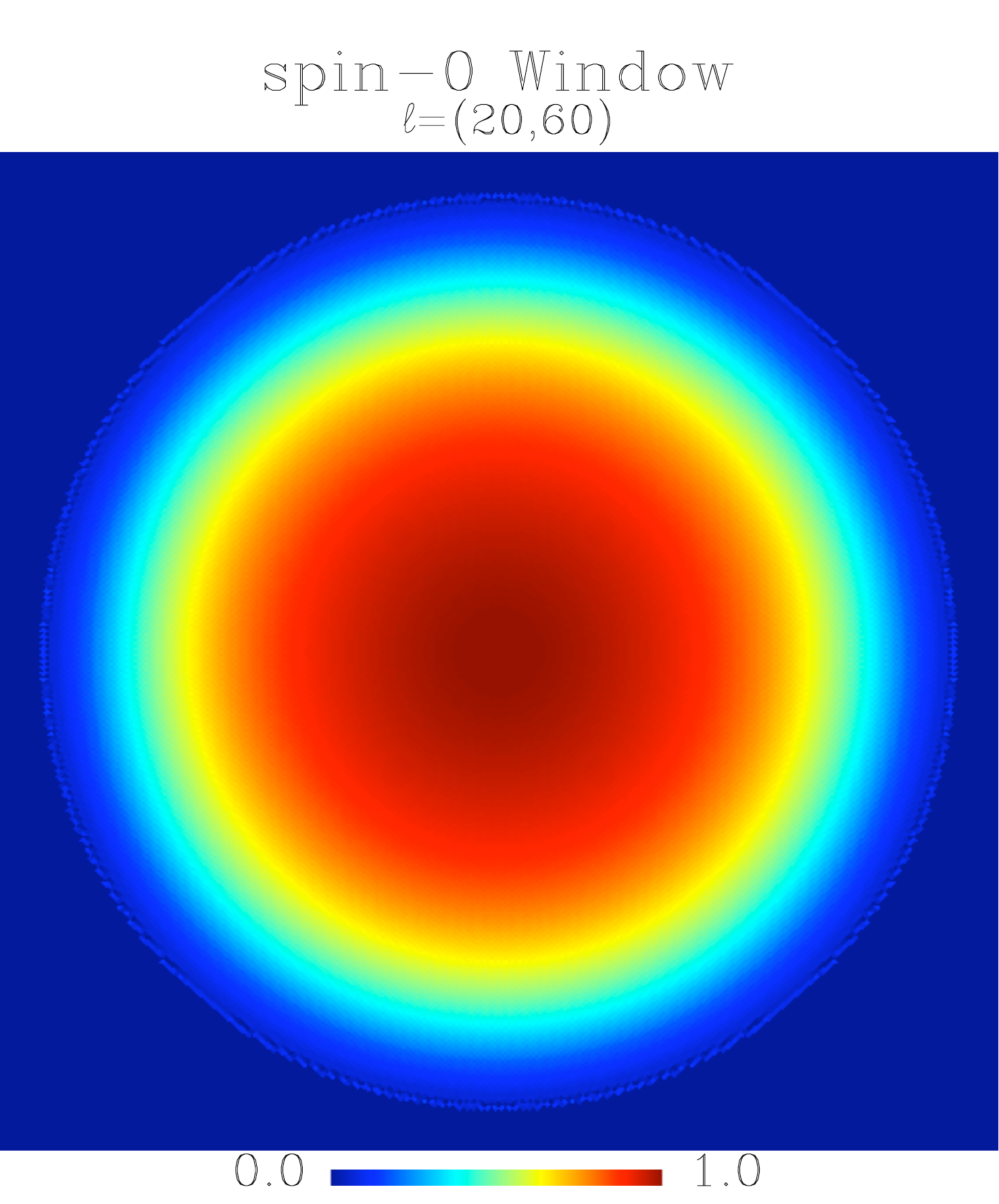}  \includegraphics[scale=0.15]{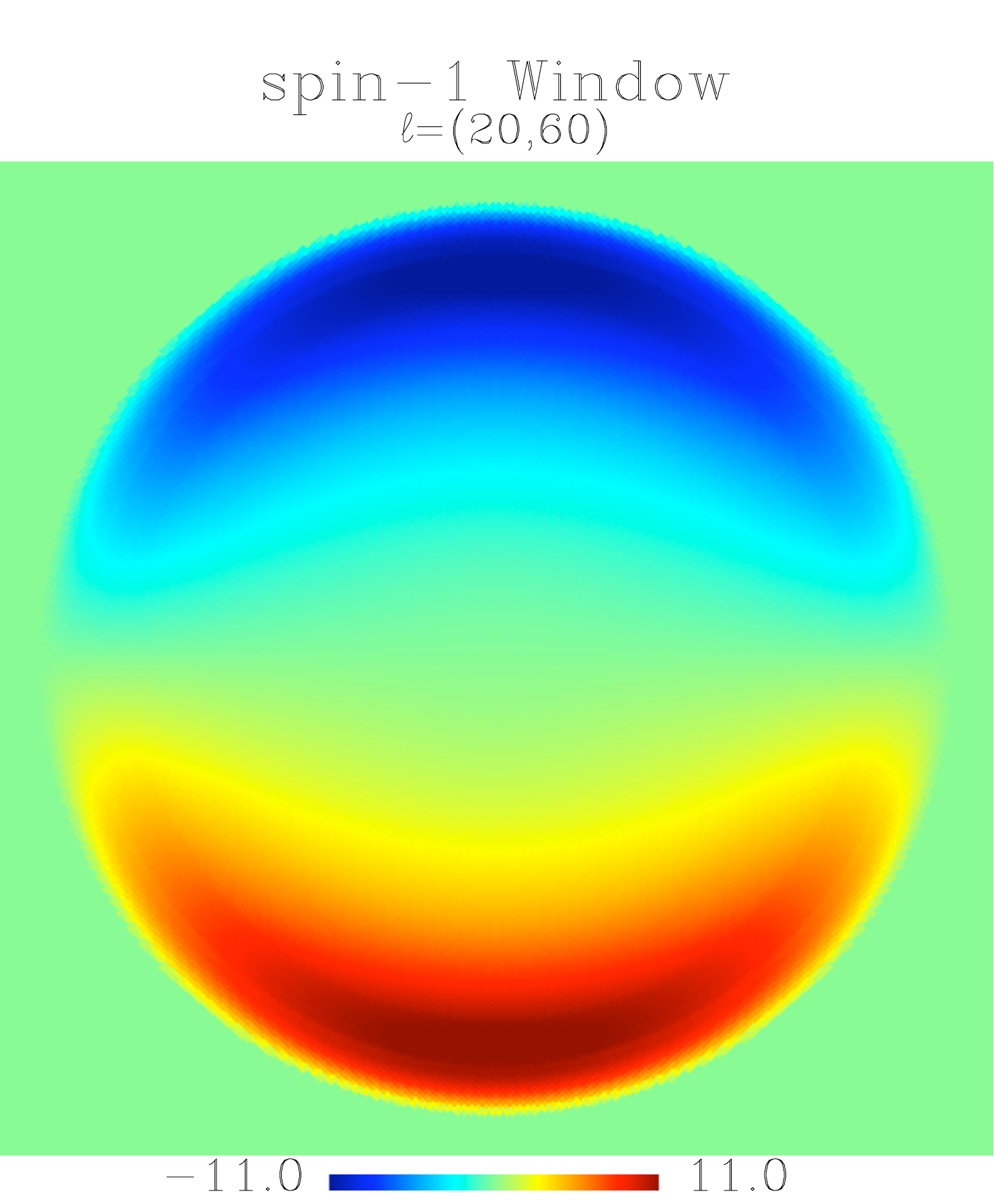} \includegraphics[scale=0.15]{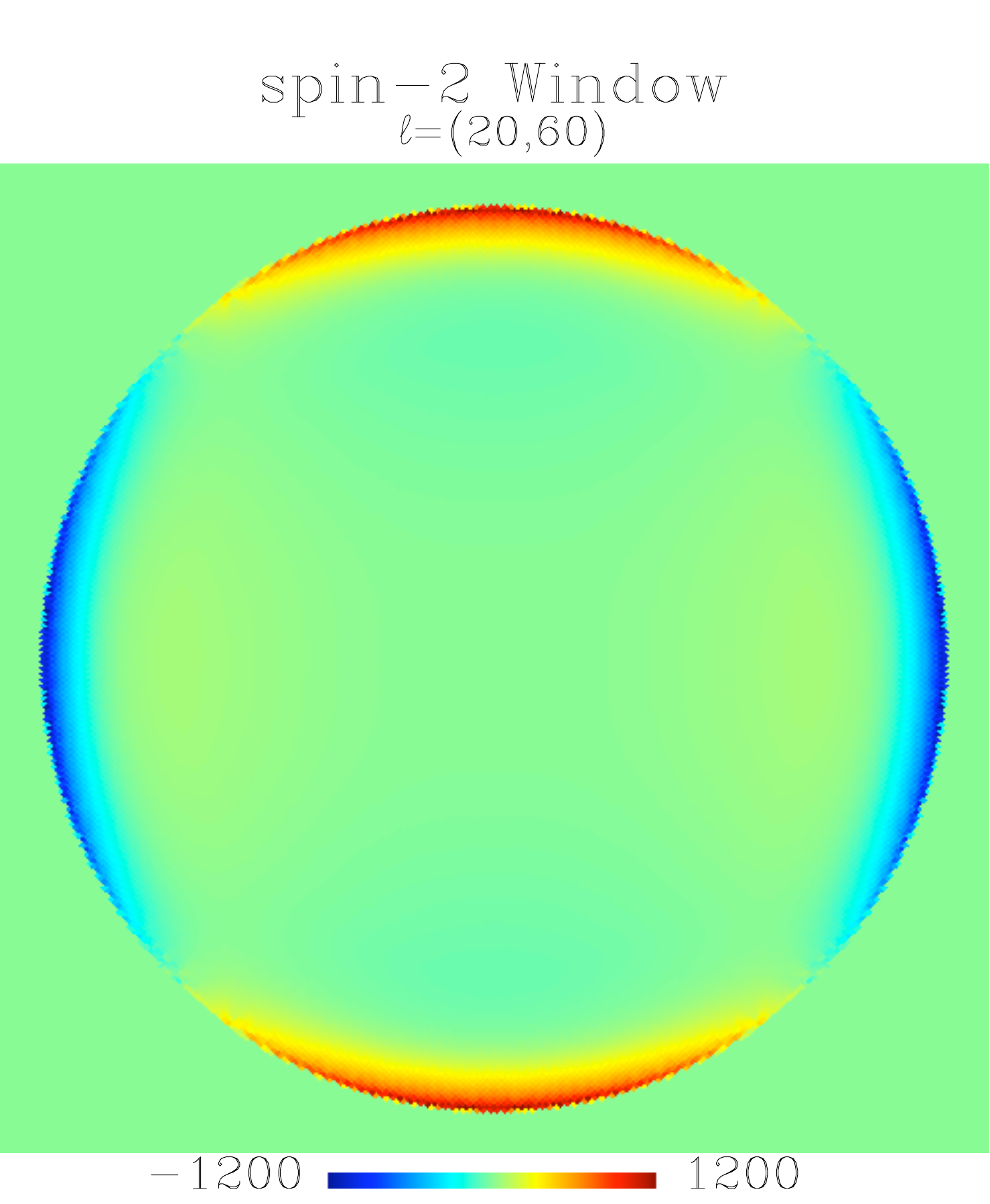}\\
\includegraphics[scale=0.15]{w0_r05_homo1.pdf}  \includegraphics[scale=0.15]{w1_r05_homo1.pdf} \includegraphics[scale=0.15]{w2_r05_homo1.pdf} \\
\includegraphics[scale=0.15]{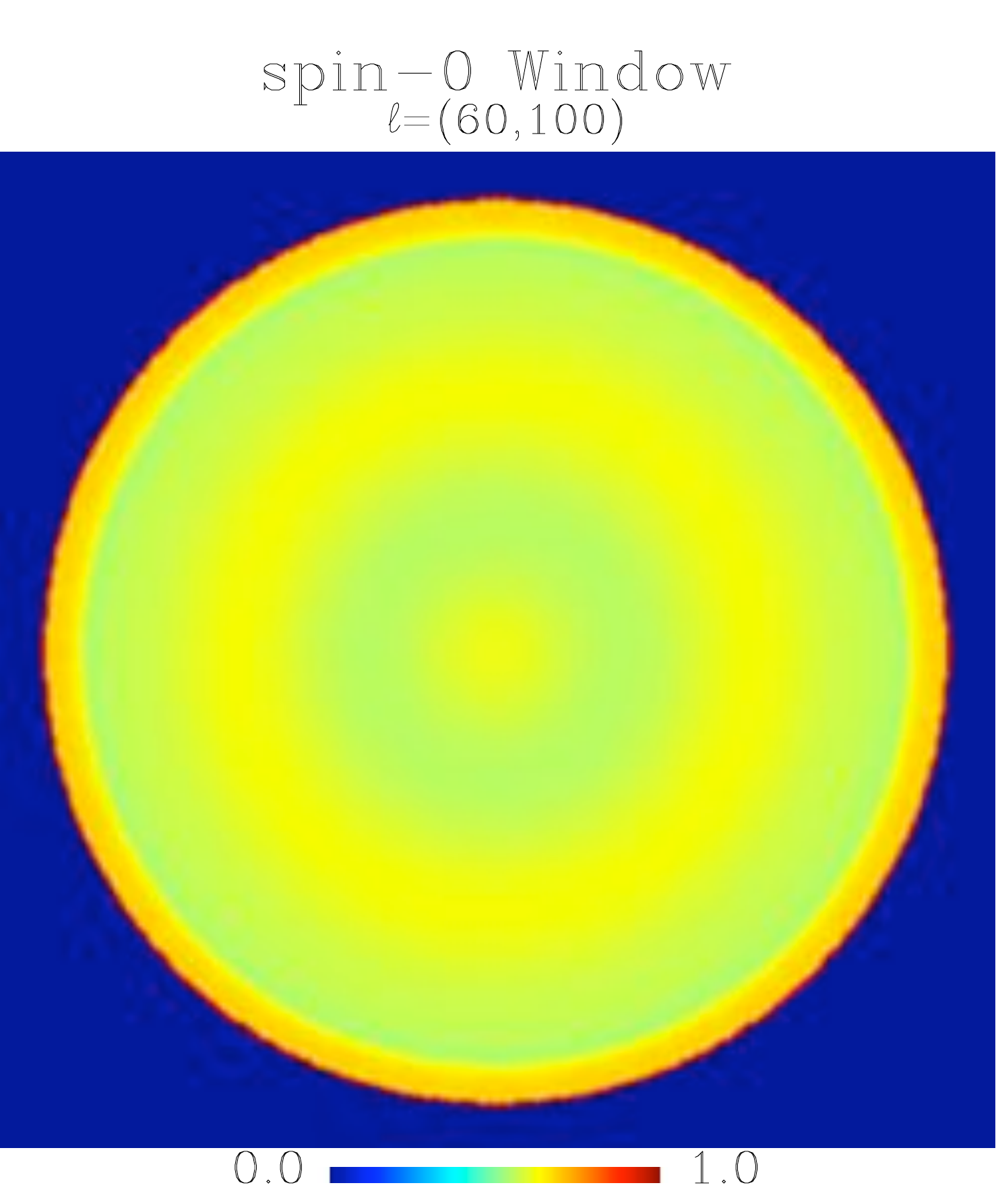}  \includegraphics[scale=0.15]{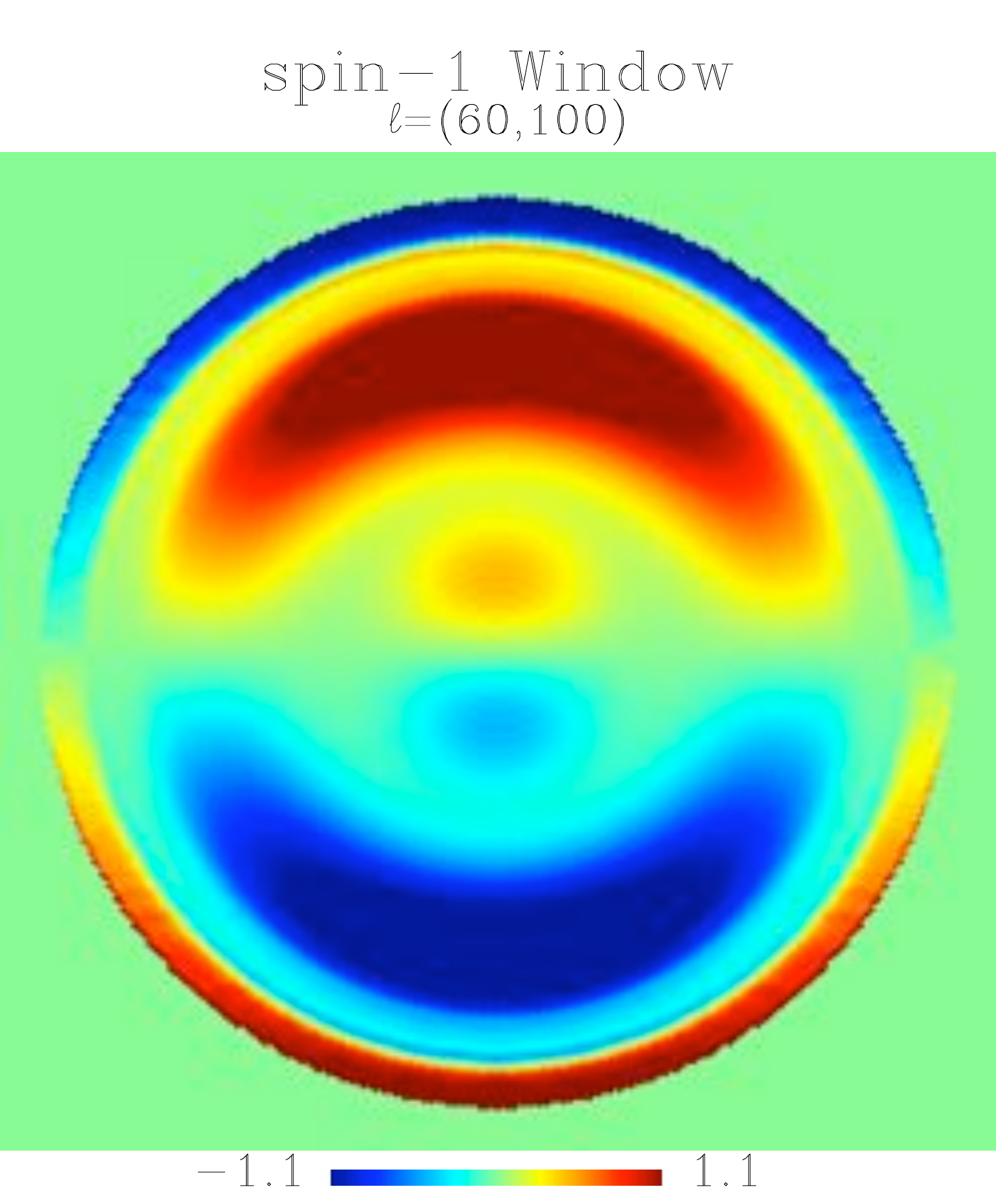} \includegraphics[scale=0.15]{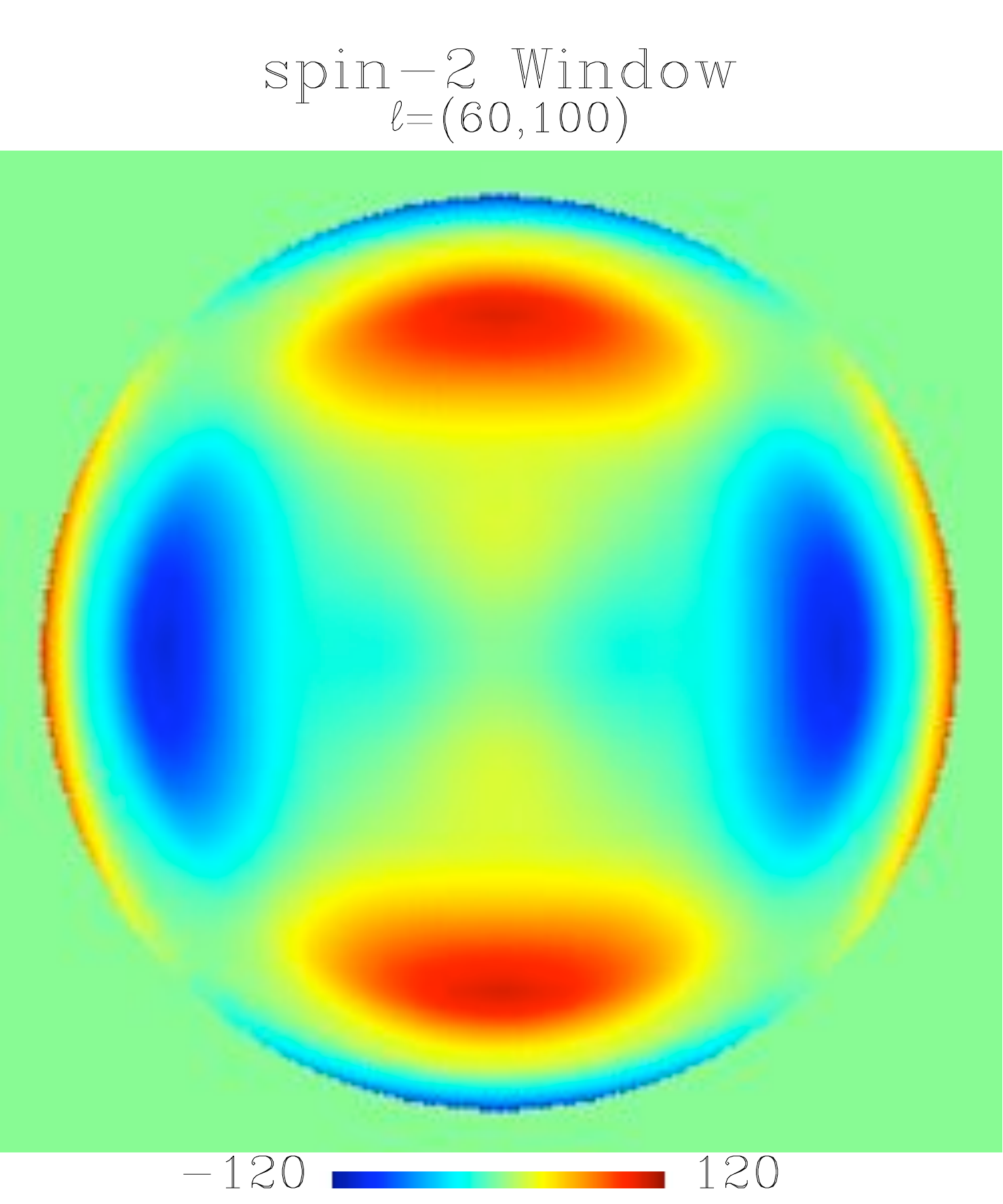}\\
\includegraphics[scale=0.15]{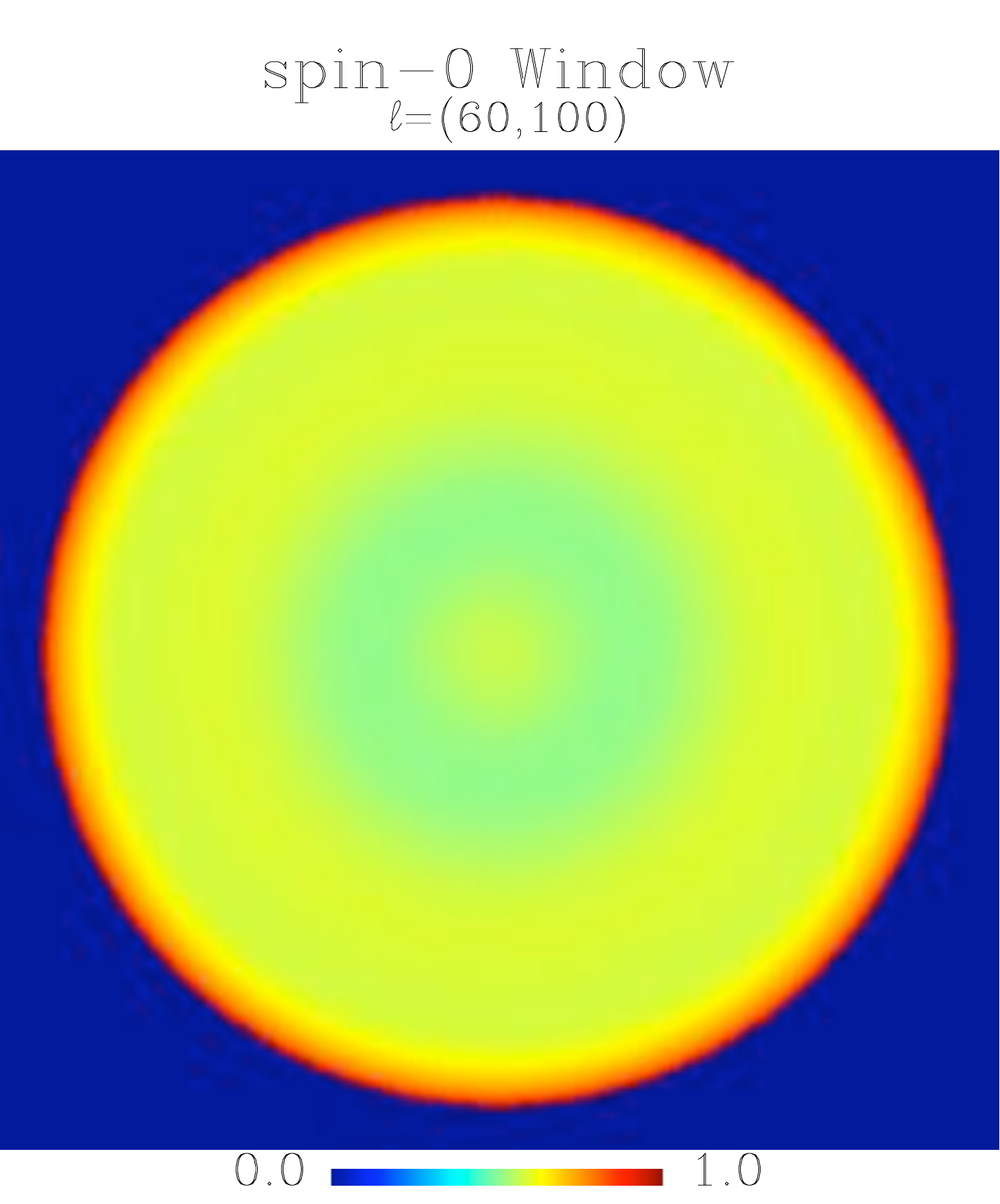}  \includegraphics[scale=0.15]{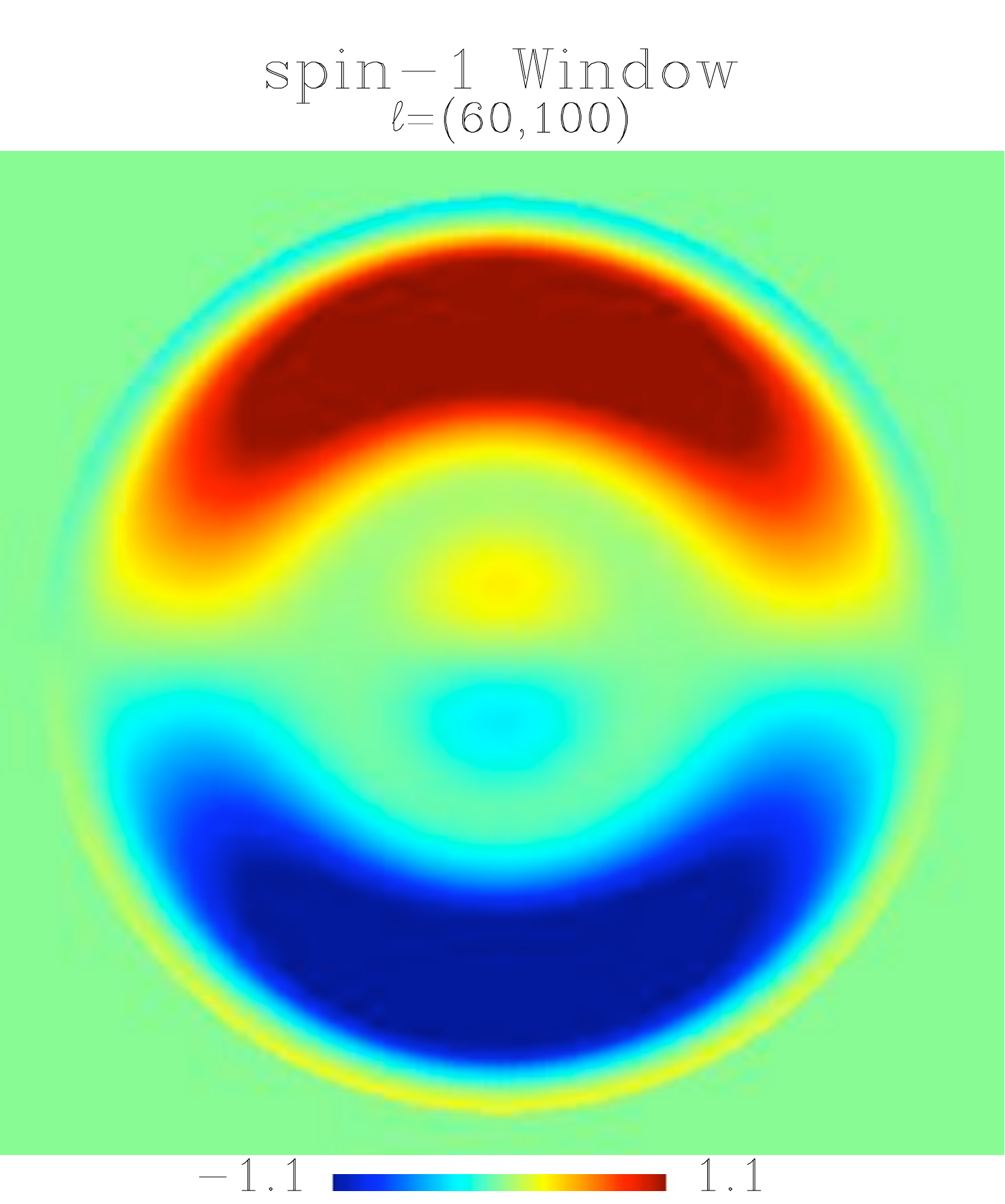} \includegraphics[scale=0.15]{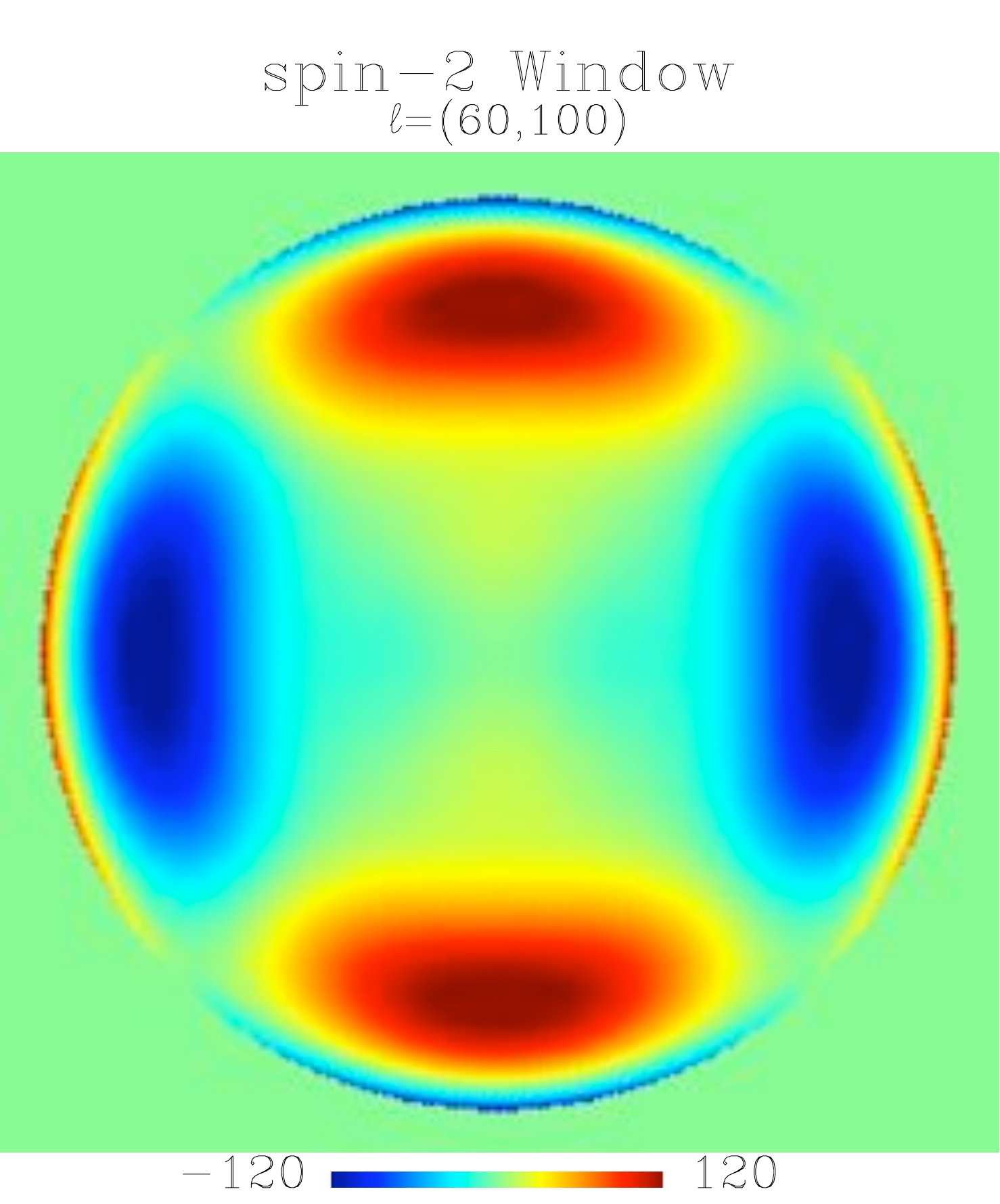}
\caption{Spin-0, spin-1, \& spin-2 optimized window functions, left to right, computed for the $\ell\in(20,60)$ bin for 
three different assumptions about the $B$-mode signal: no $B$-mode (first row), lensing-induced $B$-mode only 
(second row), lensing-induced plus primordial $B$-mode for $T/S=0.05$ (third row). In all these three cases, the 
$E$-mode signal is obtained using the cosmological parameters constrained by WMAP 5-year 
data~\cite{dunkley_etal_2008}. In the fourth row the $E$-mode signal amplitude is set to a hundredth of that value, 
i.e., on par with the $B$-mode level, and no $E$-mode signal is assumed in the fifth row. In these last two cases, 
the $B$-mode is the same as in the third row. The noise and patch properties are the same as in Figs.~\ref{pcg1} 
\&~\ref{sht1}. The shape of the windows  depends weakly on the assumed $B$-mode, but clearly more strongly on 
the $E$-mode.  For visualization purposes, in the fourth and fifth row the color stretch is 10 times smaller for the 
spin-1 and spin-2 windows than in the first three rows.}
\label{pcg-win-signal}
\end{figure}

\subsubsection{Sky signal prior and the optimized windows}
The computations of the properly optimized window functions requires some prior knowledge about the signal
expected in the data. In the context of the small-scale CMB observations at the low and intermediate $\ell$ 
the optimization is mostly between the variance of the $E$ polarization signal leaked to $B$  and the assumed 
(and presumed known) noise level. Hence this is 
the level of the $E$-mode polarization, which is of primary importance. As polarization experiments measure usually
both modes of polarization, therefore a high quality internal constraint on the $E$-mode power should be available for an experiment aiming at the $B$-mode detection. At the higher-$\ell$ end a reasonable guess about the $B$-mode power is becoming however increasingly more important. Nevertheless, the dominant $B$-mode contribution on subdegree 
scales is due to the gravitational lensing of the $E$-mode polarization and knowledge of the latter should again be 
sufficient. We elaborate more on these observations in the next Section.

Nonetheless as the prior assumptions have an impact on the final results they need to be chosen with care. 
In the case of the pixel-domain optimization we illustrate the dependence of the recovered windows on the assumed
sky signal in Fig.~\ref{pcg-win-signal} considering five different assumptions about the signal, while always keeping 
the noise level fixed at $\sigma=5.75~\mu$K-arcmin. In the first three cases we assume the $E$ mode power as in 
the best-fit WMAP 5-year model and change only the $B$ signal. The three cases correspond to (i) no $B$-mode, 
(ii) only lensing-induced $B$-mode and (iii) both lensing-induced and primordial $B$-mode 
for $T/S=0.05$. In the two remaining cases we fix the $B$-mode power as in the case (iii) and change that
of the $E$-mode to be: (iv) a hundredth of the standard value and (v) zero. In all the cases, shown in 
Fig.~\ref{pcg-win-signal} from top to bottom, the windows have been optimized for $\ell\in(20,60)$ where a high 
amount of leakage has to be corrected for. For assumptions (i) and (ii),  the $B$ signal is lower than the noise for all angular scales
whereas it
exceeds the noise level in the last three cases. The shape of the optimized windows slowly 
varies with the $B$-mode level meaning that the $E/B$ leakage from the dominant $E$-mode is the main source 
of extra variance and the optimization procedure focuses on resolving this particular source.  In the last two cases, 
the $E$-mode is first set to be comparable to (case iv) and then lower (case v) than the $B$-mode power, what
causes clearly discernible changes in the shape of the computed apodizations. In the next Section we discuss 
consequences of the observed changes in the window functions on the recovered power spectra variance.

\begin{figure}
\includegraphics[scale=0.5]{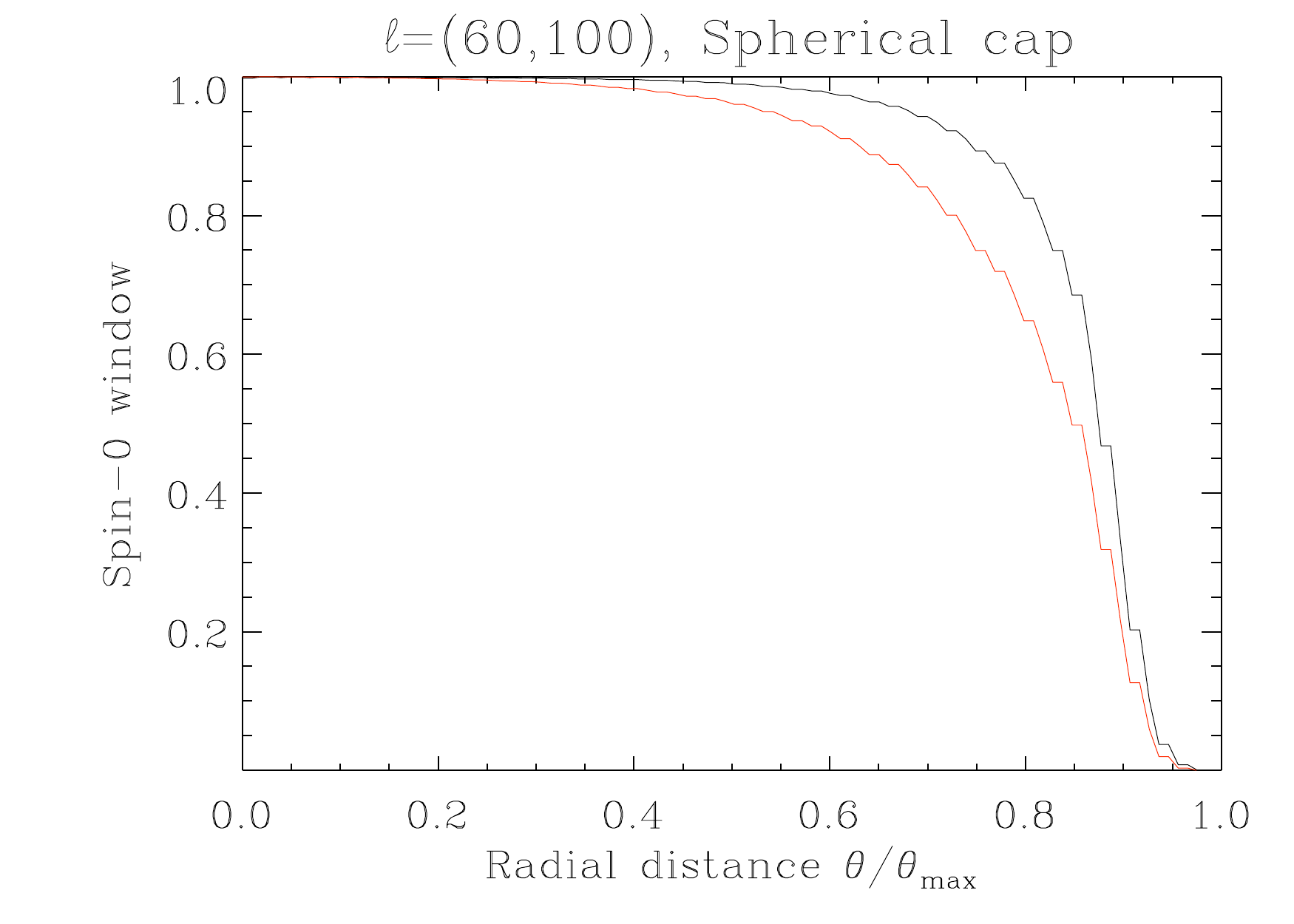}
\caption{ Radial profile of the harmonic domain optimized spin-0 window function for the noise 
dominated case (red curve) and the signal-dominated 
case (black curve). For the signal-dominated case, the primodial $B$-mode has been set to $T/S=0.5$ to 
emphasize the discrepancy between the two cases, all the other parameters are set as in Fig.~\ref{pcg-win-signal}.}
\label{sht2}
\end{figure}

In the case of the harmonic window the optimized weight functions depend on the noise and $B$-mode signal prior only, 
as emphasized by Eq.~\eref{opt-sht}. The $B$-mode power spectrum appears explicitly in ${C}^\chi_{\ell,\ell''}$ as
well as in $\mu$ and $\lambda$. Unlike in the pixel-domain case, the $E$-mode signal is not involved here as 
by design the harmonic window perfectly fulfills the Dirichlet~+~Neumann boundary conditions, ensuring that 
the $E$-mode power leaked to $B$ vanishes up to pixel effects. The dependence of such a window on the noise 
and $B$-mode prior is depicted on Fig. \ref{sht2}. The figure displays the radial profile of the spin-0 window in the 
noise-dominated (red curve) or the signal-dominated case (black curve). In the noise-dominated case, the power 
spectrum in $C^\chi_{\ell,\ell"}$ behaves as ${\ell'}^4$ and our numerical result coincides with the analytic expression given in Eq.~(33) of \cite{smith_zaldarriaga_2007} applied to a spherical cap as the survey region, small enough for the flat sky approximation used in \cite{smith_zaldarriaga_2007} to be valid. The apodization length of the window 
in the signal-dominated case is smaller due to the fact that in this case $C^B_\ell\propto\ell^{-2}$ and thus the power 
spectrum in $C^\chi_{\ell,\ell"}$ behaves roughly like ${\ell'}^2$. This conclusion corroborates
the analytic investigation of \cite{smith_zaldarriaga_2007}.

\subsection{Analytic apodizations}

Analytic window functions are in general not well suited for providing an optimal performance in the presence of 
the instrumental noise, given their limited ability to adapt to the specific properties of the data. However,
they are quick to calculate, do not require any prior knowledge about the sky signals and can 
fulfill the boundary conditions with high precision. They may therefore at least
be useful on exploratory, initial stages of any CMB data analyzes. Moreover, whenever prior information about 
the sky signals is available, the parameters of the analytic window functions can be tuned via Monte Carlo (MC) simulations 
providing a competitive performance to the optimized windows usually at much lower costs. The analytic windows 
are therefore of considerable practical interest and in this Section we will discuss two specific proposals of the 
analytic windows, elaborating and complementing the discussion presented in~\cite{smith_2006}.

First of the considered choices is a straightforward generalization of the analytic window function
proposed in~\cite{smith_2006} extended to be applicable to an arbitrary sky coverage. Calling $\delta_i$ the distance between 
the $i$-th observed pixel and the boundary, i.e., the smallest angular distance form the considered pixel to the contour of 
the mask, we can write the relevant expression as,
\begin{equation}
W_i=\left\{\begin{array}{ll}
\frac{1}{2}-\frac{1}{2}\cos\left(-\pi\frac{\delta_i}{\delta_c}\right), & \ \ \ \ \ \delta_i<\delta_c,  \\
1, &  \ \ \ \ \ \delta_i>\delta_c,
\end{array}\right.
\label{eqn:smithWinDef}
\end{equation}
and is zero outside the mask. Here, $\delta_c$ is a width of a boundary layer  of the window smoothly interpolating 
between the window core and the zero on the outside. In the following, we refer to it as an apodization length and treat 
it as an adjustable parameter. Hereafter we will refer to the window defined in Eq.~\eref{eqn:smithWinDef} as the 
$C^1$-window. One can easily check that if $\delta_i=0$ (i.e., the pixel is on the contour of the mask), the window 
function is indeed vanishing, as well as that spin-1 window which can be derived from it. Therefore, the window defined 
above indeed fulfills the conditions as required by the pure pseudo power spectrum formalism. We note however that 
the related spin-2 window is not apodized and potentially therefore could be a source of numerical problems in particular,
whenever SHT are involved.

We therefore propose a second analytic formula for the spin-0 window from which apodized spin-1 and spin-2 windows 
can be derived, reducing in principle the ringing due to a sharp spin-2 window. With the same notation as above and for
every pixel in the mask the window is defined as,
\begin{equation}
W_i=\left\{\begin{array}{ll}
-\frac{1}{2\pi}\sin{\left(2\pi\frac{\delta_i}{\delta_c}\right)}-\frac{\delta_i}{\delta_c}, &\ \ \ \ \ \delta_i<\delta_c,  \\
1, &  \ \ \ \ \ \delta_i>\delta_c,
\end{array}\right.
\label{eqn:grainWinDef}
\end{equation}
and vanishes outside the mask. We will refer to this choice as the $C^2$-window.

\begin{figure}
\includegraphics[scale=0.175]{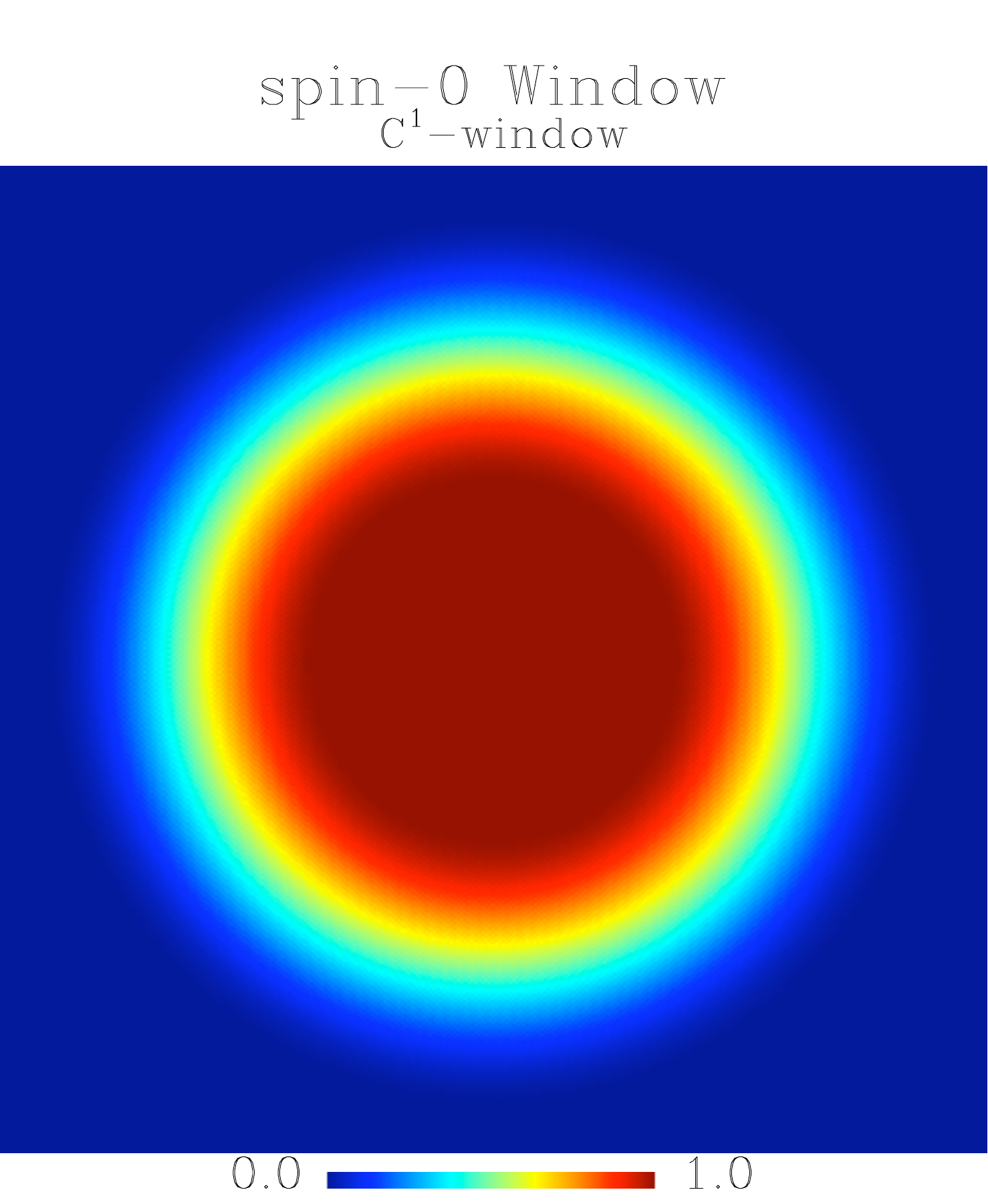}
\includegraphics[scale=0.175]{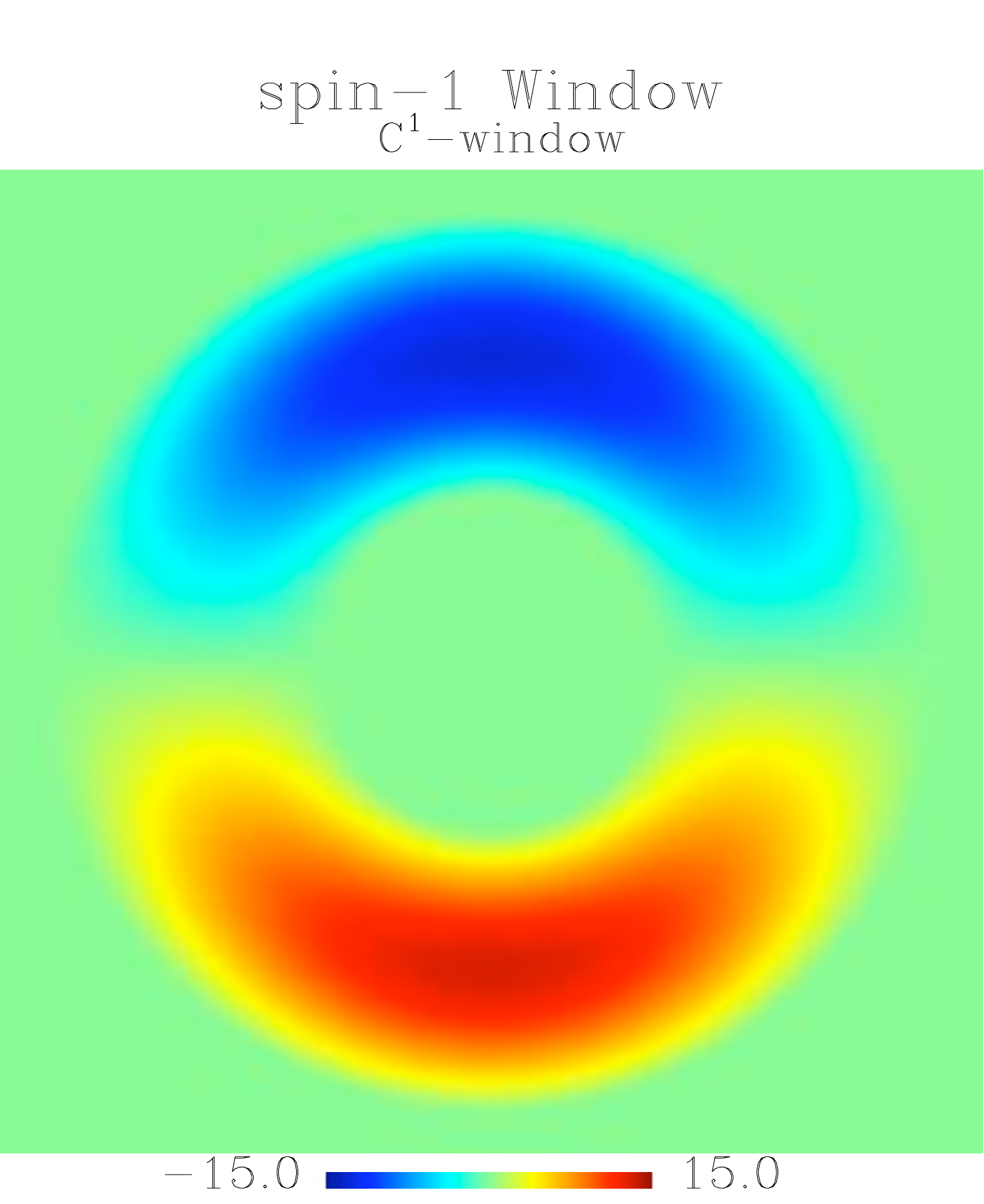}
\includegraphics[scale=0.175]{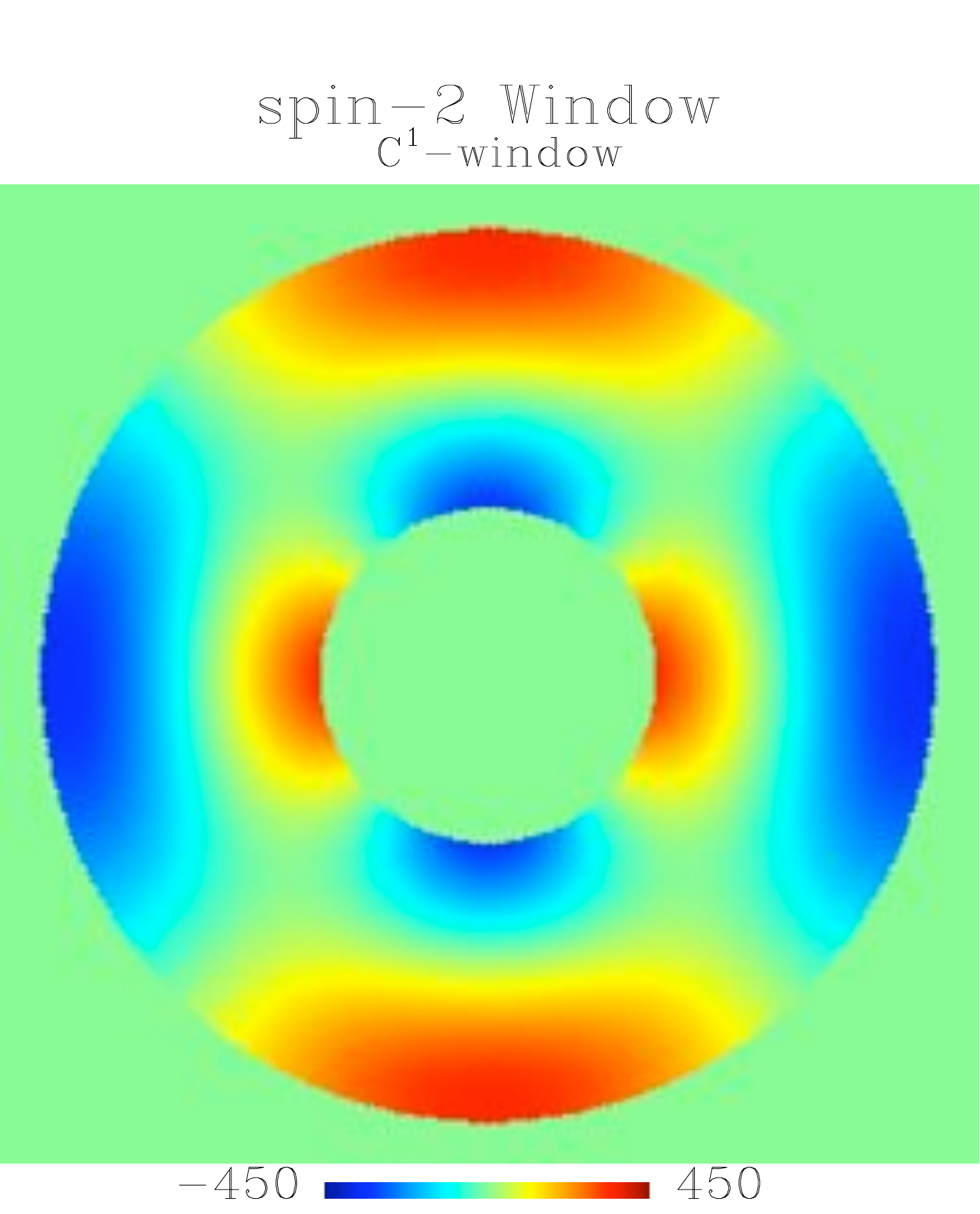} \\
\includegraphics[scale=0.175]{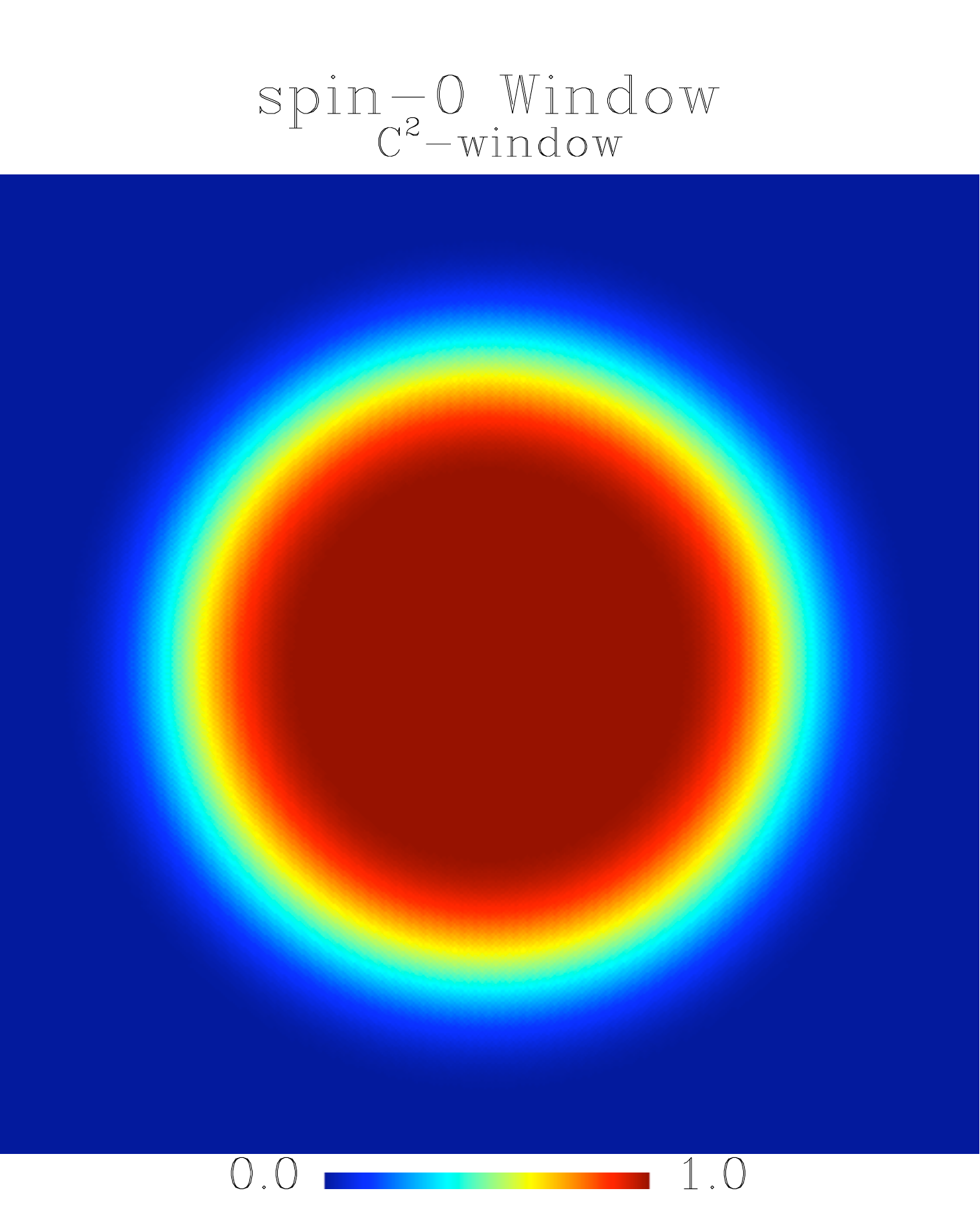}
\includegraphics[scale=0.175]{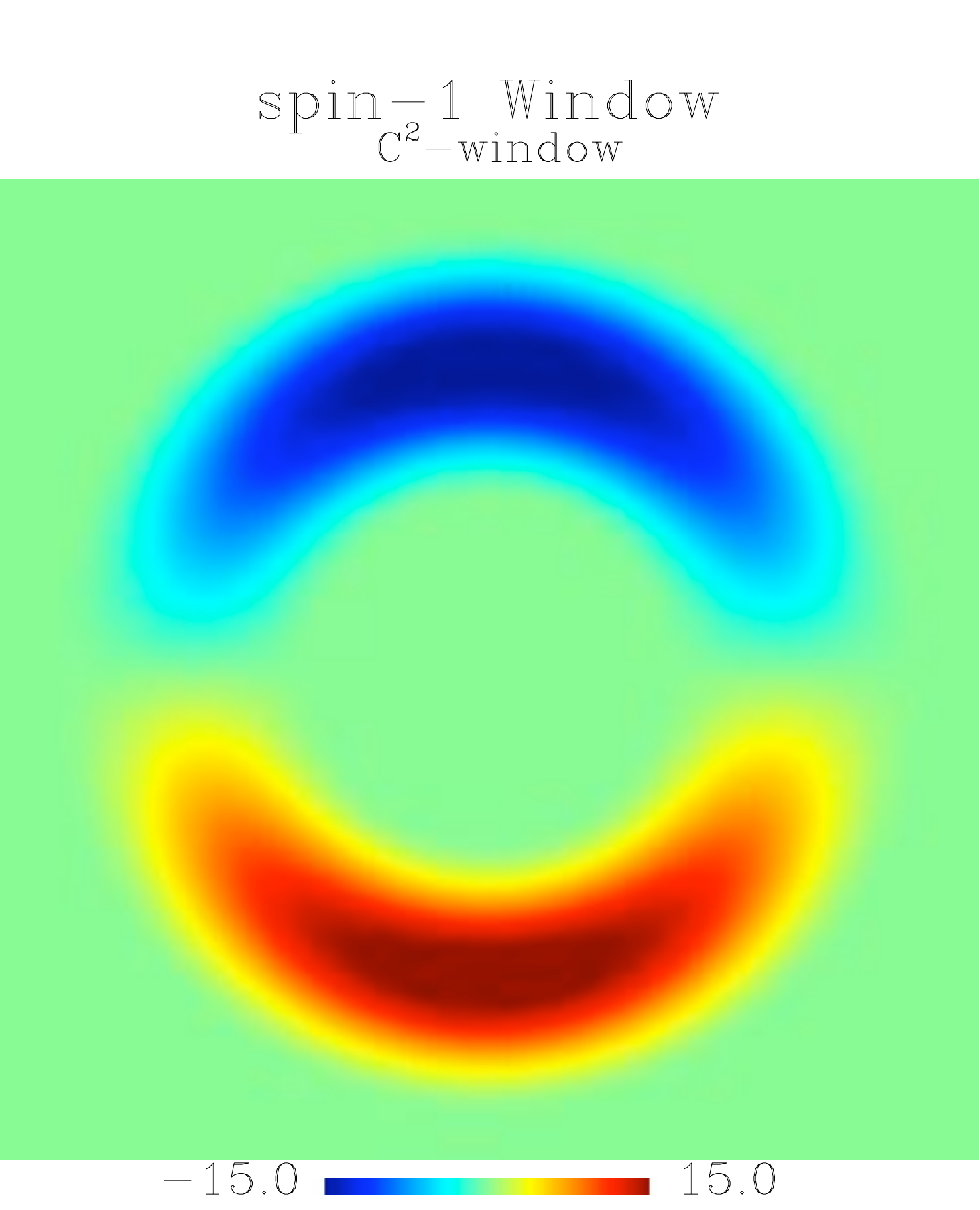}
\includegraphics[scale=0.175]{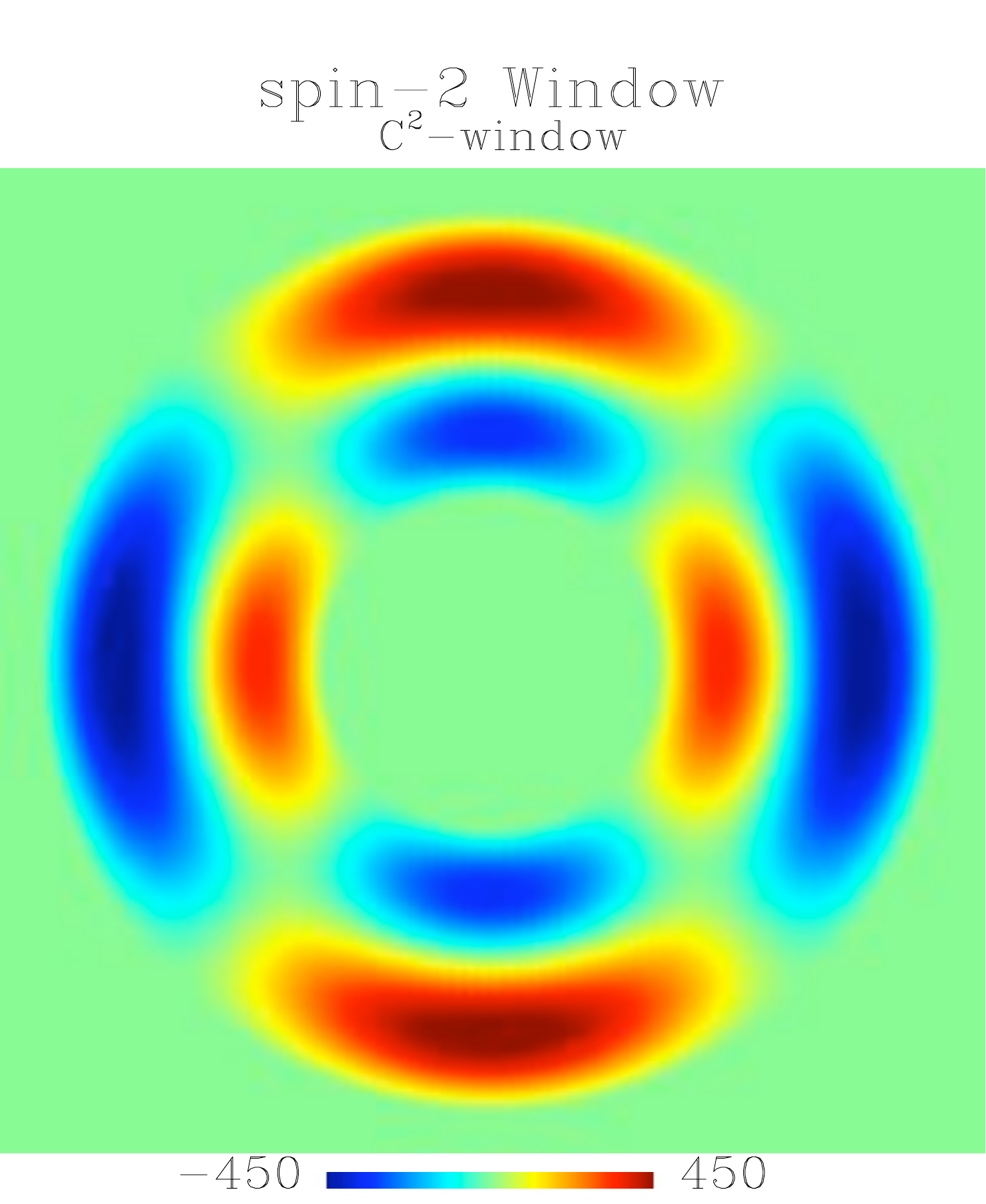}
\caption{Analytic window functions, spin-0, spin-1 and spin-2, left-to-right, respectively, 
calculated using the $C^1$ (upper panels) and $C^2$ (lower panels) functions. 
Only real parts of the spin-1 and 2 windows are shown.}
\label{smith-vs-my}
\end{figure}

The three spin windows, for each of the two cases, are shown in Fig.~\ref{smith-vs-my} 
in a case of a spherical cap. The apodization length is set to 7~degrees.
The spin-0 window indeed falls more 
rapidly to zero in the second proposal, thus leading to an apodized spin-2 window.
Nevertheless the two analytic windows are very similar with only small differences 
concentrated around the edge of the observed patch. We will discuss an impact of such 
differences in the window shapes on the $E/B$ leakage in the next Section.

\begin{figure}
\includegraphics[scale=0.175]{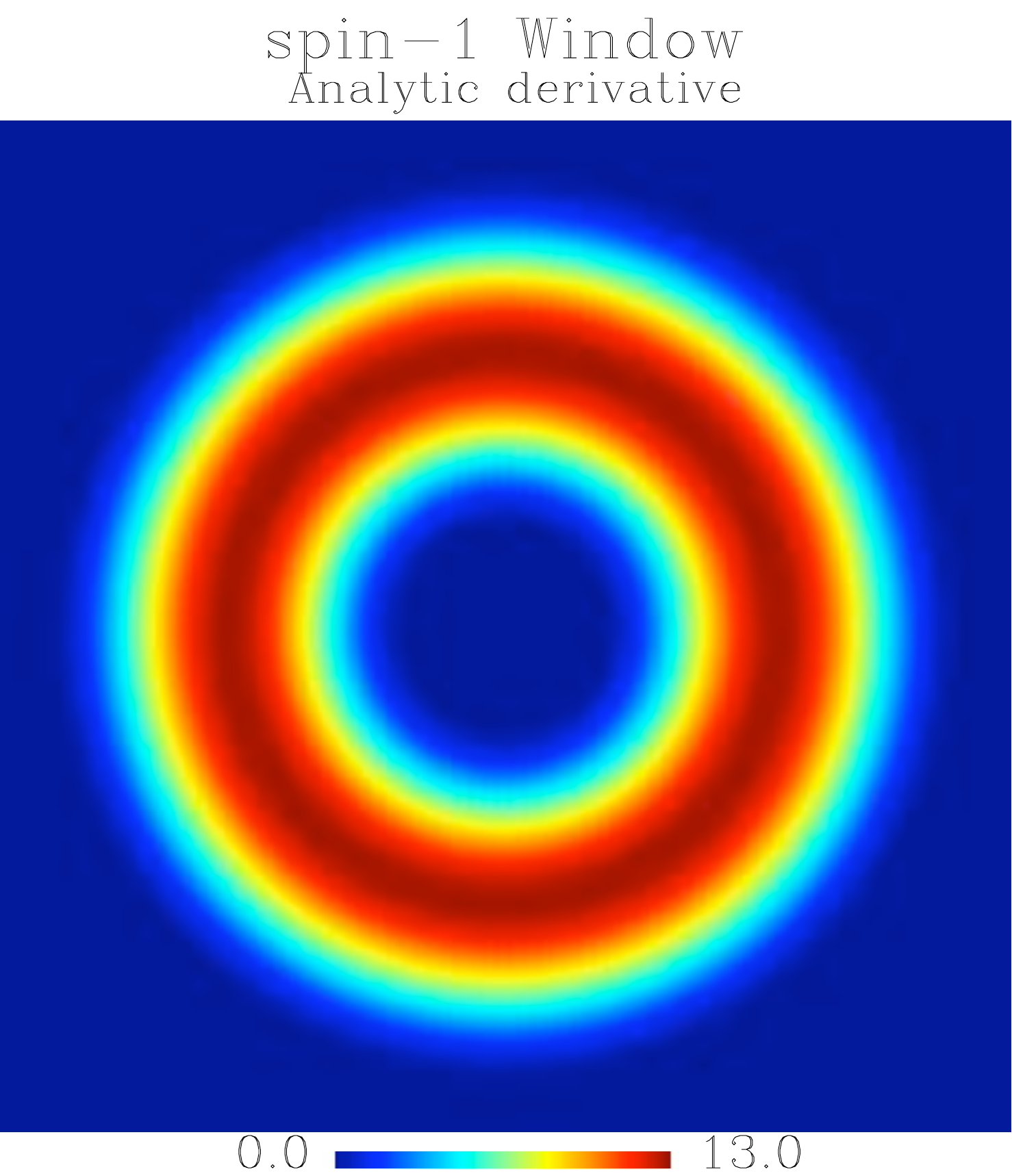} \includegraphics[scale=0.175]{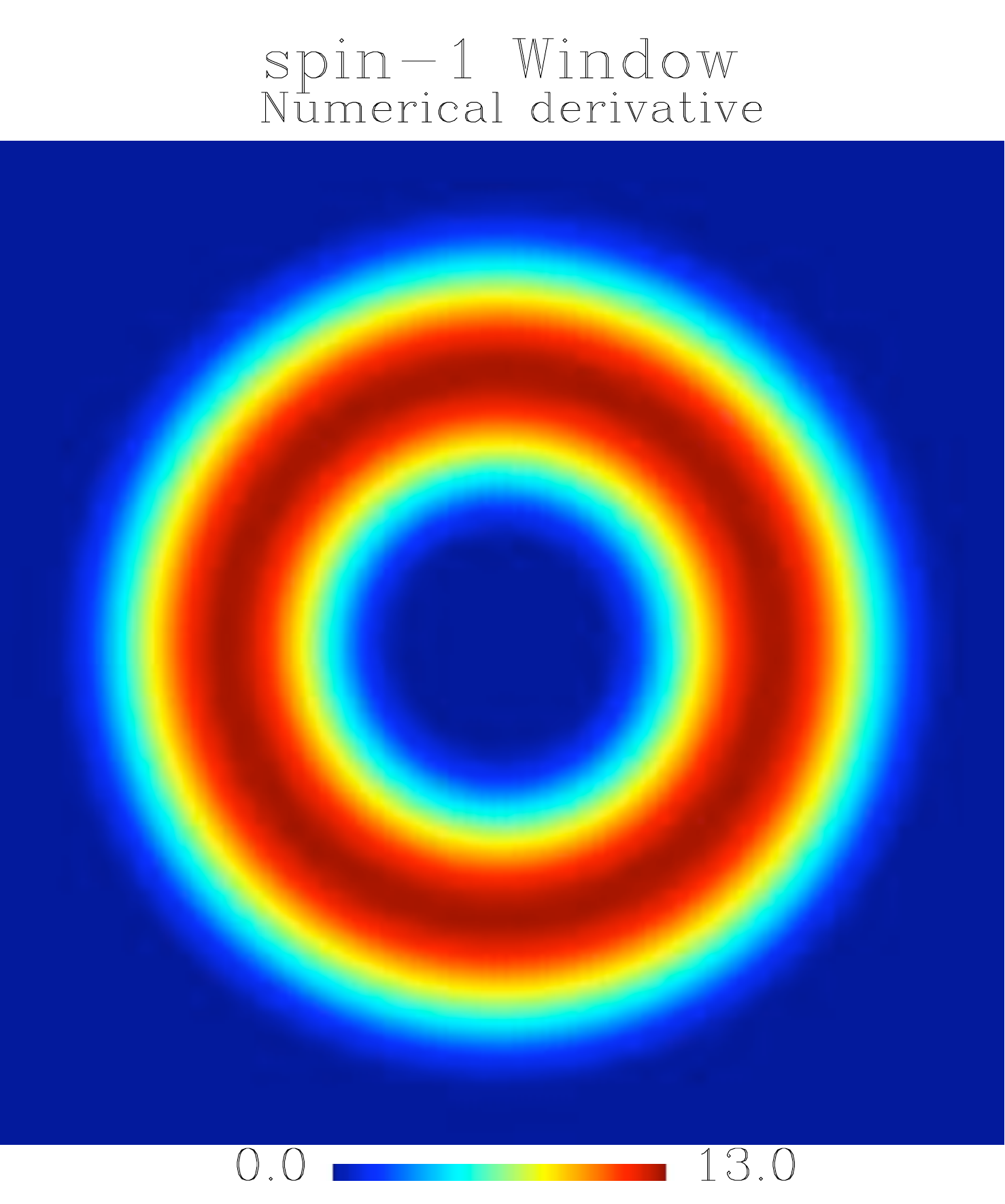} \includegraphics[scale=0.175]{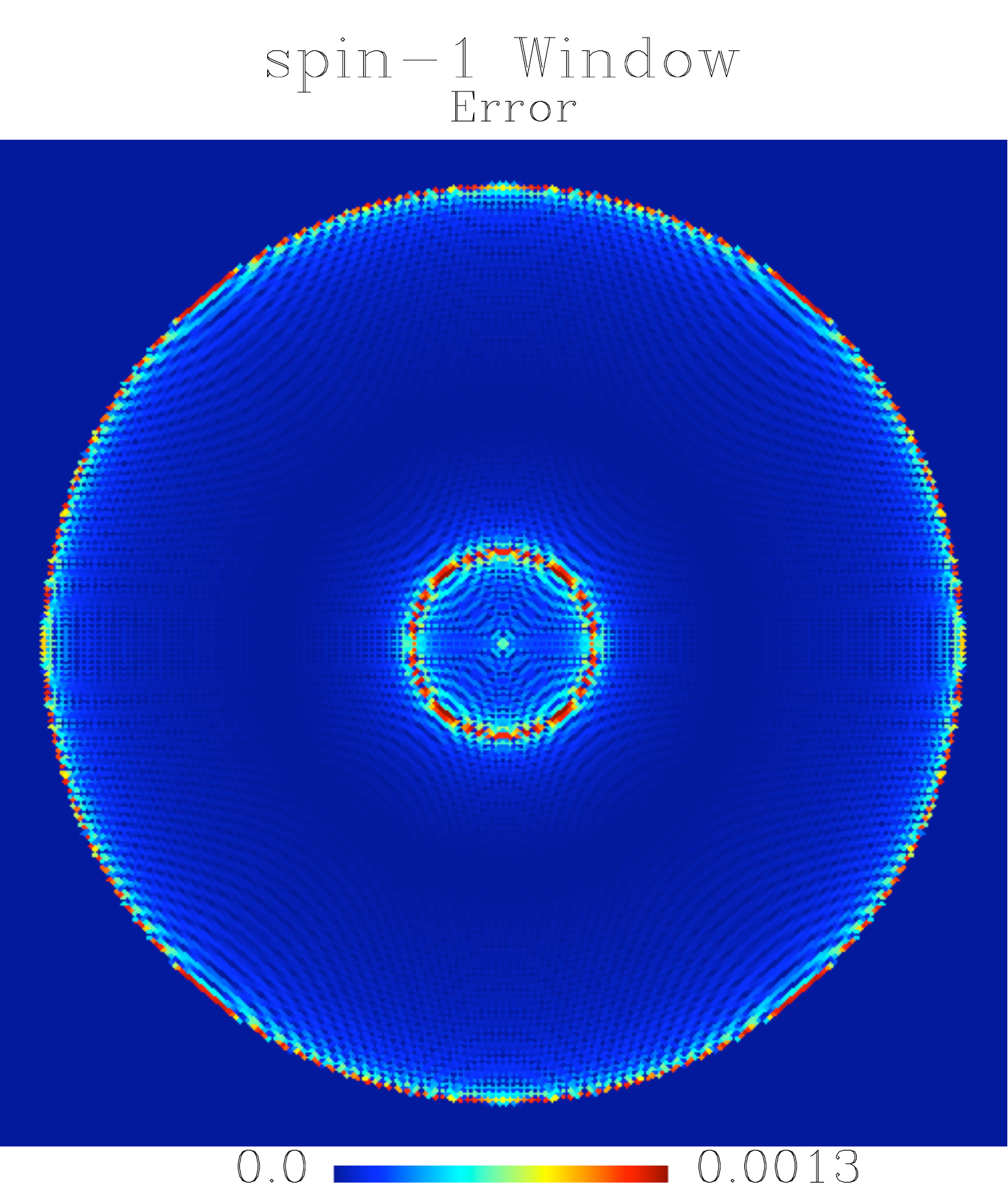} \\
\includegraphics[scale=0.175]{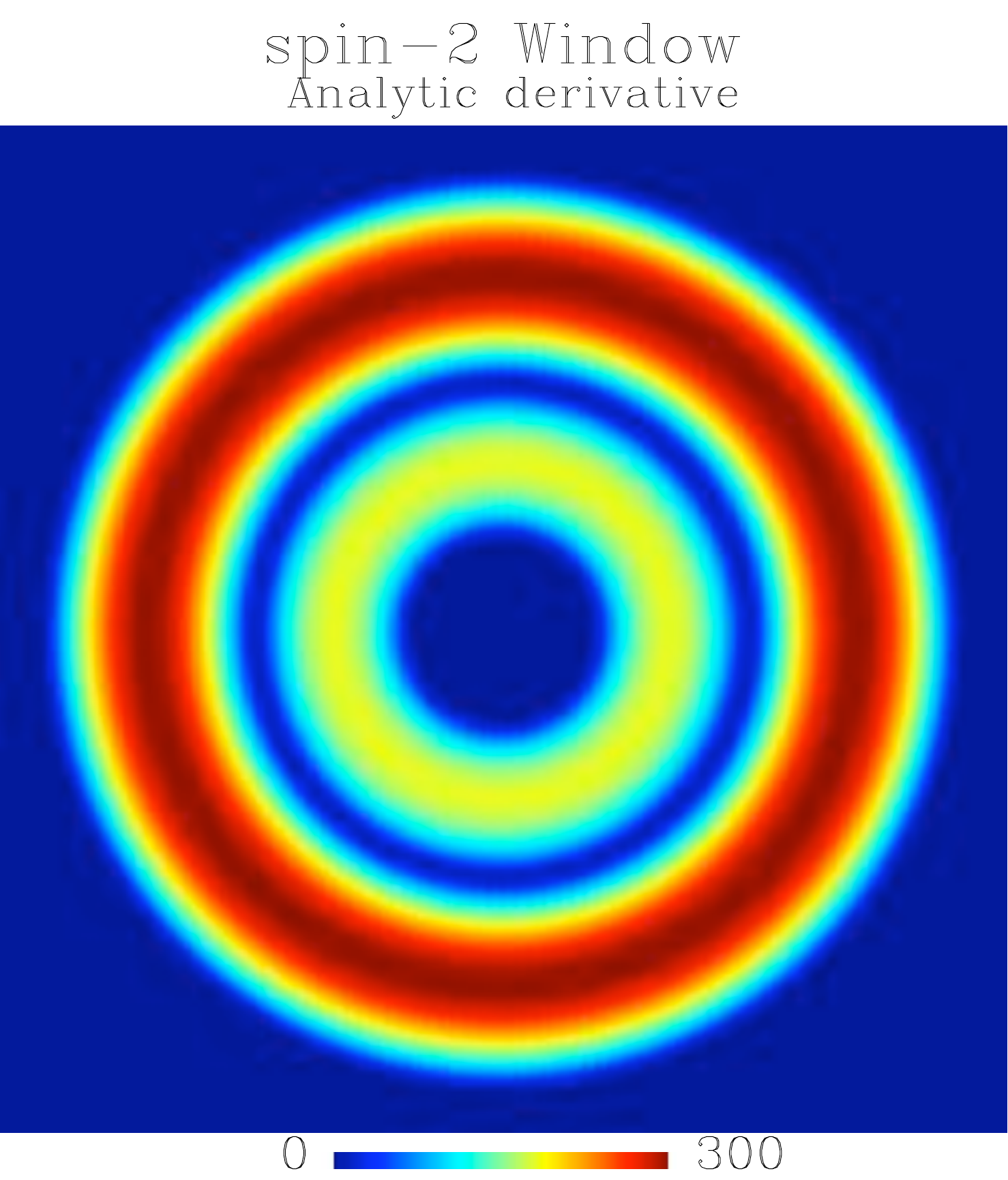} \includegraphics[scale=0.175]{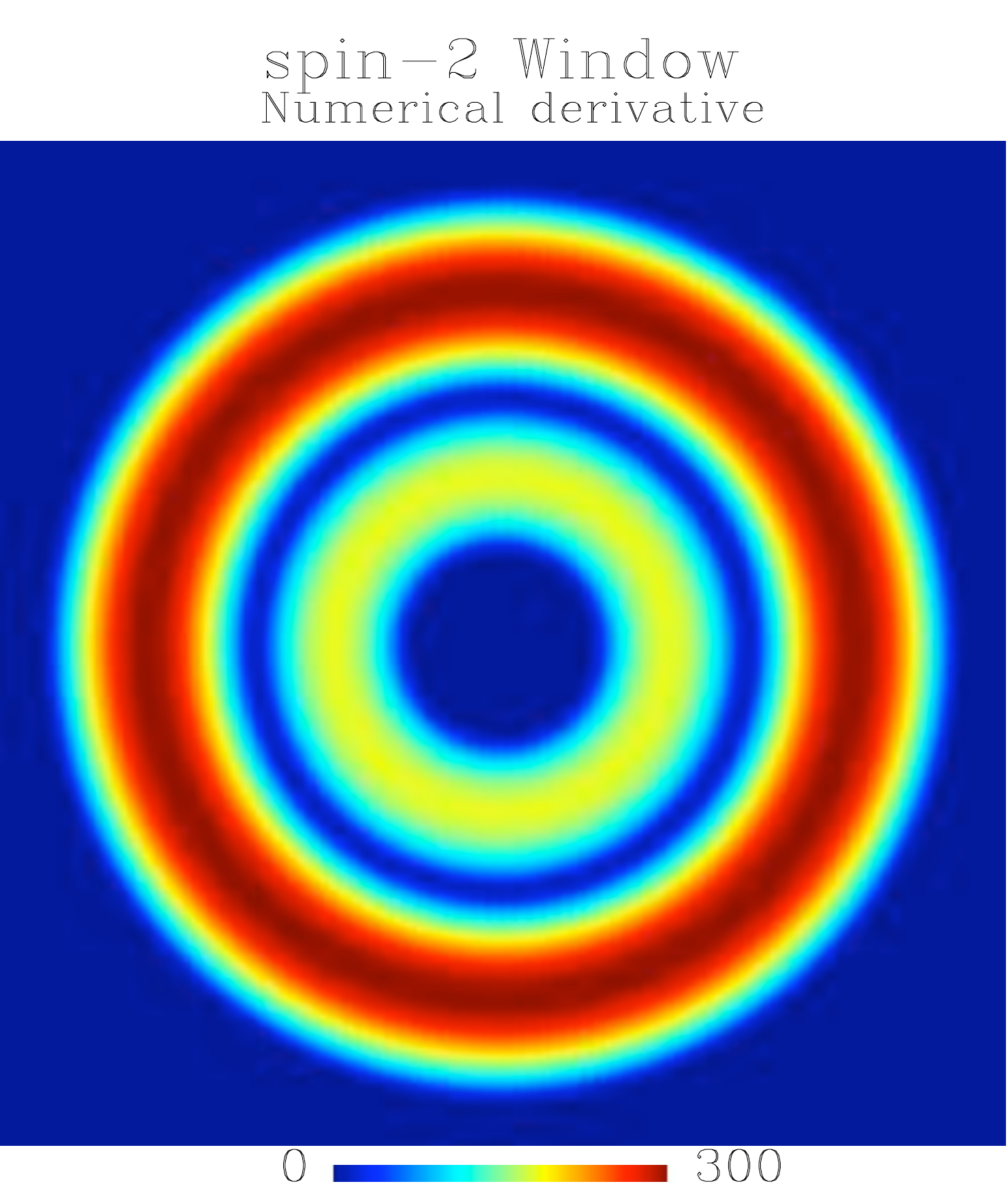} \includegraphics[scale=0.175]{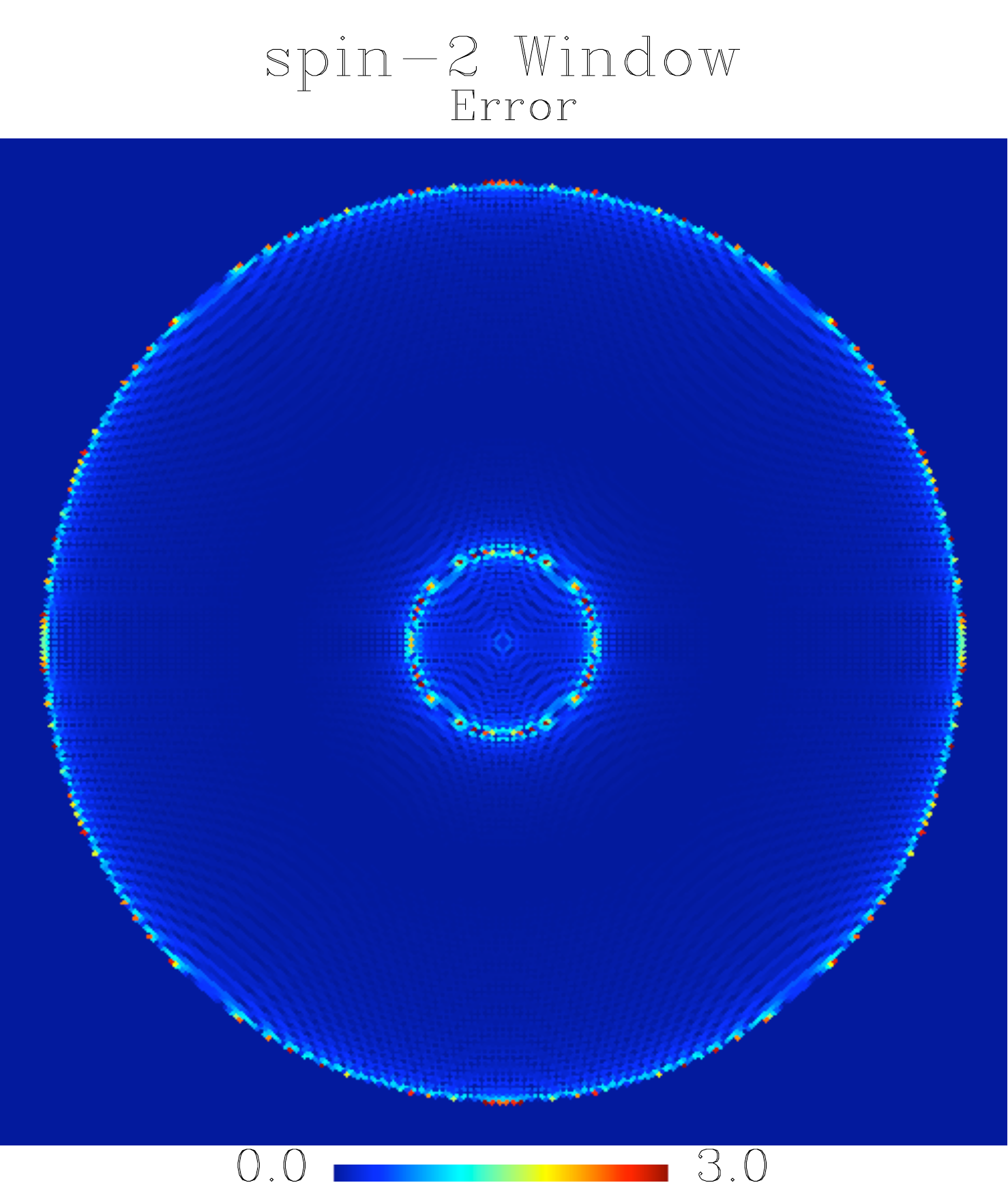} \\
\caption{Absolute value of the analytic spin-1 windows (upper panels) and spin-2 windows (lower panels) derived 
via the analytic (left line) and numerical computation (middle line). The plots in the right panels show the respective
differences. Note that in the right panels the adopted color scale spans a range of values $10^{-4}$ (up) and $10^{-2}$ 
(bottom) times smaller than in the remaining panels.}	
\label{ana-vs-num}
\end{figure}

Two remarks are in order here. First of all, the analytic windows can account for the 
presence of holes in the mask for example due to masked out point-sources or unobserved 
pixels. The distance $\delta_i$ just has to be computed between the $i$-th pixel and 
the boundary closest to it (external or internal). Second, for an arbitrary sky coverage 
(including holes), the	spin windows need to be clearly derived numerically, however, 
in some simple cases such as that of a spherical cap or square patches the spin-1 and 
spin-2 windows can be computed analytically from the spin-0 window. We can, and will, use 
that latter observation to quantify some numerical effects, involved in the discussed formalism. 
For instance we will investigate the effects of the numerical derivatives on the shape of 
the spin-weighted windows, as shown for a spherical cap case in Fig.~\ref{ana-vs-num}. 

\section{Tests and Applications}

\label{sect:applications}

Performance of the technique discussed here depends on many factors, ranging from the pixelization effects 
to assumed knowledge of the sky signals, and can be looked at from different perspectives. In this Section we discuss 
some of the most relevant effects. As the presented estimator first aims at solving the $E/B$ leakage, we start this 
Section with a discussion of a quality of the $E/B$ leakage control and follow on with a study of the estimator
efficiency expressed in terms of the $B$-mode power spectrum variance levels, which can be achieved using this 
approach in different circumstances.
Unless it is explicitly mentioned otherwise, we will work with the HEALPix pixelization
scheme adopting the resolution 
$N_{side}=512$. The input $E$-mode signal is that of the cosmological model with parameters as constrained by the
WMAP 5-year data~\cite{dunkley_etal_2008}. The $B$-mode signal includes lensing and primordial $B$-mode with 
$T/S=0.05$. For the case of homogeneous noise the noise level is set to $5.75~\mu$K-arcmin, which corresponds
to a typical level expected for the small-scale $B$-mode experiments.

To study the dependence on the patch geometry, we adopt five different patch shapes: a spherical cap, a square and 
three rectangular patches with a different elongation. Their geometrical properties are summarized in Table~\ref{tab-patch}. 
All the patches cover the same sky area (roughly $1$\%) but the length of their contour increases progressively
from the spherical cap (C) to the most elongated rectangle (R3).
\begin{table}[ht!]
\begin{center}
\begin{tabular}{ccc|ccccccc} \hline\hline
&&&& $\theta$ && $\varphi$ && perimeter &  \\
&&&& [deg.] && [deg.] && [rad.] & \\ \hline \hline
SQUARE && (S) && 20 && 20 && 1.38 & \\
RECTANGLE 1 && (R1) && 16 && 24.9 && 1.42 & \\
RECTANGLE 2 && (R2) && 12 && 33.2 && 1.57 & \\
RECTANGLE 3 && (R3) && 8 && 49.8 && 2.01 & \\ 	\hline
CAP && (C) && \multicolumn{3}{c}{radius=11.3 deg.} && 1.23 &\\ \hline\hline
\end{tabular}
\caption{Geometrical properties of the five patches used in the examples considered here: 
$\theta$ is the zenithal aperture, $\varphi$ the azimuthal aperture 
and the perimeter is the total geometrical length of the contour of the patch. The five patches cover the same sky area roughly equal to $1$\% of the entire celestial sphere.}
\label{tab-patch}
\end{center}
\end{table}

\subsection{Mode mixing kernel}
The mode mixing kernel provides the most direct measure of the $E/B$ leakage. The root-mean-square (rms) of the fraction of the power 
contained in one of the two polarized modes which is leaked to the other mode is characterized by the magnitude of 
the off-diagonal elements of the mixing kernel. For the pure estimators in principle those should be vanishing if the 
proper apodization window is used. However, in practice, the apodization is only one of the factors on which the 
level of the leakage depends and numerical effects due to approximations and simplifications involved in the calculations 
of the estimated spectra give often rise to the power leakages. In this Section we study dependence of the off-diagonal 
blocks of the mixing kernels on the specific choices of the apodization windows, pixelization scheme, pixel size, sky 
location and shape of observed sky areas, as well as errors due to the approximations involved in the numerical 
derivative calculations.

\begin{figure}
\includegraphics[scale=0.2]{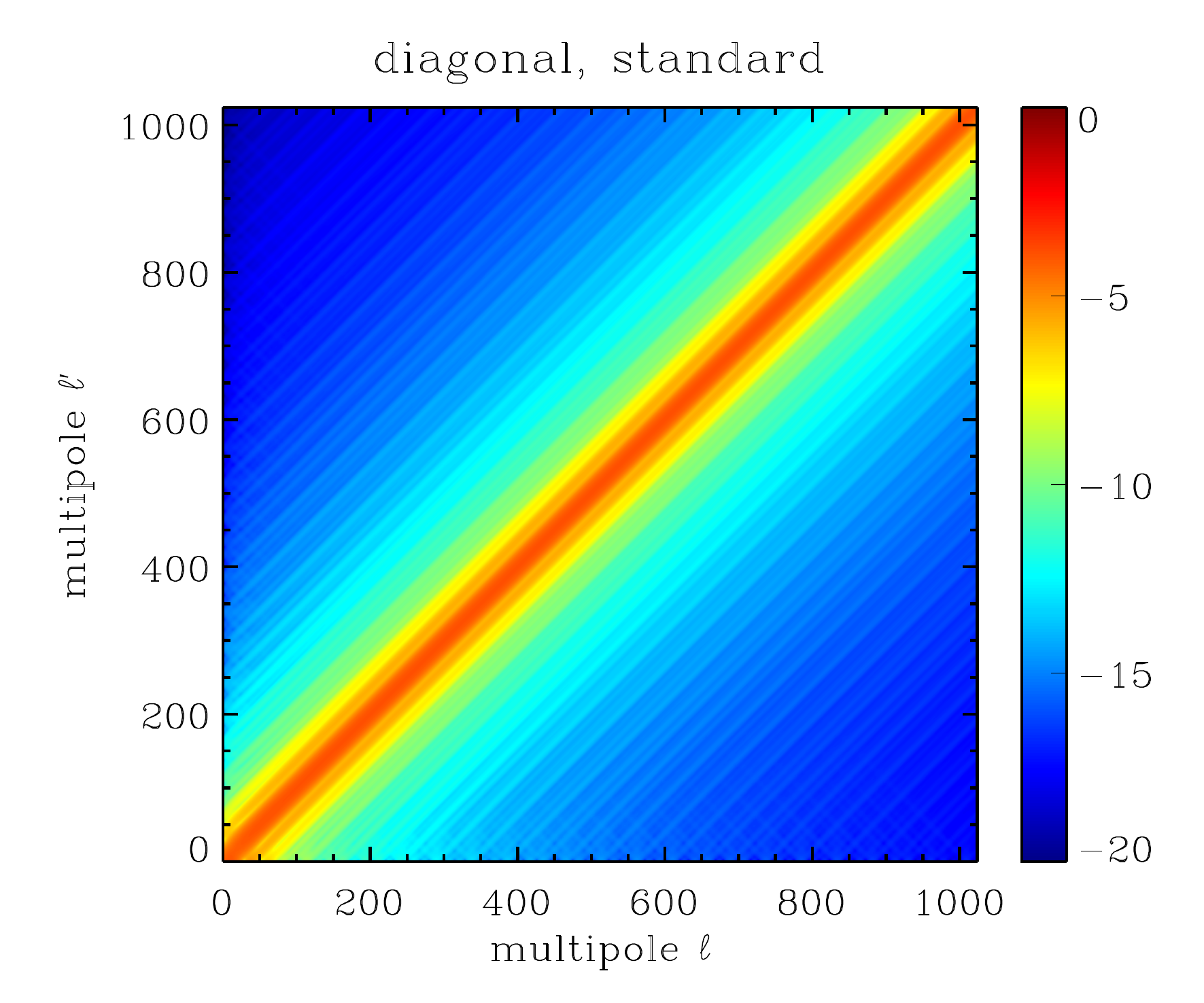} \includegraphics[scale=0.2]{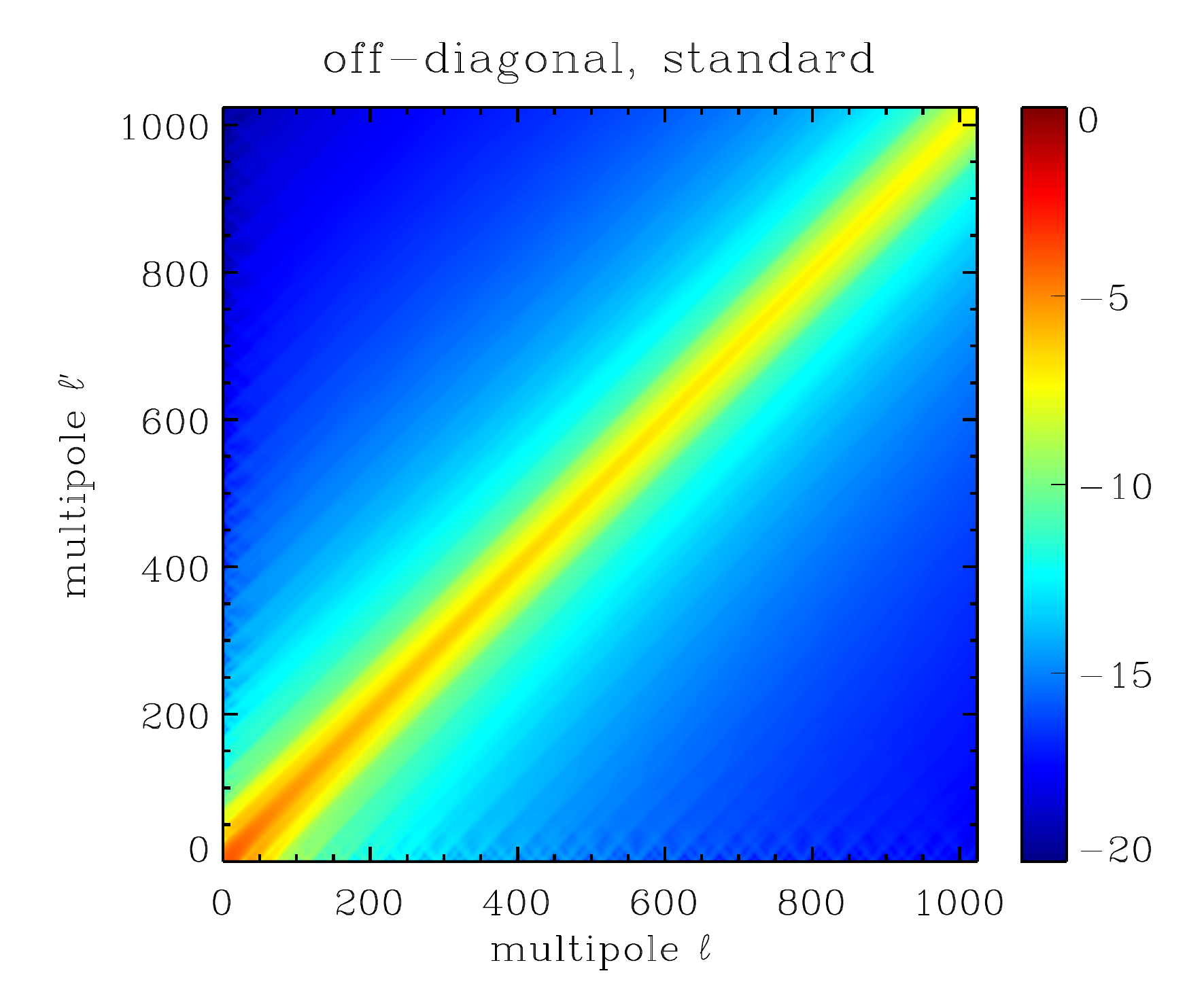} \\
\includegraphics[scale=0.2]{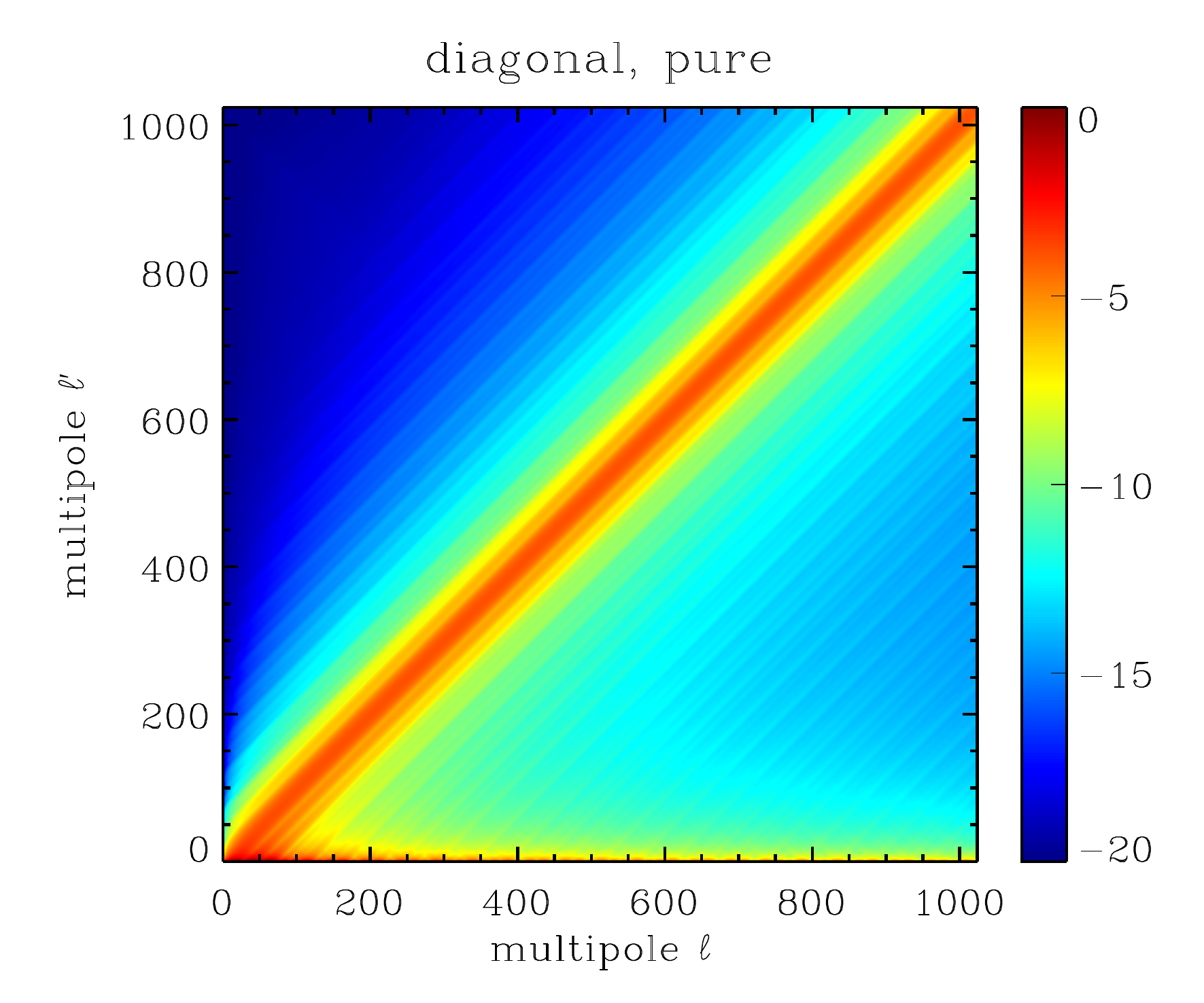} \includegraphics[scale=0.2]{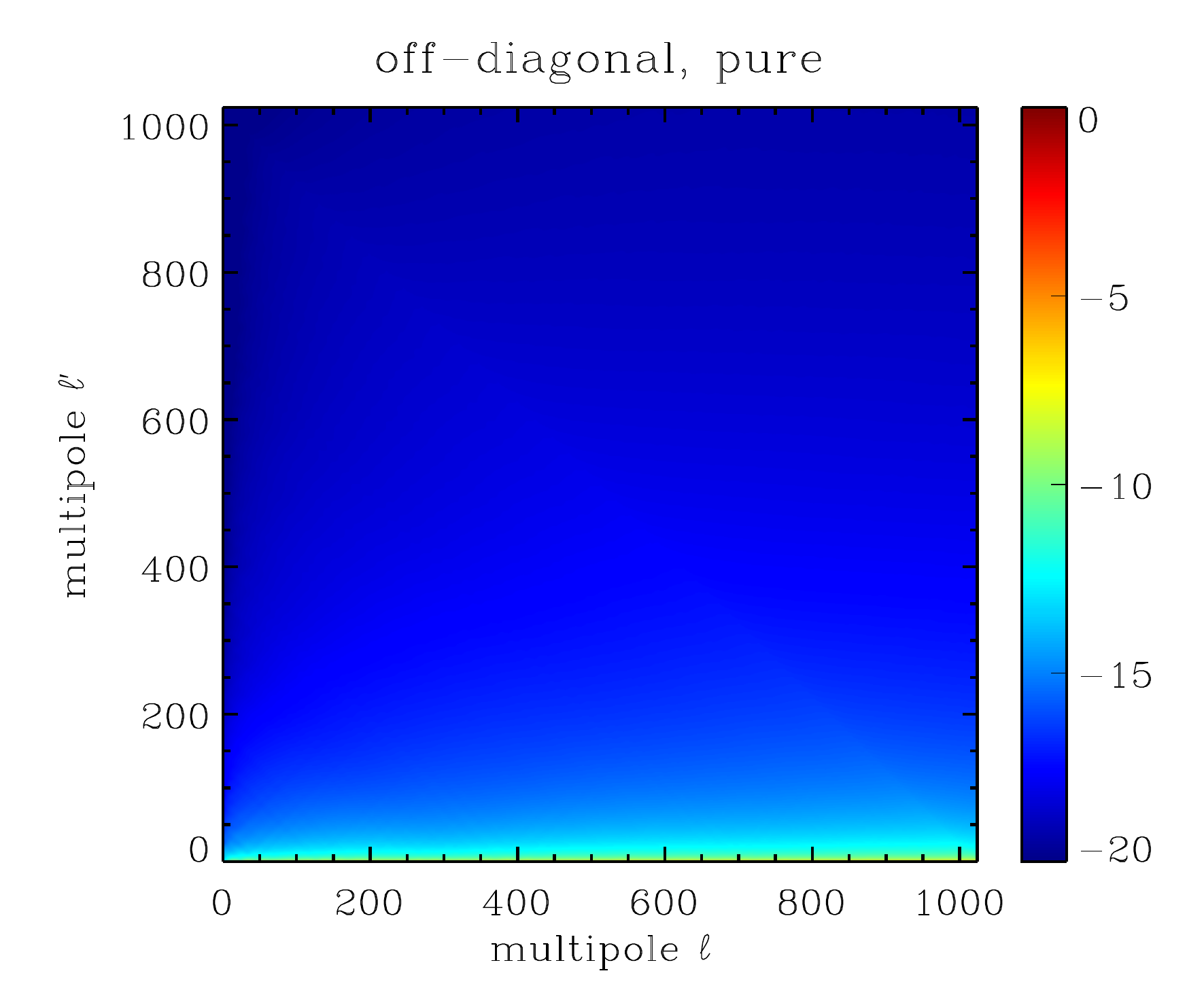} \\
\caption{{\it Upper row:} Diagonal, $BB$, ({\it left}) and off-diagonal, $EB$, ({\it right}) blocks of the mixing 
kernels for the standard estimations. {\it Lower row:} Same as above but for the pure estimator computed 
using the analytic $C^2$-window. The color scale is logarithmic and the same in all the panels. The vertical
axes correspond to the power spectrum multipoles, while the horizontal ones to the pseudo-spectrum, 
Eq.~\eref{unbiased}.
}
\label{pure-vs-std}
\end{figure}

We show an example of the mixing kernels for both standard and pure pseudo-spectra in Fig.~\ref{pure-vs-std}. 
In the standard case we note a strongly band diagonal character of both, diagonal and off-diagonal,
blocks of the mixing kernel with a magnitude 
of the diagonal  elements of  the off-diagonal block comprising a significant fraction of the corresponding elements 
in the diagonal block. This is a clear indication of the leakage. The off-diagonal block elements are manifestly 
suppressed in the case of the pure estimator, as expected given that in the displayed case we have used
a properly apodized, analytic $C^2$-window. Also as expected the diagonal block remains strongly
diagonal dominated. Nevertheless there is a substantial level of mixing between the low-$\ell$ 
multipoles of the power spectra with essentially all multipoles of the corresponding pure 
pseudo-spectra seen for both the blocks.
This extra mixing
arises due to the presence of the counterterms in the definition of the pure modes, Eq.~\eref{counterterm}.
As a consequence a special treatment on the map level aiming at reducing the power contained in the modes 
comparable and larger than the observed sky patch may be required prior to the power spectrum estimation.

\subsubsection{Pixel size}
\begin{figure}[ht!]
\includegraphics[scale=0.5]{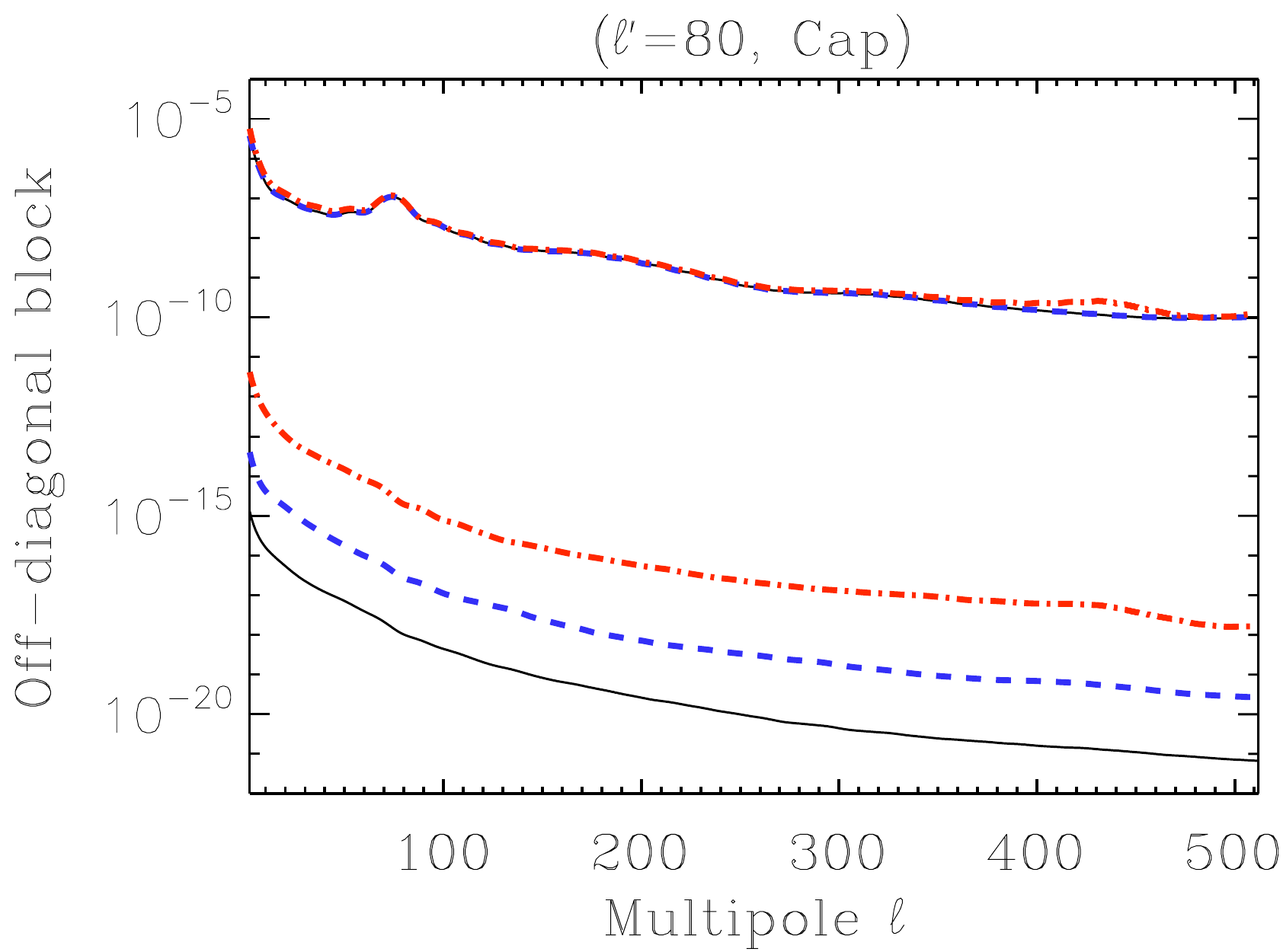} 
\caption{Off-diagonal block of the mixing kernel $M^{off}_{\ell \ell'}$ for two different classes of windows and for different 
HEALPix pixel sizes: $N_{side}=1024$ (black-solid curves), $N_{side}=512$ (blue-dashed curves) and $N_{side}=256$ 
(red-dashed-dotted curves). The three lower curves are for the $C^2$-window and the three, overlapping upper curves -- for the optimized
window computed in the pixel-domain. The apodization length has been set equal to 9 degrees for the $C^2$-window 
and the optimized windows has been optimized for $\ell\in(60,100)$. For the analytic windows, the residual leaked $E$-mode 
is induced by pixel effects and therefore decreases for the higher resolution. For the optimized window, the residual leakage 
is mainly due to a sky cut and as such does not depend strongly on the pixel size.}
\label{leakage-pix}
\end{figure}

As we already hinted at before (Sect.~\ref{subsubsect:NumImp}), pixelization of the sphere is a source of residual 
$E/B$ leakage, which will in general depend on the pixel sizes and shapes. The dependence of such a leakage on 
the HEALPix pixel size for two types of window functions is displayed in Fig.~\ref{leakage-pix}. Each of the three 
lower curves shows a single column of the off-diagonal block of the mixing kernel, $M^{off}_{\ell\ell'}$, computed 
using the $C^2$-window and corresponding to, from top to bottom, $N_{side} = 256$, $512$ \& $1024$. The residual,
leaked $E$-mode decreases for higher resolution confirming the pixel origin of this leakage.  The three upper curves 
in Fig.~\ref{leakage-pix} show the case of the optimized window computed in the pixel-domain. These windows have 
been optimized for the range of $\ell$ from $60$ to $100$ and we kept the noise per pixel area constant. In all three 
cases we find a substantial level of the residual leakage, which is independent on the adopted pixel size. This leakage
appears therefore mostly due to the cut-sky effects, which are, by design, not fully removed in the case of the optimized
windows which do not fulfill the proper boundary requirements, and are only suppressed to the levels below that of the
other uncertainties.

\subsubsection{Window choice}
\begin{figure*}
\includegraphics[scale=0.325]{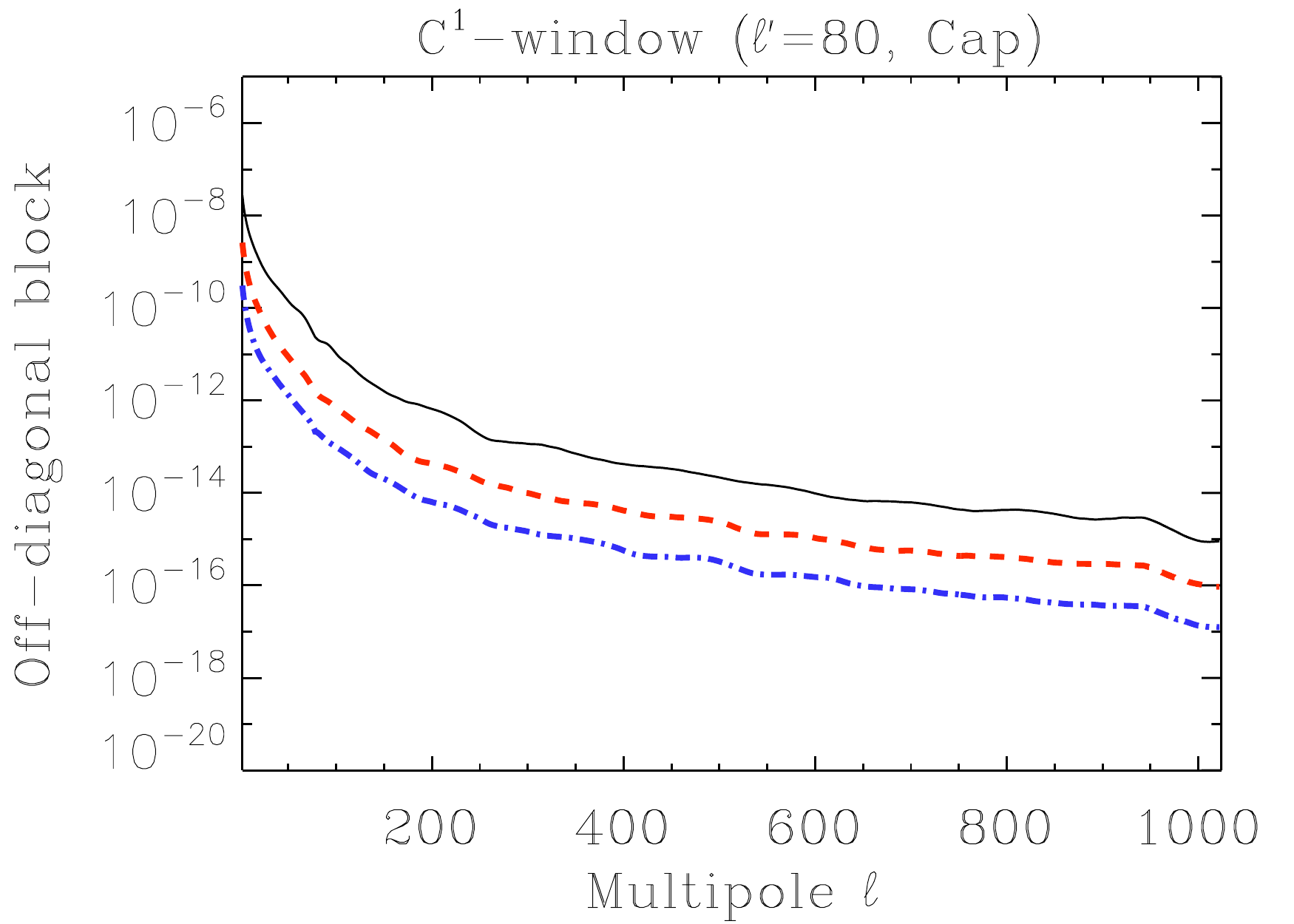} \includegraphics[scale=0.325]{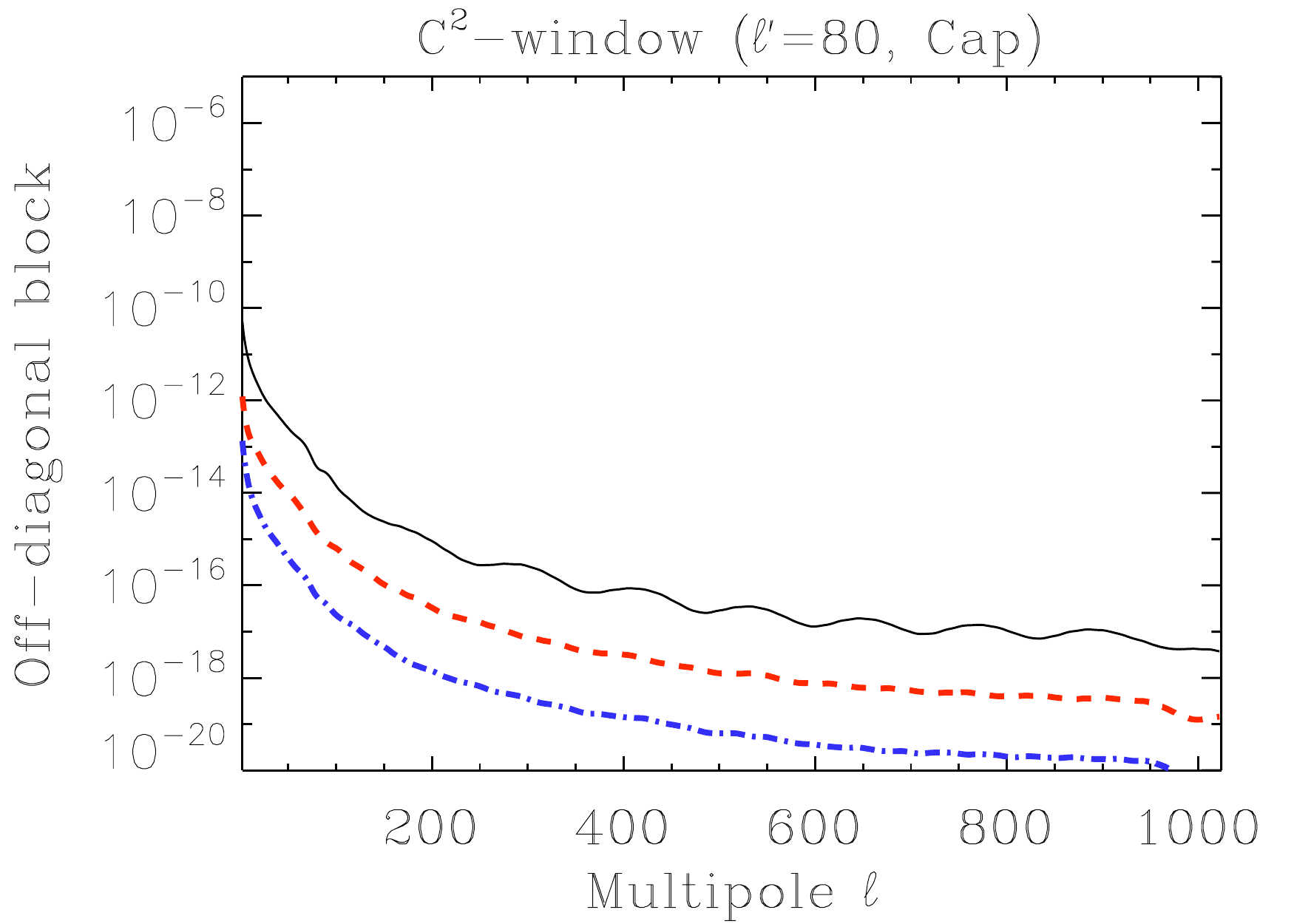} \includegraphics[scale=0.325]{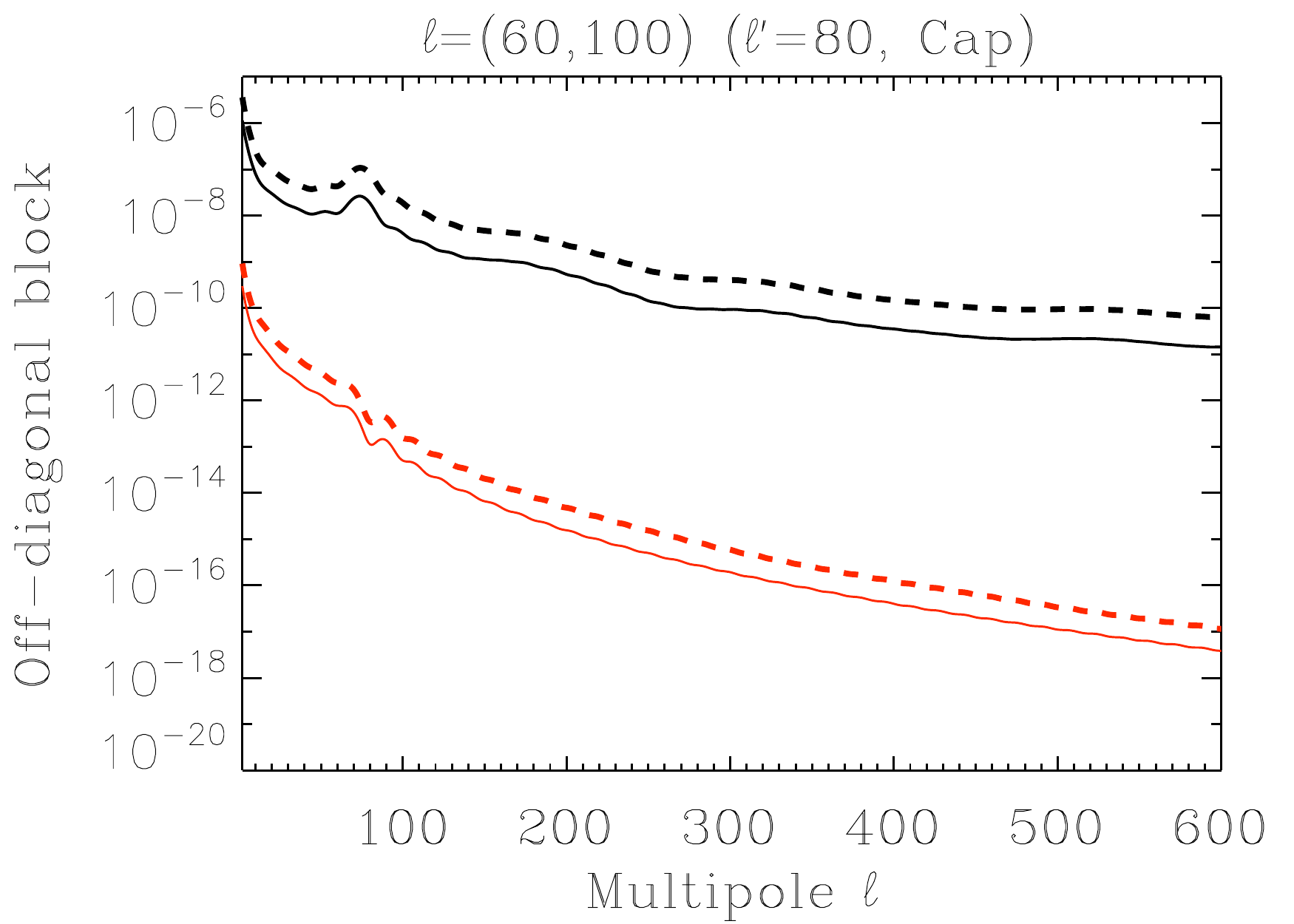}
\caption{A $\ell'=80$ column of an off-diagonal block of the mixing kernel $M^{off}_{\ell \ell'}$ for the four classes of apodized windows. 
{\it Left panel :}  $C^1$-analytic window for three different values of the apodization length : $\delta_c=3$ deg. (black-solid
curve), $\delta_c=5$ deg. (red-dashed curve) and $\delta_c=8$ deg. (blue-dashed-dotted curve). {\it Middle panel :}  Same
as left panel but for the $C^2$-window. {\it Right panel :} optimized windows computed in the pixel-domain (black curves) and in the harmonic domain (red curves). Solid curves are for optimized windows derived without $B$-mode and dashed curves 
for optimized windows with lensing-induced and primordial $B$-mode for $T/S=0.05$.}
\label{leakage-window}
\end{figure*}
A comparison of the residual levels of the $E$-to-$B$ leakage for the different sky apodizations as considered here
is presented in Fig.~\ref{leakage-window}. We plot there a single column of the off-diagonal block of the mixing kernel for 
$\ell'=80$ and computed for the four types of windows defined in Sect.~\ref{sect:apodizations} and the spherical cap area 
as defined in Table~\ref{tab-patch}. As expected the optimized window computed in the pixel-domain leads to the highest 
amount of leakage. The harmonic space window fares better in part thanks to the boundary conditions which are explicitly
imposed on it and in part due to the extra-apodization. For the two types of optimized windows, the level of leakage varies
slightly depending on a prior assumed for the $B$-mode power.

Both the analytic windows satisfy the Dirichlet and Neumann conditions and therefore lead only to a residual amount of 
$E$-to-$B$ leakage. Nevertheless, the $C^2$-window performance is superior to that of the $C^1$-window, producing 
the level of leaked $E$-mode $10^{-4}$ to $10^{-2}$ smaller. This is because the level of the leakage is reduced 
by increasing the apodization length. For the analytic windows, more apodization is achieved either by increasing the 
apodization length or by additionally forcing the spin-2 window to vanish at the edges of the survey. For the optimized 
windows, as discussed before the amplitude of the adopted $B$-mode prior leads to an increase of the effective 
apodization length, thus affecting the leakage level.  Because of pixelization, the windows are discretized and the 
Dirichlet and Neumann boundary conditions are inevitably broken. By increasing the apodization, we reduce the size 
of the discrete jump in the window values between two neighbouring pixels at the patch boundary, thus mitigating 
the extent to which that affects the boundary conditions.

\subsubsection{Derivatives}

\begin{figure}
\includegraphics[scale=0.246]{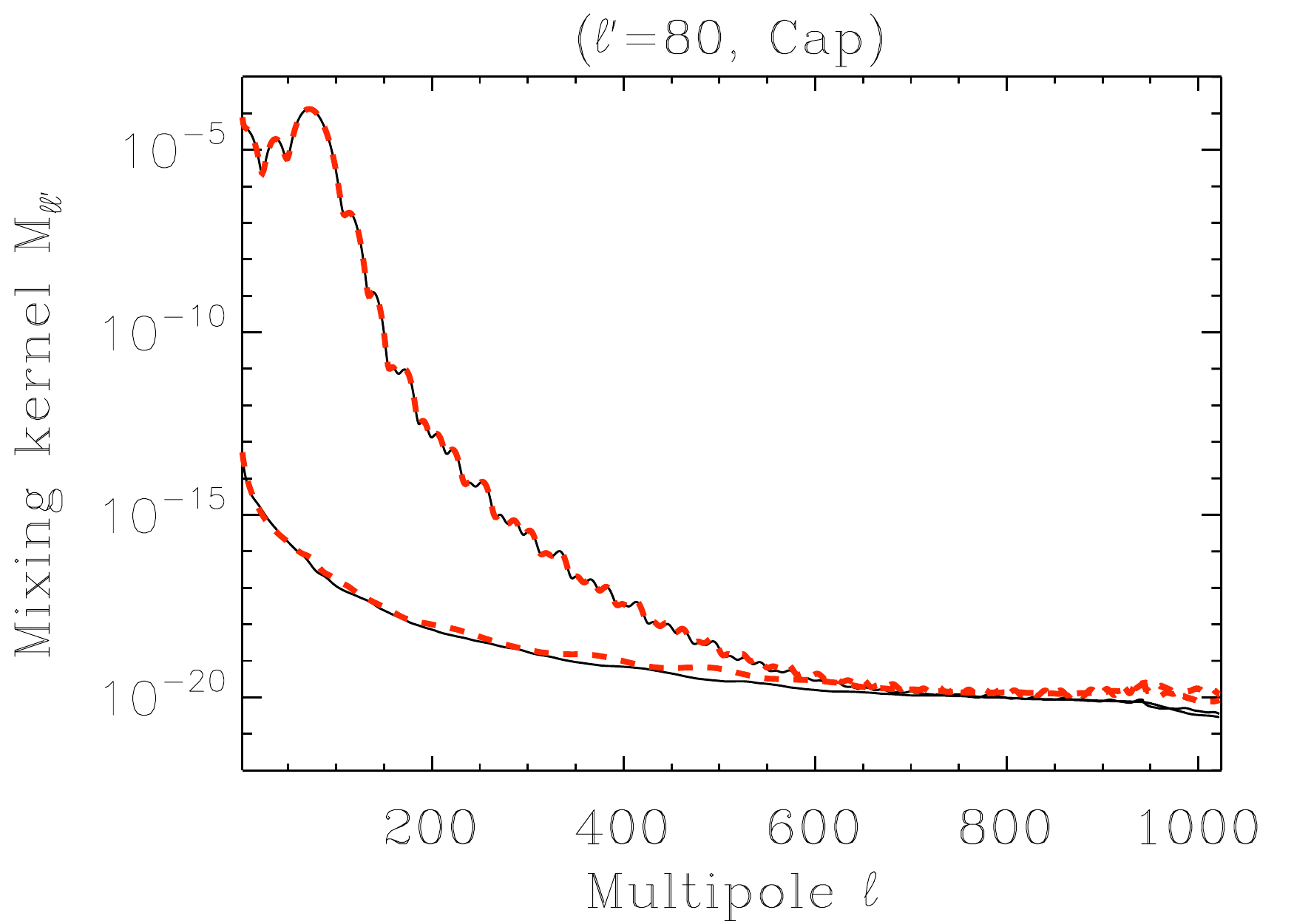} \includegraphics[scale=0.246]{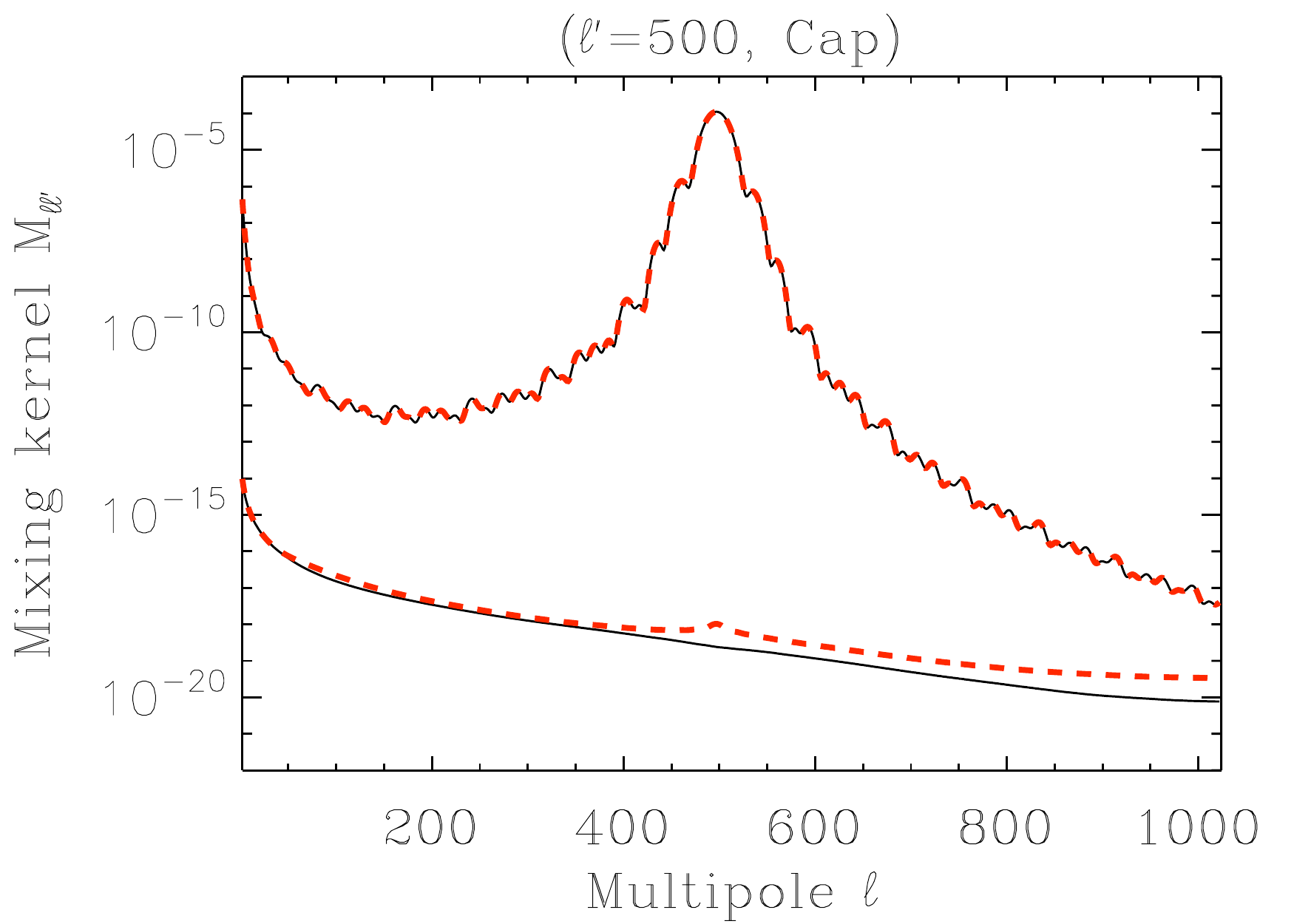}
\caption{A column of a diagonal (upper curves) and off-diagonal (lower curves) block of the mixing kernel using analytically (black-solid curves) and numerically (red-dashed curves) computed spin-1 and spin-2 windows. The dashed and solid curves overlap over most of the angular scale. Left panel shows a column calculated for $\ell' = 80$ and right -- for $\ell' = 500$. The numerical 
computation of the spin-1 and spin-2 windows leads to a slight increase of the leakage in the high $\ell$ part seen as a difference in the off-diagonal block curves.}
\label{leakage-ana-vs-num}
\end{figure}

For arbitrary patch geometries and in the case of the analytic apodizations the spin-1 and spin-2 windows 
may have to be computed  from its spin-0 counterpart numerically and therefore being often only 
approximate. In this paper, as described before, we perform the numerical derivation in the harmonic domain. The effect of using the approximate rather than exact nonzero spin windows
is depicted  in Fig.~\ref{leakage-ana-vs-num}, where again a single column of the mixing kernel is shown
for the spherical cap computed using either the analytic formula (in black) or numerical 
computations (in red) for spin-1 and spin-2 component. This shows that calculating the derivatives numerically does not affect the diagonal block elements and leads only to a slight increase of 
the $E$-to-$B$ leakage, as expressed by a small change in the off-diagonal block element magnitudes. This latter effect is indeed very small and only appears at the smallest 
scales, 
confirming thus that our prescription for the derivative computation permits us to retain very 
good control of the leakage, especially in the low-$\ell$ part, where it is of primary importance.

\subsubsection{Patch geometry}
\label{sect:patchGeom}
\begin{figure}[ht!]
\includegraphics[scale=0.246]{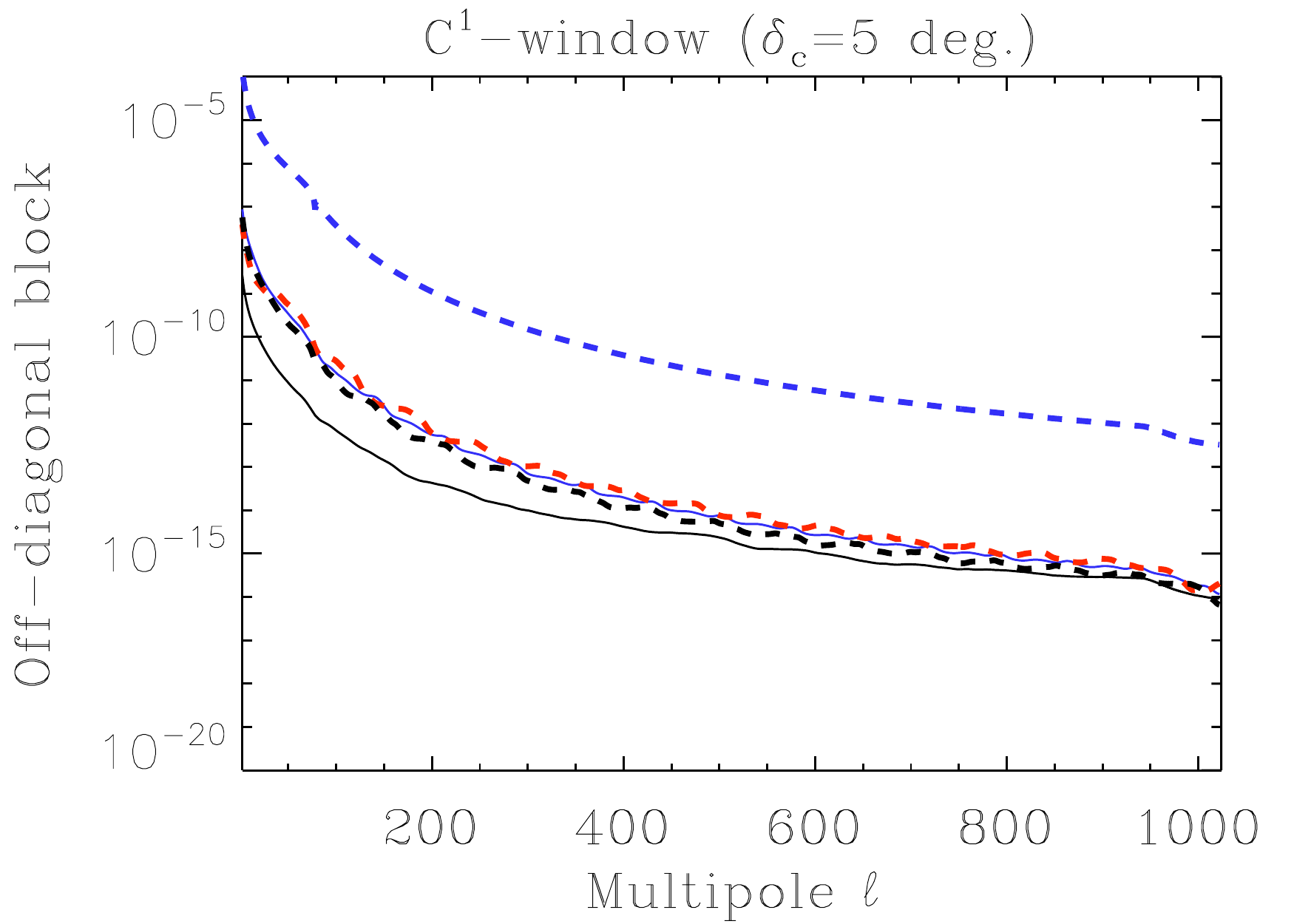} \includegraphics[scale=0.246]{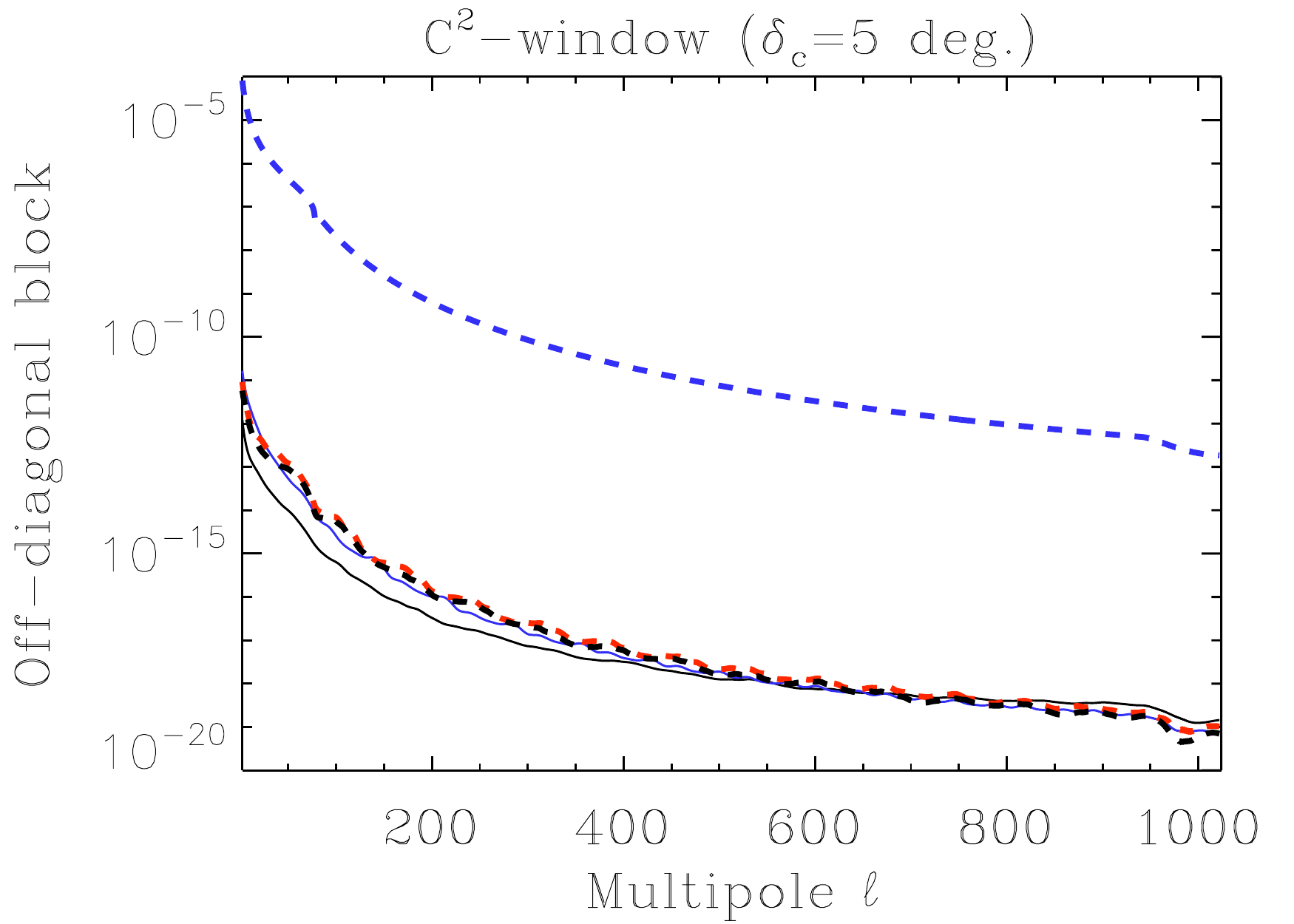} \\
\includegraphics[scale=0.246]{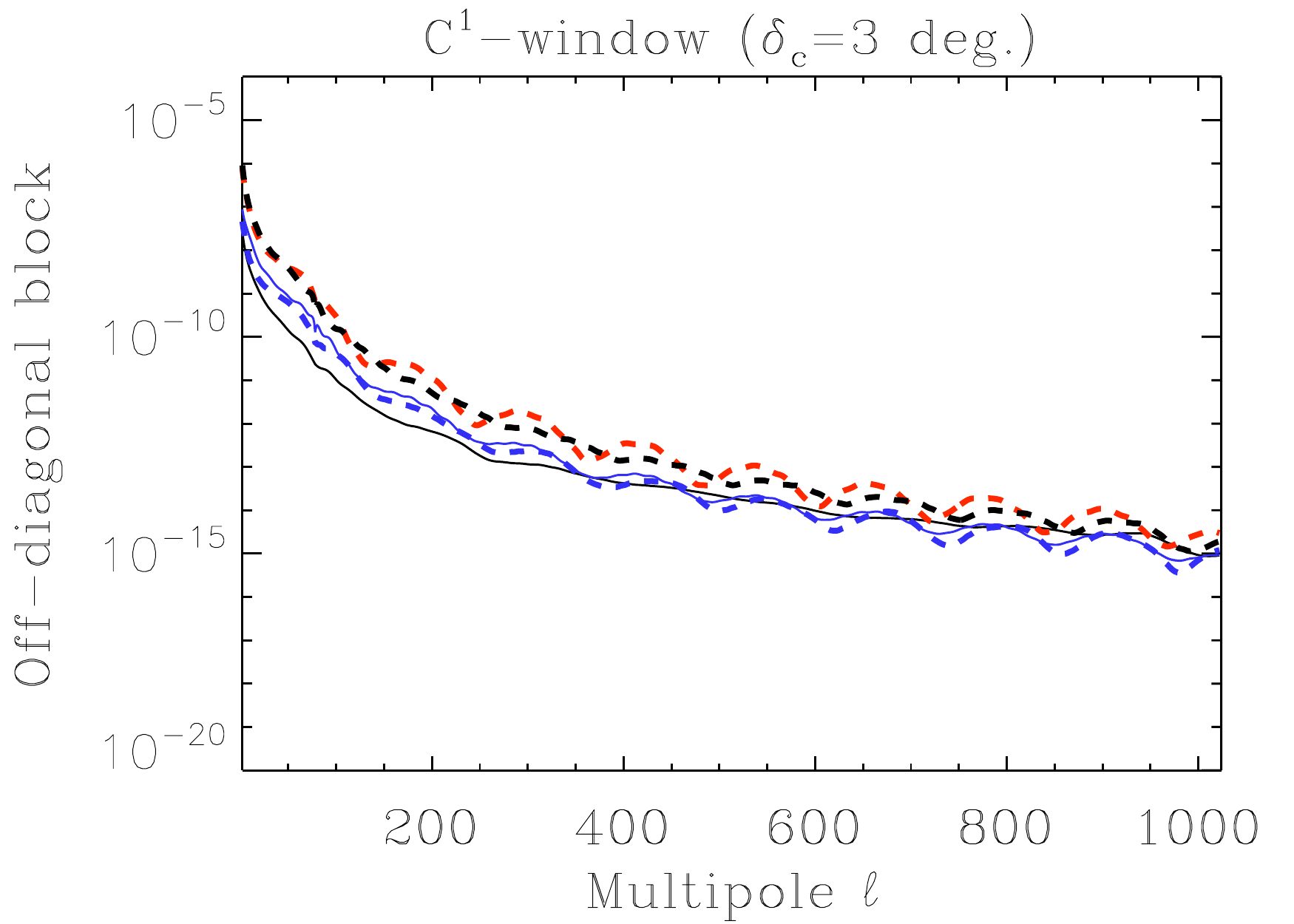} \includegraphics[scale=0.246]{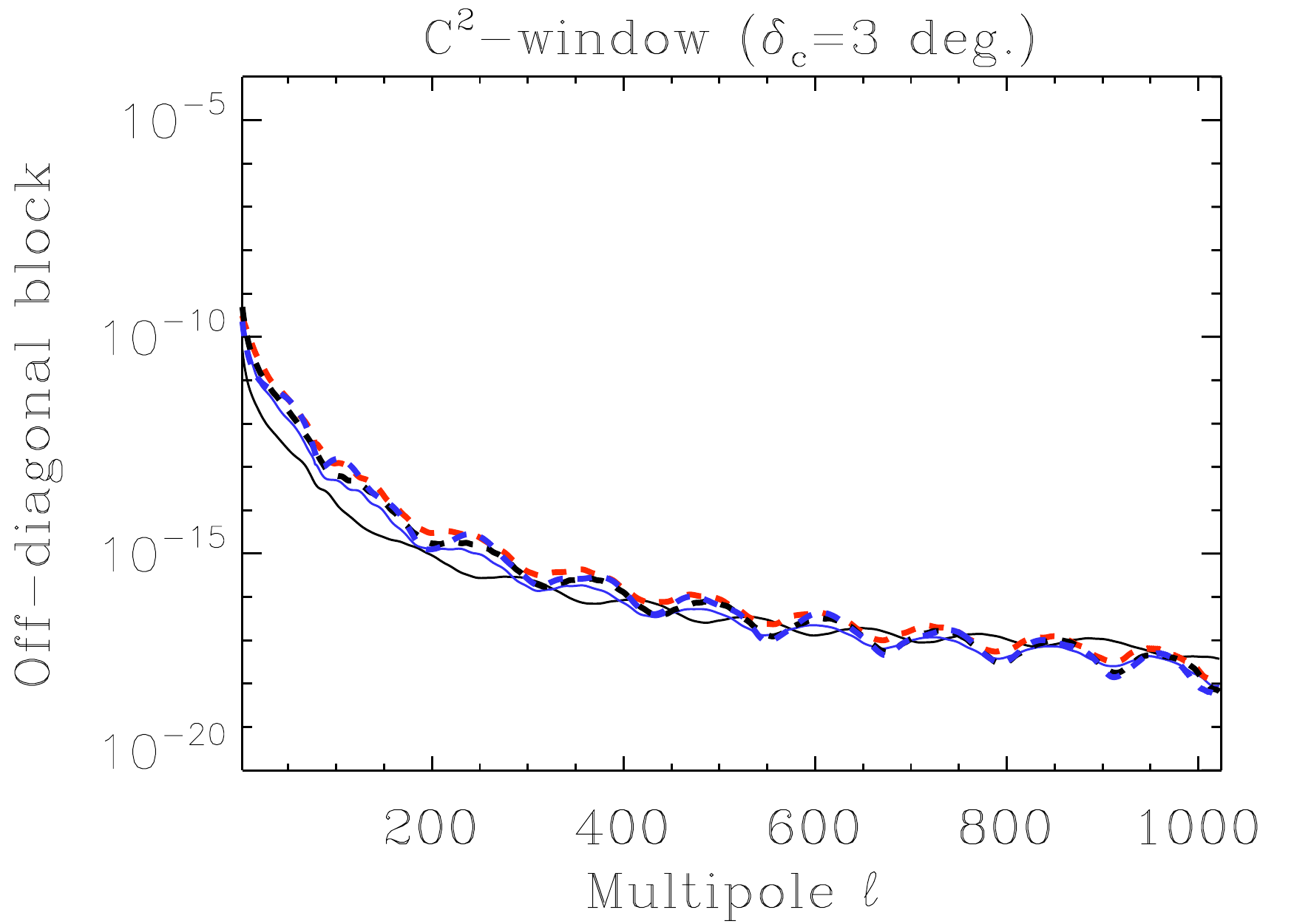}
\caption{A column of the off-diagonal block of the mixing kernel $M^{off}_{\ell \ell'}$ for $\ell'=80$ and for five different patch geometry: spherical cap (black-solid curve), square (blue-solid  curve), rectangle R1 (red-dashed curve), rectangle R2 (black-dashed-dotted curve) and rectangle R3 (blue-dashed curve). The left panels show the results for the $C^1$-window and the right panel for the $C^2$-window. Upper panels are for $\delta_c=5$ deg. and lower panels for $\delta_c=3$ deg. For the (R3) geometry, an apodization length of 5 deg. exceeds the half of the smallest size of the patch. As a consequence, the spin-1 and spin-2 windows are not continuous right in the middle of the patch, leading to a significant increase of the pixel-induced leaked $E$-mode (see blue-dashed curves of the upper panels). For all the other cases, the curves overlap proving the good control of the $E/B$ leakge irrespectively of the geometry.}
\label{leakage-geometry}
\end{figure}

The remaining leakage as a function of the geometry of the observed sky area is depicted in Fig.~\ref{leakage-geometry} for the two analytic windows and for two values of the apodization length. For an apodization length of $3$ degree the level of 
leaked $E$-mode is the same in all the considered cases, demonstrating that we can achieve a good control of the leakage 
for any elongation. For $\delta_c=5$ degree the leakage for the most elongated patch, i.e., the R3 geometry, is greatly
enhanced. This is because for such a patch, apodization lengths greater than $4$ degree exceed the half of its smallest size
causing the spin-0 window, as defined in Eqs.~\eref{eqn:smithWinDef} or \eref{eqn:grainWinDef}, to be strictly speaking nondifferentiable. 
Nevertheless the spin-1 and spin-2 windows can still be formally
introduced, using the fact that one-sided derivatives exist. That however unavoidably results in
the spin-1 and spin-2 windows, which are not always continuous, leading the leakage excess. 
This emphasizes the importance of ensuring that all three spin-$s$ windows are continuous throughout the entire
observed region -- a condition, which on occasion may require either 
trimming some irregular edges of realistic sky patches as observed by 
actual experiments or applying an extra smoothing to the spin-0 window. Nevertheless, whenever this condition is fulfilled the leakage is largely independent on the observed 
patch shape, and with the help of the presented formalism can be, at least in principle, removed down to the level as defined 
by the pixelization effects.

\begin{figure}
\includegraphics[scale=0.5]{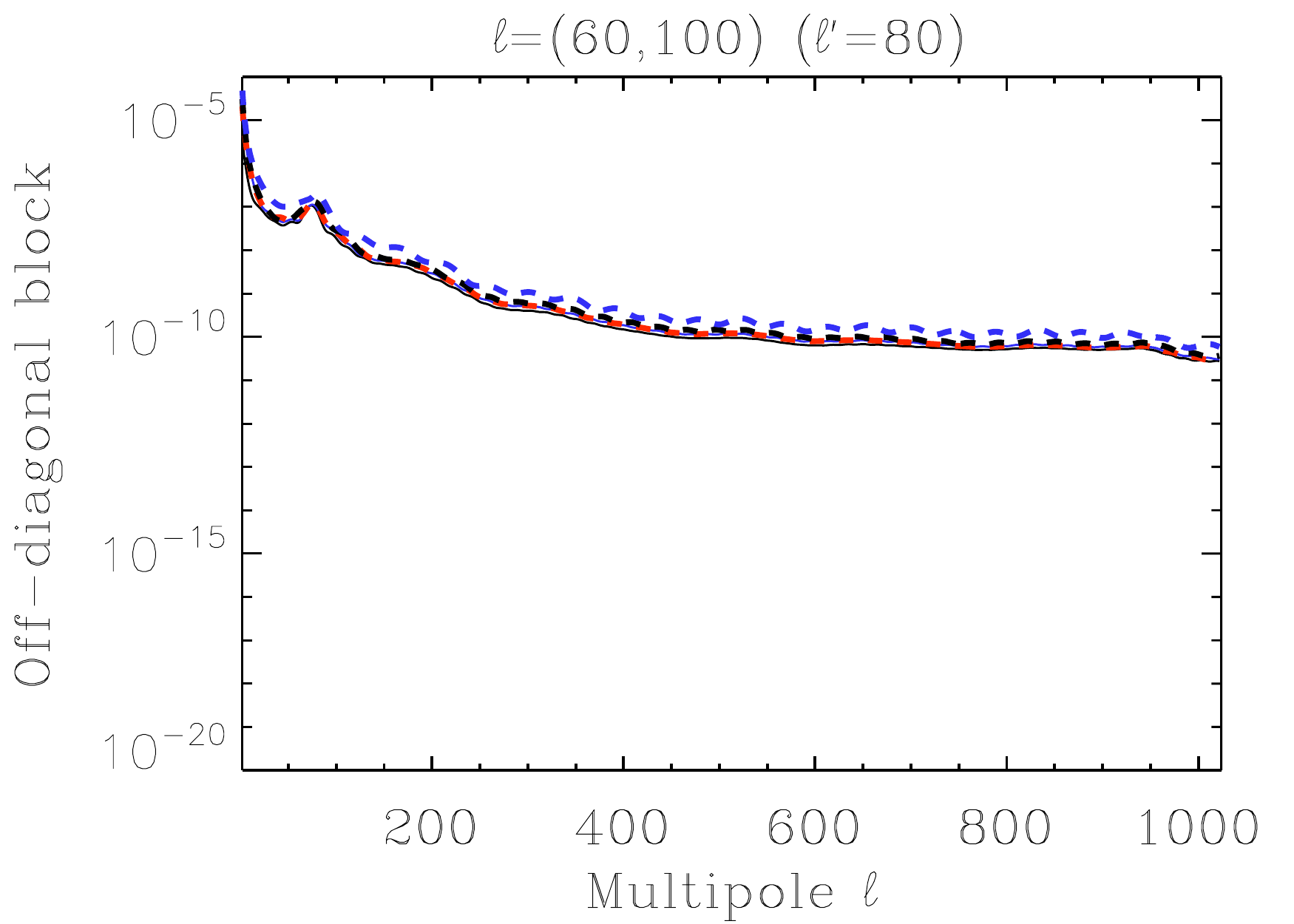}
\caption{A column of the off-diagonal block of the mixing kernel $M^{off}_{\ell \ell'}$ for $\ell'=80$ and calculated for 
windows optimized in a pixel-domain for a range $\ell\in(60,100)$ and the five different patch geometries {\bf{(the five curves overlap)}}: spherical cap
(black-solid curve), square (blue-solid  curve), rectangle R1 (red-dashed curve), rectangle R2 (black-dashed-dotted curve)
and rectangle R3 (blue-dashed curve). The axis ranges coincide with those in Fig.~\ref{leakage-geometry}.}
\label{leakage-geom-opt}
\end{figure}

The case of optimized windows computed in the pixel-domain is displayed on Fig.~\ref{leakage-geom-opt} where the level 
of leakage is shown for the windows optimized in the $\ell\in(60,100)$ bin. The noise and signal prior is the same for all 
the patches. As for the analytical sky apodizations, the level of leakage only marginally depends on the patch geometry. 
This means that the optimization procedure manage to lower the level of leaked $E$ mode below the noise irrespectively 
of the shape of the boundaries (we remind that the five surveys cover the same sky area, which roughly leads to the 
same amount of noise in $\ell$-space). We note that as expected the level of leakage allowed by the optimized windows 
is a few orders of magnitude higher than for the analytic windows.

\subsubsection{Dependence on a sky position/pixelization}

\begin{figure*}[ht!]
\includegraphics[scale=0.34]{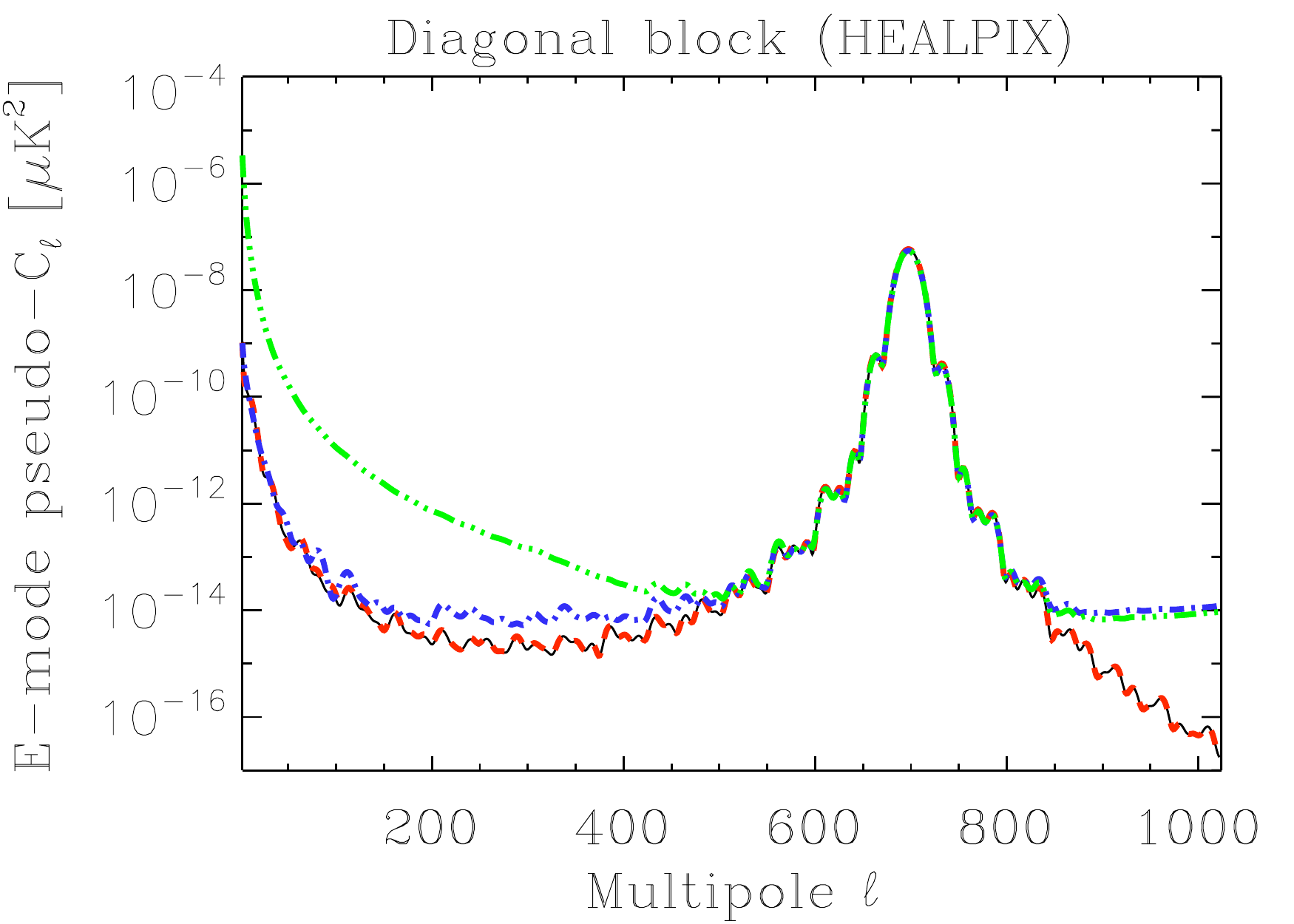} \includegraphics[scale=0.34]{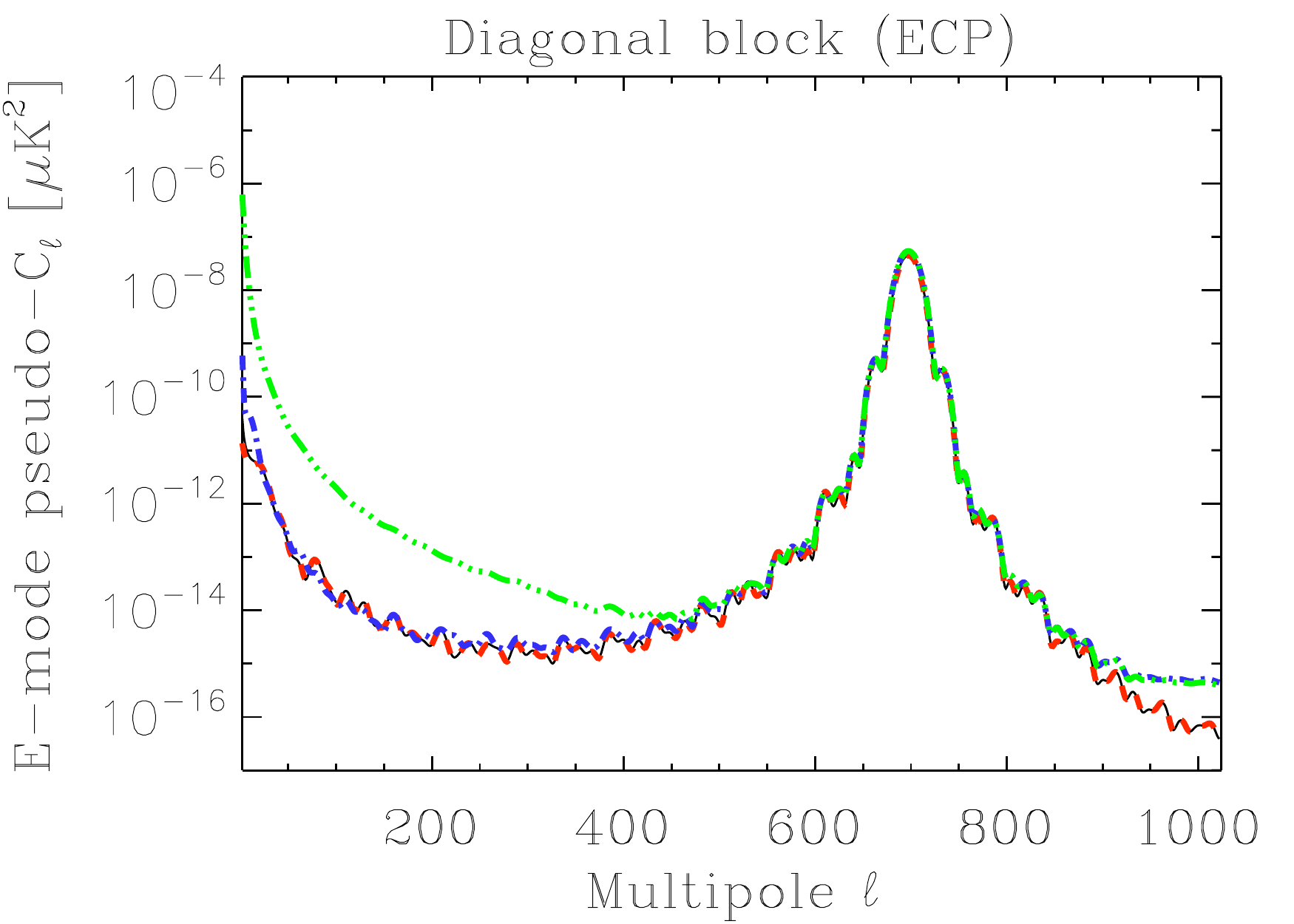} \includegraphics[scale=0.34]{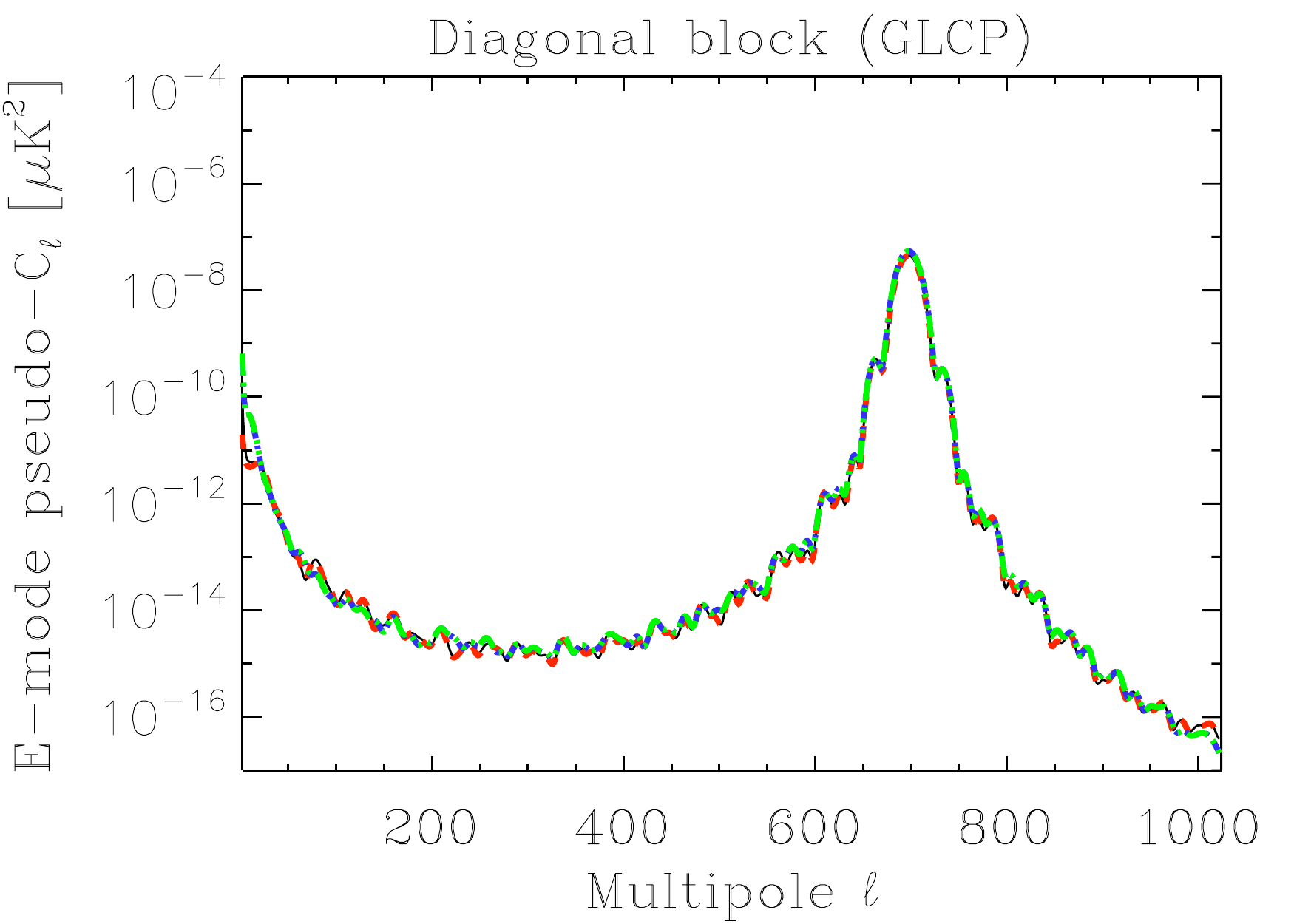} \\
\includegraphics[scale=0.34]{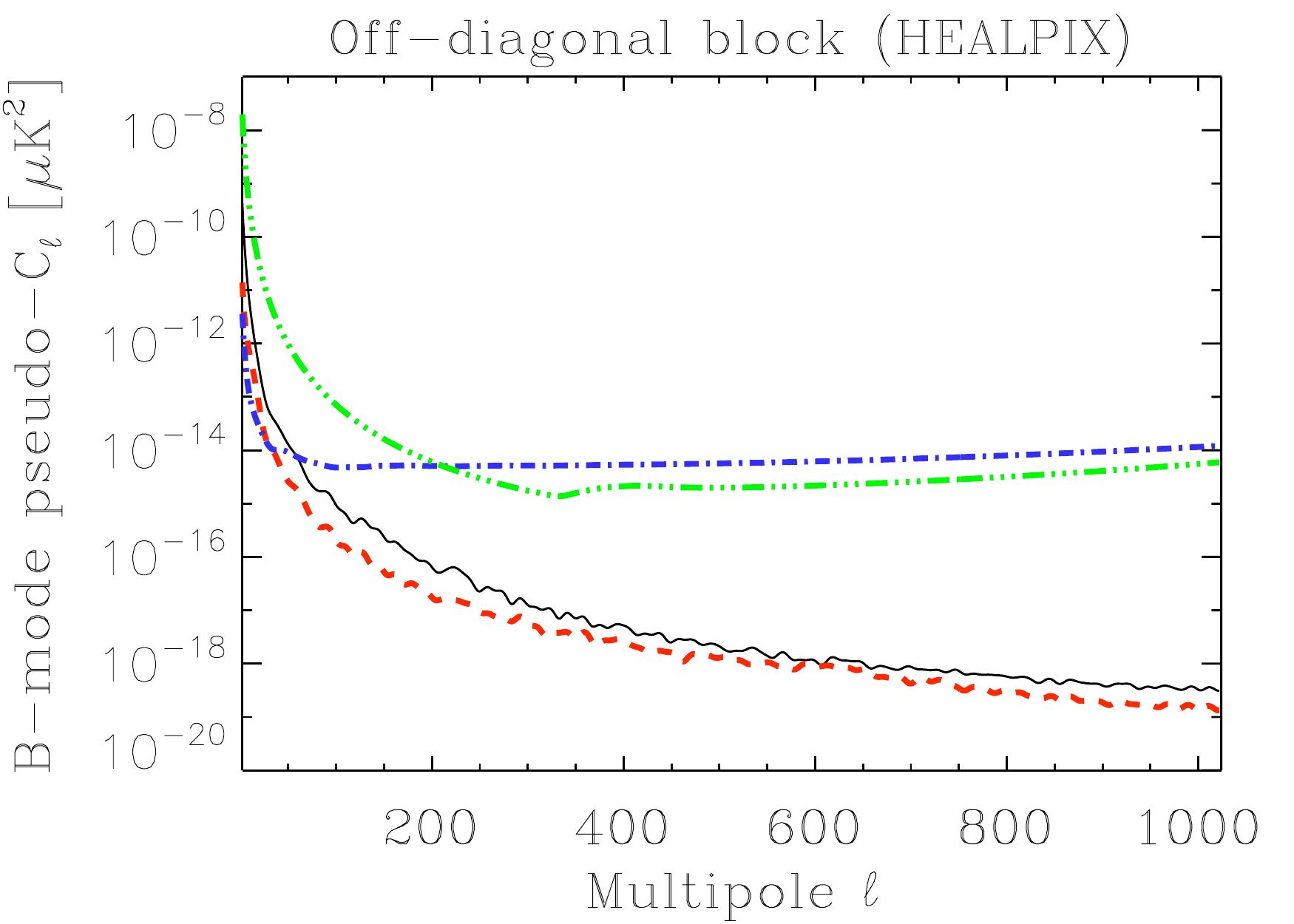} \includegraphics[scale=0.34]{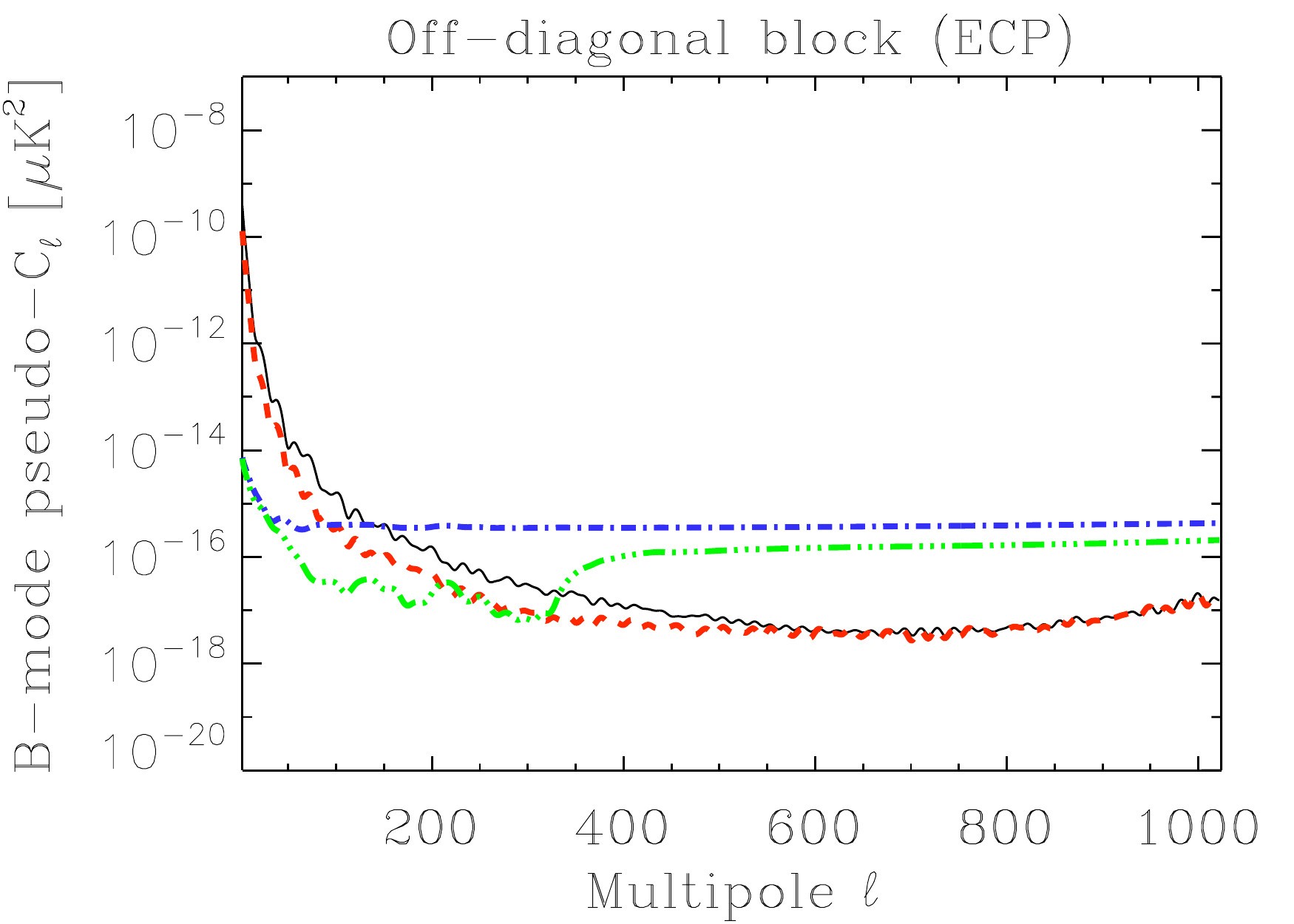} \includegraphics[scale=0.34]{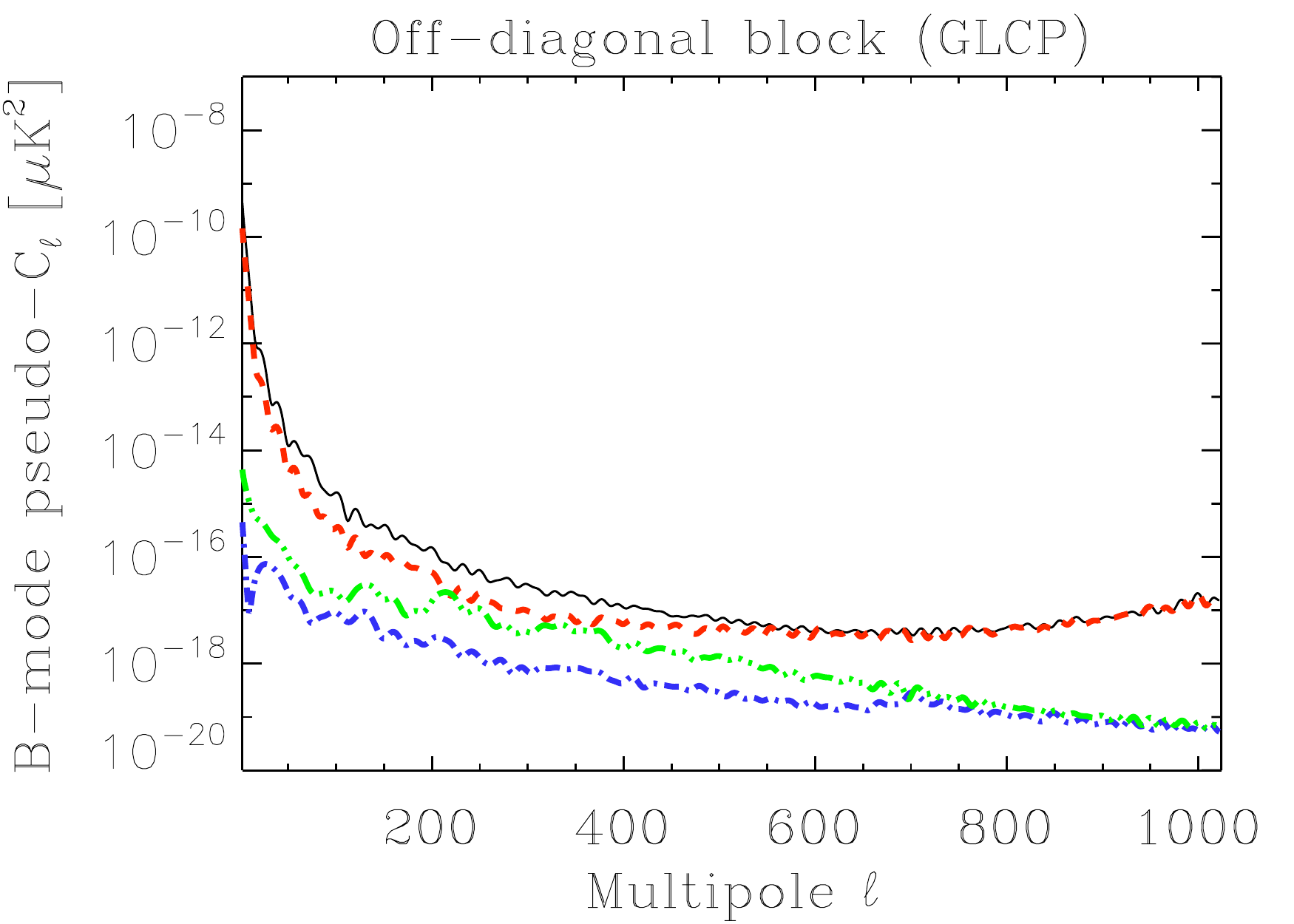} 
\caption{An impact of the sky location and the pixelization scheme on the diagonal block (upper panels) and off-diagonal block (lower panels) of the mixing kernel $M_{\ell\ell'}$. The different curves partially overlap for the diagonal block. {\it Left panels :} Computation in the HEALPix scheme for a cap centered at the equatorial plane using analytic spin-1 and spin-2 windows (black-solid curve) or numerical spin-1 and spin-2 windows (red-dashed curve) and for a cap centered at the north pole using analytic spin-1 and spin-2 windows (blue-dashed-dotted curve) or numerical spin-1 and spin-2 windows (green-dashed-dotted curve). {\it Middle panels :}  Same as the left panels for ECP pixelization scheme. {\it Right panels :}  Same as the left panels for GLCP pixelization scheme. The window function used here is the $C^1$-window with the apodization length set $\delta_c=8$~degrees. In all three cases the pixelization parameters are chosen in such a way that there is the identical number of the constant declination rings within the the sky cap for each of them. For the HEALPix scheme, the bad sampling of the sphere 
 close to the north pole (particularly along the azimuthal direction) leads to strong pixelization effects becoming dominant even for the diagonal block. A similar problem is also seen in the ECP case, but is resolved in the case of the GLCP scheme, which properly samples the sphere all the way to the poles in both polar and azimuthal directions.  The effect can be also alleviated by shifting the cap closer to the equatorial plane.
}
\label{fig:skypos}
\end{figure*}

Numerical computations of derivatives of spin-0 windows will depend in general on the sky location and on the adopted
pixelization scheme, even if the windows are defined analytically. For instance, in the case of a spherical cap centered at 
the north pole, the precision of a calculation of the azimuthal part of the SHT quickly deteriorates in the HEALPix scheme
because of the decreasing number of constant declination pixels present in the polar areas. As a consequence, the spin-1
and spin-2 windows computed in the harmonic space can, sometimes drastically, depart from their true behavior often exaggerating the level of the leakage. Such a problem can be cured either by rotating the cap close to the 
equatorial plane, as often plausible in cases with  a small sky coverage, or by using a different pixelization/gridding scheme such as GLCP (i.e., Gauss-Legendre Celestial Pixelization -- ECP-like but with the constant declination rows distributed 
as the zeros of the Legendre polynomial of some order as supported by the S$^2$HAT software~\cite{s2hat}), for which 
the number of azimuthal pixels (grid-points) remains the same for each constant declination row, and which provide a nearly exact quadrature 
over the polar angle. We provide examples of 
the potential effects in Fig.~\ref{fig:skypos}. We note that the choice of the pixelization used for the derivative computations
can be different from that used for the analyzed maps, though care needs to be taken while performing interpolations required in such 
a case. In addition, similar effects may also play a role whenever any SHT is to be done, as for example, in 
the computation of the pure multipoles, and therefore choice of the pixelization and/or its orientation are of primary
importance for techniques as the one discussed here.

\subsection{Power spectra uncertainty}

The errors of the estimated $B$-mode power spectra include contributions due to both the signal and the noise. 
For any $\ell$-bin the signal sample variance is due to the variance of the $B$-mode signal modes from that bin but 
also from other multipoles aliased to it. It is also due to power of the leaked $E$-mode into $B$, i.e., the 
$E/B$ leakage. The window functions discussed before lead to different trade-offs between these different sources 
of the uncertainties. In this Section, we will assess the power spectra uncertainty using MC simulations
comparing the performances of the pure pseudo cross power spectrum approach for different choices of the window functions. 
We will also consider the effects due to patch geometry and signal priors, complementing the discussion in the previous
Section.

Hereafter, we estimate the variance as the standard deviation of 1,000 MC simulations. Unless specified explicitly
otherwise we will employ the same set of parameters as before, i.e., we will work with the HEALPix scheme with the resolution $N_{side}=512$ and assuming the WMAP 5-year data cosmological model~\cite{dunkley_etal_2008} for the 
input $E$-mode signal. The $B$-mode signal includes lensing and primordial $B$-mode with $T/S=0.05$. We first 
focus on the homogeneous noise case with a level of $5.75~\mu$K-arcmin. The estimated power spectrum are 
binned with the bins of a width of $\Delta\ell=40$ and with the lowest-$\ell$ bin starting at $\ell = 20$. Except for the 
cases investigating effects of the patch geometry, the simulated maps cover a spherical cap (C) area centered at the 
equator as described in Table~\ref{tab-patch}.

\begin{figure}
\includegraphics[scale=0.375]{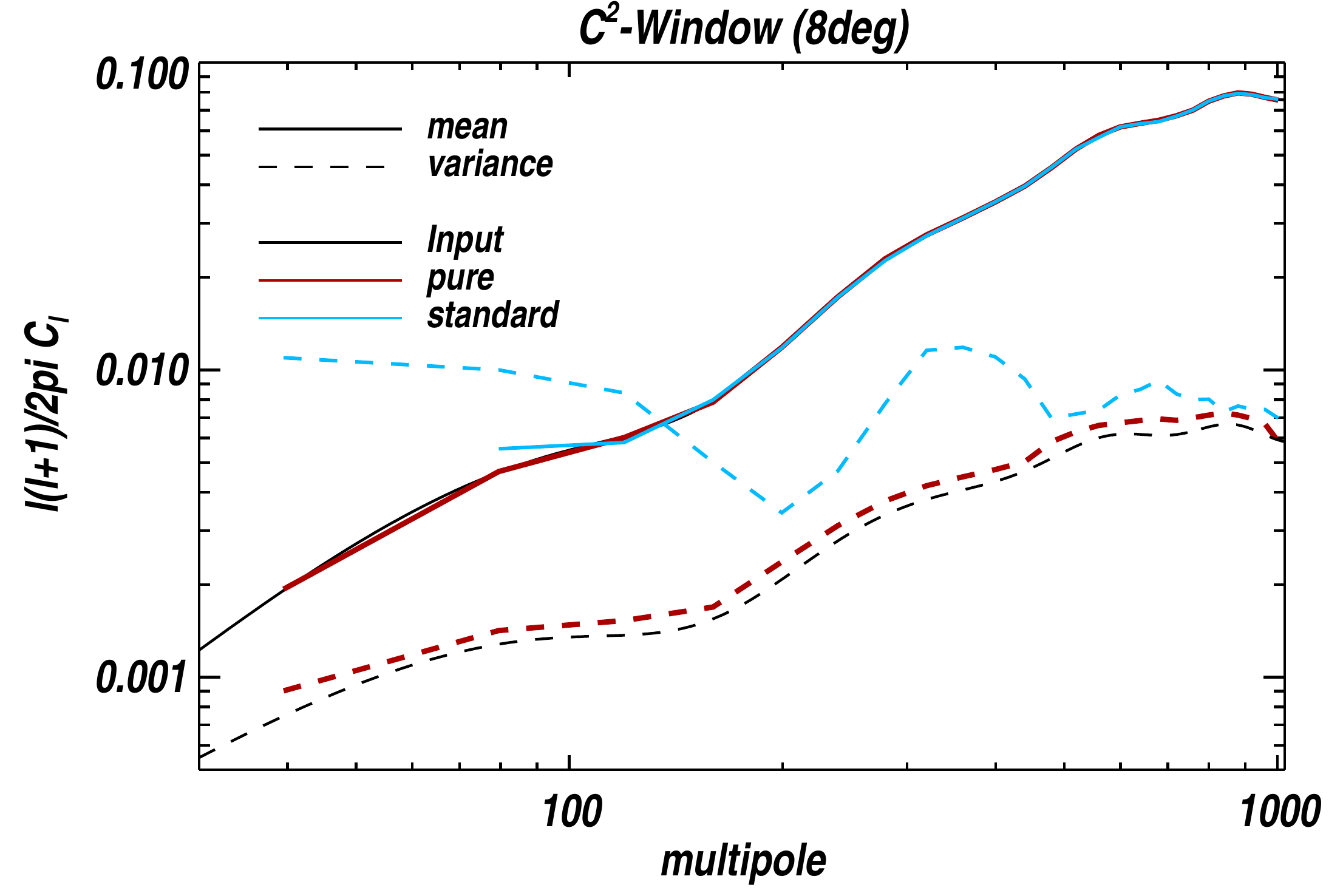}
\caption{ Average $B$-mode spectra (solid curves) and their sample variances (dashed curves) obtained using standard, 
in blue, or pure estimators, in red, and computed from $1000$ signal-only MC simulations done on a spherical cap covering $1$\% of the sky.
Also shown are the input power spectrum (black-solid) and the Fisher estimate of the variance (black-dashed). The extra-variance for the standard estimators 
comes from the $E$-mode power leaked into $B$, while the sample variance for 
the pure estimators is close to optimal.
}
\label{fig:psVarNoNoise}
\end{figure}

As a benchmark for the derived numerical results we will use the simplified, theoretical variance derived for a cross-spectra from 
the Fisher consideration~\cite{hivon_etal_2002,jaffe_etal_2000},
\begin{equation}
\Delta\mathcal{C}^B_{\ell}=\sqrt{\frac{2f_{sky}^{-1}}{(2\ell+1)}\left(C^{B\;2}_\ell+C^B_\ell\frac{\sigma^2}{B_\ell^2}+\frac{\sigma^4}{B^4_\ell}\right)},
\label{fisher}
\end{equation}
where $\sigma^2$ is the noise power spectrum and $f_{sky}$ will be hereafter taken to be an actual covered sky area. As such this uncertainty provides a lower limit 
on any power spectrum estimation error, including the one derived using the pure estimator discussed here, and therefore 
a useful reference to gauge the performance of the pure method against. We emphasize that we do not expect that our 
estimator will be ever  able to reach this limit even in the cases when no $E/B$ leakage is allowed for, i.e., when
$C_\ell^{E}$ is set to $0$. In fact a tighter lower limit could be readily obtained, if an effective, and thus usually smaller, rather than
actual sky area is used as $f_{sky}$ in Eq.~\eref{fisher} accounting for the presence of the apodizations in the pure formalism.
However, even then the excess variance is usually seen owing to the presence of the counter terms in 
Eq.~\eref{counterterm}.

We illustrate some of these effects in Fig.~\ref{fig:psVarNoNoise}, where we present the results of the signal-only MC
simulations for both standard and pure pseudo-spectra approaches and contrast them with the semianalytic predictions 
as calculated using the Fisher approach, Eq.~\eref{fisher}.

\begin{figure*}
\includegraphics[scale=0.375]{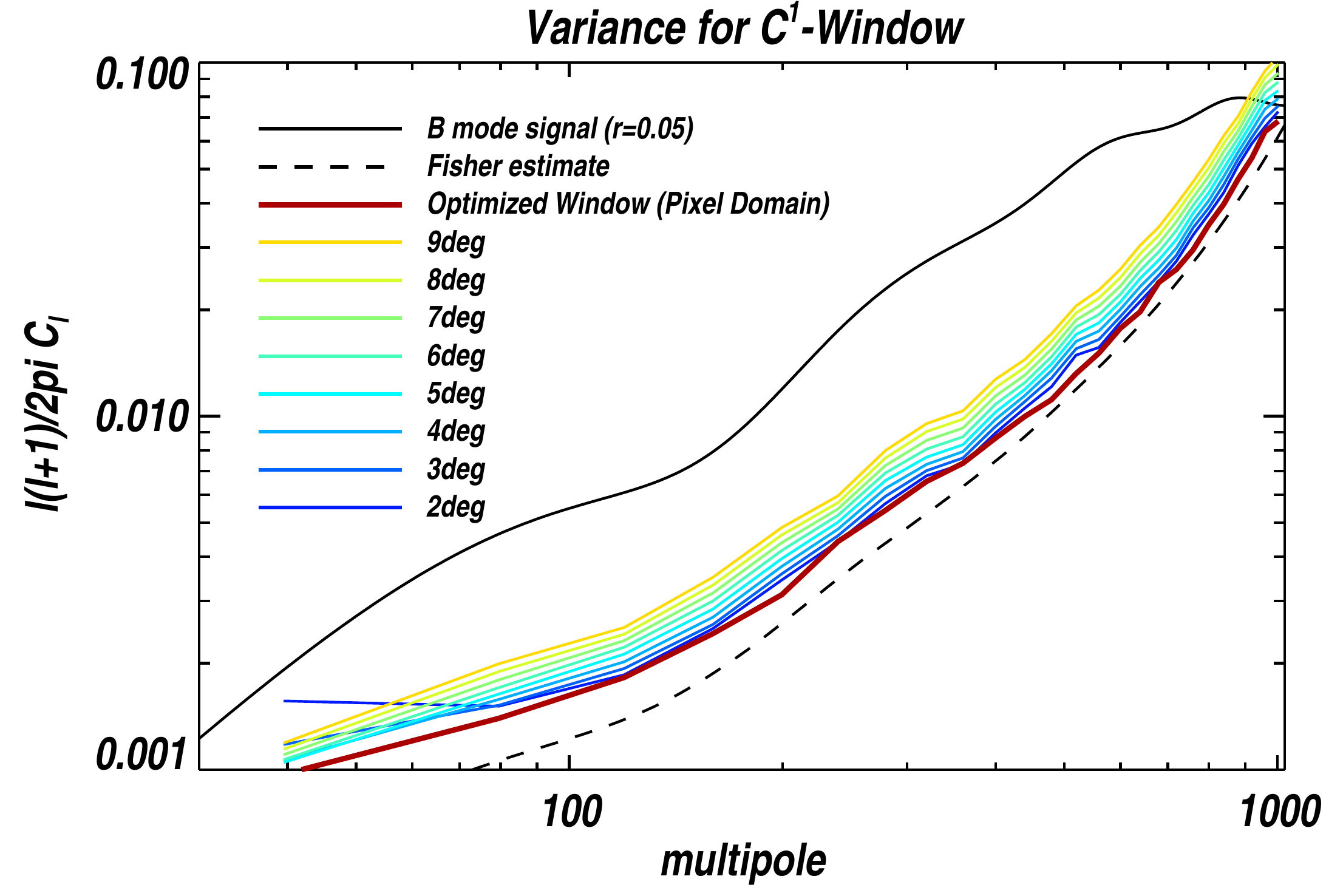} 
\includegraphics[scale=0.375]{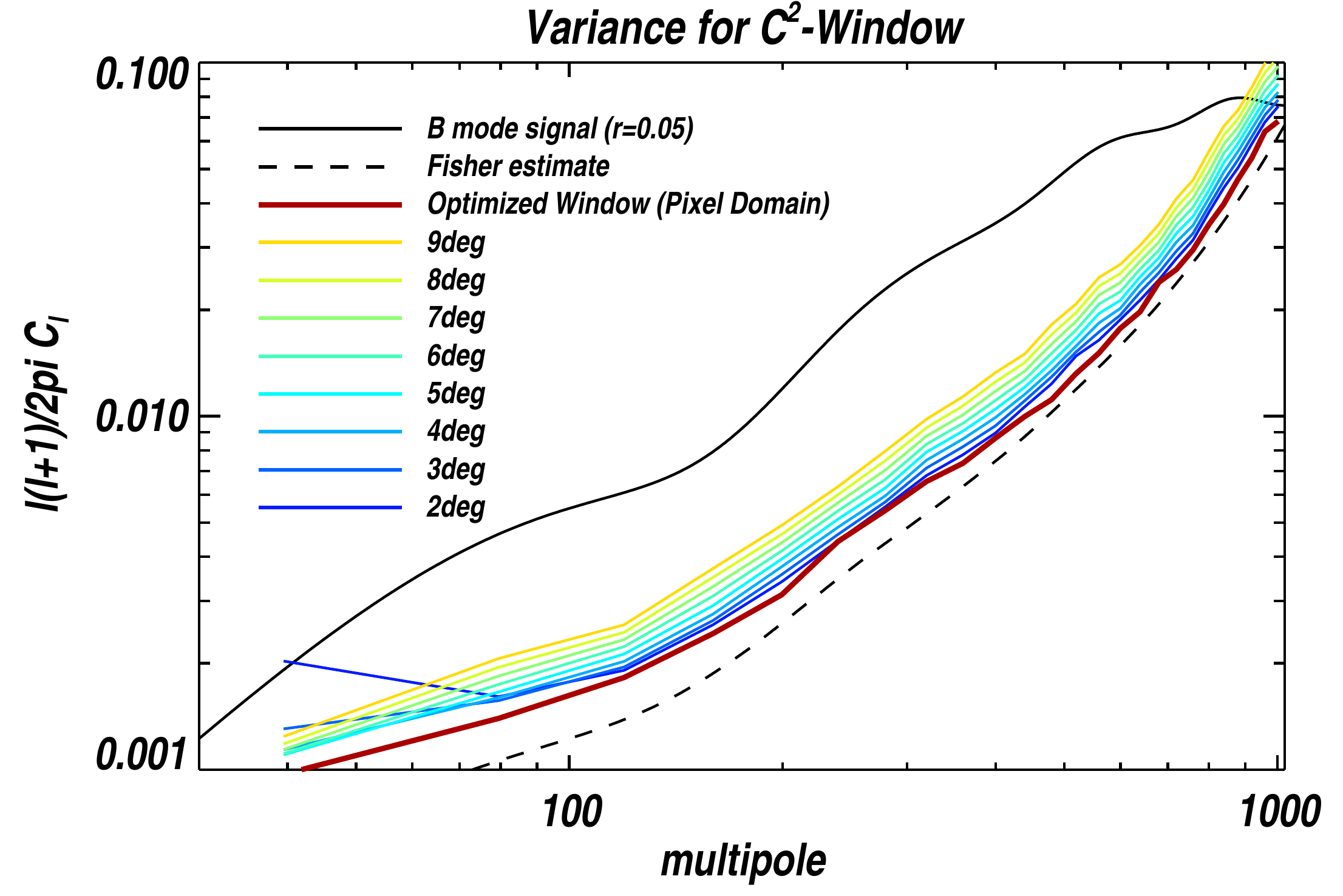}
\caption{Variance of the estimated $B$-mode power spectrum using the $C^1$-window, {\it left panel}, and the 
$C^2$-window, {\it right panel}, for different apodization lengths: from $2$ (blue) to $9$ (yellow). The red curve 
shows the variance obtained using the optimized windows computed in the pixel-domain and optimized bin-by-bin.
The observed survey is the spherical cap (C) and the cosmological model and the level of homogeneous noise 
are described in the text. The input theoretical $B$-mode is shown with a black-solid line and a Fisher estimate 
of the variance with a black-dashed curve.
}
\label{variance-length}
\end{figure*}

\subsubsection{Effect of the window function}

The variance of the estimated $B$-mode power spectra for a number of the apodization lengths and the two analytic proposals is displayed in Fig.~\ref{variance-length} and contrasted against the variance derived using the optimized 
windows computed in the pixel-domain (red curve) and its Fisher estimate (black-dashed curve). As a reference, 
the input theoretical $B$-mode used in the simulations is plotted as black-solid lines. Given the adopted here sky 
coverage (and hence bin-width) as well as the noise level, both in a ball-park of the anticipated experiments, we find 
that in the range of $\ell$ from $20$ to $1000$ the variance is always few times smaller than the signal itself thus in 
principle allowing for a good reconstruction of the $B$-mode power spectrum.

The apodization length has clearly an effect on the performance of the method. For the $C^1$-windows, the lowest 
variance is reached using a $5$~degree apodization in the first bin and $2$~degrees in all the others. For the 
$C^2$-windows, $4$ and $3$~degrees should be used in the first and second bins respectively, whereas the 
$2$~degree window gives the lowest variance for all the other bins. This is due to the fact that noise in the 
subsequent bins is bigger for the higher  $\ell$ values and thus though the leakage level somewhat increases for 
the smaller apodizations, boosting up the variance. This is compensated by the simultaneous decrease of the noise 
variance. At least for the homogeneous noise the best optimization length can be clearly tuned for every analytic 
window and each bin if some prior information about the sky power is available. This can be done determining via 
MC simulations the total estimator variance for a range of considered apodizations and selecting the one leading to 
the minimal scatter.

\begin{figure}
\includegraphics[scale=0.375]{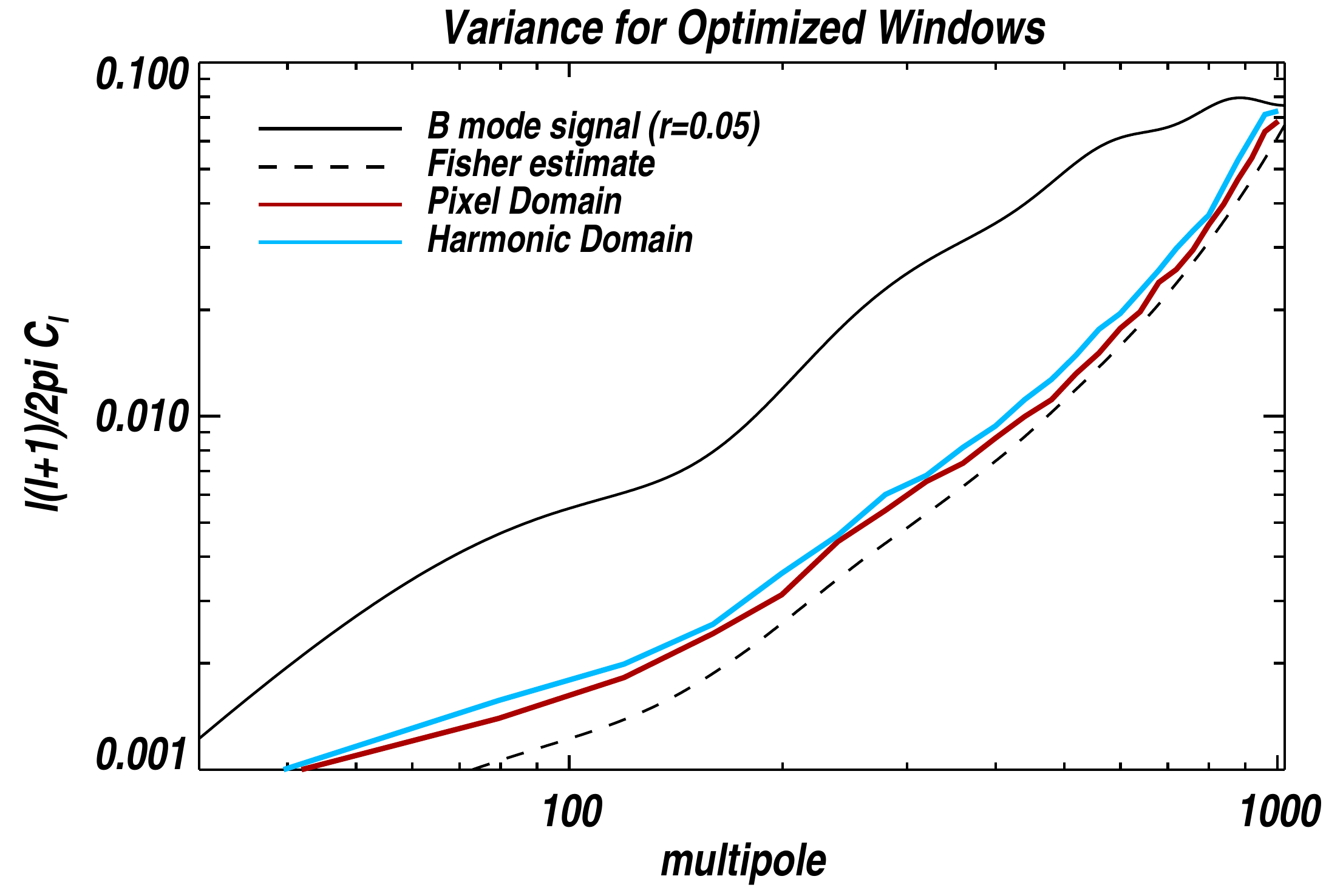}
\caption{Variance of the estimated $B$-mode power spectrum using the optimized windows computed in the pixel 
(blue) and harmonic space (red curve). The black-dashed curve shows the Fisher estimate of the variance and 
the black-solid line -- the input theoretical $B$-mode. The same performance is achieved using the two implementations 
of the optimized window and is close to optimal for $\ell>200$.}
\label{variance-optimal}
\end{figure}

The variance derived using the optimized windows computed either in the pixel-domain (red curve) or in harmonic 
space (blue curve) is displayed in Fig.~\ref{variance-optimal}. The two windows lead to a similar level of variance, 
which is also on par with the Fisher forecast at least for $\ell>200$. Clearly, in the cases studied here the use of pure 
pseudo-spectrum technique yields results close to the best achievable for angular scales smaller than a degree. 
Moreover, the derived variance is indeed at least no larger than the one obtained with the analytical windows as can be seen in Fig.~\ref{variance-length}.
The very good agreement between the harmonic space computation and the pixel-domain computation of the 
optimized windows shows that the extra apodization used during the harmonic computation does not cause any 
significant departure from optimality. This justifies the use of the harmonic window in the homogeneous, or nearly 
so, noise cases as it is much faster, and better controls the $E/B$ leakage across the entire range of $\ell$ and 
does it without any apparent loss of performance for the final precision of the estimated power spectrum.

\subsubsection{Effect of the signal prior}

As we have emphasized before the optimization procedure requires knowledge of both the noise and signal 
properties, including the $B$-mode. In most of the cases of the forthcoming CMB polarization experiments
noise and $E$-mode properties will be probably sufficiently constrained for high and intermediate $\ell$ modes, so the major source of uncertainties 
will result from our ignorance of the primordial $B$-mode and low-$\ell$ $E$-mode power.

\begin{figure*}
\includegraphics[scale=0.375]{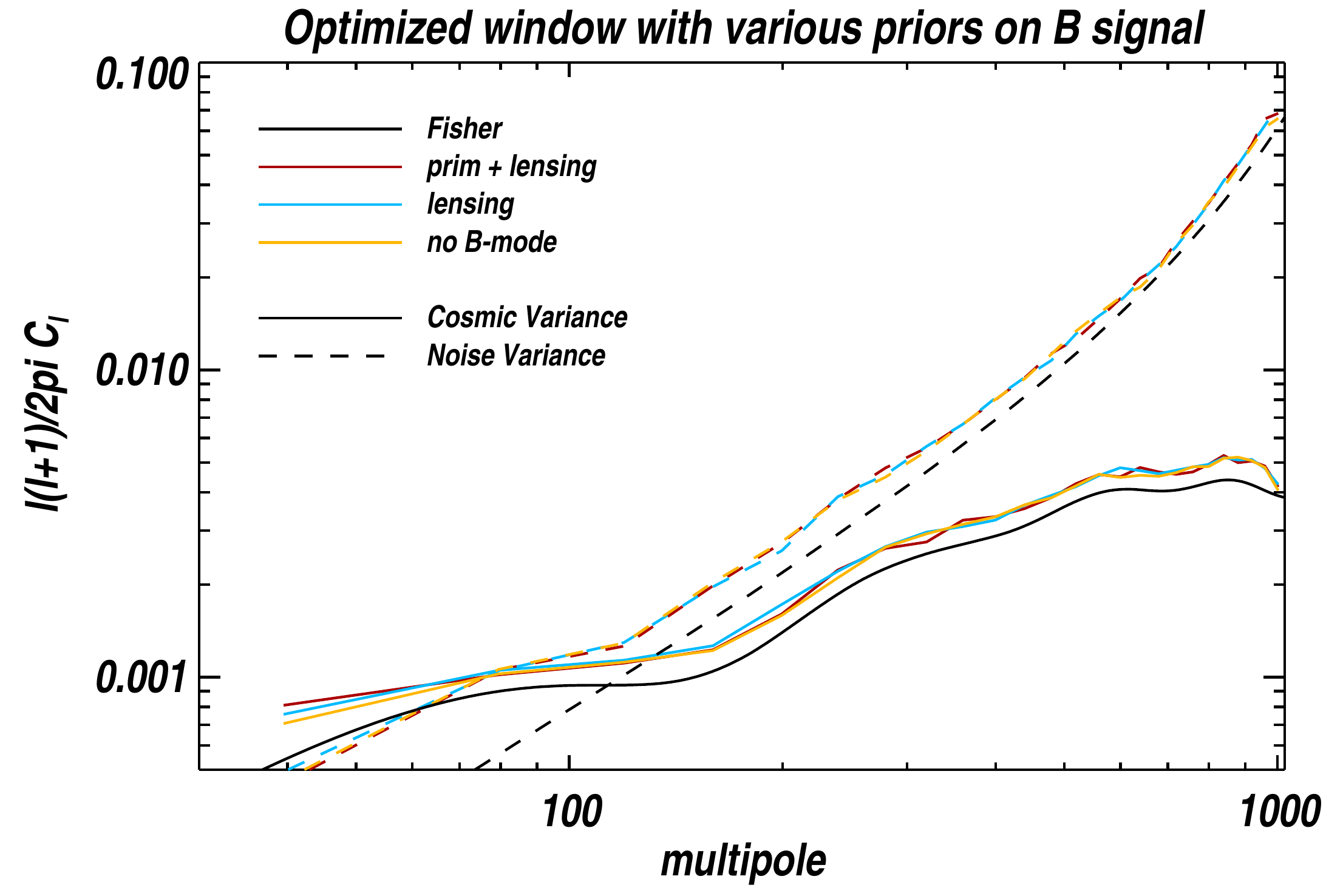} \includegraphics[scale=0.375]{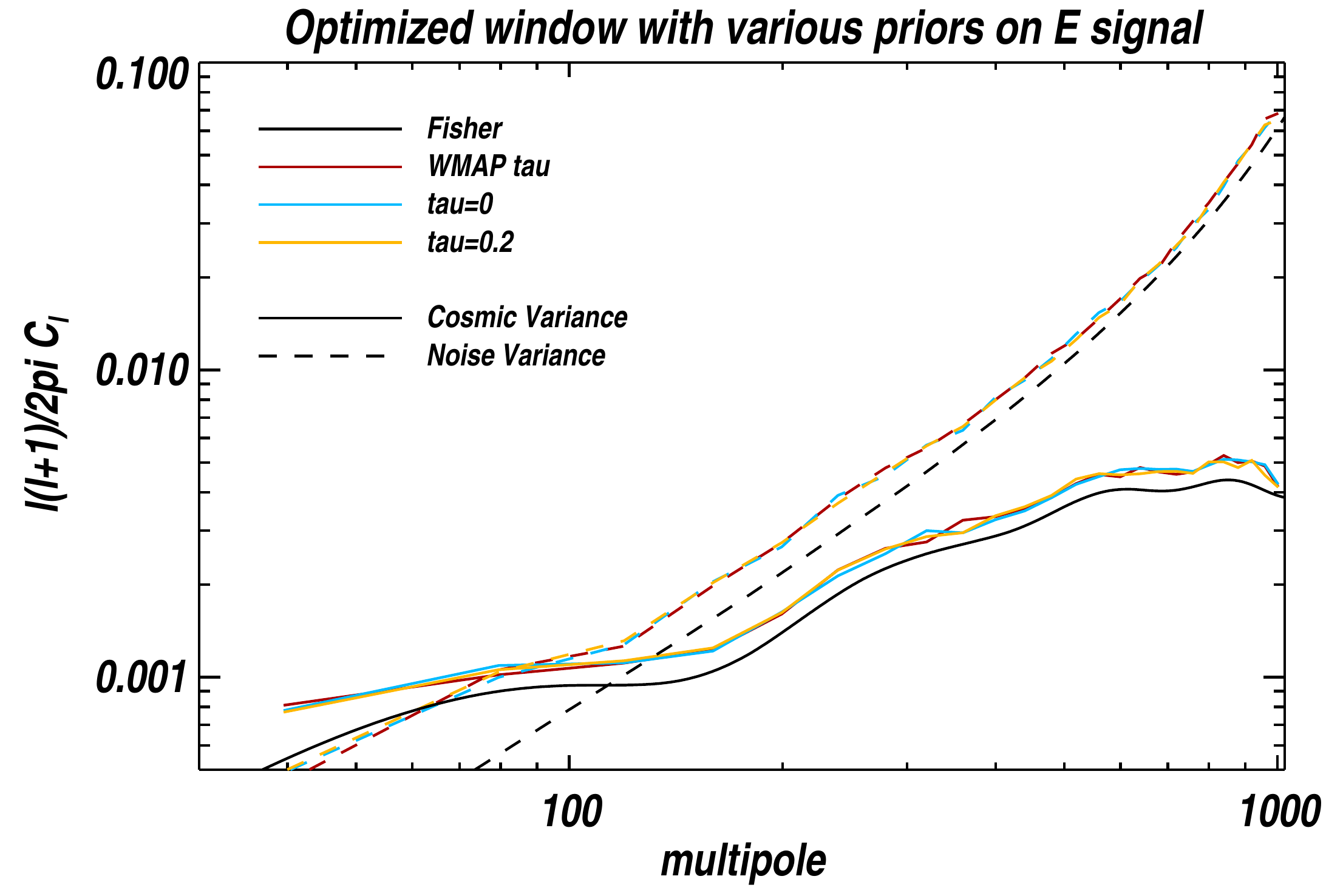}
\caption{{\it{Left panel:}} Noise (dashed curves) and sampling (solid) variance of the estimated $B$-mode power spectrum using pixel-domain optimized windows calculated for three different assumptions about the $B$-mode signal: no $B$-mode (blue curves), only lensing-induced $B$-mode (red curves) and lensing-induced plus primordial $B$-mode for $T/S=0.05$ (yellow curves). The Fisher estimates of the noise and signal variances are shown in black. This is the 3rd case model, which has been used to generate the signal maps analyzed
in all three cases.  Though the assumed sky priors for $B$-modes in two first cases are wrong, that does not seem to have a strong
impact on either the signal or the noise variance. {\it{Right panel:}} Same as left panel but using windows optimized for three different assumption of the $E$-mode signal via a varying reionization optical depth: $\tau=0$ (red curves), $\tau=0.2$ (yellow curves) and $\tau=\tau_{WMAP}=0.087$ (blue curves). This last model was also used to generate the signal maps analyzed
in all three cases. As before no strong impact of the priors on either the recovered signal 
or the noise variance is seen in any of the considered cases.
}
\label{variance-signal}
\end{figure*}

To assess the influence of the $B$-mode prior on 
the performance of the optimized windows, we have performed simulations using lensing-induced and primordial 
$B$-mode ($T/S=0.05$) as input signal but estimating the power spectrum using three different sets of windows 
optimized for different $B$-mode signal : (1) no $B$-mode, (2) lensing-induced $B$-mode and (3) lensing-induced 
plus primordial $B$-mode for $T/S=0.05$. As the two first assumptions differ from the $B$-mode used to synthesize 
the map in the simulations, we expect that the optimized windows derived under those hypotheses could occur
to be suboptimal.
The results of such simulations are shown in Fig.~\ref{variance-signal}, left panel, where the signal variance (solid curves) and 
the noise variance (dashed curves) for the three sets of optimized windows are displayed. The black curves correspond to 
their Fisher estimates. It clearly shows that the prior on the $B$-mode used in the optimization procedure does not 
significantly affect the performance of the pure pseudo-spectrum estimation, as long as the postulated power of the 
$B$-mode remains much smaller than that of assumed in the $E$-mode prior. This last condition will be always fulfilled 
taking into account the current upper limits on the $B$-mode at least as long as CMB is the only sky signal considered.
As a consequence, a conservative, but sufficient, i.e., nearly optimal approach would consist in assuming lensing-induced 
$B$-mode for optimization, as its level and shape will be largely known from the modeling of the $E$-mode spectrum.

Though in general we expect to have a sufficiently well constrained $E$-mode power spectrum for the future polarization-sensitive
experiments, that may not be always so at least as far as the very lowest multipoles are considered. To test the impact 
of those on the variance of the recovered power spectra we have computed the optimized windows with a fixed $B$-mode
power spectrum, including lensing-induced and primordial $B$-mode for $T/S=0.05$, but using three different $E$-mode priors. 
The three $E$-mode power spectra used in this test were calculated for three different values of the reionization optical 
depth $\tau$: (1) $\tau=0$, (2) $\tau=0.2$ and (3) $\tau=\tau_{WMAP}=0.087$ as constrained by the WMAP 5-year results.
These three power spectra have been used to first compute the corresponding optimized windows and later to perform
MC simulations in order to obtain the variance estimates. The 'true' sky used to produce the maps to-be-analyzed here
assumed $\tau=\tau_{WMAP}$ and thus coincided with the third prior as considered above.
The resulting noise (dashed-colored  curves) and sample (solid-colored curves) for the three prior are displayed on 
Fig.~\ref{variance-signal}, right panel, alongside the Fisher estimate (black-dashed curve) and the input $C^B_\ell$ (black-solid 
curve). We find that the variance is only marginally affected by the $E$-prior, even at large angular scales where a 
varying optical depth significantly changes the shape and amplitude of the $E$-mode. We note that our knowledge of
the $E$-mode spectrum will be soon much more precise than the uncertainty permitted in the test described here,
providing therefore an external prior certainly sufficiently accurate for the purpose of the method discussed in this 
paper.

\subsubsection{Effect of the patch geometry}

\begin{figure}
\includegraphics[scale=0.375]{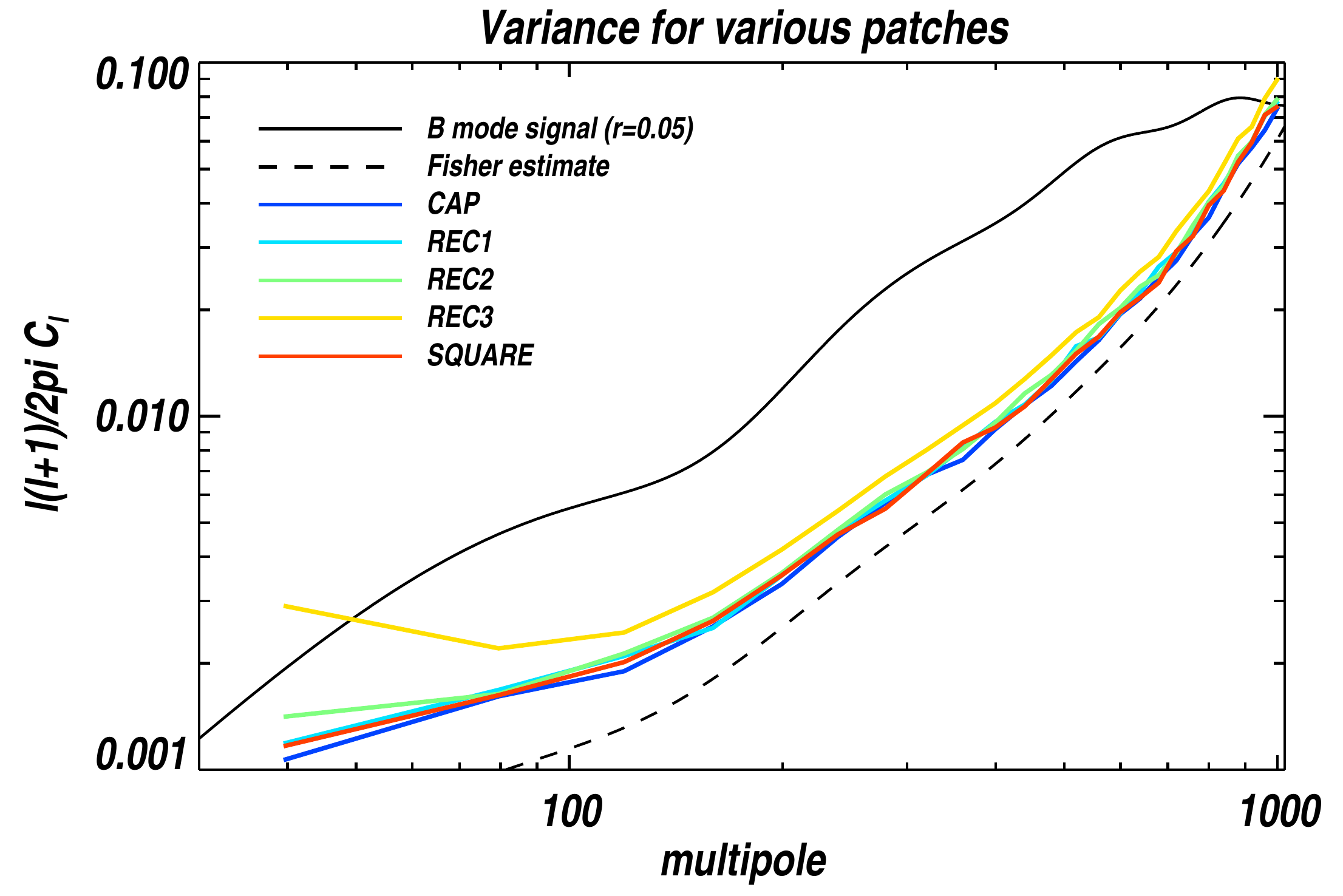}
\caption{Variance of the estimated $B$-mode power spectrum with respect to the patch geometry using 
the $C^2$-window with an apodization length optimized via MC simulations for each bin and each considered patch. 
The black-dashed curve shows the Fisher estimate of the variance and the black-solid line -- the theoretical input 
$B$-mode spectrum. For $\ell>200$, the estimation is close to optimal and a good reconstruction of the $B$-mode 
can be obtained. For smaller multipoles, the performance deteriorates for more elongated patches though a 
variance smaller than the expected power level is recovered in most of the studied cases.}
\label{variance-patch}
\end{figure}

From the previous discussion, we have already learned that the shape of the observed sky area does not affect our
ability to control the $E/B$ leakage with the help of a proper apodization and pure formalism. We can therefore anticipate 
that the dependence of the variance of the estimated $B$-mode power spectrum on the patch geometry will follow
the same pattern as for the standard pseudo-spectrum estimator.

We show the numerical results in Fig.~\ref{variance-patch}, alongside the Fisher estimate, which is the same for 
all considered sky patches as they all have the same sky area (Table~\ref{tab-patch}).
The power spectra have been estimated using the $C^2$-window function with an apodization length optimized for 
each bin and every patch separately. Except for the REC3 patch, for which the variance of the estimated power 
spectrum exceeds the power level in the first bin, the variance is small enough to ensure a good reconstruction of 
the B-mode power spectrum. Moreover, for all the patches, it is roughly the same and on the level comparable to that 
of the Fisher variance at least down to $\ell \sim 200$.  For larger angular scales however, the patch geometry starts
affecting the performance of the estimation of the power spectrum as the variance increases for more elongated patches. 
A similar behavior can be seen also in the case of the optimized windows, as expected given that the patch geometry 
does not affect the level to which the $E/B$ leakage is controlled.

\subsection{More realistic example}
\begin{figure}
\includegraphics[scale=0.25]{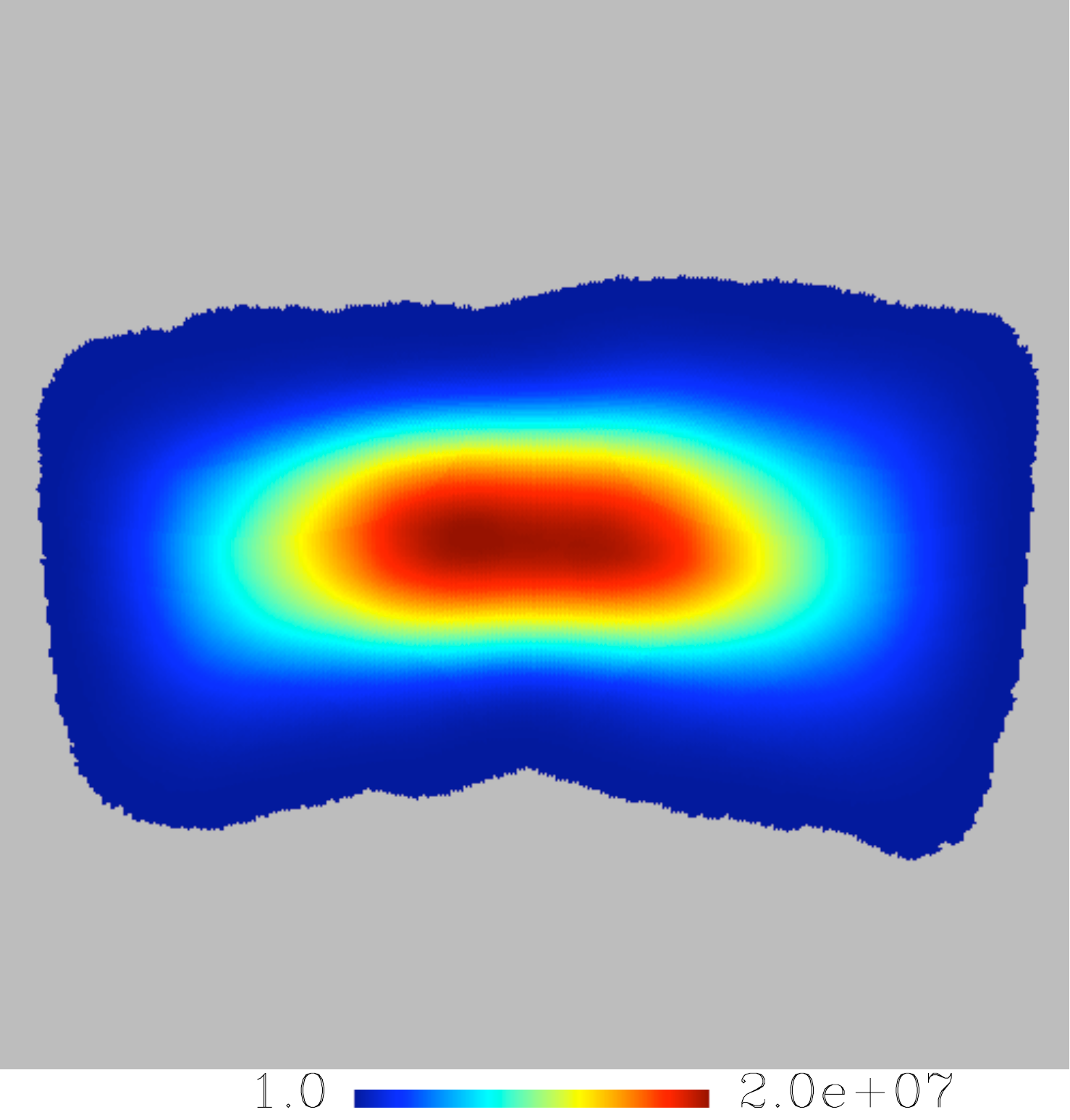}
\includegraphics[scale=0.25]{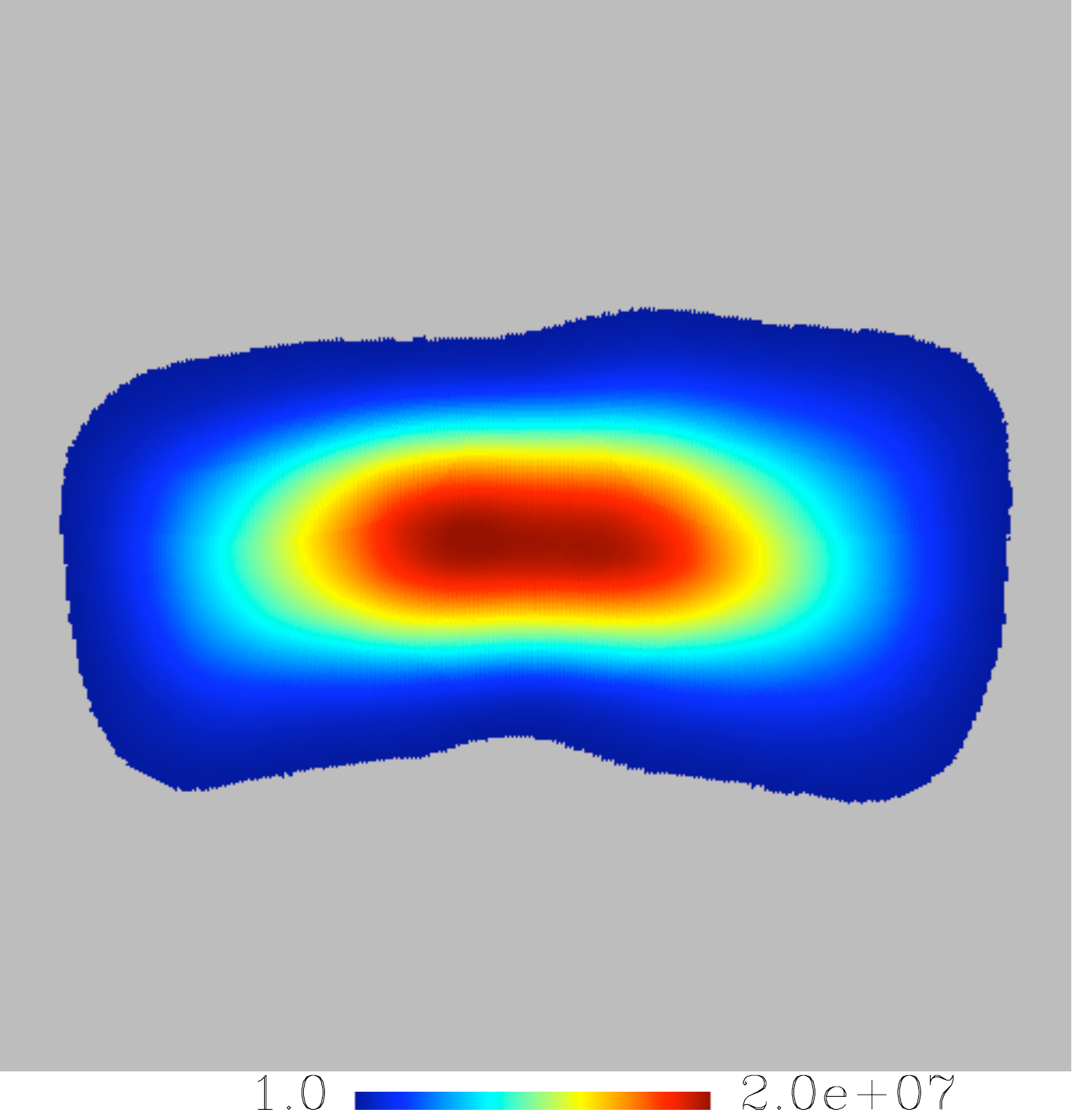}
\caption{A distribution of sky observation for an observational strategy mimicking that of the EBEX balloon-borne 
experiment. The left panel shows the full observed sky patch, while the right one -- its well-observed part. The
latter contains pixels for which a number of the observations is not smaller than $5\times10^{-3}$ of the best 
observed pixel. The density of observations per pixel ranges from $\sim 2\times10^7$ in the center down to 
$1$  or $10^5$ at the edge in the full and reduced sky cases respectively. The pixel size is roughly 7~arcmin, corresponding to $N_{side}=512$. The size of the shown panels 
is roughly  $20^\circ \times 20^\circ$ each. The color stretch adopted here is linear.}
\label{nhit}
\end{figure}

The sky coverage of the realistic experiments is clearly more complex than any of the cases studied so far.
The observed patches are usually irregular and a density of observations per sky area is pixel-dependent 
giving rise to significant noise inhomogeneity of sky maps recovered from data of such experiments. 
\begin{figure*}
\includegraphics[scale=0.375]{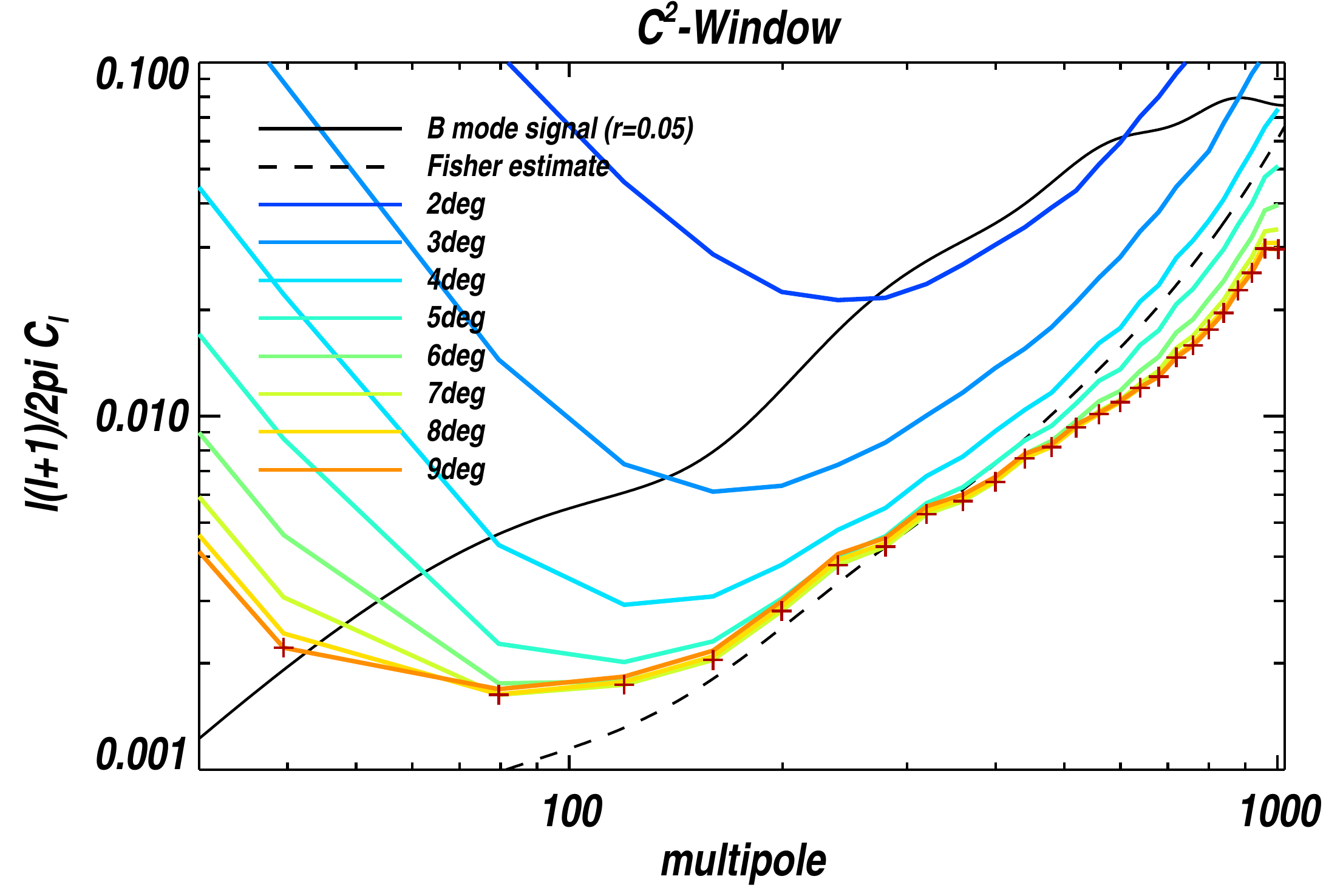} \includegraphics[scale=0.375]{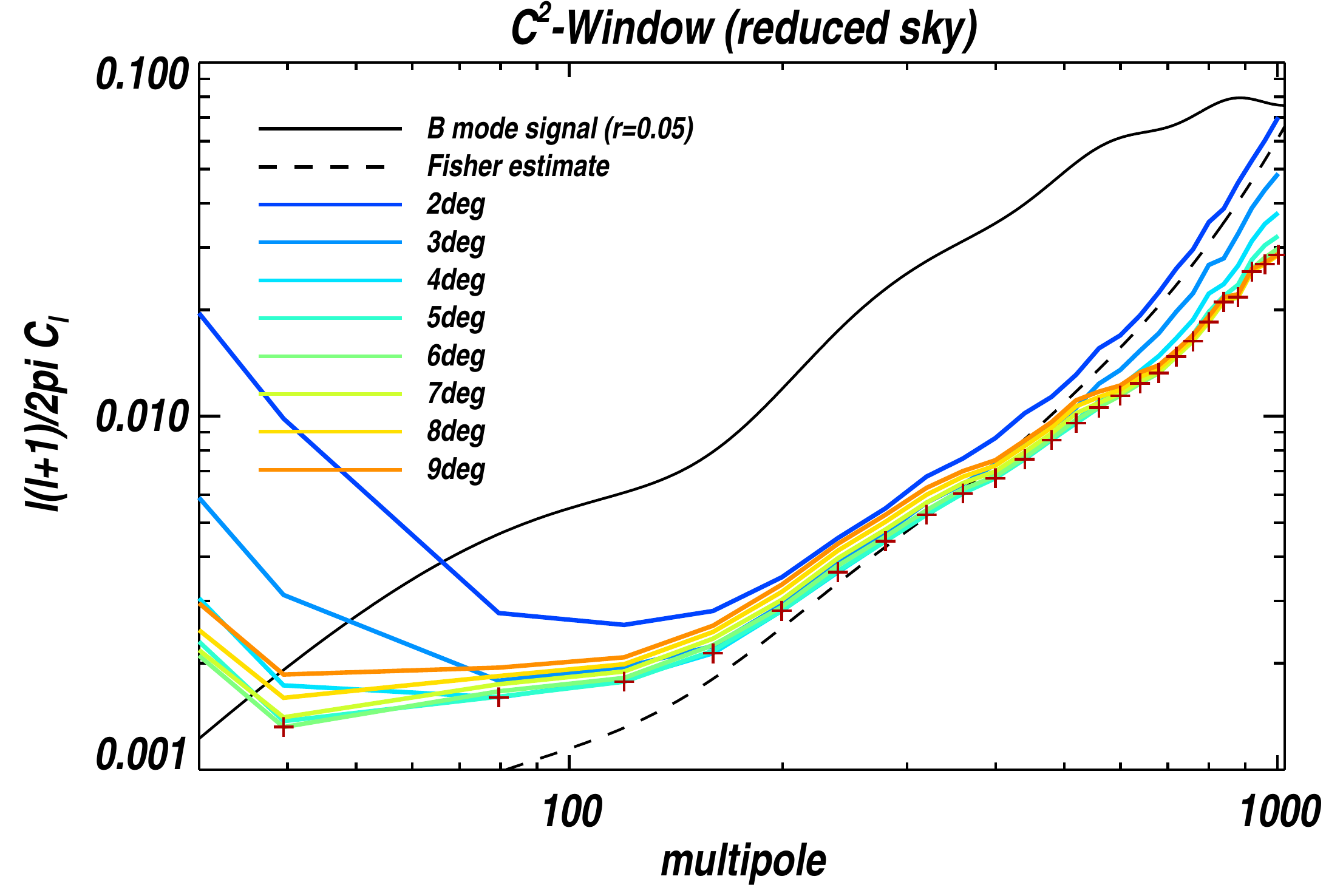} \\
\includegraphics[scale=0.375]{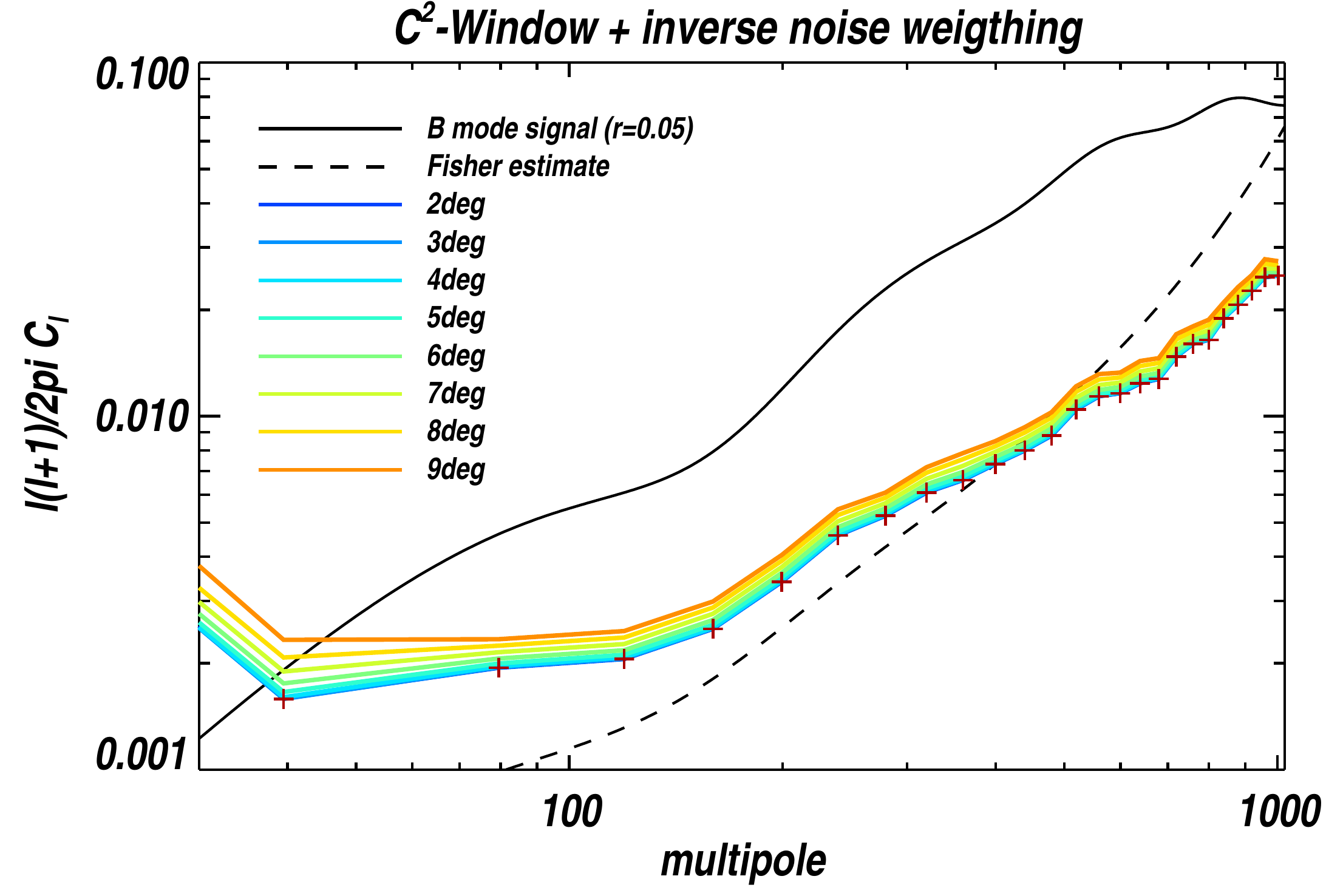} \includegraphics[scale=0.375]{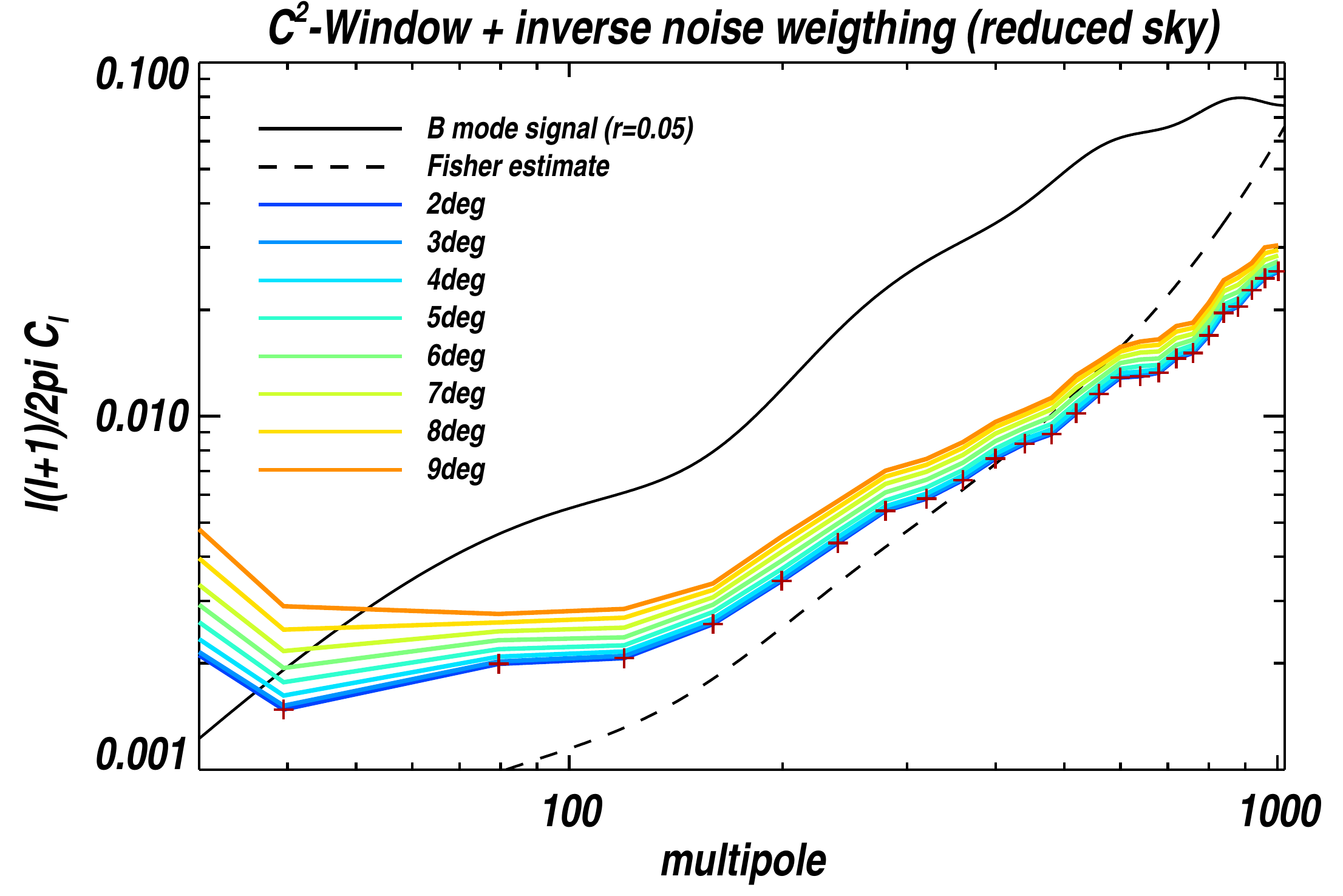}
\caption{{\it Upper panels:} Variance of the estimated $B$-mode power spectrum computed using $C^2$-windows 
with different apodization length (color curves) for the full observed (left panel) and the well-observed
(right panel) sky area. The black-dashed curves show the Fisher estimate of the 
variance calculated assuming the average noise level (see text) and are only shown as a reference here.
The black solid curve represents the input $B$-mode power spectrum. The red crosses indicate the lowest variance
found for each bin, thus defining the optimized apodization length selected via the MC optimization procedure. 
For a given apodization length, removing the noisiest pixel significantly reduces the noise. Nevertheless, except for the 
first two bins the optimized variance is nearly the same for the two 
considered sky patches.
{\it Lower panels:} Same as above but using a window function given as a product of the analytic $C^2$-window and the inverse square-noise variance. This extra noise weighting largely alleviates the problem of the noisiest pixels resulting in similar variance estimates for the full and reduced observed sky patches. Moreover the optimized apodization length, marked with red crosses, is independent of the bin number.
}
\label{inhomo-analytic}
\end{figure*}	
In this Section we therefore assess the performance of the pure pseudo-power-spectrum approach in the case 
of an observation mimicking a long-duration CMB balloon-borne experiment, such as for instance EBEX~\cite{oxley_etal_2004}. The adopted here sky coverage and its sampling, i.e., a number of observations per pixel,
 corresponds to one of the possible scanning strategies for that experiment and is shown in  Fig.~\ref{nhit}.
As it is typical of the small-scale experiments the realistic distribution of the observations 
is very inhomogeneous.
The number of samples per pixel ranges from $1$ at the patch edge to roughly $2\times10^7$ 
in the center. Though in principle  including all the pixels could be advantageous, if they are properly apodized, 
in practice it turns out not to be always the case. 
Moreover, we find that due the dynamical range for the noise levels the PCG-based, iterative calculation of 
the optimized windows fails to converge within a reasonable number of iterations, while a direct inversion of the
involved matrix is clearly too costly to be performed.
For this reason in the following examples together with the full observed sky patch we consider
also its well-observed
part as shown in the right panel of Fig.~\ref{nhit}. It contains all the pixels for which a number of observations is 
not smaller than $10^5$, reducing the dynamical range of the noise levels from roughly $3000$ to $15$. This has 
an unavoidable and unfortunate impact on a observed sky area, which in the studied case  decreases from 
$1.1$\% down to $0.89$\% of the total sky. We set the noise level per sample in such a way that average noise 
level for the full patch corresponds to that used in the homogeneous noise cases studied earlier.

In Fig.~\ref{inhomo-analytic} we show the variances of the $B$-mode spectra computed using full (left panel) and only 
well-observed part of the patch (right panel). The gain with respect to the noise variance largely compensates for the 
loss in terms of sample variance due to removal of the very noisy pixels close to the edge of the full survey area. This is especially apparent 
for $\ell<150$. When analyzing the full sky survey, long apodization lengths are preferred 
as underlined by the red crosses displaying the minimal variance achieved for each bin. This is 
because a long apodization length allows us better to mitigate the influence of the noisiest pixels at the edge and thus to lower 
the noise variance. This also explains why small apodization lengths are suboptimal for the 
entire range of considered $\ell$-modes.
However, once the noisiest pixels are removed (top-right panel), the situation changes drastically. Moderate apodizations are 
preferred at large angular scales ($4$ degrees for $\ell<100$). 
This value of the optimized apodization length is the same as for the homogeneous case. Nevertheless, at small angular scales, the behavior 
of the variance with respect to the apodization length for the inhomogeneous noise differs from the homogeneous 
noise case. For the latter, small, $\sim2$ degrees, apodizations are preferred at small angular scales whereas long apodizations should be used at small scales, $\sim8$ degrees for $\ell>800$, for the former. Such a modification is due 
to the noise variance (we remind that with analytic windows, the $E$-to-$B$ leakage is sufficently lowered below the 
noise level). In this part of the spectrum, windowing should converge towards inverse-noise weighting for the estimation 
to be close to optimal. In the homogeneous noise case, this requires to lower as much as possible the apodization length. 
In the inhomogeneous case, however, this means that windowing should be as close as possible to the actual noise
distribution. For our realistic example, the noise distribution as shown on Fig. \ref{nhit}, exhibits a relatively high 
apodization length, which translates into a large apodization needed for the window to be optimal. 
\begin{figure*}
\includegraphics[scale=0.375]{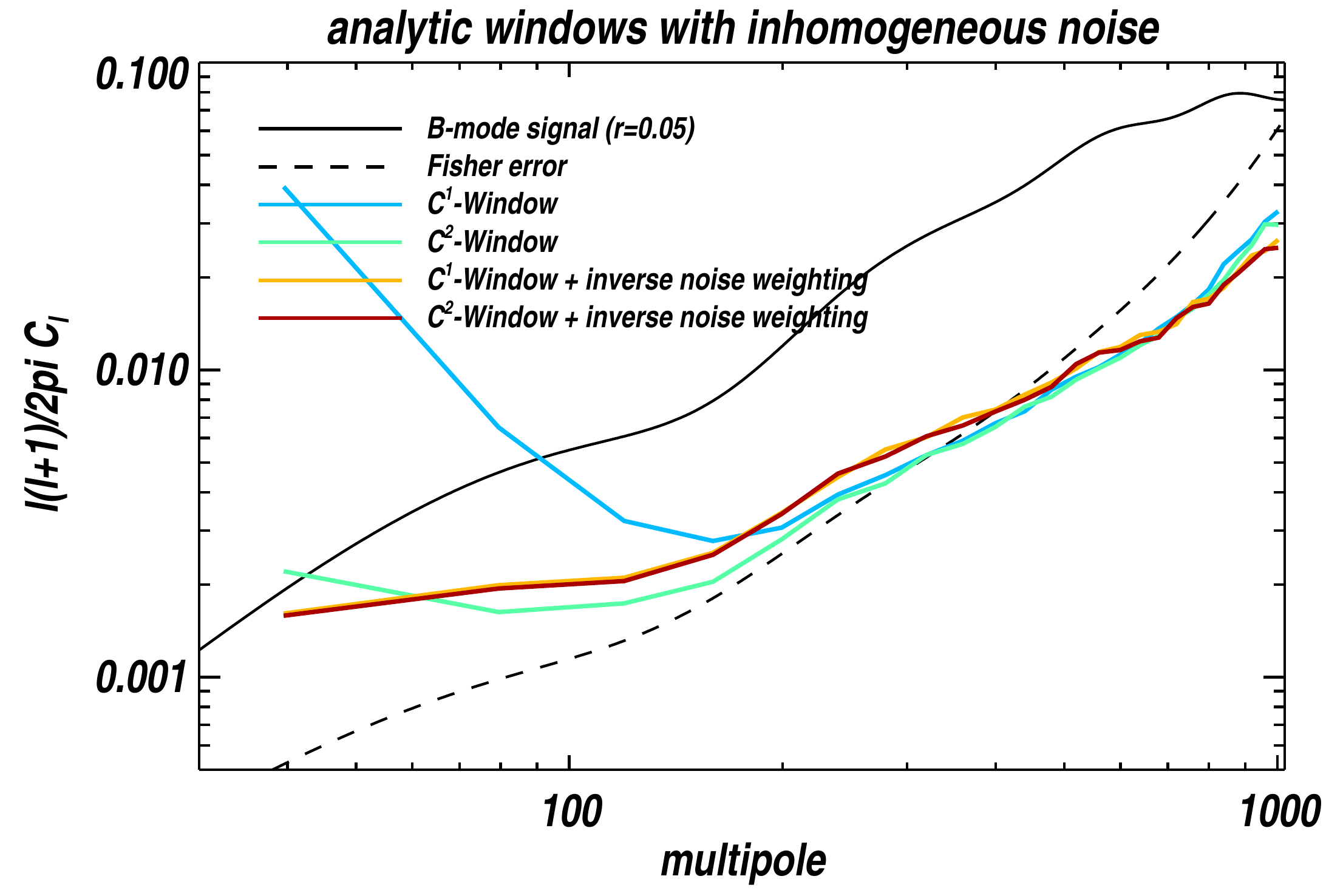} \includegraphics[scale=0.375]{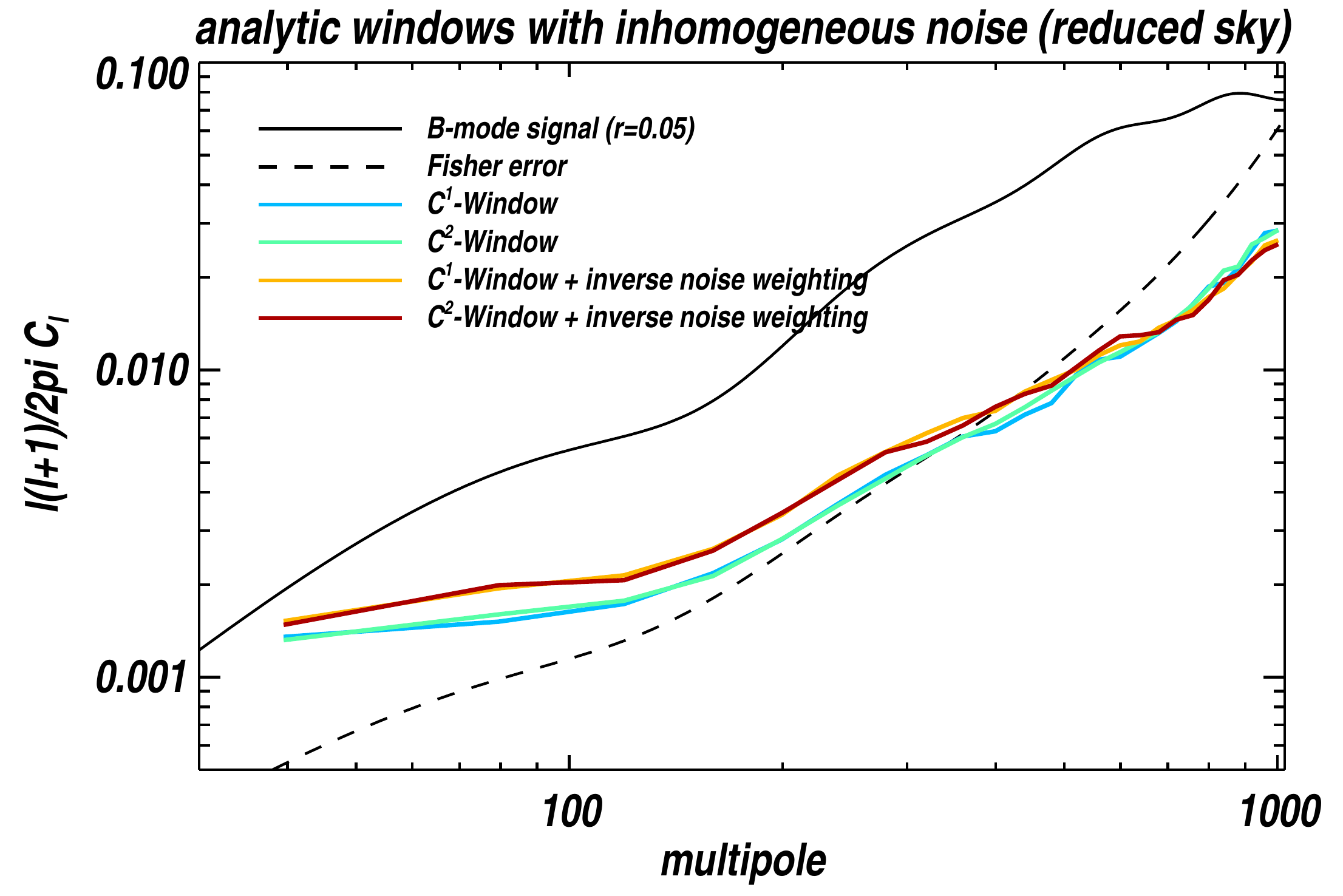}
\caption{Variance of the estimated $B$-mode cross-spectra using the $C^1$- and $C^2$-windows as well as their inverse-noise weighted extensions for the full observed (left panel) and well-observed sky (right panel) areas. The black-dashed curves show the Fisher estimate of the variance derived for the average noise level and the black solid curve depicts the input $B$-mode power spectrum. If the full observed map is to be analyzed, adding inverse-noise weighting suppresses the variance below the estimated power spectrum level at the low-$\ell$ regime. With the noisiest pixels removed, adding inverse-noise weighting does not change much except at large angular scale where it results in some excess variance, when compared with the simple analytic windowing.
}
\label{inhomo-analytic-resume}
\end{figure*}

In a more general case the question if any given analytic window can produce the results close to the optimal is more complex and will depend on both the window and the noise distribution.
In fact, at high $\ell$, the near optimality can be reached only if the adopted window function approximates the inverse noise weighting.

Our discussion and results derived above show that the analytic windows introduced in this paper can indeed be appropriate for noise patterns similar to EBEX.
A more general approach, applicable to other noise inhomogeneity patterns is also however possible.
We first note that multiplying any of the analytic apodization studied here by some smooth function of the sky position will
result in a proper apodization, which suppresses the $E/B$ leakage. We emphasize that this observation is indeed strictly
true only for nonpixelized skies and as usual a care has to be exercised whenever pixelization is introduced. In general,
for pixelized skies the smooth function needs also to be sufficiently flat at the boundary not to lead to some uncontrolled
level of leakage. In the case at hand we derive the new spin-0 window by multipling the analytic windows by a 
pixel-dependent weight corresponding to the inverse squared noise weighting, $\propto1/\sigma^2(p)$, where, 
$\sigma^2(p)$ is the estimated noise variance at pixel $p$.
We then numerically compute the spin-1 and spin-2 components from it.
This allows us to mimic inverse-noise weighting while fulfilling the proper boundary constraints.
The effect of such inverse-noise weighting is displayed in the lower panels of Fig.~\ref{inhomo-analytic} for the full survey (left panel) and its well-observed part (right panel), where colors refer to the value of the apodization length of $W_0$. 
The results show that using this inverse-noise weighting significanlty reduces the total variance, particularly for small apodization length. This explains the origin of the extra-variance at low-$\ell$ end seen in the upper panels of Fig.~\ref{inhomo-analytic} relating it to the
noise of the noisy pixels present at the map boundary. 

Interestingly whatever survey we use (full or reduced), the lowest variance is always 
reached for a small apodization length ($\delta_c=2$ degrees) apparently alleviating the 
need of going through the optimization procedure of the apodization length. 
We note however that this 
fact is related to the specific noise distribution considered here and will be applicable 
in general only to the cases with a sufficient, noise-induced apodization present at the patch
edges, which, as shown here, are however of a practical interest.
At the high-$\ell$ part, this is because the near optimality is reached thanks to the 
inverse-noise weighting and therefore the small apodization length of the analytic part of the 
window is preferred, as it affects the total window shape to the least extent.
At the low-$\ell$ part of the spectrum, without the extra noise-weighting an apodization length 
of 4 degrees was needed for the analytic weighting alone. 
The effective, noise-induced apodization is larger than that and the apodization length for the
analytic part of the window can be therefore smaller as the latter is needed only to fulfill the
boundary conditions and reduce the $E/B$ leakage.

Somewhat counter-intuitively the numerically derived estimates of the power spectrum uncertainties are seen in Fig.~\ref{inhomo-analytic} to be superior to the forecasts 
based on the 
Fisher formalism. This is due to the fact that the latter are based on the average noise level computed assuming the same number of observed samples as in the
actual case, but now homogeneously 
distributed over the entire observed sky area, and the numerical
calculations benefit, thanks to either noise-weighting or appropriate $\ell$-bin dependent apodization, from the very low noise core of the patch, which 
dominates the constraints at the high-$\ell$ end of the spectrum.

A comparison of the best estimates of the spectral variance derived with different analytic
windows is 
shown in Fig.~\ref{inhomo-analytic-resume}. It demonstrates the importance of the down-weighting 
of the very noisy pixels for the full patch analysis (left panel), which is needed to achieve a 
variance level lower than the input $B$-mode at large angular scales. This effect is particularly prominent for the $C^1$-window, which by construction is less smooth towards the patch boundary than the $C^2$-window, and adding inverse noise weighting is here mandatory.  If the noisiest pixels are removed from the analysis (right panel), using inverse-noise weighting performs the same as analytic windows alone except for $\ell<200$, where inverse-noise weighting is somewhat suboptimal as compared to analytic windowing alone. This effect is analogous to the usual loss of precision
in the signal-dominated regime also seen in the standard pseudo power spectrum cases.
However, as noted previously, the results derived with the explicit inverse-noise weighting have not required any MC-based optimization in the case studied here. Therefore though suboptimal 
for $\ell<200$ they can certainly be used to provide a first and reliable estimate of the $B$-mode 
power spectrum, at least in the cases whenever the noise distribution complies with the 
assumptions as already spelled out earlier.

\begin{figure}
\includegraphics[scale=0.375]{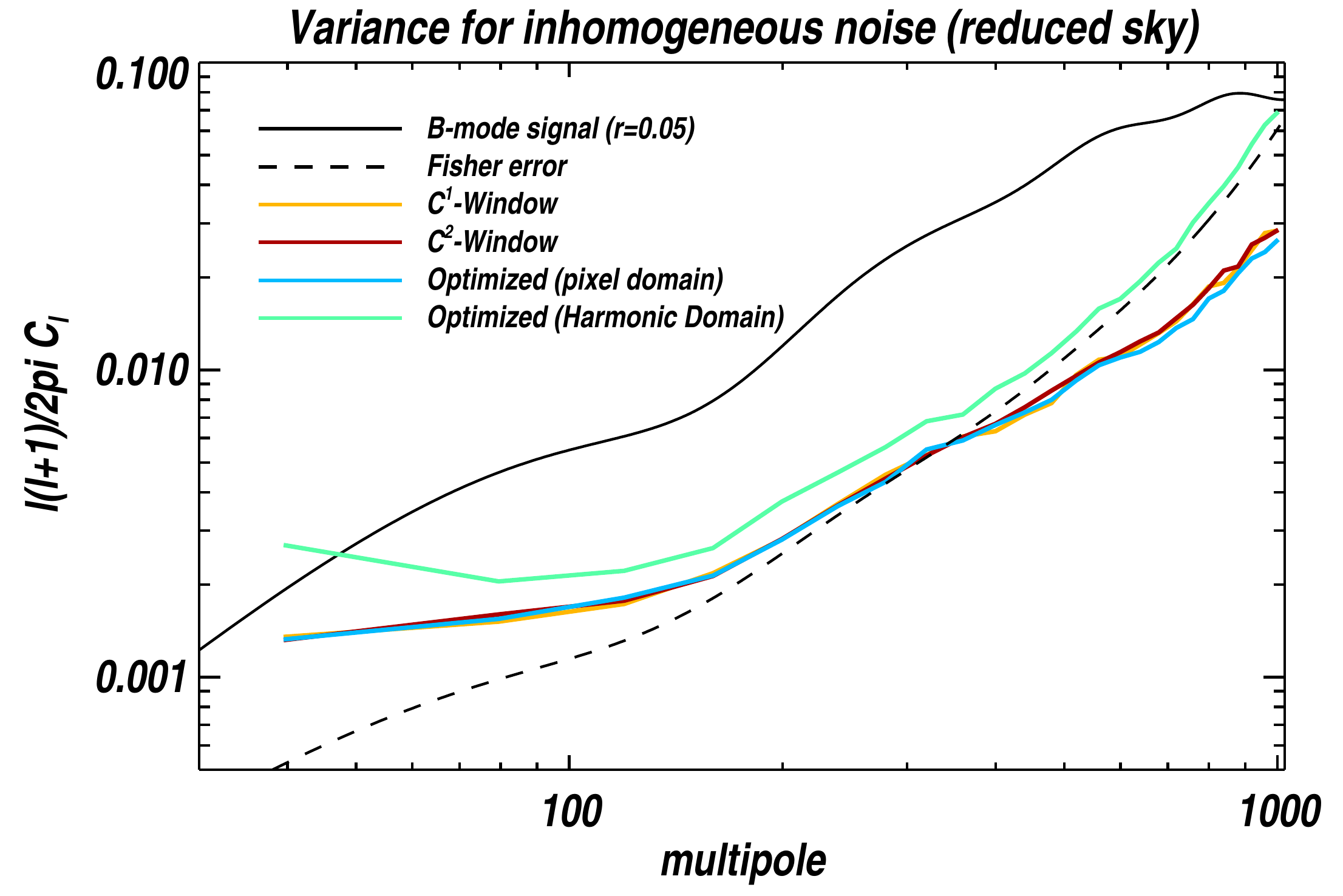}
\caption{Variance of the estimated $B$-mode power spectrum for the realistic study case considered here. Four 
different windows have been used : $C^1$-window (yellow curve), $C^2$-window (red curve), optimized window computed in the harmonic domain (green curve) and the optimized window computed in the pixel-domain (blue curve). The apodization length of the analytic windows has been optimized in each bin. The black-dashed curve stands for the Fisher estimates of the variance and the black-solid one for the input theoretical $B$-mode. The Fisher estimate has been calculated assuming the average noise level. All the windows show comparable performance with an exception of the harmonic domain optimized one. This departure is due to the homogeneous approximation 
of the noise making the effective apodization length of the harmonic optimized window suboptimal.
\label{variance-ebex}
}
\end{figure}

In Fig.~\ref{variance-ebex} we show the variances derived using the numerically optimized 
windows and compare them against those obtained with the help of the analytic ones discussed already
before. Only the reduced, well-observed sky area has been used here, due to the numerical
convergence problems experienced by the PCG procedure as mentioned before.
The apodization length of the analytic windows (which include no noise-weighting here) has been 
optimized in each bin in order to minimize the variance. Clearly, the windows optimized 
numerically in the pixel-domain display very similar levels of the uncertainty on the recovered 
power spectrum level as derived with the analytic apodizations.
The harmonic space optimized window shows however an excess of variance in the entire range of
considered $\ell$-modes. This is because it has been calculated using the average noise level,
which in all bins has incorporated all the pixels, including those close to the patch boundary,
therefore it does not benefit from the low-variance of the pixels in the very center of the patch.
At the high-$\ell$ end the variance obtained with the harmonic domain window follows closely
the Fisher matrix estimate, which has been derived using the same approximation for the adopted
noise level.

\section{Conclusions}

\label{sect:concl}

We have implemented the pure pseudo-spectrum formalism of \cite{smith_2006, smith_zaldarriaga_2007} extending it 
to incorporate the cross-spectrum approach and discussed practical issues related to its implementation. In particular, we have presented a quick, efficient way to compute pure multipoles and their respective mode mixing kernels utilizing harmonic domain derivatives, and demonstrated its overall consistency. 
We have also considered a number of apodization proposals. Those have included analytical, semianalytical and numerically-derived functions. We have studied the performance of the implemented formalism from the point of view 
of a level of the residual $E/B$ leakage as well as final variance of the estimated spectra, and investigated their
dependences on a number of factors such as an observed sky geometry, window choice, pixelization type and pixel size. We have considered the
role of the sky priors in the apodization optimization procedures, as well as proposed an iterative, prior-free approach to the $B$-mode power spectrum estimation.

Our results support the assertion of ~\cite{smith_2006, smith_zaldarriaga_2007} and show that the pure pseudo 
spectrum approach not only can successfully resolve the $E/B$ leakage problem but, if properly applied, it also allows to
bring down the total variance of the estimated power spectrum to levels comparable with those derived from the optimal
approaches. And it can do that at the significantly lower computational costs. In fact, we usually find that the power
spectrum variance derived with the pure approach is within a factor of $2$ away from the reference variance
based on the Fisher matrix approach, which neglects the $E/B$ leakage and assumes no apodization 
(but a binary mask).

We have shown that the analytic apodizations, in particular the $C^2$-window proposed here, 
can deliver a competitive performance, particularly if optimized using the MC simulations, 
at the low calculational cost. The applicability of the analytic windows can be greatly
extended by combining them with the inverse square-noise weighting. Such weighting may 
sometimes produce somewhat suboptimal results at the intermediate and low end of the 
$\ell$-range, where it can/should be replaced by the unweighted analytic or numerically 
optimized windows. 
We have however found that in some cases of practical interest the noise weighted analytic
windows may not require any optimization procedure. They therefore provide a suitable first
guess useful in particular in a preliminary quick-look analysis.

In the cases with a nearly homogeneous or mildly inhomogeneous noise distribution, we have 
derived a semianalytic, quick-to-compute harmonic domain window, which though requires some 
prior information about the sky signal, provides yet another cheap and efficient alternative 
to be used in the pure spectrum calculation. However, in the highly inhomogeneous cases such a 
procedure may be suboptimal and further work is needed to bring its performance in-line with the 
other methods discussed here.

The numerical window optimization as proposed in~\cite{smith_zaldarriaga_2007} have been shown
to provide an efficient framework for the optimized window computation applicable to 
a wide variety of the problems. It can however be numerically prohibitive and iterative solvers 
needed to overcome this problem may suffer due to lack of convergence. We have found that
a particularly common case is whenever a high dynamic range of expected noise levels is present. 
In such circumstances a conservative tresholding procedure may have to be applied in 
practice to avoid such difficulties.

We note that although these conclusions have been derived from the cross-spectrum analysis they should be equally
applicable to autospectra cases. 

In this work we have not considered issues related to bin-bin correlations of the power spectrum estimates. This is
an important topic for the further applications of the results of the method discussed here, in particular for the
cosmological parameter estimation, and will be treated in the forthcoming work. 

This paper focuses only on the statistical uncertainty incurred due to either the noise
and/or sampling variance. Other potential sources of error, for instance,
resulting from a foreground component separation and/or systematic effect
treatment are not included. Clearly, such effects will have to be incorporated in a final
error budget of $B$-mode power spectrum estimated from any realistic CMB polarization data.
The results presented here set a stage for an analysis of those issues in
future work.

In this work, we have not discussed the role and a potential impact of pixels missing from the main body of the observed map.
Such "holes" in the maps could result from, for example, a point-sources removal or masking.
We note that their presence does not lead to any fundamental problems in the pure formalism, as has been shown in the case of the pixel-domain
optimized windows in \cite{smith_zaldarriaga_2007}. However, the practical issues exist. In particular in the cases with the analytic windowing, whenever  
the missing pixels are closer to each other and/or observed patch edge than the considered apodization length a special care must be taken to avoid
windows discontinuities (Sect.~\ref{sect:patchGeom}). 
Such effects may cause some additional $E/B$ leakage, however they are
unlikely to lead to any appreciable increase of the total variance of the recovered $B$-mode power spectrum.
A more detailed study of the effect of holes in $B$-mode reconstruction will be presented in a future work.

The results derived in this paper were obtained with the help of a publicly available software package S$^2$HAT~\cite{s2hat}. The set of MPI-parallel routines permitting quick computation of the pure multipoles as well as spin-weighted apodizations given an input spin-0 window, developed in 
the course of this work, can be downloaded from \cite{s2hat_pure}.

\acknowledgements

We acknowledge the use of the HEALpix~\cite{gorski_etal_2005}, CMBfast~\cite{cmbfast}, \&\ CAMB~\cite{camb} software packages. We also thank Marc Betoule for his sky apodization routine~\cite{betoule}. We thank Sam Leach, Will
Grainger, Chris Cantalupo and Ted Kisner and the EBEX team for providing the information and tools to simulate 
the realistic experiment study case and Shaul Hanany for comments.
JG and RS acknowledge partial support of the European Commission Marie Curie IR Grant, MIRG-CT-2006-036614.
\\
\\
\begin{appendix}

\section{Mixing kernel}
\label{app:mixKernel}
	The mixing kernel is implicitly defined by Eq.~\eref{unbiased}. It can be explicitly derived in the harmonic
	space by expressing the pseudo-$C_\ell$'s, averaged over CMB realizations, as functions of the CMB power spectrum. The derivation is very similar to the one described in \cite{hivon_etal_2002}. 
	We first recall that the pure pseudo-$a_{\ell m}$'s are given by the following expression:
\begin{eqnarray}
	{}_E\tilde{a}_{\ell{m}}&=&{E}_{2,\ell{m}}+2\frac{N_{\ell,1}}{N_{\ell,2}}{E}_{1,\ell{m}}+\frac{1}{N_{\ell,2}}{E}_{0,\ell{m}}, \label{aeapp}\\
	{}_B\tilde{a}_{\ell{m}}&=&{B}_{2,\ell{m}}+2\frac{N_{\ell,1}}{N_{\ell,2}}{B}_{1,\ell{m}}+\frac{1}{N_{\ell,2}}{B}_{0,\ell{m}},\label{abapp}
\end{eqnarray}
where $N_{\ell,s}\equiv\sqrt{(\ell+s)!/(\ell-s)!}$. Employing now harmonic space decompositions for the polarization field $Q+iU$ and the different spin-$s$ window functions, we can show that,
\begin{widetext}
\begin{eqnarray}
	E_{s,\ell m}&=&\frac{(-1)^{m}}{4}\displaystyle\sum_{\ell'm'}\sum_{\ell''m''}N_{\ell'',(2-s)}\sqrt{\frac{(2\ell+1)(2\ell'+1)(2\ell''+1)}{4\pi}}w_{\ell''m''}\left(\begin{array}{ccc}
			\ell & \ell' & \ell'' \\
			-m & m' & m''
		\end{array}\right) \nonumber \\
	&&\times\left\{\left[
		\left(\begin{array}{ccc}
			\ell & \ell' & \ell'' \\
			-s & 2 & -2+s
		\end{array}\right)+
		\left(\begin{array}{ccc}
			\ell & \ell' & \ell'' \\
			s & -2 & 2-s
		\end{array}\right)\right]a^E_{\ell'm'}+i\left[
		\left(\begin{array}{ccc}
			\ell & \ell' & \ell'' \\
			-s & 2 & -2+s
		\end{array}\right)-
		\left(\begin{array}{ccc}
			\ell & \ell' & \ell'' \\
			s & -2 & 2-s
		\end{array}\right)\right]a^B_{\ell'm'}\right\}, \nonumber \\
	B_{s,\ell m}&=&\frac{(-1)^{m}}{4}\displaystyle\sum_{\ell'm'}\sum_{\ell''m''}N_{\ell'',2-s}\sqrt{\frac{(2\ell+1)(2\ell'+1)(2\ell''+1)}{4\pi}}w_{\ell''m''}\left(\begin{array}{ccc}
			\ell & \ell' & \ell'' \\
			-m & m' & m''
		\end{array}\right) \nonumber \\
	&&\times\left\{\left[
		\left(\begin{array}{ccc}
			\ell & \ell' & \ell'' \\
			-s & 2 & -2+s
		\end{array}\right)+
		\left(\begin{array}{ccc}
			\ell & \ell' & \ell'' \\
			s & -2 & 2-s
		\end{array}\right)\right]a^B_{\ell'm'}-i\left[
		\left(\begin{array}{ccc}
			\ell & \ell' & \ell'' \\
			-s & 2 & -2+s
		\end{array}\right)-
		\left(\begin{array}{ccc}
			\ell & \ell' & \ell'' \\
			s & -2 & 2-s
		\end{array}\right)\right]a^E_{\ell'm'}\right\}. \nonumber
\end{eqnarray}
The above expressions can be then inserted into Eqs.~\eref{aeapp} and \eref{abapp}, which we can use to define the pure pseudo-$C_\ell$'s. Taking then the average over CMB realizations and making use of the orthogonality relations,
\begin{displaymath}
	(2\ell''+1)\displaystyle\sum_{mm'}\left(\begin{array}{ccc}
			\ell & \ell' & \ell'' \\
			-m & m' & m''
		\end{array}\right)\left(\begin{array}{ccc}
			\ell & \ell' & \ell''' \\
			-m & m' & m'''
		\end{array}\right)=\delta_{\ell''\ell'''}\delta_{m''m'''},
\end{displaymath}
we arrive at,
\begin{equation}
	\left(\begin{array}{c}
		\tilde{\mathcal{C}}^E_\ell \\
		\tilde{\mathcal{C}}^B_\ell
	\end{array}\right)=\displaystyle\sum_{\ell'}
		\left(\begin{array}{cc}
			M^{diag}_{\ell\ell'} & M^{off}_{\ell\ell'} \\
			M^{off}_{\ell\ell'} & M^{diag}_{\ell\ell'}
		\end{array}\right)	
			\left(\begin{array}{c}
				{C}^E_{\ell'} \\
				{C}^B_{\ell'}
			\end{array}\right).
\end{equation}
On defining the power spectrum of the spin-0 window function by
\begin{displaymath}
	\mathcal{W}_{\ell''}=\frac{1}{2\ell''+1}\displaystyle\sum_{m''}w_{\ell''m''}w^\dag_{\ell''m''},
\end{displaymath}
and introducing the following notation
\begin{displaymath}
	J^{\pm}_{s}(\ell,\ell',\ell'')=\left(\begin{array}{ccc}
		\ell & \ell' & \ell'' \\
		-2+s & 2 & -s
		\end{array}\right)\pm\left(\begin{array}{ccc}
		\ell & \ell' & \ell'' \\
		2-s & -2 & s
		\end{array}\right),
\end{displaymath}
we can express the diagonal and off-diagonal blocks (in the $E/B$ subspaces) of the mixing matrix as:
\begin{eqnarray}
	M^{diag}_{\ell\ell'}&=&\frac{2\ell'+1}{16\pi}\displaystyle\sum_{\ell''}(2\ell''+1)\mathcal{W}_{\ell''}\left[J^{+}_0(\ell,\ell',\ell'')+2\sqrt{\frac{(\ell+1)!(\ell-2)!(\ell''+1)!}{(\ell-1)!(\ell+2)!(\ell''-1)!}}J^{+}_1(\ell,\ell',\ell'')\right.
	\label{eqn:mlldiaganal}
	\\
	&&\left.+\sqrt{\frac{(\ell-2)!(\ell''+2)!}{(\ell+2)!(\ell''-2)!}}J^{+}_2(\ell,\ell',\ell'')\right]^2, \nonumber \\
	M^{off}_{\ell\ell'}&=&\frac{2\ell'+1}{16\pi}\displaystyle\sum_{\ell''}(2\ell''+1)\mathcal{W}_{\ell''}\left[J^{-}_0(\ell,\ell',\ell'')+2\sqrt{\frac{(\ell+1)!(\ell-2)!(\ell''+1)!}{(\ell-1)!(\ell+2)!(\ell''-1)!}}J^{-}_1(\ell,\ell',\ell'')\right.
	\label{eqn:mlloffanal}
	\\
	&&\left.+\sqrt{\frac{(\ell-2)!(\ell''+2)!}{(\ell+2)!(\ell''-2)!}}J^{-}_2(\ell,\ell',\ell'')\right]^2. \nonumber
\end{eqnarray}
\end{widetext}
We note that in the standard pseudo-spectrum approach, the mixing kernels are obtained by removing the $J^{\pm}_1$ and $J^\pm_2$ terms in the above expressions as the latter encode the contribution of the counterterms.

If the window function is apodized up to its first derivative, the off-diagonal block of the mixing matrix should be equal to zero. We have checked numerically that this is indeed the case, when the above formulae are
applied.

For the binned power spectrum estimation, the mixing matrix as implicitly defined in Eq.~\eref{unbiased}, is given by,
\begin{equation}
M^{XX'}_{\alpha\alpha'}=\displaystyle\sum_{\ell\in\alpha}\sum_{\ell'\in\alpha'}\frac{\ell(\ell+1)}{\ell'(\ell'+1)\Delta\ell}M^{diag}_{\ell\ell'}\delta_{XX'}.
\end{equation}

In the above formulae we have assumed that the derivative constraints between the three spin-$s$ windows are exactly fulfilled. However, in practice they may be only approximately satisfied and the mixing kernel should be 
therefore expressed not only as a function of $W$, but also of $W_1$ and $W_2$. 
If such extensions are incorporated, the mixing kernel reads,
\begin{widetext}
\begin{eqnarray}
	M^{diag}_{\ell\ell'}&=&\frac{2\ell'+1}{16\pi}\displaystyle\sum_{\ell''m''}\left|w_{0,\ell''m''}\left(\begin{array}{ccc}
		\ell & \ell' & \ell'' \\
		-2 & 2 & 0
		\end{array}\right)+w_{0,\ell''m''}\left(\begin{array}{ccc}
		\ell & \ell' & \ell'' \\
		2 & -2 & 0
		\end{array}\right)\right. 
		\label{eqn:mlldiaganalspin}
		\\
	&&+2\sqrt{\frac{(\ell+1)!(\ell-2)!}{(\ell-1)!(\ell+2)!}}\left[-w_{-1,\ell''m''}\left(\begin{array}{ccc}
		\ell & \ell' & \ell'' \\
		-1 & 2 & -1
		\end{array}\right)+w_{1,\ell''m''}\left(\begin{array}{ccc}
		\ell & \ell' & \ell'' \\
		1 & -2 & 1
		\end{array}\right)\right] \nonumber \\
	&&\left.+\sqrt{\frac{(\ell-2)!}{(\ell+2)!}}\left[w_{-2,\ell''m''}\left(\begin{array}{ccc}
		\ell & \ell' & \ell'' \\
		0 & 2 & -2
		\end{array}\right)+w_{2,\ell''m''}\left(\begin{array}{ccc}
		\ell & \ell' & \ell'' \\
		0 & -2 & 2
		\end{array}\right)\right]\right|^2 \nonumber
\end{eqnarray}
\begin{eqnarray}
	M^{off}_{\ell\ell'}&=&\frac{2\ell'+1}{16\pi}\displaystyle\sum_{\ell''m''}\left|w_{0,\ell''m''}\left(\begin{array}{ccc}
		\ell & \ell' & \ell'' \\
		-2 & 2 & 0
		\end{array}\right)-w_{0,\ell''m''}\left(\begin{array}{ccc}
		\ell & \ell' & \ell'' \\
		2 & -2 & 0
		\end{array}\right)\right. 
		\label{eqn:mlloffanalspin}
		\\
	&&+2\sqrt{\frac{(\ell+1)!(\ell-2)!}{(\ell-1)!(\ell+2)!}}\left[-w_{-1,\ell''m''}\left(\begin{array}{ccc}
		\ell & \ell' & \ell'' \\
		-1 & 2 & -1
		\end{array}\right)-w_{1,\ell''m''}\left(\begin{array}{ccc}
		\ell & \ell' & \ell'' \\
		1 & -2 & 1
		\end{array}\right)\right] \nonumber \\
	&&\left.+\sqrt{\frac{(\ell-2)!}{(\ell+2)!}}\left[w_{-2,\ell''m''}\left(\begin{array}{ccc}
		\ell & \ell' & \ell'' \\
		0 & 2 & -2
		\end{array}\right)-w_{2,\ell''m''}\left(\begin{array}{ccc}
		\ell & \ell' & \ell'' \\
		0 & -2 & 2
		\end{array}\right)\right]\right|^2, \nonumber
\end{eqnarray}
\end{widetext}
where
\begin{equation}
	w_{\pm s,\ell''m''}=\displaystyle\int_{4\pi}W_{\pm s}(\Omega){}_{\pm s}Y^\dag_{\ell''m''}(\Omega)d^2\Omega.
\end{equation}
These last expressions can be rewritten using the $E$ and $B$ decomposition of the spin-weighted window functions, defined as,
\begin{eqnarray}
w^{(E)}_{0,\ell m}&=&-w_{0,\ell m},
\end{eqnarray}
\begin{eqnarray}
w^{(E)}_{s,\ell m}&=&-\frac{1}{2}\left(w_{s,\ell m}+(-1)^sw_{-s,\ell m}\right),
\end{eqnarray}
\begin{eqnarray}
w^{(B)}_{s,\ell m}&=&\frac{i}{2}\left(w_{s,\ell m}-(-1)^sw_{-s,\ell m}\right),
\end{eqnarray}
with $s=1$ or $2$. Using this decomposition in the expression for the mixing kernel leads to,
\begin{widetext}
\begin{eqnarray}
	M^{diag}_{\ell\ell'}&=&\frac{2\ell'+1}{16\pi}\displaystyle\sum_{\ell''m''}\left|w^{(E)}_{0,\ell''m''}J^+_{0}+2\sqrt{\frac{(\ell+1)!(\ell-2)!}{(\ell-1)!(\ell+2)!}}w^{(E)}_{1,\ell''m''}J^+_{1}+\sqrt{\frac{(\ell-2)!}{(\ell+2)!}}w^{(E)}_{2,\ell''m''}J^+_{2}\right|^2
	\label{eqn:mlldiaganalEB}
	 \\
	&&+\left|2\sqrt{\frac{(\ell+1)!(\ell-2)!}{(\ell-1)!(\ell+2)!}}w^{(B)}_{1,\ell''m''}J^-_{1}+\sqrt{\frac{(\ell-2)!}{(\ell+2)!}}w^{(B)}_{2,\ell''m''}J^-_{2}\right|^2, \nonumber \\
	M^{off}_{\ell\ell'}&=&\frac{2\ell'+1}{16\pi}\displaystyle\sum_{\ell''m''}\left|w^{(E)}_{0,\ell''m''}J^-_{0}+2\sqrt{\frac{(\ell+1)!(\ell-2)!}{(\ell-1)!(\ell+2)!}}w^{(E)}_{1,\ell''m''}J^-_{1}+\sqrt{\frac{(\ell-2)!}{(\ell+2)!}}w^{(E)}_{2,\ell''m''}J^-_{2}\right|^2	
	\label{eqn:mlloffanalEB}
	 \\
	&&+\left|2\sqrt{\frac{(\ell+1)!(\ell-2)!}{(\ell-1)!(\ell+2)!}}w^{(B)}_{1,\ell''m''}J^+_{1}+\sqrt{\frac{(\ell-2)!}{(\ell+2)!}}w^{(B)}_{2,\ell''m''}J^+_{2}\right|^2. \nonumber
\end{eqnarray}
\end{widetext}
We point out that because the spin-0 window is real, it does not have any $B$ component. As a consequence, when the derivative relation is completely fulfilled, the spin-1 and spin-2 windows have also a vanishing $B$ part and the second line in Eqs. \eref{eqn:mlldiaganalEB} and \eref{eqn:mlloffanalEB} are equal to zero. This is the case when using analytic windows or optimized windows computed in the harmonic domain. However, because the derivative relation is relaxed during the pixel implementation of the optimized windows, the $B$ part of the two spin-weighted windows is not zero anymore in this special case, though it remains much smaller than their $E$ part. 

Equations \eref{eqn:mlldiaganalEB} and~\eref{eqn:mlloffanalEB} are directly applicable to
the mixing kernel computations for both auto- and cross-spectra, but in the latter
case only when the same apodization is adopted for all the maps. However two 
different apodizations are often required in the cross-spectrum case, for example,x
because noise properties of the two maps differ, and then the above equations have to be 
appropriately 
generalized. This can be done by representing each of the moduli present on the right hand 
sides of Eqs.~\eref{eqn:mlldiaganalEB} and~\eref{eqn:mlloffanalEB} as a product of a
respective expression, of which the modulus is taken, and its complex conjugate and 
subsequently using one of the two
different apodizations for calculations of the expression and the other -- its conjugate.
For example, the square modulus
\begin{displaymath}
	\left|2\frac{N_{\ell,1}}{N_{\ell,2}}J^-_1w^{(B)}_{1,\ell''m''}+\frac{1}{N_{\ell,2}}J^-_2w^{(B)}_{2,\ell''m''}\right|^2
\end{displaymath}
in Eq. \eref{eqn:mlldiaganalEB}, is replaced by
\begin{eqnarray}
\left(2\frac{N_{\ell,1}}{N_{\ell,2}}J^-_1{}_{(0)}w^{(B)}_{1,\ell''m''}+\frac{1}{N_{\ell,2}}J^-_2{}_{(0)}w^{(B)}_{2,\ell''m''}\right)^* 
\nonumber
\\ 
\times\left(2\frac{N_{\ell,1}}{N_{\ell,2}}J^-_1{}_{(1)}w^{(B)}_{1,\ell''m''}+\frac{1}{N_{\ell,2}}J^-_2{}_{(1)}w^{(B)}_{2,\ell''m''}\right),
\nonumber
\end{eqnarray}
where ${}_{(0)}w^{(B)}_{s,\ell''m''}$ and ${}_{(1)}w^{(B)}_{s,\ell''m''}$ are the window functions applied to the first and second map respectively.

In this paper we use Eqs.~\eref{eqn:mlldiaganalEB} and~\eref{eqn:mlloffanalEB}, or their generalization as described above, for the calculations of all the mixing kernels.

\section{Optimized window for cross-spectra}

\label{app:CrossOptWin}

In this Appendix, we show how the optimized ansatz proposed in \cite{smith_zaldarriaga_2007} in the context of the autospectra can be extended to incorporate the case of cross-spectra. Depending on the general context, different choices can be adopted for such an extension. We will present three options, focusing on the case of two maps, $\mathbf{d}^{(0)}$ and $\mathbf{d}^{(1)}$. In the following, $i,~j$ indices will run over pixels and $A,~B$ indices over maps.

The first and simplest choice is to optimize with respect to the auto pseudo-spectrum of each of the maps. The optimized windows are then directly given by the ansatz of \cite{smith_zaldarriaga_2007} applied to each of the maps separately,
\begin{eqnarray}
\displaystyle\sum_{j=1}^{N_{obs}}\mathbf{C}^{(00)}_{ij}{\mathbf{P}}^{(\alpha)}_{ij}W^{(0)}_i&=&1, \label{x-opt1}
\\
\displaystyle\sum_{j=1}^{N_{obs}}\mathbf{C}^{(11)}_{ij}{\mathbf{P}}^{(\alpha)}_{ij}W^{(1)}_i&=&1, \label{x-opt1p}
\end{eqnarray}
where $\mathbf{C}^{AA}_{ij}=\left<\mathbf{d}^A_i\mathbf{d}^A_j\right>$ is the covariance matrix of the $A$-th map and ${\mathbf{P}}^{(\alpha)}_{ij}$ the geometrical matrix projecting in the $\ell,~m$ space (see \cite{smith_zaldarriaga_2007}).

The second option optimizes with respect to the autospectrum of the two maps combined together in the optimal way. In this case the optimal, quadratic estimator and the pseudo-spectrum one can be written as,
\begin{eqnarray}
\mathcal{O}_\alpha&=&\displaystyle\sum_{i,j=1}^{N_{obs}}\sum_{A,B=0}^1\mathbf{d}^A_{i}\left(\mathbf{C}^{-1}{\mathbf{P}}^{(\alpha)}\mathbf{C}^{-1}\right)^{AB}_{ij}\mathbf{d}^B_{j}, \\
\mathcal{C}_\alpha&=&\displaystyle\sum_{i,j=1}^{N_{obs}}\sum_{A,B=0}^1\mathbf{d}^A_{i}W^A_i{\mathbf{P}}^{(\alpha)}_{ij}W^B_j\mathbf{d}^B_{j},
\end{eqnarray}
where 
$\mathbf{C}\equiv\left<\mathbf{d}\mathbf{d}^t\right>$ 
is the full covariance matrix of the data consisting of the two maps concatenated together, i.e.,
\begin{eqnarray}
\mathbf{C}
&=&\left(\begin{array}{cc}
\mathbf{S}_{ij}+\mathbf{N}^{(00)}_{ij} & \mathbf{S}_{ij} \\
\mathbf{S}_{ij} & \mathbf{S}_{ij}+\mathbf{N}^{(11)}_{ij}
\end{array}\right), \nonumber
\end{eqnarray}
with $\mathbf{S}$ and $\mathbf{N}$ the signal and noise covariance matrices respectively. From the above definitions of the estimators, the formalism of \cite{smith_zaldarriaga_2007} is easily generalized to get,
\begin{equation}
\displaystyle\sum_{i,j=0}^{N_{obs}}\sum_{B=0}^1\mathbf{C}^{AB}_{ij}{\mathbf{P}}^{(\alpha)}_{ij}W^B_j=1,
\label{x-opt2}
\end{equation}
This is easily generalized to an arbitrary amount of maps by summing over the entire set of maps.

To compare the above optimized window to the one given by Eqs. \eref{x-opt1} \& \eref{x-opt1p}, we assume the noise properties to be the same for the two maps. As a consequence, optimal windowing will be identical for each map. By plugging this into Eq. \eref{x-opt2}, one can easily show that optimal windows coming from this last equation differs from optimal windows coming from Eqs. \eref{x-opt1} \& \eref{x-opt1p} by the way the noise is accounted for: with Eq. \eref{x-opt2}, the noise part is divided by a factor 2 as compared to Eqs. \eref{x-opt1} \& \eref{x-opt1p}. If $N_{map}$ maps are to be analyzed, then the noise contribution would be divided by a factor $N_{map}$ instead of $2$.

The third and final option we consider consists of optimizing the weighting with respect to each cross-spectrum. For two maps, the optimal, quadratic and pseudo-sepctrum estimators read,
\begin{eqnarray}
	\mathcal{O}_\alpha&=&\displaystyle\sum_{i,j=1}^{N_{obs}}\mathbf{d}^{(0)}_{i}\left({\mathbf{C}}^{-1}{\mathbf{P}}^{(\alpha)}{\mathbf{C}}^{-1}\right)^{(01)}_{ij}\mathbf{d}^{(1)}_{j}, \\
	\mathcal{C}_\alpha&=&\displaystyle\sum_{i,j=1}^{N_{obs}}\mathbf{d}^{(0)}_{i}W^{(01)}_i{\mathbf{P}}^{(\alpha)}_{ij}W^{(01)}_j\mathbf{d}^{(1)}_{j}. \\
\end{eqnarray}
The optimal, quadratic estimators can be recasted as a function of the inverse of the covariance matrix $\mathbf{D}=\mathbf{C}^{-1}$,
\begin{widetext}
\begin{equation}
	\mathcal{O}_\alpha=\displaystyle\sum_{i,j=1}^{N_{obs}}\sum_{m,n=1}^{N_{obs}}\mathbf{d}^{(0)}_{i}\left(D^{(00)}+D^{(10)}\right)^T_{im}{\mathbf{P}}^{(\alpha)}_{mn}\left(D^{(01)}+D^{(11)}\right)_{nj}\mathbf{d}^{(1)}_{j}.
\end{equation}
\end{widetext}
The explicit expression of $\mathbf{D}$ are given by,
\begin{eqnarray}
D^{(00)}&=&\left[\mathbf{C}^{(00)}-{\mathbf{C}^{(01)}} {\mathbf{C}^{(11)}}^{-1}{\mathbf{C}^{(01)}}\right]^{-1}, \\
D^{(11)}&=&\left[\mathbf{C}^{(11)}-{\mathbf{C}^{(01)}}{\mathbf{C}^{(00)}}^{-1} {\mathbf{C}^{(01)}}\right]^{-1}, \\
D^{(01)}&=&-{\mathbf{C}^{(00)}}^{-1}{\mathbf{C}^{(01)}}D^{(11)}, \\
D^{(10)}&=&-{\mathbf{C}^{(11)}}^{-1}{\mathbf{C}^{(01)}}D^{(00)}.
\end{eqnarray}
Following \cite{smith_zaldarriaga_2007}, the optimized window is obtained by minimizing the distance between the optimal, quadratic estimator and the cross pseudo-spectrum estimators with respect to $W^{(01)}$. This leads to the following linear system,
\begin{eqnarray}
&&\displaystyle\sum_{i=1}^{N_{obs}}\mathbf{C}^{(01)}_{ik}\mathbf{P}^{(\alpha)}_{ik}W^{(01)}_i = \ \ 
\label{x-opt3}
\\
&&\frac{1}{2}\displaystyle\sum_{i,j=1}^{N_{obs}}\mathbf{C}^{(01)}_{jk}\mathbf{P}^{(\alpha)}_{ik}\left(D^{(00)}_{ij}+D^{(10)}_{ij}+D^{(01)}_{ij}+D^{(11)}_{ij}\right).
\nonumber
\end{eqnarray}
This last approach is probably the more appropriate as the resulting windows are optimized for each cross-spectra. However, due to the rather complicated right hand side in Eq. \eref{x-opt3}, it is also more numerically involved than the two first approaches.

\section{Harmonic computation of optimized windows}

\label{app:harmOptWin}

In this Appendix, we describe the major steps to compute the optimized windows in the case of white and homogeneous noise. As such noise is completely described by its power spectrum, we perform
all the computations in the harmonic space. We start from Eqs.~\eref{eqn:chiEdef} \&~\eref{eqn:chiBdef} 
and thus choose to use the $\chi$-fields rather than the spin-weighted approach in parlance of \cite{smith_zaldarriaga_2007} .
 Consequently, we will optimize only the spin-0 window and compute the spin-1 and spin-2 windows as its numerical derivatives. We remind that the $\chi^B$ field is a scalar field defined as \cite{smith_zaldarriaga_2007}
 \begin{equation}
 	\chi^B\equiv\sqrt{\frac{(\ell+2)!}{(\ell-2)!}}\mathbf{D}^B_2\cdot\mathbf{P},
\end{equation}
where the $\ell$-dependent prefactor compensates for the $\ell$-dependant normalization of the $\mathbf{D}^B_2$ operator as defined in Eq.~\eref{eqn:bYsDef}. The harmonic decomposition of the $\chi^B$ field reads
\begin{equation}
	\chi^B=\displaystyle\sum_{\ell,m} \sqrt{\frac{(\ell+2)!}{(\ell-2)!}}a^B_{\ell m}Y_{\ell m},
\end{equation}
where $a^B_{\ell m}$ denotes a type $B$ multipole of the polarization vector $\mathbf{P}$.

Following the variational interpretation as proposed in \cite{smith_zaldarriaga_2007}, the optimized windows are the one which minimize the total aliased power in the pseudo-$C_\ell$'s.  The $B$ pseudo-spectrum averaged over CMB realization reads then,
\begin{equation}
	\left<\tilde{C}_\ell\right>=\displaystyle\sum_{\ell''m''}{C}^\chi_{\ell,\ell''}w_{0,\ell''m''}w^*_{0,\ell''m''},
	\label{equ-chi-app}
\end{equation}
where,
\begin{eqnarray}
	{C}^\chi_{\ell,\ell''}&=&\displaystyle\sum_{\ell'}\frac{2\ell'+1}{4\pi}{\frac{(\ell-2)!(\ell'+2)!}{(\ell+2)!(\ell'-2)!}}\left(\begin{array}{ccc}
			\ell & \ell' & \ell'' \\
			0 & 0 & 0
		\end{array}\right)^2\nonumber \\
		&\times& \left(C^B_{\ell'}B^2_{\ell'}+\sigma^2\right),
		\label{cell-chi}
\end{eqnarray} 
and $w_{0,\ell m}$ stands for the harmonic representation of the spin-0 window function.  This window has to 
be apodized up to its first derivative to make Eqs.~\eref{eqn:chiEdef} \&~\eref{eqn:chiBdef} and 
Eqs.~\eref{aetilde} and \eref{abtilde} (i.e., the $\chi$-fields and the spin-weighted approaches) equivalent.

As a consequence, we have to minimize $\left<\tilde{C}_\ell\right>$ with respect to $w_{0,\ell m}$ under the three constraints. First of all, the spin-0  window has to be normalized within the observed region,\begin{equation}
	\displaystyle\sum_{\ell m}w_{0,\ell m}\mathcal{M}^\dag_{\ell m}=1,
\end{equation}
where $\mathcal{M}_{\ell m}$ is the harmonic representation of the mask. (The normalization constant can be set to unity without loss of generality.) Second, the spin-0 window and its first derivative, the spin-1 window, have to vanish at the contour of the observed region, denoted $\mathcal{P}$ in the following. This translates into $2N_c$ constrains, where $N_c$ is the number of pixels on the contour
\begin{equation}
	W(i)=0~\mathrm{and}~W_1(i)=0,~~~~~\mathrm{for~all}~i~\in~\mathcal{P}.
\end{equation}
With such a large number of external constraints, the standard Lagrange multiplier technique used to derive the optimized spin-0 window function becomes hard to solve. 
The complexity of the problem can be reduced by assuming that the integral of the spin-0 and the spin-1 windows on the contour have to vanish, leading to only two additional constraints 
\begin{eqnarray}
	&&\displaystyle\sum_{\ell m}w_{0,\ell m}\mathcal{P}^\dag_{\ell m}=0, \\
	&&\displaystyle\sum_{\ell m}\sqrt{\ell(\ell+1)}w_{0,\ell m}\mathcal{P}^\dag_{\ell m}=0.
\end{eqnarray}
Here $\mathcal{P}_{\ell m}$ is the harmonic representation of the contour. The second equation above corresponds to the spin-1 constraint as $w_{1,\ell m}=\sqrt{\ell(\ell+1)}w_{0,\ell m}$. The optimized spin-0 window can be
then derived by applying the standard Lagrange multiplier techniques, resulting in,
\begin{equation}
	w_{0,\ell''m''}=\frac{1}{{C}^\chi_{\ell,\ell''}}\left(\lambda \mathcal{M}_{\ell''m''}+\mu \mathcal{P}_{\ell''m''}\right).
	\label{opt-sht-app}
\end{equation}
The two constants, $\lambda$ and $\mu$, are derived by plugging the above solution into the constraints. Defining the following total power,
\begin{eqnarray}
	N_{A,0}&=&\displaystyle\sum_{\ell'' m''}\frac{1}{{C}^\chi_{\ell,\ell''}}\left|\mathcal{M}_{\ell'' m''}\right|^2, 
	\label{eqn:Na0}
	\\
	N_{A,1}&=&\displaystyle\sum_{\ell'' m''}\frac{1}{{C}^\chi_{\ell,\ell''}}\left|\mathcal{P}_{\ell'' m''}\right|^2, \\
	N_{A,2}&=&\displaystyle\sum_{\ell'' m''}\frac{\ell(\ell+1)}{{C}^\chi_{\ell,\ell''}}\left|\mathcal{P}_{\ell'' m''}\right|^2, \\
	N_{X,0}&=&\displaystyle\sum_{\ell'' m''}\frac{1}{{C}^\chi_{\ell,\ell''}}\left|\mathcal{M}_{\ell'' m''}\mathcal{P}^\dag_{\ell'' m''}\right|, \\
	N_{X,1}&=&\displaystyle\sum_{\ell'' m''}\frac{\sqrt{\ell(\ell+1)}}{{C}^\chi_{\ell,\ell''}}\left|\mathcal{M}_{\ell'' m''}\mathcal{P}^\dag_{\ell'' m''}\right|, \\
	N_{X,2}&=&\displaystyle\sum_{\ell'' m''}\frac{\sqrt{\ell(\ell+1)}}{{C}^\chi_{\ell,\ell''}}\left|\mathcal{P}_{\ell'' m''}\right|^2,
		\label{eqn:NX2}
\end{eqnarray}
the two constants read,
\begin{eqnarray}
	\lambda^{-1}&\equiv&N_{A,0} \nonumber \\
	&-&\frac{N^2_{X,0}N_{A,2}+N^2_{X,1}N_{A,1}-2N_{X,0}N_{X,2}^2}{N_{A,1}N_{A,2}-N^2_{X,2}}, 
	\label{eqn:lambdaDef}
	\\
	\mu&\equiv&-\frac{\lambda}{N_{A,1}N_{A,2}-N^2_{X,2}}\left[
	\left(N_{X,0}N_{A,2}-N_{X,1}N_{X,2}\right)\r.\nonumber\\
	&-&
	\l.\sqrt{\ell(\ell+1)}\left(N_{X,0}N_{X,2}-N_{X,1}N_{A,1}\right)\right].
	\label{eqn:muDef}
\end{eqnarray}

In this derivation the boundary constraints have been partially relaxed, making the window only partially optimize. The numerical experiments, however, show
that the resulting power spectra uncertainties are at the same level as the ones obtained using the exact, pixel-domain computation of the optimized window (see Fig. \ref{variance-optimal}). This demonstrates that the proposed boundary constraint simplification does not compromise the performances of this fast computed, optimized window.

We point out that the above approach is general and does not apply only to the $B$-mode fields. For instance,
we can use it to calculate optimized windows for a temperature (scalar) map.
It suffices then to replace
\begin{displaymath}
	{\frac{(\ell-2)!(\ell'+2)!}{(\ell+2)!(\ell'-2)!}}\left(C^B_{\ell'}B^2_{\ell'}+\sigma^2\right)
\end{displaymath}
in Eq.~\eref{cell-chi} by the total (signal plus noise) power spectra, i.e., $C^T_{\ell'}B^2_{\ell'}+\sigma^2$, of 
the map. Moreover, we can straightforwardly relax some of the boundary requirements. For example, 
if no boundary conditions are to be imposed, we just need to replace Eqs.~\eref{eqn:lambdaDef}-\eref{eqn:muDef} with,
\begin{eqnarray}
\lambda & = & N_{A,0}^{-1}\\
\mu & = & 0,
\end{eqnarray}
while in a case when only the Dirichlet boundary condition is to be imposed, the two constants are given by:
\begin{eqnarray}
\lambda & = & \frac{N_{A,1}}{N_{A,0}N_{A,1}-N^2_{X,0}}\\
\mu & = & \frac{N_{X,0}}{N^2_{X,0}-N_{A,0}N_{A,1}}.
\end{eqnarray}
\vfill

\end{appendix}


\end{document}